\documentstyle[10pt,epsf,epsfig,dp_delphititle,cite,mcite]{dp_delphi}
\makeindex
\def\DpPaperGroup{PH-EP}
\def\DpPaperRef{2005-045}
\def\DpDate{4 October 2005}
\def\DpAuthors{DELPHI Collaboration}
\def\DpSubmit{(Accepted by Eur. Phys. J. C)}
\def\DpTitle{{Measurement and Interpretation of Fermion-Pair Production \\
 at LEP energies above the \boldmath{\Zzero} Resonance}}
\def\DpComment{}
\def\DpEMail{}
\newcommand {\capsty}{\it}
\newcommand {\ssize}{\scriptsize}

\newcommand {\TeV} {\mbox{$\mathrm{TeV}$}}
\newcommand {\GeV} {\mbox{$\mathrm{GeV}$}}

\newcommand {\MeV} {\mbox{$\mathrm{MeV}$}}

\newcommand {\pbarn} {\mbox{$\mathrm{pb}$}}
\newcommand {\fbarn} {\mbox{$\mathrm{fb}$}}

\newcommand {\invpbarn} {\mbox{$\mathrm{pb}^{-1}$}}

\newcommand{\ZFITTER}{\mbox{ZFITTER}}

\newcommand{\PYTHIA}{\mbox{PYTHIA}}
\newcommand{\KK}{\mbox{KK}}

\newcommand{\TAUOLA}{\mbox{TAUOLA}}

\newcommand{\BHWIDE}{\mbox{BHWIDE}}
\newcommand{\WPHACT}{\mbox{WPHACT}}

\newcommand{\BDKRC}{\mbox{BDKRC}}

\newcommand{\lt}{\raisebox{0.2ex}{$<$}}
\newcommand{\gt}{\raisebox{0.2ex}{$>$}}
\newcommand{\chisq}{\mbox{$\chi^{2}$}}

\newcommand {\Zzero} {\mbox{$\mathrm{Z}$}}
\newcommand {\ffbar} {\mbox{$\mathrm{f\bar{f}}$}}

\newcommand {\lplm} {\mbox{$\mathrm{l^{+}l^{-}}$}}

\newcommand {\ee} {\mbox{$\mathrm{e^{+}e^{-}}$}}

\newcommand {\mumu} {\mbox{$\mu^{+}\mu^{-}$}}

\newcommand {\tautau} {\mbox{$\tau^{+}\tau^{-}$}}

\newcommand {\qqbar} {\mbox{$\mathrm{q\bar{q}}$}}
\newcommand {\qqbarg} {\mbox{$\mathrm{q\bar{q}}(\gamma)$}}
\newcommand {\ZZpair} {\mbox{$\mathrm{ZZ}$}}
\newcommand {\WWpair} {\mbox{$\mathrm{W^{+}W^{-}}$}}

\newcommand {\eeff} {\mbox{$\ee \rightarrow \ffbar$}}

\newcommand {\eell} {\mbox{$\ee \rightarrow \lplm$}}

\newcommand {\eeee} {\mbox{$\ee \rightarrow \ee$}}

\newcommand {\eemm} {\mbox{$\ee \rightarrow \mumu$}}

\newcommand {\eett} {\mbox{$\ee \rightarrow \tautau$}}

\newcommand {\eeqq} {\mbox{$\ee \rightarrow \qqbar$}}
\newcommand {\eeqqg} {\mbox{$\ee \rightarrow \qqbarg$}}

\newcommand{\ggee}{\gamma\gamma \rightarrow \ee}
\newcommand{\ggmm}{\gamma\gamma \rightarrow \mumu}
\newcommand{\ggtt}{\gamma\gamma \rightarrow \tautau}
\newcommand{\ggqq}{\gamma\gamma \rightarrow \qqbar}

\newcommand {\eeWW} {\mbox{$\ee \rightarrow \WWpair $}}
\newcommand {\eeZZ} {\mbox{$\ee \rightarrow \ZZpair $}}

\newcommand {\MZ} {\mbox{$\mathrm{M}_{\mbox{\ssize{Z}}}$}}

\newcommand {\MZbar} {\mbox{$\overline{\mathrm{M}}_{\mbox{\ssize{Z}}}$}}
\newcommand {\MW} {\mbox{$\mathrm{M_W}$}}

\newcommand {\MH} {\mbox{$\mathrm{M_H}$}}
\newcommand {\MT} {\mbox{$\mathrm{m_{t}}$}}
\newcommand {\GZ} {\mbox{$\Gamma_{\mbox{\ssize{Z}}}$}}
\newcommand {\GZbar} {\mbox{$\overline{\Gamma}_{\mbox{\ssize{Z}}}$}}

\newcommand{\sqq}{\mbox{$\sigma_{qq}$}}
\newcommand{\see}{\mbox{$\sigma_{ee}$}}
\newcommand{\smu}{\mbox{$\sigma_{\mu\mu}$}}
\newcommand{\stau}{\mbox{$\sigma_{\tau\tau}$}}
\newcommand {\Afb} {\mbox{$\mathrm{A_{FB}}$}}
\newcommand {\Afbe} {\mbox{$\mathrm{A^{e}_{\mbox{\ssize FB}}}$}}
\newcommand {\Afbm} {\mbox{$\mathrm{A^{\mu}_{\mbox{\ssize FB}}}$}}
\newcommand {\Afbt} {\mbox{$\mathrm{A^{\tau}_{\mbox{\ssize FB}}}$}}

\newcommand{\dsdcth}{\mbox{${{\mathrm{d}}\sigma}/{{\mathrm{d}}\cos\theta}$}}
\newcommand {\roots} {\mbox{$\sqrt{s}$}}
\newcommand {\rootsp} {\mbox{$\sqrt{s^{\prime}}$}}
\newcommand {\sqsps} {\mbox{$\rootsp/\roots$}}
\newcommand {\Mff} {\mbox{${\mathrm{M}}_{\small{\ffbar}}$}}

\newcommand {\acol} {\theta_{acol}}

\newcommand {\prad} {\mathrm{P_{\mbox{\small{RAD}}}}}

\newcommand {\Evis} {\mathrm{E_{\mbox{\small{VIS}}}}}
\newcommand {\mydeg} {\mbox{$^\circ$}}

\newcommand{\rtote}{\mathrm{r^{tot}_{e}}}
\newcommand{\rtotm}{\mathrm{r^{tot}_{\mu}}}
\newcommand{\rtott}{\mathrm{r^{tot}_{\tau}}}
\newcommand{\rtotl}{\mathrm{r^{tot}_{l}}}
\newcommand{\rtoth}{\mathrm{r^{tot}_{had}}}

\newcommand{\jtote}{\mathrm{j^{tot}_{e}}}
\newcommand{\jtotm}{\mathrm{j^{tot}_{\mu}}}
\newcommand{\jtott}{\mathrm{j^{tot}_{\tau}}}
\newcommand{\jtotl}{\mathrm{j^{tot}_{l}}}
\newcommand{\jtoth}{\mathrm{j^{tot}_{had}}}

\newcommand{\rfbe}{\mathrm{r^{fb}_{e}}}
\newcommand{\rfbm}{\mathrm{r^{fb}_{\mu}}}
\newcommand{\rfbt}{\mathrm{r^{fb}_{\tau}}}
\newcommand{\rfbl}{\mathrm{r^{fb}_{l}}}

\newcommand{\jfbe}{\mathrm{j^{fb}_{e}}}
\newcommand{\jfbm}{\mathrm{j^{fb}_{\mu}}}
\newcommand{\jfbt}{\mathrm{j^{fb}_{\tau}}}
\newcommand{\jfbl}{\mathrm{j^{fb}_{l}}}

\newcommand {\mco}{\multicolumn{1}{|c|}}
\newcommand{\rf}{\mathrm{r_{f}}}
\newcommand{\jf}{\mathrm{j_{f}}}
\newcommand{\gf}{\mathrm{g_{f}}}
\newcommand{\rfa}{\mathrm{r_{f}^{a}}}
\newcommand{\jfa}{\mathrm{j_{f}^{a}}}
\newcommand{\gfa}{\mathrm{g_{f}^{a}}}

\newcommand{\susyfy}[1]{\mbox{$\stackrel{\sim}{#1}$}}
\newcommand{\msneut}{\mbox{$m_{\tiny{\susyfy{\nu}}}$}}
\newcommand{\sneut} {\susyfy{\nu}}
\newcommand{\snul} {\susyfy{\nu}_{l}}
\newcommand{\snue} {\susyfy{\nu}_{\mathrm{e}}}
\newcommand{\snumu} {\susyfy{\nu}_{\mu}}
\newcommand{\snutau}{\susyfy{\nu}_{\tau}}
\newcommand {\Zprime} {\mbox{$\mathrm{Z}^{'}$}}
\newcommand {\thtzzp} {\mbox{$\Theta_{\mbox{\ssize{Z}}\mbox{\ssize{Z}$^{'}$}}$}}
\newcommand {\MZp} {\mbox{$\mathrm{M}_{\mbox{\ssize{Z}$^{'}$}}$}}
\newcommand {\MZpsq} {\mbox{$\mathrm{M}^{2}_{\mbox{\ssize{Z}$^{'}$}}$}}
\newcommand {\thtzzplim} {\mbox{$\Theta_{\mathrm{Z} \mathrm{Z}^{'}}^{limit}$}}
\newcommand {\MZplim} {\mbox{$\mathrm{M}_{\mathrm{Z}^{'}}^{limit}$}}
\begin{document}
%%%%%%%%%%%%%%%%%%%%%%%%%% They are a problem with Coll.Sty ?
\makeatletter
\makeatother
%%%%%%%%%%%%%%%%%%%%%%%%%% ??????????????????????????????????
% Generate the title page
\begin{titlepage}
\pagenumbering{roman}
\CERNpreprint{\DpPaperGroup}{\DpPaperRef} % Reference of the paper
\date{{\small\DpDate}} % Date of the paper
\title{\DpTitle} % Title of the paper
\address{\DpAuthors} % General name of the author(s)
\begin{shortabs} % Start the abstract
\noindent
%%--------------------------------------------------------------------
%%-- ABSTRACT --------------------------------------------------------
%%--------------------------------------------------------------------
\noindent
This paper presents DELPHI measurements and interpretations of cross-sections, 
forward-backward asymmetries, and angular distributions, 
%$\dsdcth$, 
for the \eeff\ process for centre-of-mass energies above the 
\Zzero\ resonance, from $\roots \sim 130 - 207$ $\GeV$ at the LEP collider. 

The measurements are consistent with the predictions of the Standard Model
and are used to study a variety of models including the S-Matrix ansatz for
\eeff\ scattering and several models which include physics beyond 
the Standard Model: the exchange of \Zprime\ bosons, contact interactions
between fermions, the exchange of gravitons in large extra dimensions
and the exchange of \sneut\ in R-parity violating supersymmetry. 

\vskip 2cm

\centerline
{\it This paper is dedicated to the memory of Alan Segar.}

\end{shortabs}
\vfill
\begin{center}
\DpSubmit \ \\ % Horrible hack to allow to have DpSubmit empty
\DpComment \ \\
\DpEMail \ \\
\end{center}
\vfill
\clearpage
\headsep 10.0pt
\addtolength{\textheight}{10mm}
\addtolength{\footskip}{-5mm}
\begingroup
% Commands to process the author names
%
\newcommand{\DpName}[2]{\hbox{#1$^{\ref{#2}}$},\hfill}
\newcommand{\DpNameTwo}[3]{\hbox{#1$^{\ref{#2},\ref{#3}}$},\hfill}
\newcommand{\DpNameThree}[4]{\hbox{#1$^{\ref{#2},\ref{#3},\ref{#4}}$},\hfill}
\newskip\Bigfill \Bigfill = 0pt plus 1000fill
\newcommand{\DpNameLast}[2]{\hbox{#1$^{\ref{#2}}$}\hspace{\Bigfill}}
%
%\small
\footnotesize
\noindent
\DpName{J.Abdallah}{LPNHE}
\DpName{P.Abreu}{LIP}
\DpName{W.Adam}{VIENNA}
\DpName{P.Adzic}{DEMOKRITOS}
\DpName{T.Albrecht}{KARLSRUHE}
\DpName{T.Alderweireld}{AIM}
\DpName{R.Alemany-Fernandez}{CERN}
\DpName{T.Allmendinger}{KARLSRUHE}
\DpName{P.P.Allport}{LIVERPOOL}
\DpName{U.Amaldi}{MILANO2}
% ...
\normalsize
\endgroup
\newpage
\titlefoot{Department of Physics and Astronomy, Iowa State
     University, Ames IA 50011-3160, USA
    \label{AMES}}
\titlefoot{Physics Department, Universiteit Antwerpen,
     Universiteitsplein 1, B-2610 Antwerpen, Belgium \\
     \indent~~and IIHE, ULB-VUB,
     Pleinlaan 2, B-1050 Brussels, Belgium \\
     \indent~~and Facult\'e des Sciences,
     Univ. de l'Etat Mons, Av. Maistriau 19, B-7000 Mons, Belgium
    \label{AIM}}
\titlefoot{Physics Laboratory, University of Athens, Solonos Str.
     104, GR-10680 Athens, Greece
    \label{ATHENS}}
\titlefoot{Department of Physics, University of Bergen,
     All\'egaten 55, NO-5007 Bergen, Norway
    \label{BERGEN}}
\titlefoot{Dipartimento di Fisica, Universit\`a di Bologna and INFN,
     Via Irnerio 46, IT-40126 Bologna, Italy
    \label{BOLOGNA}}
\titlefoot{Centro Brasileiro de Pesquisas F\'{\i}sicas, rua Xavier Sigaud 150,
     BR-22290 Rio de Janeiro, Brazil \\
     \indent~~and Depto. de F\'{\i}sica, Pont. Univ. Cat\'olica,
     C.P. 38071 BR-22453 Rio de Janeiro, Brazil \\
     \indent~~and Inst. de F\'{\i}sica, Univ. Estadual do Rio de Janeiro,
     rua S\~{a}o Francisco Xavier 524, Rio de Janeiro, Brazil
    \label{BRASIL}}
\titlefoot{Coll\`ege de France, Lab. de Physique Corpusculaire, IN2P3-CNRS,
     FR-75231 Paris Cedex 05, France
    \label{CDF}}
\titlefoot{CERN, CH-1211 Geneva 23, Switzerland
    \label{CERN}}
\titlefoot{Institut de Recherches Subatomiques, IN2P3 - CNRS/ULP - BP20,
     FR-67037 Strasbourg Cedex, France
    \label{CRN}}
\titlefoot{Now at DESY-Zeuthen, Platanenallee 6, D-15735 Zeuthen, Germany
    \label{DESY}}
\titlefoot{Institute of Nuclear Physics, N.C.S.R. Demokritos,
     P.O. Box 60228, GR-15310 Athens, Greece
    \label{DEMOKRITOS}}
\titlefoot{FZU, Inst. of Phys. of the C.A.S. High Energy Physics Division,
     Na Slovance 2, CZ-180 40, Praha 8, Czech Republic
    \label{FZU}}
\titlefoot{Dipartimento di Fisica, Universit\`a di Genova and INFN,
     Via Dodecaneso 33, IT-16146 Genova, Italy
    \label{GENOVA}}
\titlefoot{Institut des Sciences Nucl\'eaires, IN2P3-CNRS, Universit\'e
     de Grenoble 1, FR-38026 Grenoble Cedex, France
    \label{GRENOBLE}}
\titlefoot{Helsinki Institute of Physics and Department of Physical Sciences,
     P.O. Box 64, FIN-00014 University of Helsinki, 
     \indent~~Finland
    \label{HELSINKI}}
\titlefoot{Joint Institute for Nuclear Research, Dubna, Head Post
     Office, P.O. Box 79, RU-101 000 Moscow, Russian Federation
    \label{JINR}}
\titlefoot{Institut f\"ur Experimentelle Kernphysik,
     Universit\"at Karlsruhe, Postfach 6980, DE-76128 Karlsruhe,
     Germany
    \label{KARLSRUHE}}
\titlefoot{Institute of Nuclear Physics PAN,Ul. Radzikowskiego 152,
     PL-31142 Krakow, Poland
    \label{KRAKOW1}}
\titlefoot{Faculty of Physics and Nuclear Techniques, University of Mining
     and Metallurgy, PL-30055 Krakow, Poland
    \label{KRAKOW2}}
\titlefoot{Universit\'e de Paris-Sud, Lab. de l'Acc\'el\'erateur
     Lin\'eaire, IN2P3-CNRS, B\^{a}t. 200, FR-91405 Orsay Cedex, France
    \label{LAL}}
\titlefoot{School of Physics and Chemistry, University of Lancaster,
     Lancaster LA1 4YB, UK
    \label{LANCASTER}}
\titlefoot{LIP, IST, FCUL - Av. Elias Garcia, 14-$1^{o}$,
     PT-1000 Lisboa Codex, Portugal
    \label{LIP}}
\titlefoot{Department of Physics, University of Liverpool, P.O.
     Box 147, Liverpool L69 3BX, UK
    \label{LIVERPOOL}}
\titlefoot{Dept. of Physics and Astronomy, Kelvin Building,
     University of Glasgow, Glasgow G12 8QQ
    \label{GLASGOW}}
\titlefoot{LPNHE, IN2P3-CNRS, Univ.~Paris VI et VII, Tour 33 (RdC),
     4 place Jussieu, FR-75252 Paris Cedex 05, France
    \label{LPNHE}}
\titlefoot{Department of Physics, University of Lund,
     S\"olvegatan 14, SE-223 63 Lund, Sweden
    \label{LUND}}
\titlefoot{Universit\'e Claude Bernard de Lyon, IPNL, IN2P3-CNRS,
     FR-69622 Villeurbanne Cedex, France
    \label{LYON}}
\titlefoot{Dipartimento di Fisica, Universit\`a di Milano and INFN-MILANO,
     Via Celoria 16, IT-20133 Milan, Italy
    \label{MILANO}}
\titlefoot{Dipartimento di Fisica, Univ. di Milano-Bicocca and
     INFN-MILANO, Piazza della Scienza 2, IT-20126 Milan, Italy
    \label{MILANO2}}
\titlefoot{IPNP of MFF, Charles Univ., Areal MFF,
     V Holesovickach 2, CZ-180 00, Praha 8, Czech Republic
    \label{NC}}
\titlefoot{NIKHEF, Postbus 41882, NL-1009 DB
     Amsterdam, The Netherlands
    \label{NIKHEF}}
\titlefoot{National Technical University, Physics Department,
     Zografou Campus, GR-15773 Athens, Greece
    \label{NTU-ATHENS}}
\titlefoot{Physics Department, University of Oslo, Blindern,
     NO-0316 Oslo, Norway
    \label{OSLO}}
\titlefoot{Dpto. Fisica, Univ. Oviedo, Avda. Calvo Sotelo
     s/n, ES-33007 Oviedo, Spain
    \label{OVIEDO}}
\titlefoot{Department of Physics, University of Oxford,
     Keble Road, Oxford OX1 3RH, UK
    \label{OXFORD}}
\titlefoot{Dipartimento di Fisica, Universit\`a di Padova and
     INFN, Via Marzolo 8, IT-35131 Padua, Italy
    \label{PADOVA}}
\titlefoot{Rutherford Appleton Laboratory, Chilton, Didcot
     OX11 OQX, UK
    \label{RAL}}
\titlefoot{Dipartimento di Fisica, Universit\`a di Roma II and
     INFN, Tor Vergata, IT-00173 Rome, Italy
    \label{ROMA2}}
\titlefoot{Dipartimento di Fisica, Universit\`a di Roma III and
     INFN, Via della Vasca Navale 84, IT-00146 Rome, Italy
    \label{ROMA3}}
\titlefoot{DAPNIA/Service de Physique des Particules,
     CEA-Saclay, FR-91191 Gif-sur-Yvette Cedex, France
    \label{SACLAY}}
\titlefoot{Instituto de Fisica de Cantabria (CSIC-UC), Avda.
     los Castros s/n, ES-39006 Santander, Spain
    \label{SANTANDER}}
\titlefoot{Inst. for High Energy Physics, Serpukov
     P.O. Box 35, Protvino, (Moscow Region), Russian Federation
    \label{SERPUKHOV}}
\titlefoot{J. Stefan Institute, Jamova 39, SI-1000 Ljubljana, Slovenia
     and Laboratory for Astroparticle Physics,\\
     \indent~~Nova Gorica Polytechnic, Kostanjeviska 16a, SI-5000 Nova Gorica, Slovenia, \\
     \indent~~and Department of Physics, University of Ljubljana,
     SI-1000 Ljubljana, Slovenia
    \label{SLOVENIJA}}
\titlefoot{Fysikum, Stockholm University,
     Box 6730, SE-113 85 Stockholm, Sweden
    \label{STOCKHOLM}}
\titlefoot{Dipartimento di Fisica Sperimentale, Universit\`a di
     Torino and INFN, Via P. Giuria 1, IT-10125 Turin, Italy
    \label{TORINO}}
%\titlefoot{INFN,Sezione di Torino, and Dipartimento di Fisica Teorica,
%     Universit\`a di Torino, Via P. Giuria 1,\\
%     \indent~~IT-10125 Turin, Italy
\titlefoot{INFN,Sezione di Torino and Dipartimento di Fisica Teorica,
     Universit\`a di Torino, Via Giuria 1,
     IT-10125 Turin, Italy
    \label{TORINOTH}}
\titlefoot{Dipartimento di Fisica, Universit\`a di Trieste and
     INFN, Via A. Valerio 2, IT-34127 Trieste, Italy \\
     \indent~~and Istituto di Fisica, Universit\`a di Udine,
     IT-33100 Udine, Italy
    \label{TU}}
\titlefoot{Univ. Federal do Rio de Janeiro, C.P. 68528
     Cidade Univ., Ilha do Fund\~ao
     BR-21945-970 Rio de Janeiro, Brazil
    \label{UFRJ}}
\titlefoot{Department of Radiation Sciences, University of
     Uppsala, P.O. Box 535, SE-751 21 Uppsala, Sweden
    \label{UPPSALA}}
\titlefoot{IFIC, Valencia-CSIC, and D.F.A.M.N., U. de Valencia,
     Avda. Dr. Moliner 50, ES-46100 Burjassot (Valencia), Spain
    \label{VALENCIA}}
\titlefoot{Institut f\"ur Hochenergiephysik, \"Osterr. Akad.
     d. Wissensch., Nikolsdorfergasse 18, AT-1050 Vienna, Austria
    \label{VIENNA}}
\titlefoot{Inst. Nuclear Studies and University of Warsaw, Ul.
     Hoza 69, PL-00681 Warsaw, Poland
    \label{WARSZAWA}}
\titlefoot{Fachbereich Physik, University of Wuppertal, Postfach
     100 127, DE-42097 Wuppertal, Germany \\
\noindent
{$^\dagger$~deceased}
    \label{WUPPERTAL}}
\addtolength{\textheight}{-10mm}
\addtolength{\footskip}{5mm}
\clearpage
\headsep 30.0pt
\end{titlepage}
%%%%%%%%%%%%%%%%%%%%%%%%%
% Change for the document body
%%\pagestyle{heading} % for page numbering
\pagenumbering{arabic} % page numbering in number
\setcounter{footnote}{0} %
\large
\section{Introduction}
\label{sec:intro}

%%....................................................................
%%.. Summary .........................................................
%%....................................................................
%\fbox{
%\begin{minipage}{0.87\textwidth}
%\begin{itemize}
% \item General overview of LEP II program
% \begin{itemize}
%   \item range of energies
% 
%   \item total luminosity
% \end{itemize}
%
% \item References to previous literature:
%
% \begin{itemize}
%  \item LEP I Lineshape publications
%  \item Previous DELPHI LEP II \ffbar\ publications
%  \item Other LEP Experiments
% \end{itemize}
%
% \item Organization of paper
%\end{itemize}
%\end{minipage}
%}
%\vskip 0.2cm

%%....................................................................
%%.. Contents ........................................................
%%....................................................................

This paper presents measurements and interpretations of cross-sections, 
$\sigma$, forward-backward asymmetries, $\Afb$,
%\footnote{The forward-backward 
%asymmetry is defined as 
%$\Afb = \frac{\int_{0}^{c} d\sigma/d\cos\theta d\cos\theta - 
%               \int_{-c}^{0} d\sigma/d\cos\theta d\cos\theta}
%             {\int_{0}^{c} d\sigma/d\cos\theta d\cos\theta + 
%                \int_{-c}^{0} d\sigma/d\cos\theta d\cos\theta}$ 
%where $\theta$ is the polar angle of the final state fermion with respect to 
%the direction of the incoming electron and $c$ is the angular acceptance.}
and angular distributions, 
$\dsdcth$, for \eeff\ processes for centre-of-mass energies above the 
\Zzero\ resonance, as measured in the DELPHI experiment at the LEP collider. 
Measurements of flavour tagged \qqbar\ samples will be included in an 
additional publication.

For the first part of the LEP programme, LEP I, \ee\ collisions were provided 
at centre-of-mass energies close to the \Zzero\ resonance. Measurements of 
the process \eeff\ were used to determine properties of the \Zzero\ and 
electroweak parameters of the Standard 
Model~\cite{ref:delphils:91,*ref:delphils:95,ref:delphils:00}.
In 1995 the operation of LEP moved into the LEP II programme. The collision 
energy was raised to significantly above the \Zzero\ resonance, and a 
total of approximately $0.7 \: \fbarn^{-1}$ of integrated luminosity was 
delivered to the 
DELPHI experiment at energies ranging from 130 \GeV\ 
to a maximum collision energy of 209 \GeV\ 
during LEP II operations. A breakdown of the centre-of-mass energies,
and integrated luminosities is given in Table~\ref{tab:intro:ecmlum_aprox}.
Measurements of the process \eeff\ from LEP II are less sensitive to the 
electroweak parameters of the Standard Model. Nevertheless, taken together, 
the measurements constitute a test of the Standard Model at the 
${\mathcal{O}}(1\%)$ level, at the highest \ee\ collision 
energies to date. Furthermore, the \eeff\ measurements at LEP II are 
predicted to be more sensitive to a variety of models of physics beyond the 
Standard Model than the LEP I measurements. Having determined many of the
parameters of the Standard Model, largely from LEP I data, studies of the 
process \eeff\ at LEP II are, therefore, a suitable place to look for physics 
beyond the Standard Model.

%%%%%%%%%%%%%%%%%%%%%%%%%%%
\begin{table}[p]
\begin{center}
 \renewcommand{\arraystretch}{1.5}
 \begin{tabular}{|c|c|c|c|}
 \hline
 Year      & Nominal   & Mean     & Integrated \vspace{-1.0ex} \\ 
           & Energy    & Energy   & Luminosity \vspace{-1.0ex} \\
           & ($\GeV$)  & ($\GeV$) & ($\invpbarn$)              \\
 \hline\hline
 1995      & 130       & 130.3    &     3                      \\
           & 136       & 136.3    &     3                      \\ 
 \hline
 1996      & 161       & 161.5    &    10                      \\ 
           & 172       & 172.4    &    10                      \\ 
 \hline
 1997      & 183       & 182.7    &    53                      \\ 
%          & 130       & 130.2    &     3                      \\
%          & 136       & 136.2    &     3                      \\
 \hline
 1998      & 189       & 188.6    &   155                      \\ 
 \hline
 1999      & 192       & 191.8    &    25                      \\
           & 196       & 195.7    &    76                      \\
           & 200       & 199.7    &    83                      \\
           & 202       & 201.8    &    40                      \\
 \hline
 2000      & 205       & 204.9    &    82                      \\
           & 207       & 206.6    &   135                      \\
 \hline\hline
 Total     &           & 195.6    &   675                      \\
 \hline
\end{tabular}
\end{center}
\caption{\capsty{Nominal centre-of-mass energies with approximate 
         centre-of-mass energy and integrated luminosity collected by DELPHI. 
         The values used in the final analyses differ due to data collection 
         efficiencies and selection criteria.}}
\label{tab:intro:ecmlum_aprox}
\end{table}
%%%%%%%%%%%%%%%%%%%%%%%%%%%

Results from the analysis of data at centre-of-mass energies from 
130--189~\GeV\ have already been 
published~\cite{ref:delphiff:130-172,ref:delphiff:183-189}.  
Results at energies from 192--207 \GeV\ are published here for the first time.
Some previously published results at energies of  183 and 189
\GeV\ \cite{ref:delphiff:183-189} have been reanalysed and are presented 
again here.
For completeness we present again the results from 130--172~\GeV\ data which 
have not been reanalysed.
Results on \ffbar\ production from the other LEP experiments can be found 
in~\cite{ref:alephff:1996,*ref:alephff:1997,*ref:alephff:2000,ref:l3ff:1996,
*ref:l3ff:1997,*ref:l3ff:1998,*ref:l3ff:1999a,*ref:l3ff:1999b,*ref:l3ff:2000a,
*ref:l3ff:2000b,ref:opalff:1996,*ref:opalff:1997,*ref:opalff:1998,
*ref:opalff:1999,*ref:opalff:2000,*ref:opalff:2003}.

%%%%%%%%%%%%%%%%%%%%
%\fbox{
%\begin{minipage}{0.87\textwidth}
%
% Point out any changes to any previous results
%
%\end{minipage}
%}
%\vskip 0.2cm
%%%%%%%%%%%%%%%%%%%%

The operation of LEP during the LEP II programme is discussed in 
Section~\ref{sec:lep}. The DELPHI detector itself is described in 
Section~\ref{sec:detector}, and the measurement of the integrated luminosity
is described in Section~\ref{sec:lumi}. Features of \ffbar\ production at
LEP II, details of the event simulation and theoretical predictions
are given in Section~\ref{sec:analysis}. Sections~\ref{sec:ee} to~\ref{sec:qq}
cover the analysis of the individual channels \eeee,
\eemm, \eett\ and \eeqq. In each section the criteria for selecting events 
are described and the methods for evaluating the efficiency and backgrounds are
discussed. Results of the individual measurements are provided and the 
principal sources of systematic error are described. The sets of measurements
are summarised in Section~\ref{sec:results}, where the results are compared
to the predictions of the Standard Model. Further interpretations of the 
results are presented in Section~\ref{sec:interp}. In particular the results,
together with LEP I data, are interpreted with the S-Matrix formalism in
Section~\ref{sec:smat}, and also in a variety of models which include 
explicit forms of physics beyond the Standard Model: 
models with \Zprime\ bosons in Section~\ref{sec:zp}, 
contact interactions (Section~\ref{sec:cntc}),
%the exchange of \sneut\ in \rpviol\ SUSY (Section~\ref{sec:sneut})
models which include the exchange of gravitons in large extra dimensions
(Section~\ref{sec:grav}) and models which consider possible $s$ or $t$ channel 
sneutrino $\snul$ exchange in R-parity violating supersymmetry 
(Section~\ref{sec:sneutrino}). 
In each case, a resum{\'e} of the model, the method of comparing predictions of
the model to the data  and the results of the interpretation are provided.
Conclusions drawn from the DELPHI analyses of the \eeff\ processes at LEP II 
are given in Section \ref{sec:conclusions}.

%%--------------------------------------------------------------------
%%-- LEP -------------------------------------------------------------
%%--------------------------------------------------------------------
\section{LEP}
\label{sec:lep}

%%....................................................................
%%.. Summary .........................................................
%%....................................................................
%\fbox{
%\begin{minipage}{0.87\textwidth}
%\begin{itemize}
% \item The operation of LEP during LEP II
%  \begin{itemize}
%   \item LEP II energy-time profile
%   \item \Zzero\ runs
%  \end{itemize}
%
% \item Energy calibration
%
% \item Other beam parameter
%  \begin{itemize}
%   \item Beam Acollinearity
%   \item Beam Energy spread
%   \item Beam Energy difference
%  \end{itemize}
%\end{itemize}
%\end{minipage}
%}
%\vskip 0.2cm

%%....................................................................
%%.. Contents ........................................................
%%....................................................................

The LEP collider was upgraded from its original configuration, used for running
at energies around the \Zzero\ resonance, by the 
addition of superconducting RF cavities. This then allowed operations at 
energies well above the \Zzero\ resonance.
In the years 1995 to 1999 LEP delivered \ee\ collisions at one or more
discrete energies, each LEP fill corresponding to a particular 
energy. 
%Generally throughout a year, the fills at the end of a year 
%were at higher energies than those at the start of the year.
In 2000 the luminosity at any given time was delivered at the maximum 
energy available from the LEP RF system, within a margin of safety which 
allowed for the trip of either two, one or no RF units before the beam was 
lost. In a given fill the 
energy would usually be increased from the limit set by two RF trips, to
the limit set by one trip and eventually no RF trips. The luminosity was, 
therefore, delivered
more or less continuously over a range of energies. For analysis purposes
the data were grouped into two energy points: data taken at centre-of-mass
energies between 202.5 and 205.5~\GeV\ and data taken at energies above
205.5 \GeV. Data collected during the time in which the energy was 
increasing are not analysed here, which represents a loss of 
approximately $1\%$ of the delivered luminosity. 

As well as providing collisions at energies above the \Zzero\ resonance,
LEP also ran at a centre-of-mass energy close to the pole of the \Zzero\ 
resonance in each year. The data gathered at this energy were used by 
DELPHI for detector calibration purposes. Typically 2.5 $\pbarn^{-1}$
were delivered at the start of the year in 1996-2000, with additional
luminosity delivered on several other occasions when requested by
the experiments. The data collected in 1995 at centre-of-mass energies 
of 130 and 136 \GeV\ followed a long run at energies close 
to the \Zzero\ resonance.

The energy of the \ee\ collisions was determined by the LEP energy group. 
During LEP I this was based primarily on the resonant depolarisation
technique~\cite{ref:lep:lep1lepenergyrdp,*ref:lep:lep1lepenergy1995,*ref:lep:lep1lepenergy1999}, 
where the energy was determined
with very high precision at the end of a large number of fills 
- making the measurements at close to the actual collision energies. 
The measurements were used to normalise a model of the
RF system, from which the energies during actual collisions were determined. 
At LEP II the energy was again calibrated using the
resonant depolarisation technique, but it was not possible to obtain
polarisation at the actual collision energies, so the RF model
had to be used to extrapolate from the calibration energies 
(up to 60~\mbox{$\mathrm{GeV}$} per beam) 
to the collision energies. The accuracy of this 
extrapolation was checked using a number of techniques:
\begin{itemize}
 \item dedicated LEP runs, at collision energies, in which
       the energy of the electron and positron beams were determined from 
       the deflection of the beams in a magnetic spectrometer whose 
       magnetic field was known to high precision;
 \item measurements of the synchrotron tune as a function of RF voltage;

 \item measurements of the total magnetic field seen by the beams using the
       flux loop system of LEP.
\end{itemize}
The precision that was obtained was $\pm (20-40)$ \MeV\ on the centre-of-mass 
energy~\cite{ref:lep:lep2lepenergy}. The instantaneous difference in energy 
between the electron and positron beams was less than $\pm 100$ \MeV.

The beam energy spread at LEP II is larger than at LEP I, due to the
increased synchrotron energy loss at the higher beam energies. At LEP I
the beam energy spread was $\sim 40$ \MeV, at LEP II this increases
to $\sim180$ \MeV\ at the highest centre-of-mass energies. At LEP I the 
beam energy spread had to be taken into account in the estimation of the
total cross-section. For example, at the peak of the \Zzero\ resonance, 
the expected cross-section was significantly lower than would have been 
observed for a monochromatic beam. At LEP II the corresponding correction 
is not significant for \eeff\ since the
cross-sections are nearly linear over a small range of energies. However,
for certain measurements at LEP II it is necessary to account for the beam 
energy spread for the determination of the event kinematics.

%%--------------------------------------------------------------------
%%-- The DELPHI Detector ---------------------------------------------
%%--------------------------------------------------------------------
\section{The DELPHI Detector}
\label{sec:detector}

%%....................................................................
%%.. Summary .........................................................
%%....................................................................
%\fbox{
%\begin{minipage}{0.87\textwidth}
%\begin{itemize}
% \item DELPHI detector~\cite{ref:det:delphidet,ref:det:lep1perf}
%  \begin{itemize}
%   \item Changes of hardware w.r.t LEP I
%   \item Changes of running which improved collection eff for
%         highest energies
%  \end{itemize}
%
% \item Include comment about TPC S6 in 2000
%
%\end{itemize}
%\end{minipage}
%}
%\vskip 0.2cm

A detailed description of the DELPHI apparatus as used at LEP I 
and its performance can be found in 
refs.~\cite{ref:det:delphidet,ref:det:lep1perf}. For
the present analysis the following parts of the detector are relevant:
\begin{itemize}

\item for the measurement of charged particles the Microvertex
Detector (VD), the Inner Detector (ID), the Time Projection Chamber
(TPC), the Outer Detector(OD) and the Forward Chambers A and B (FCA
and FCB). For the running from 1995 onwards a lengthened Inner Detector was
installed. The polar angle\footnote{The DELPHI coordinate system is a
RH system with the $z$-axis collinear with the incoming electron beam,
the $x$ axis pointing to the centre of the LEP accelerator and the $y$
axis vertical. The polar angle $\theta$ is with reference to the
$z$-axis, and $\phi$ is the azimuthal angle in the $x,y$ plane. r $=
\sqrt{x^2+y^2}$.} coverage was thereby extended from
$23^\circ<\theta<157^\circ$ to $15^\circ<\theta<165^\circ$ with a
corresponding increase in forward tracking efficiency. For a period in
2000 when part of the TPC was not operational (see later), space
points from the Barrel RICH detector were included in the track fit;
\item for the measurement of electromagnetic energy the High-density Projection
Chamber (HPC) and the Forward Electromagnetic Calorimeter (FEMC); these
detectors were also used for identifying minimum ionising particles;
\item for the measurement of the hadronic energy and muon identification the
Hadron Calorimeter (HCAL), which covered both the barrel and endcap regions; 
\item for muon identification the barrel (MUB), the endcap (MUF), and
the surround muon chambers (SMC), which completed the polar angle
coverage between barrel and endcap;
\item for the trigger, besides the detectors mentioned above, the
barrel Time of Flight counters (TOF), the endcap scintillators (HOF) and a
scintillator layer embedded in the HPC;
\item for the measurement of luminosity the Small Angle Tile 
Calorimeter (STIC).

\end{itemize}

The DELPHI detector was upgraded for LEP II 
data taking.
Changes were made to the subdetectors, the trigger system, 
the run control and the algorithms used in the offline reconstruction
of tracks, which improved the performance compared 
to LEP I~\cite{ref:det:lep1perf}.

The major change was the inclusion of the Very Forward Tracker 
(VFT)~\cite{ref:det:vft}, which extended the coverage of the innermost 
silicon tracker out to $11\mydeg < \theta < 169\mydeg$. Together with improved tracking 
algorithms
%%~\cite{ref:det:lep2tracking} 
and alignment and calibration procedures
%%~\cite{ref:det:lep2track_align_calib} 
optimised for LEP II, these changes led to an improved track reconstruction 
efficiency in the forward regions of DELPHI.

A smaller change was the removal of the tungsten nose cone in front of
the Small Angle Tile Calorimeter, to increase the forward
electromagnetic coverage.
%%its angular
%%acceptance.  
This change had consequences for the luminosity
determination discussed in Section~\ref{sec:lumi}.

%%
%% HCAL upgrades - not relevant here
%% Taggers - not relevant
%%

Changes were made to the electronics of the trigger and timing system 
which improved the
stability of the running during data taking~\cite{ref:det:lep2trigger}. The
trigger conditions were optimised for LEP II running, to give high
efficiency for Standard Model 2- and 4-fermion processes and also to give 
sensitivity to events which might be signatures of new physics.
In addition, improvements were made to the operation of the detector during
the LEP cycle, to prepare the detector for data taking at the very start
of stable collisions of the \ee\ beams, and to respond to adverse backgrounds
from LEP, if they occurred. These changes led to an overall improvement
in the efficiency for collecting the delivered luminosity from 
$\sim85\%$ at the start of LEP II in 1995 to $\sim95\%$ by the end in 2000.

During the operation of the DELPHI detector in 2000 one of the 12 sectors of
the central tracking chamber, the TPC, failed. After
September $1^{\mathrm{st}}$ it was not possible to detect tracks left 
by charged particles inside the broken sector. The data affected corresponds
to $\sim 1/4$ of that collected in 2000. Nevertheless, the redundancy
of the tracking system of DELPHI meant that tracks of charged particles 
passing through that sector
could still be reconstructed from signals in other tracking 
detectors. A modified track reconstruction algorithm was used in
this sector, which included space points reconstructed in the Barrel RICH
detector, these points having a significant role in the determination of the
polar angle of tracks.
%%~\cite{ref:det:lep2track_s6}.
As a result, the track reconstruction efficiency was only slightly reduced in 
the region covered by the broken sector, but on average the resolution on the 
perigee parameters of the tracks was worse than prior to the failure of 
the sector. The impact of the failure of this part of the detector
on the different analyses is discussed further in Section~\ref{sec:analysis}.

%%--------------------------------------------------------------------
%%-- Luminosity ------------------------------------------------------
%%--------------------------------------------------------------------
\section{Luminosity determination}
\label{sec:lumi}

The luminosity measurement followed closely the analysis described 
in~\cite{ref:delphils:00}. It was based on counting the number of
Bhabha events in the Small Angle Tile Calorimeter of DELPHI, which covered the region between 29 and 185 mrad with respect
to the beam line. A detailed description of this detector and its performance 
can be found in \cite{ref:lumi:stic}.
It provided a very uniform energy response and an accurate energy resolution of 
$2.0 \%$ for 100 GeV electrons.
The tungsten nose, which was used at LEP I to define the inner edge of the 
calorimeter, was removed between data taking in 1995 and 1996. After 1995 the 
definition of the geometrical acceptance was entirely based on the 
reconstructed radii of the electron and positron showers.

The trigger for Bhabha events was prescaled by a factor 3 to 4 to reduce the 
overall trigger rate. A comparison between the measured luminosity and the 
scalers of the trigger shows that the prescaling 
had no effect on the luminosity measurement~\cite{ref:det:lep2trigger}.
Furthermore a prescaled single arm trigger was used to monitor possible 
trigger inefficiencies, which were found to be smaller than 
2 $\times$ 10$^{-4}$. 

In the selection of the Bhabha events, only the most energetic electromagnetic 
cluster in each arm of the STIC was used.
To remove the background due to off-momentum particles, the following cuts 
were applied:
\begin{itemize}
\item in each arm, the energy of the cluster was required to be 
      larger than $65 \%$ of the beam energy;
\item the acoplanarity\footnote{In general the acoplanarity is defined
      as the complement of the angle, in the plane transverse to the
      beam, between two tracks. In this case the ``tracks'' are lines
      joining the two clusters to the interaction point.}  between the
      two clusters was required to be less than 20$^\circ$.
\end{itemize}
A special trigger, requiring a coincidence between the signal from one arm 
of the STIC and a delayed signal from the other, measured the residual 
background due to off-momentum particles. The measurement showed that it was 
smaller than 10$^{-4}$ of the accepted events.

In order to minimize the sensitivity of the accepted cross-section to the 
transverse position of the interaction point, an asymmetric acceptance was 
defined, with a narrow side and a wide side. 
The following cuts were applied to define the geometrical acceptance:
\begin{itemize}
\item the radial position of the reconstructed shower was required to be 
      between 10 and 25 cm on the narrow side;
\item the radial position of the reconstructed shower was required to be 
      between 9.1 and 28 cm on the wide side.
\end{itemize}
The cuts were chosen at the borders between rings, where the best spatial 
resolution is achieved. 
The narrow side was alternated between the forward 
and the backward hemispheres in successive events, in order to minimize the 
sensitivity to the longitudinal position of the interaction point.

The calculation of the accepted cross-section was based on the event generator 
BHLUMI 4.04 \cite{ref:lumi:bhlumi}, which includes the full 
$\mathcal{O}$($\alpha^{2})$ calculation.
The generated events were passed through a detailed simulation of the detector,
and analysed in the same way as the real data.  
The total accepted cross-section was estimated to be 10.712 $\pm$ 0.010 nb, 
at a centre-of-mass energy of 200 GeV. The contribution of the process 
$e^+ e^- \rightarrow \gamma \gamma $ in the selected sample of Bhabha events
 was 
calculated to be $0.06 \%$.  The theoretical uncertainty on the estimation
of the cross-section is estimated to be 
$\pm0.12 \%$~\cite{ref:lumi:bhlumi_unc}. There is a factor of two improvement on the error 
quoted in~\cite{ref:delphiff:130-172,ref:delphiff:183-189}; in the analysis of 
the results and interpretations made in the paper these smaller errors
have been applied to all data, including those data sets which have not been
reanalysed in full.

The number of selected Bhabhas collected at LEP II energies during
the different years of data taking are reported in Table \ref{tab:lumi:nev}.

%%---------------------------------
\begin{table}
\begin{center}
\renewcommand{\arraystretch}{1.2}
\begin{tabular}{|c||c|c|c|c|c|c|}
\hline 
Year & 1995 & 1996 & 1997 & 1998 & 1999 & 2000 \\
\hline 
\hline 
Number of Bhabha events ($\times 10^3$) & 156 & 144 & 384 & 1195 & 856 & 757 \\
\hline
\end{tabular}
\caption{\capsty{Number of Bhabha events at LEP II.}}
\label{tab:lumi:nev}
\end{center}
\end{table}
%%---------------------------------

The experimental systematics in the luminosity measurement were
dominated by the uncertainty in the radial cut which defines the inner
border of the acceptance.  The bias of the radius reconstruction was
carefully studied by comparing the STIC measurement with the
independent measurement from the two planes of the Silicon Shower
Maximum Detectors, embedded inside the STIC. The difference between
the two measurements was monitored throughout data taking and the bias
was evaluated to be 250 $\mu$m, corresponding to a systematic uncertainty of
$\pm0.5 \%$. The other sources of uncertainties in the event selection
were estimated to contribute $\pm0.04 \%$ to the systematics.  The
luminosity had to be corrected for the average displacement of the
interaction point from its nominal position, as measured by the DELPHI
tracking system. The residual systematic uncertainty, connected with
the fill-to-fill fluctuations, was evaluated to be $\pm0.03 \%$.  The
uncertainty in the beam energy at LEP II propagated into a systematic uncertainty
of $\pm0.04 \%$.  A detailed list of the contributions to the systematic
uncertainty is given in Table \ref{tab:lumi:syst}.  The overall
systematic uncertainty  is evaluated to be $\pm0.5 \%$ and it is common to all LEP~II
data taking, except for 1995. For 1995, the presence of the tungsten
mask made the luminosity determination more accurate, giving an
experimental systematic uncertainty of $\pm0.09 \%$ on the luminosity
determination, for data taken during that year.

%%---------------------------------
\begin{table}
\begin{center}
\renewcommand{\arraystretch}{1.0}
\begin{tabular}{|l|c|}
\hline 
Source  & ${\Delta \mathcal{L} / \mathcal{L}}$ (\%) \\
\hline
\hline 
Bias on inner edge of the acceptance &  0.5  \\
Beam energy                          &  0.04 \\
Position of the IP                   &  0.03 \\
Event selection                      &  0.04 \\
Background subtraction               &  0.01 \\
Trigger inefficiency                 &  0.02 \\
\hline
\hline 
Total experimental                   &  0.5  \\
\hline
\hline 
Total theoretical                    &  0.12 \\
\hline
\end{tabular}
\caption{\capsty{Contributions to the systematic uncertainty of the luminosity 
         measurement at LEP II energies from 1996 onwards.}}
\label{tab:lumi:syst}
\end{center}
\end{table}
%%---------------------------------

%%--------------------------------------------------------------------
%%-- ANALYSIS --------------------------------------------------------
%%--------------------------------------------------------------------
\section{Analysis of $\boldmath{\ffbar}$ final states}
\label{sec:analysis}

%%....................................................................
%%.. Summary .........................................................
%%....................................................................
%\fbox{
%\begin{minipage}{0.87\textwidth}
%\begin{itemize}
% \item
% Discussion of general features of \ffbar\ production at LEP II
% \begin{itemize}
%  \item size of cross-sections asymmetries compared to LEP I
%  \item importance of radiative return
%  \item typically radiative return topologies - $\rootsp$
%  \item backgrounds
% \end{itemize}
%
% \item Generators
%
% \item Theory, semi-analytical calculations and theoretical uncertainty
%
% \item Common issues in signal definition
% \begin{itemize}
%  \item $\rootsp$ definition
%  \item ISR-FSR Interference
%  \item pairs
%  \item angular acceptance
%  \item $\rootsp$ acceptance
% \end{itemize}
%
%\end{itemize}
%\end{minipage}
%}
%\vskip 0.2cm

%%....................................................................
%%.. Contents ........................................................
%%....................................................................

\begin{figure}[tp]
\begin{center}
 \begin{tabular}{c}
  \mbox{\epsfig{file=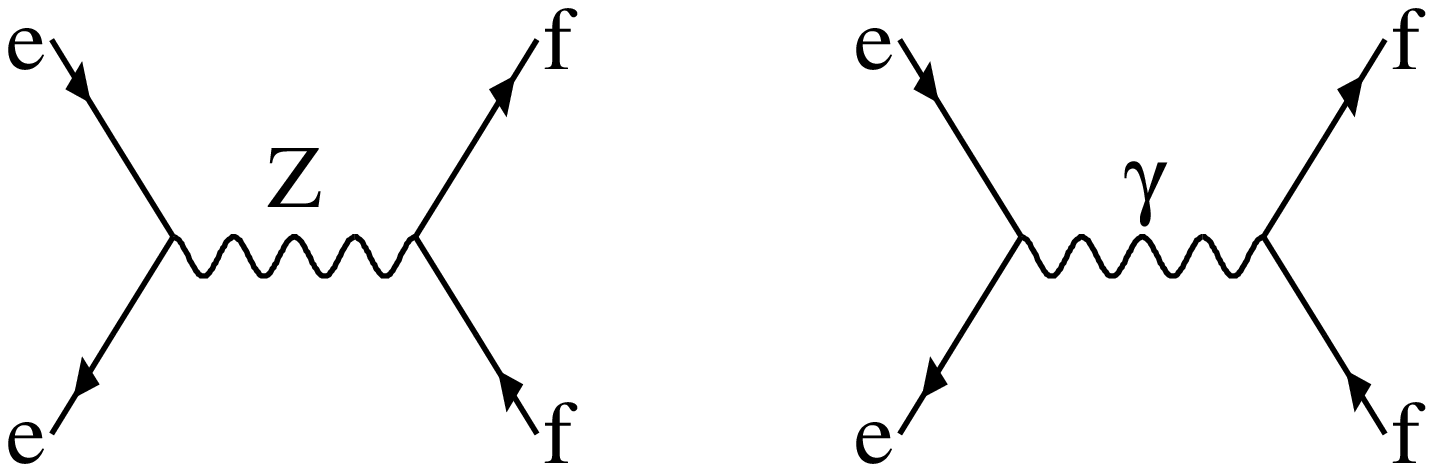,width=0.49\textwidth}} \\
  \mbox{\epsfig{file=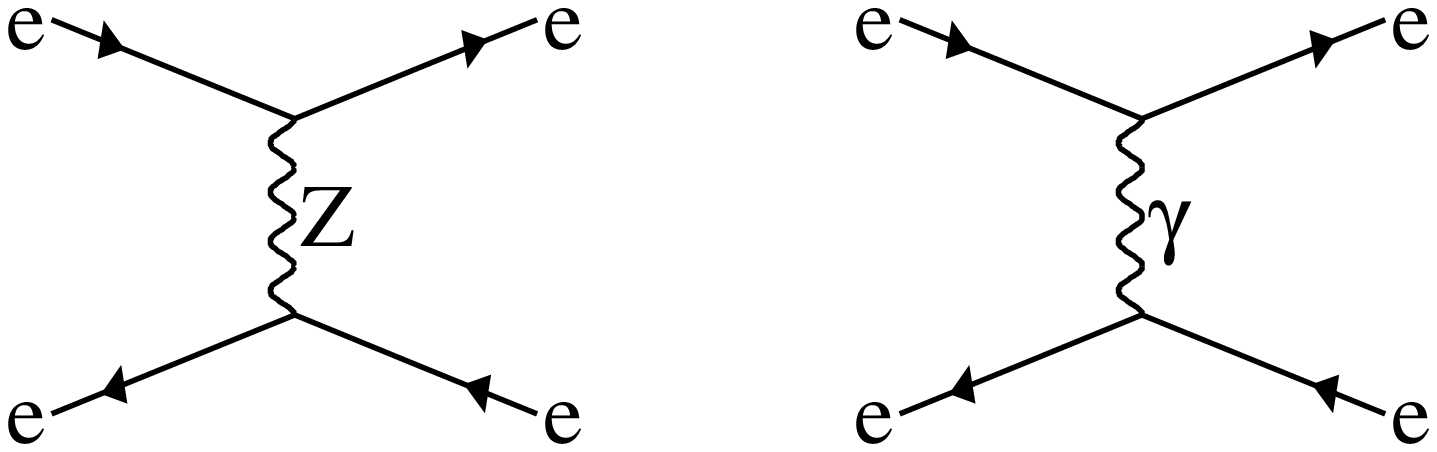,width=0.49\textwidth}} \\
 \end{tabular}
\end{center}
\caption{\capsty{The Standard Model $s$-channel processes
         \eeff\ (upper) and $t$-channel processes in \eeee\ (lower).}}
\label{fig:feyn:eeff}
\end{figure}

The process \eeff\ at LEP II is very similar to that at LEP I. In the
Standard Model the $s$-channel processes \eemm, \eett\ and \eeqq\ are 
described by the exchange of virtual photons and \Zzero\ bosons, shown
as Feynman diagrams in Figure~\ref{fig:feyn:eeff}. At LEP I 
the process is dominated by the formation and decay of the \Zzero. 
At LEP II, the exchange of the photon and the interference between 
the \Zzero\ and the photon (which is highly suppressed at the \Zzero\ 
pole due to the phases of the \Zzero\ and photon exchange amplitudes) 
are more important. Moreover, the predicted Born level cross-sections for 
the $s$-channel process are some 2-3 orders of magnitude smaller than
at the \Zzero\ resonance, and the forward-backward asymmetries are
larger. 

%%%%%%%%%%%%%%%%%%%%%%%%%%%%%%%%%%%%%%%
%%%%%%%%%%%%%%%%%%%%%%%%%%%%%%%%%%%%%%%

At the peak of the \Zzero\ resonance QED radiative corrections are significant,
leading to a $\sim 30\%$ reduction of the cross-section relative to
the Born level prediction. At LEP II energies, QED radiative corrections are
again significant, but here they lead to an increase of a factor of 
${\mathcal{O}}(2-3)$
in the total cross-section above the Born level predictions. This is
predominantly through the process of {\it{radiative return}}, in which a photon
is radiated from the incoming electron or positron, reducing the
centre-of-mass energy of the hard scattering from the full centre-of-mass 
energy, $\roots$, to close to the \Zzero\ resonance. 
The photon is typically emitted along the direction of the
incoming beams, and usually goes undetected down the LEP beampipe.
The final-state fermion pairs in {\it{radiative return}} events are,
therefore, usually acollinear, though they are typically produced 
back-to-back in the plane transverse to the incoming beams.
An important step of the analysis for each channel is to determine 
the reduced centre-of-mass energy of the collisions, $\rootsp$. This
is used to separate events with high collision energies from {\it{radiative
return}} events. Measured cross-sections and forward-backward asymmetries
are quoted for {\itshape{non-radiative}} samples, for which 
$\rootsp \sim \roots$ and for {\itshape{inclusive}} samples of events, 
which also include {\it{radiative return}} events.
The algorithms employed to determine $\rootsp$ are different in each channel. 

The process \eeee\ is dominated by the $t$-channel exchange of virtual
photons.  The $t$-channel processes are shown in
Figure~\ref{fig:feyn:eeff}.  At the \Zzero\ resonance the
cross-sections are sensitive to $s$-channel \Zzero\
exchange. Therefore, at LEP I it was reasonable to subtract off the
predicted $t$-channel contributions to measured quantities for \eeee\
collisions in order to extract \Zzero\ parameters, such as the \Zzero\
mass and coupling to electrons . Above the \Zzero\ resonance, this
subtraction is less appropriate. Physics beyond the Standard Model may
manifest itself through $t$-channel processes. No subtraction of the
$t$-channel contributions to the measured cross-sections, asymmetries
or differential cross-sections has been made.  A cut on the
acollinearity of the electron-positron pair in the final state is used
to separate events with high invariant masses from {\it{radiative
return}} events.

In addition to falling cross-sections for the signal \eeff\ processes
compared to LEP I, the backgrounds from other processes increase
at LEP II. The cross-section for two-photon collision processes increases 
as $\log s$ and new channels such as $\eeWW$ and $\eeZZ$
open at $\roots \sim 2\MW$ and $\roots \sim 2\MZ$ respectively.

With some small modifications to take into account the new 
{\it{radiative return}} topologies and the smaller signal over background 
ratio, the selection of \eeff\ events at LEP~II is very similar to that at 
LEP I. Classification into \ee, \mumu, \tautau\ and \qqbar\ final states is
based on the the multiplicity of final state particles, the responses of the 
electromagnetic and hadronic calorimeters and the muon chambers, and the 
momenta of charged particles measured in the tracking system.

%-.-.-.-.-.-.-.-.-.-.-.-.-.-.-.-.-.-.-.-.-.-.-.-.-.-.-.-.-.-.-.-.-.-.-.-.-
\subsubsection*{Event simulation}

To determine selection efficiencies and backgrounds for each analysis,
events were simulated using a variety of event generators and the DELPHI 
detector simulation~\cite{ref:det:lep1perf} and passed through the full data 
analysis chain. To allow studies of the data taken after 
September $1^{\mathrm{st}}$ 2000, samples of events were simulated 
dropping information from the broken sector of the TPC.

The $s$-channel \eeff\ processes were simulated with \KK~4.14~\cite{ref:mc:kk},
while events in the \eeee\ channel were simulated with 
\BHWIDE~1.01~\cite{ref:mc:bhwide}. The fragmentation of \qqbar\ events into 
hadrons 
was simulated using \PYTHIA~6.156~\cite{ref:mc:pythia,ref:mc:pythia_orig},
%%%new text after EPJ%%%%
 with parameters tuned to DELPHI data~\cite{ref:mc:tuning}.
%%%
Spin effects in $\tau$-lepton 
decays were handled by \TAUOLA~2.6~\cite{ref:mc:tauola_1st,*ref:mc:tauola_2nd} 
using the helicity approximation. There were typically 
1 million simulated events per energy for the \eeee\ and \eeqq\ channels, 
with 100,000 simulated events for the \eemm\ channel and 200,000 simulated 
events for the \eett\ channel at each energy.

Four-fermion background events, including high-mass two-photon 
collision events, were simulated with the generator 
\WPHACT~2.0~\cite{ref:mc:wphact1997,*ref:mc:wphact2003}. The generated events 
were divided into 3 samples~\cite{ref:mc:delphi4f}; the first dominated by 
charged current processes, \eeWW; the second dominated by neutral current 
processes with topologies similar to \eeZZ\ events; and the third sample
being neutral current samples dominated by multi-peripheral processes, 
$\ee \rightarrow \ee\ffbar$.
%%\eeZee.
Low mass two-photon collisions were simulated with \BDKRC~\cite{ref:mc:bdkrc}
for leptonic final states and with \PYTHIA\ for hadronic final states. Since 
the samples of 4-fermion events were generated with certain kinematic cuts, 
the background subtraction involved an extrapolation to estimate the 
backgrounds coming from events which were not simulated. Theoretical 
uncertainties amount to $\pm5\%$ on the 
$\ee \rightarrow \ee\ffbar$
%\eeZee\ 
samples, $\pm0.5\%$ on the
\eeWW\ and $\pm2\%$ on the \eeZZ\ samples.
Where possible the real data were used to control the simulation predictions
for the backgrounds in the selected samples.

%-.-.-.-.-.-.-.-.-.-.-.-.-.-.-.-.-.-.-.-.-.-.-.-.-.-.-.-.-.-.-.-.-.-.-.-.-
\subsubsection*{Theoretical predictions and signal definition}

The measurements reported in this paper are compared to theoretical 
predictions, from the Standard Model and from models which include physics
beyond the Standard Model. Throughout this paper Standard Model predictions
are taken from the semi-analytical QED corrected, Improved Born Approximation 
computations of \ZFITTER~\cite{ref:th:zfitter} for \eemm, \eett\ and \eeqq, 
and from \BHWIDE~\cite{ref:mc:bhwide} for \eeee.
ZFITTER version 6.36~\footnote{The following \mbox{ZFITTER}\ flags were
used:
 AFBC:~1 SCAL:~0 SCRE:~0 AMT4:~4 BORN:~0
 BOXD:~2 CONV:~2 FOT2:~3 GAMS:~1
 DIAG:~1 BARB:~2 PART:~0 POWR:~1
 PRNT:~0 ALEM:~2 QCDC:~3 VPOL:~1 WEAK:~1
 FTJR:~1 EXPR:~0 EXPF:~0 HIGS:~0 AFMT:~1
 CZAK:~0 PREC:10 HIG2:~0 ALE2:~3 GFER:~2
 ISPP:~2 FSRS:~1 MISC:~0 MISD:~1 IPFC:~5
 IPSC:~0 IPTO:-1 FBHO:~0 FSPP:~0 FUNA:~0
 ASCR:~1 SFSR:~1 ENUE:~1 TUPV:~1.
For \eeqq~FINR:~0  INTF:~0, while for \eemm\ and \eett\ FINR:~1 INTF:~2
For {\it{non-radiative}} samples FSRS:~0, while for {\it{inclusive}} samples
FSRS:~1.}
was used with the following central values for input parameters
\begin{eqnarray*}
   \MZ                    & = & 91.1875 ~\GeV/c^{2}, \\
   \MT                    & = & 175.0 ~\GeV/c^{2},   \\
   \MH                    & = & 150.0 ~\GeV/c^{2},   \\          
   \alpha_{s}(\MZ)        & = & 0.118,              \\
   \Delta\alpha^{(5)}_{had} & = & 0.02761,            \\
\end{eqnarray*}
where $\Delta\alpha^{(5)}_{had}$ is the contribution to the running of 
the electromagnetic coupling constant, $\alpha$, due to contributions from
hadronic loops containing 5 quark flavours.
To make the comparison it was necessary to match the signal definitions in the
data, simulation and semi-analytical computations. There were several choices
to be made: definition of $\rootsp$; subtraction or inclusion of QED 
corrections from the interference between Initial State Radiation (ISR) and 
Final Sate Radiation (FSR); 
inclusion of \ffbar\ events with additional radiated 
pairs of fermions; angular acceptance and $\rootsp$ acceptance. The signal 
definition adopted here was chosen to make analysis of the {\it{non-radiative}}
events as straightforward as possible:
\begin{itemize}
 \item for \eemm\ and \eett, $\rootsp$ is taken to be the invariant mass of 
       the fermion pair, \Mff. For \eeqq, $\rootsp$ is taken to be the 
invariant mass
       of the fermion pair with any FSR included - for $s$-channel processes
       this corresponds to computing the invariant mass, $Q$, of the virtual 
       propagator;

 \item QED corrections from the interference between ISR and FSR are included 
       for \eemm\ and \eett. These corrections are included in the simulated 
       events. For \eeqq\ these corrections are not included in the simulated 
       events, furthermore, the definition $\rootsp=Q$ is ambiguous in the 
       presence of interference between ISR and FSR. Therefore, for the \eeqq\
       channel, the signal is defined as having no ISR-FSR interference. 
       Corrections are applied to the data to remove the effects of the
       interference;

 \item events with additional radiated fermion pairs are subtracted as part
       of the 4-fermion background. The bulk of the pairs come from internal
       conversion of ISR photons into \ee\ pairs, which are lost in the
       beampipe, and which are topologically equivalent to 
       {\it{radiative return}} events. However, for the {\it{non-radiative}}
       samples this background is small compared to the signal;

 \item the total cross-sections and forward-backward asymmetries are quoted 
       in the full $4\pi$ acceptance for \eemm, \eett and \eeqq. 
       For \eemm\ and \eett\ this involves an extrapolation from the fiducial 
       volume of the detector using generated events. 
       For the electrons the measurements are made within the acceptance 
       $44\mydeg<\theta<136\mydeg$ - which corresponds to the acceptance of 
       the HPC. Measurements in the FEMC region, $12\mydeg<\theta<35\mydeg$ and  $145\mydeg<\theta<168\mydeg$, are not reported here, since they are subject to greater experimental uncertainties and the cross-sections are dominated by the $t$-channel process.
The differential cross-sections for \eemm\
       and \eett\ are quoted within the fiducial volume of the detector,
       with a cut at $|\cos\theta| = 0.97$ for \eemm\ and 0.96 for \tautau
       final states;
 
 \item for the {\it{inclusive}} samples the $\rootsp$ acceptances for \mumu, \tautau\ final states are  
       $\rootsp \gt 75~\GeV$ and $\rootsp \gt 0.10~\roots$ for \qqbar\ 
       final states. For the {\itshape{non-radiative}} samples 
       $\sqsps \gt 0.85$ for all 
       these processes.
       For \ee\ final states the cut on $\rootsp$ is replaced by a cut on the
       acollinearity, $\acol < 20\mydeg$; both $s$ and $t$ channel processes
       leading to \ee\ final states are considered as signal.
\end{itemize}

In the following sections the analyses of the different final states are
discussed and results of the measurements are presented. In all cases the 
latest theoretical predictions have been used, updating values given
in previous publications~\cite{ref:delphiff:130-172,ref:delphiff:183-189}.

%-.-.-.-.-.-.-.-.-.-.-.-.-.-.-.-.-.-.-.-.-.-.-.-.-.-.-.-.-.-.-.-.-.-.-.-.-
\subsubsection*{Experimental uncertainties}

Although a large number of possible biases and sources of uncertainty were
investigated for all final states, only those sources of bias and uncertainty
which lead to significant systematic errors for each particular analysis
are described in the sections below. Sources of bias and uncertainty which 
are negligible for a particular channel are not described.

For measurements of the differential cross-sections for the \eemm\ and \eett\ 
channels, some of the bins used for the analysis contain only a small number of
observed events. For these measurements statistical errors were computed both
from the square root of the number of observed events and of the
number of events expected from the Standard Model. The second error provides
a suitable weight which can be used to combine measurements from different
energies and can also be used when combining data from different LEP
experiments or fitting small deviations from the Standard Model to the 
measurements. In all other cases the statistical errors were computed solely
from the square root of the number of observed events.

%-.-.-.-.-.-.-.-.-.-.-.-.-.-.-.-.-.-.-.-.-.-.-.-.-.-.-.-.-.-.-.-.-.-.-.-.-
\subsubsection*{Analyses of the various final states}

%%--------------------------------------------------------------------
%%-- ANALYSIS OF ee FINAL STATES -------------------------------------
%%--------------------------------------------------------------------
\subsection{$\boldmath{\ee}$ final states}
\label{sec:ee}
%%%Insert Paolo's new text, with a few corrections 091203
%%%%%%%%%%%%%%%%%%%%%%%%%%%%%%%%%%%%%%%%%%%%%%%%%%%%%%%%%%%%%%%%%%%%%%

An analysis of \ee\ final states at \ee\ collision energies
of $189$~GeV and above is presented. This updates the analysis of data taken
at $189$~GeV \cite{ref:delphiff:183-189}.
Compared to the previous analysis, an explicit correction is applied for
charge misidentification in the measurement of the forward-backward 
asymmetry.
New  results are presented for differential cross-sections.
Data taken at collision energies below
$189$~GeV~\cite{ref:delphiff:130-172,ref:delphiff:183-189}, have not
been reanalysed.

%% -------------------------------------------------------------------
\subsubsection{Analysis}
\label{sec:ee:analysis}

%% -------------------------------------------------------------------
\subsubsection*{Run Selection}
Runs were selected requiring that the Vertex Detector (VD), the TPC and 
the electromagnetic calorimeters in the barrel of DELPHI (HPC) were
operative. For the 2000 data fraction with a TPC sector dead,
the requirement for the TPC was restricted to the remaining sectors of the 
detector.

%% -------------------------------------------------------------------
\subsubsection*{Event selection}

Events were selected with two almost-independent
sets of
experimental cuts, chosen in such a way as to minimize the correlations
between the two sets 
\cite{ref:delphils:91,*ref:delphils:95,ref:delphils:00}.
Only the barrel region of DELPHI was used for this analysis.
In each selection, both the electron and the positron
were required to be within the range $44^{\circ}<\theta<136^{\circ}$
and the acollinearity was required to be smaller than $20^\circ$.
A cut in polar angle at $90\pm2^\circ$ was applied to remove the region
with neither TPC nor HPC coverage.

The first  set of cuts  ({\bf{selection A}}) is based on requirements
of calorimetric-energy clusters  associated to hits in the Vertex Detector.
In particular events were selected if they had:

\begin{itemize}
\item at least two electromagnetic clusters in the HPC
      one with energy above $75 \%$ of the beam energy and another
      above $55 \%$ and with an acollinearity angle between the clusters 
      less than $20^{\circ}$;
\item at least two track segments in opposite hemispheres seen by the
VD and no more than four in total;
events with a 2-versus-2 topology were excluded;
%compatible within $\pm xx^\circ$
%with the electromagnetic clusters ;
\item no energy in the last two layers of the hadron calorimeter (HCAL) 
in correspondence with electromagnetic clusters detected at large distance
($> \pm 1 ^\circ$ in $\phi$) from the HPC sector boundaries.
\end{itemize}

The second selection ({\bf{selection B}}) is based on the charged-particle
momentum as measured by the tracking system.
In particular events were selected if they had:

\begin{itemize}
  \item at least 2 charged-particle tracks, of momentum greater than
$1.5~\GeV /c $
and distance of closest approach to the nominal vertex
position less than 5~cm, seen by
the DELPHI tracking system outside the  VD, with acollinearity less than
$20^{\circ}$ and no more than four tracks in total;
the 2-versus-2 track  topology was excluded;
\item the quadratic sum ($p_{rad}$) of the two momenta of the
highest-momentum charged particle in each hemisphere greater than
$0.99\sqrt{s}/2$;
\item no energy observed in the last three layers of HCAL associated to
the impact points of the two highest-momentum charged particles;
\item the OD hit pattern associated to the impact points
of the tracks compatible with the pattern of a particle showering in or
before the OD, or giving back-scattering from the HPC calorimeter;
\item no tracks identified as a muon.
\end{itemize}

%% -------------------------------------------------------------------
\subsubsection*{Estimation of the selection efficiency and background}

%%
%% NB. 
%%  - The trigger efficiency is taken to be (100+-0)%
%%  - The contamination from feed through is taken into account in the
%%    selection correlation correction, coming from MC.
%%

%%%%%%%%%%%%%%%%%%%%%%%%%%%%%%%%%%
\begin{figure}[tp]
\begin{center}
\begin{tabular}{cc}
\mbox{\epsfig{file=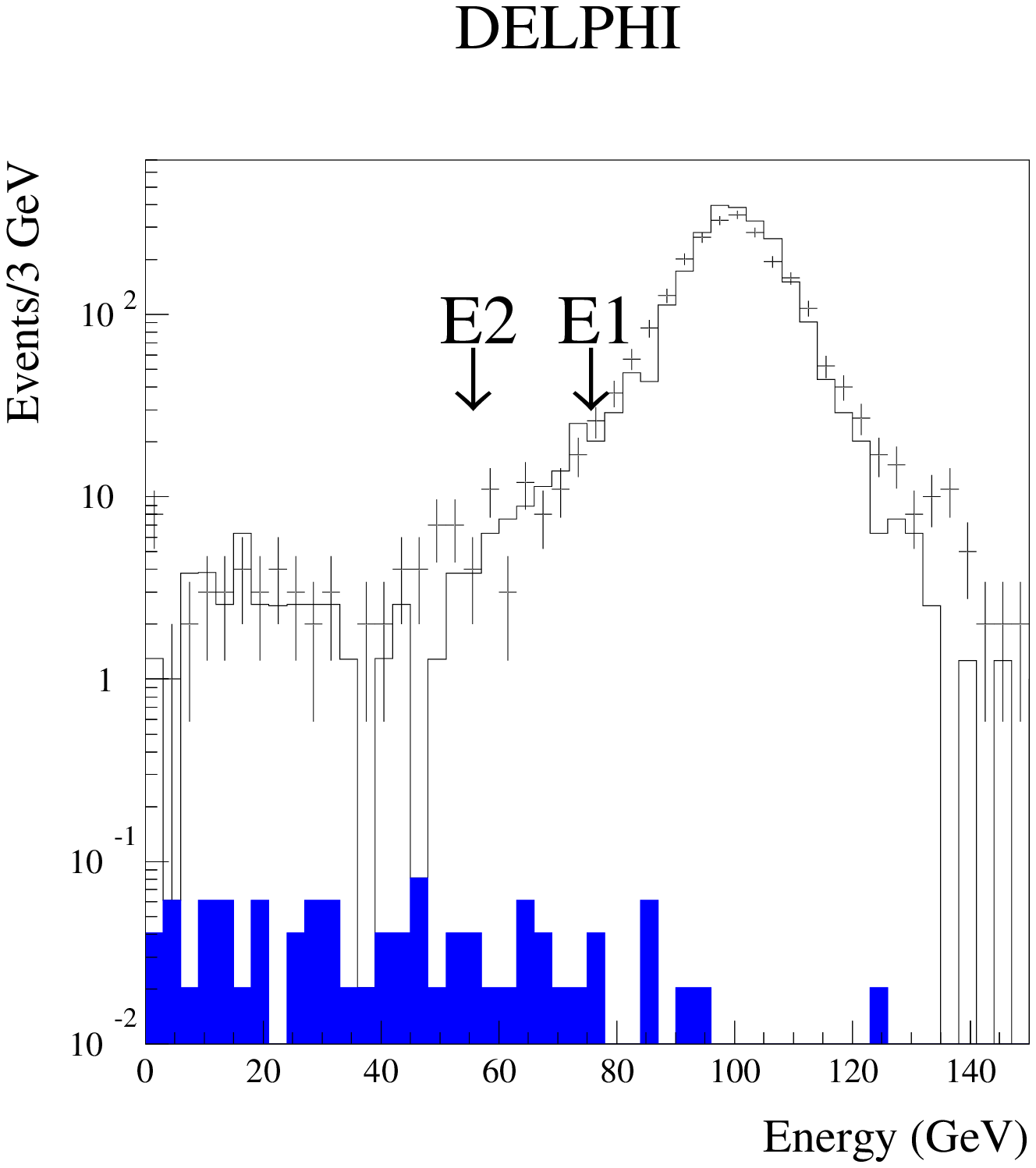,width=0.50\textwidth}} &
\mbox{\epsfig{file=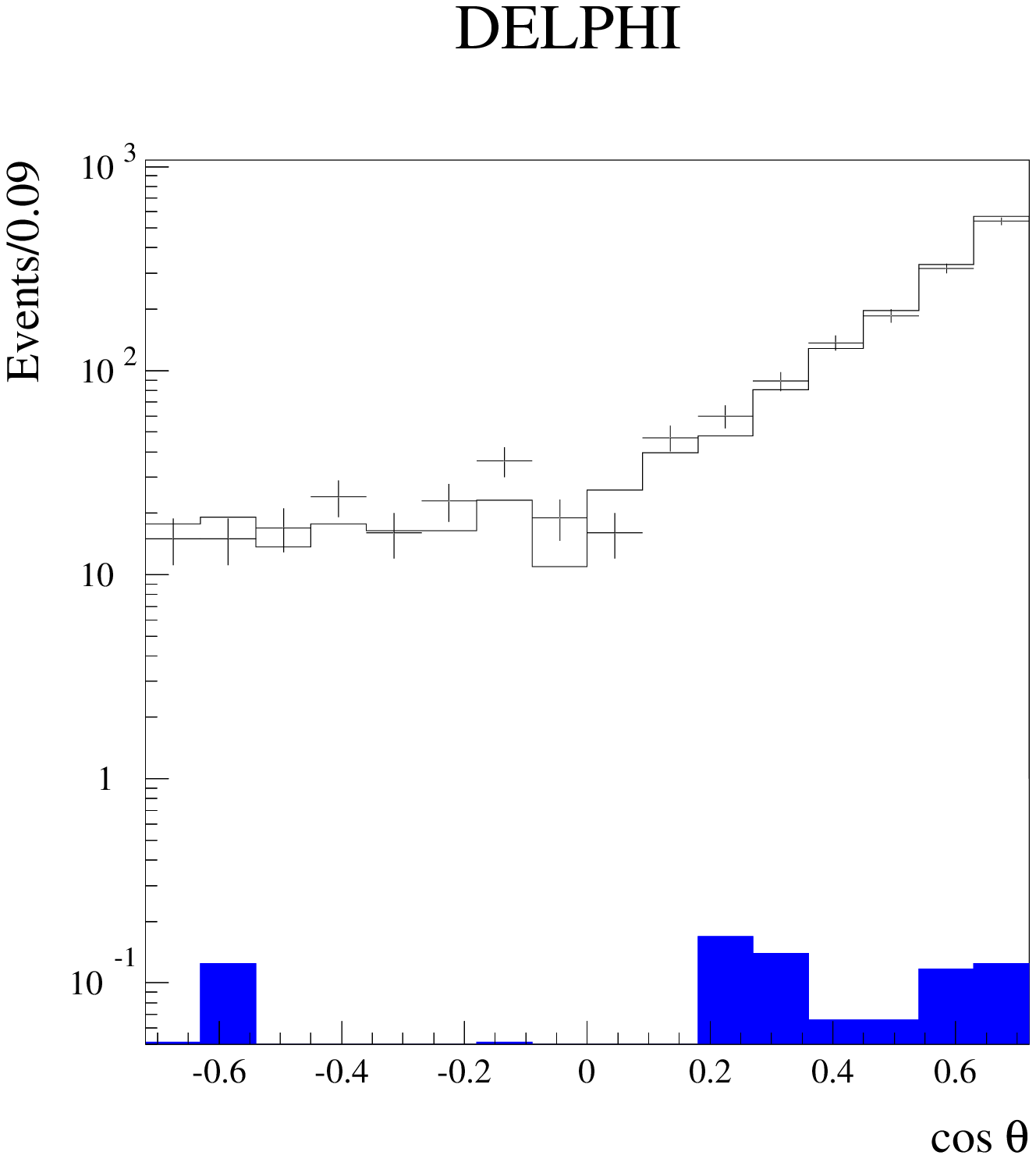,width=0.50\textwidth}}
\end{tabular}
\end{center}
\caption{\capsty{The energy of the two most energetic clusters (left) and
the uncorrected $\cos\theta$ distribution (right).
Both distributions correspond to data taken at $\sqrt{s}\sim$200 GeV and
the left plot does not include the energy cuts (indicated by the arrows) applied in
selection A. The histogram corresponds to the simulated
events and the dark area shows the $\tau$ background. }}
\label{fig:ee:ana}
\end{figure}
%%%%%%%%%%%%%%%%%%%%%%%%%%%%%%%%%%

Considering the selections A and B as independent, the efficiency of each
of them  could be easily computed by a comparison
of the number of events selected by each one separately or by both
simultaneously. This is much easier at LEP II energies, 
given the $t$-channel dominance, compared to the 
analysis
at the $\Zzero$ resonance because of the much smaller background 
( $\tau $ events  estimated from simulation) as shown in
Figure~\ref{fig:ee:ana}(left).

The efficiency of the two selections at the different energy
points is given in  Table~\ref{tab:ee:ana}.
In both selections the measured efficiencies did not include the loss
due to the exclusion of the polar angle region around $90^\circ$.
The total number of events selected by selection A at the different centre-of-mass
energies is shown in Table~\ref{tab:ee:ana}.

The simulated $\ee$ events were
used to estimate and remove the bias caused by
the correlation between the two selections due to the detector
structure or to the kinematics of the events. The bias on the
efficiency was found to be $(0.4 \pm 0.7)\%$.
However in the 2000 data new algorithms were used for track reconstruction
based on VD hits. An increased correlation between the two
selections was expected. In order to measure it, 1999 data reconstructed 
with
the new  and the old algorithm were compared. This gave a correction factor
of $1.016 \pm 0.003 $ to be applied to the cross-section measurements of the
2000 data.

\begin{table}[tp]
  {
%%\small
  \begin{center}
  \renewcommand{\arraystretch}{1.2}
  \begin{tabular}{|l|c|c|c|c|c|c|c|}
  \hline
  \multicolumn{8}{|c|}{\eeee} \\
  \hline
  \hline
                     & \multicolumn{7}{|c|}{Energy point (GeV)} \\
  \cline{2-8}
                     & 189    & 192    & 196    & 200    & 202    & 205    & 207    \\
  \hline
  \hline
  Energy (GeV)       & 188.63 & 192.17 & 196.10 & 200.12 & 202.07 & 204.88 & 206.59 \\
  Lumi ($\invpbarn$) & 155.11 &  25.12 &  76.16 &  83.07 & 40.05  &  88.55 & 128.39  \\
  No. Events (A)     &   3179 &    518 &   1568 &   1554 &   778  &  1500  &   2126 \\
%No. Events (B)      &    970 &   2710 &    448 &   1286 &   1290  &   631  
% &  ??  2878 ??    \\
  Efficiency (A) (\%)&   91.9 &   92.3 &   93.7 &   92.8 &  93.6 &   92.6 &   92.6 \\
  Efficiency (B) (\%)&   77.4 &   76.6 &   76.4 &   76.7 &  75.1 &   72.6 &   72.6 \\
  Background     (\%)&    0.2 &    0.2 &    0.2 &    0.2 &   0.2 &    0.2 &    0.2 \\
%  Stat. error ($\%$)&   1.72 &   4.31 &   2.50 &   2.48 &  3.60 &   2.53 &   2.10 \\
%  Syst. error ($\%$)&   0.85 &   0.85 &   0.85 &   0.85 &  0.85 &   0.92 &   0.92 \\
\hline
  \end{tabular}
  \end{center}
  }
  \caption{\capsty{Details of LEP II analysis for \eeee\ channel. The table 
  shows the actual centre-of-mass energy and luminosity analysed at each energy
  point, the number of events selected by selection A and the efficiencies of 
  selecting \eeee\ events with selections A and B and the background selected 
  with selection A.}} 
  \label{tab:ee:ana}
\end{table}
%%%%%%%%%%%%%%%%%%%%%%%%%%%%%%%%%%

%% -------------------------------------------------------------------
\subsubsection{Results}
\label{sec:ee:res}

%% -------------------------------------------------------------------
\subsubsection*{Cross-sections}
The cross-section was obtained as:
\begin{equation}
   \see = \frac{N_A}{{\cal{L}} \epsilon_A} \cdot c_f +\sigma^{90},
   \label{eq:csee}
\end{equation}
\noindent
where $N_A$ is the number of events selected by selection A and
$\epsilon_A$ is the efficiency of this selection, $\mathcal{L}$ is the
luminosity, $c_f$ is a correction factor including the bias in the
evaluation of $\epsilon_A$ due to the correlation between the two selection
methods and also the background subtraction, and $\sigma^{90}$ is the
correction for the central region $(90\pm2)^{\circ}$. The value of
$\sigma^{90}$ was computed by TOPAZ0~\cite{ref:th:topazzero} and
ALIBABA~\cite{ref:th:alibaba} and its value ranges from $0.43\pm 0.04$~pb 
to $0.36 \pm0.04$~pb.

The statistical error on the cross-section includes the statistical
uncertainty on the determination of the efficiency, taking into 
account the statistical uncertainties on correlated and uncorrelated
subsamples from selections A and B.

All the values of the measured cross-sections are given in
Table \ref{tab:ee:res} and shown in Figure~\ref{fig:ana:sig-cmp}.

%%%%%%%%%%%%%%%%%%%%%%%%%%%%%%%%%%
\begin{table}[tp]
 %%%%%%%%%%%%%%%% ee %%%%%%%%%%%%%%%%
 {\small
 \begin{center}
 \renewcommand{\arraystretch}{1.2}
 \begin{tabular}{|c|c|c||c|c|c|}
 \hline
 \multicolumn{6}{|c|}{\eeee} \\
 \hline
 \hline
 \multicolumn{1}{|c|}{$\roots$} &
 \multicolumn{2}{|c||}{$\acol < 20\mydeg$} &
 \multicolumn{1}{|c|}{$\roots$} &
 \multicolumn{2}{|c|}{$\acol < 20\mydeg$} \\
 \cline{2-3}
 \cline{5-6}
 \multicolumn{1}{|c|}{(GeV)} &
 $\see$ (pb) & $\Afbe$ &
 \multicolumn{1}{|c|}{(GeV)} &
 $\see$ (pb) & $\Afbe$ \\
 \hline
 \hline
 130 & 
 $\begin{array}{c}
 42.00 \pm  4.00 \pm  0.78 \\ (48.70)
 \end{array}$ &
 $\begin{array}{c}
 0.810 \pm 0.060 \pm 0.003 \\ (0.810)
 \end{array}$ &
 192 & 
 $\begin{array}{c}
 22.71 \pm  0.98 \pm  0.24 \\ (22.13)
 \end{array}$ &
 $\begin{array}{c}
 0.831 \pm 0.024 \pm 0.003 \\ (0.820)
 \end{array}$ \\
 \hline
 136 & 
 $\begin{array}{c}
 47.10 \pm  4.20 \pm  0.73 \\ (44.60)
 \end{array}$ &
 $\begin{array}{c}
 0.890 \pm 0.040 \pm 0.003 \\ (0.810)
 \end{array}$ &
 196 & 
 $\begin{array}{c}
 22.33 \pm  0.55 \pm  0.23 \\ (21.24)
 \end{array}$ &
 $\begin{array}{c}
 0.823 \pm 0.014 \pm 0.003 \\ (0.821)
 \end{array}$ \\
 \hline
 161 & 
 $\begin{array}{c}
 27.10 \pm  1.80 \pm  0.43 \\ (31.90)
 \end{array}$ &
 $\begin{array}{c}
 0.820 \pm 0.040 \pm 0.003 \\ (0.830)
 \end{array}$ &
 200 & 
 $\begin{array}{c}
 20.52 \pm  0.51 \pm  0.21 \\ (20.36)
 \end{array}$ &
 $\begin{array}{c}
 0.788 \pm 0.016 \pm 0.003 \\ (0.823)
 \end{array}$ \\
 \hline
 172 & 
 $\begin{array}{c}
 30.30 \pm  1.90 \pm  0.45 \\ (28.00)
 \end{array}$ &
 $\begin{array}{c}
 0.810 \pm 0.040 \pm 0.003 \\ (0.830)
 \end{array}$ &
 202 & 
 $\begin{array}{c}
 21.11 \pm  0.74 \pm  0.22 \\ (19.97)
 \end{array}$ &
 $\begin{array}{c}
 0.831 \pm 0.020 \pm 0.003 \\ (0.822)
 \end{array}$ \\
 \hline
 183 & 
 $\begin{array}{c}
 25.63 \pm  0.76 \pm  0.26 \\ (24.54)
 \end{array}$ &
 $\begin{array}{c}
 0.814 \pm 0.017 \pm 0.003 \\ (0.817)
 \end{array}$ &
 205 & 
 $\begin{array}{c}
 18.94 \pm  0.48 \pm  0.21 \\ (19.33)
 \end{array}$ &
 $\begin{array}{c}
 0.797 \pm 0.016 \pm 0.004 \\ (0.820)
 \end{array}$ \\
 \hline
 189 & 
 $\begin{array}{c}
 22.73 \pm  0.40 \pm  0.23 \\ (22.93)
 \end{array}$ &
 $\begin{array}{c}
 0.804 \pm 0.010 \pm 0.003 \\ (0.820)
 \end{array}$ &
 207 & 
 $\begin{array}{c}
 18.52 \pm  0.39 \pm  0.20 \\ (19.07)
 \end{array}$ &
 $\begin{array}{c}
 0.820 \pm 0.012 \pm 0.004 \\ (0.822)
 \end{array}$ \\
 \hline
 \end{tabular}
 \end{center}
 }
 %%%%%%%%%%%%%%%%%%%%%%%%%%%%%%%%%%%%
  \caption{\capsty{Measured cross-sections and forward-backward asymmetries
           for {\it{non-radiative}} \eeee\ events. The statistical error is followed by 
	     the total systematic error. 
           In parentheses the expected values as computed by BHWIDE,
           which have a precision of $\pm$2\% on \see\ and $\pm$0.02 on 
           \Afbe, are given. Results are quoted for an acceptance of 
           44$^\circ<\theta<$136$^\circ$.}}
  \label{tab:ee:res}
\end{table}
%%%%%%%%%%%%%%%%%%%%%%%%%%%%%%%%%%

%% -------------------------------------------------------------------
\subsubsection*{Forward-backward asymmetries}

In order to measure the forward-backward asymmetry the charge of the event
was defined as positive when the positron was in the forward hemisphere 
with
respect to the incoming positron direction, and negative in the opposite 
case.
In the $\eeee$ events, in addition to the canonical charge definition from
reconstructed tracks, it is possible to correlate the charges of an event by
looking at the effects of the bending due to the magnetic field on the 
impact position of HPC clusters, giving a high redundancy on the charge  
determination. The latter method was used to determine the charge of
an event in cases where the reconstructed charges of the tracks were
equal.

Given the  high expected asymmetry the measurements have to be corrected
for the residual wrong charge assignments. The asymmetry was corrected by a
factor $1+2\xi_{\pm}$ with $\xi_{\pm}=(7.2 \pm 1.4)\cdot 10^{-3}$ 
corresponding to
the charge misassignment which was determined from simulated events:
\begin{equation}
  A^0_{FB}=\frac{N_+ -N_-}{N_+ + N_-}\cdot(1+2 \xi_{\pm}),
\end{equation}
\noindent
where $N_+$ and $N_-$ are the number of events with positive and negative
charge, respectively.
In the previous publications this correction was not applied and an 
asymmetric
error was given.
The forward-backward asymmetry is corrected also for the missing central
region:
\begin{equation}
  A_{FB}=A^0_{FB}\cdot \frac{\see-\sigma^{90}}{\see}
+\frac{\sigma^{90}_+-\sigma^{90}_-}{\see},
\label{afbee}
\end{equation}
\noindent
where $\sigma^{90}_+$ and $\sigma^{90}_-$ are the computed cross-sections for
the regions $[88^{\circ}, 90^{\circ}]$ and $[90^{\circ},92^{\circ}]$ 
respectively.

All the values of the measured forward-backward asymmetries are given in
Table \ref{tab:ee:res} and shown in Figure~\ref{fig:ana:afb-cmp}.

%% -------------------------------------------------------------------
\subsubsection*{Differential cross-sections}

%. . . . . . . . . . . . . . . . . . . . . . . . . . . . . . . . . . . 
\begin{table}[p]
 %%%%%%%%%%%%%%%% ee %%%%%%%%%%%%%%%%
 %%%%%%%%%%%%%%%% ee %%%%%%%%%%%%%%%% This from John 28/1/04
 {\footnotesize
 \begin{center}
 \setlength{\tabcolsep}{1.0mm}
 \begin{tabular}{ccc}
 \multicolumn{3}{r}{\fbox{\hspace{6.38cm}
 $\eeee$
 \hspace{6.38cm}}} \\
 \\
 \renewcommand{\arraystretch}{1.2}
 \begin{tabular}
 {|@{[}r@{,}r@{]}|c|r@{$\pm$}c@{$\pm$}c|}
 \hline
\multicolumn{6}{|c|}{$\roots \sim 189$} \\
 \hline
 \hline
 \multicolumn{2}{|c|}{} &
 \multicolumn{4}{|c|}{$\dsdcth$ (pb)} \\
 \cline{3-6}
 \multicolumn{2}{|c|}{$\cos\theta$} &
 \multicolumn{1}{|c|}{SM} &
 \multicolumn{3}{|c|}{Measurement} \\
 \hline
 \hline
 -0.72 & -0.54 &  1.89 &  2.69 &  0.31 &  0.09 \\
 -0.54 & -0.36 &  2.36 &  2.31 &  0.32 &  0.04 \\
 -0.36 & -0.18 &  2.98 &  2.53 &  0.34 &  0.03 \\
 -0.18 &  0.00 &  4.23 &  4.73 &  0.36 &  0.23 \\
  0.00 &  0.09 &  5.86 &  5.27 &  0.49 &  0.34 \\
  0.09 &  0.18 &  7.55 &  8.18 &  0.77 &  0.07 \\
  0.18 &  0.27 & 10.16 &  9.79 &  0.88 &  0.10 \\
  0.27 &  0.36 & 14.44 & 12.58 &  1.06 &  0.14 \\
  0.36 &  0.45 & 21.15 & 21.48 &  1.31 &  0.21 \\
  0.45 &  0.54 & 32.75 & 31.21 &  1.60 &  0.32 \\
  0.54 &  0.63 & 52.28 & 52.76 &  2.04 &  0.51 \\
  0.63 &  0.72 & 87.65 & 87.85 &  2.59 &  0.86 \\
 \hline
 \end{tabular}
 &
 \renewcommand{\arraystretch}{1.2}
 \begin{tabular}
 {|@{[}r@{,}r@{]}|c|r@{$\pm$}c@{$\pm$}c|}
 \hline
\multicolumn{6}{|c|}{$\roots \sim 192$} \\
 \hline
 \hline
 \multicolumn{2}{|c|}{} &
 \multicolumn{4}{|c|}{$\dsdcth$ (pb)} \\
 \cline{3-6}
 \multicolumn{2}{|c|}{$\cos\theta$} &
 \multicolumn{1}{|c|}{SM} &
 \multicolumn{3}{|c|}{Measurement} \\
 \hline
 \hline
 -0.72 & -0.54 &  1.81 &  2.03 &  0.73 &  0.09 \\
 -0.54 & -0.36 &  2.29 &  1.87 &  0.80 &  0.04 \\
 -0.36 & -0.18 &  2.93 &  2.40 &  0.87 &  0.03 \\
 -0.18 &  0.00 &  4.05 &  5.36 &  1.00 &  0.23 \\
  0.00 &  0.09 &  5.64 &  4.05 &  1.40 &  0.34 \\
  0.09 &  0.18 &  7.44 & 11.39 &  2.18 &  0.08 \\
  0.18 &  0.27 &  9.83 &  9.74 &  2.20 &  0.10 \\
  0.27 &  0.36 & 13.80 & 12.03 &  2.64 &  0.14 \\
  0.36 &  0.45 & 20.35 & 20.72 &  3.34 &  0.21 \\
  0.45 &  0.54 & 31.32 & 29.61 &  3.88 &  0.32 \\
  0.54 &  0.63 & 50.21 & 54.58 &  4.79 &  0.51 \\
  0.63 &  0.72 & 85.17 & 87.93 &  6.28 &  0.87 \\
 \hline
 \end{tabular}
 &
 \renewcommand{\arraystretch}{1.2}
 \begin{tabular}
 {|@{[}r@{,}r@{]}|c|r@{$\pm$}c@{$\pm$}c|}
 \hline
\multicolumn{6}{|c|}{$\roots \sim 196$} \\
 \hline
 \hline
 \multicolumn{2}{|c|}{} &
 \multicolumn{4}{|c|}{$\dsdcth$ (pb)} \\
 \cline{3-6}
 \multicolumn{2}{|c|}{$\cos\theta$} &
 \multicolumn{1}{|c|}{SM} &
 \multicolumn{3}{|c|}{Measurement} \\
 \hline
 \hline
 -0.72 & -0.54 &  1.68 &  1.73 &  0.41 &  0.09 \\
 -0.54 & -0.36 &  2.21 &  2.37 &  0.43 &  0.04 \\
 -0.36 & -0.18 &  2.77 &  3.03 &  0.48 &  0.03 \\
 -0.18 &  0.00 &  3.89 &  4.13 &  0.50 &  0.23 \\
  0.00 &  0.09 &  5.52 &  7.23 &  0.75 &  0.34 \\
  0.09 &  0.18 &  6.97 &  7.01 &  1.04 &  0.07 \\
  0.18 &  0.27 &  9.53 & 11.27 &  1.23 &  0.09 \\
  0.27 &  0.36 & 13.38 & 14.29 &  1.47 &  0.13 \\
  0.36 &  0.45 & 19.39 & 23.39 &  1.72 &  0.19 \\
  0.45 &  0.54 & 29.77 & 30.14 &  2.14 &  0.29 \\
  0.54 &  0.63 & 48.46 & 51.75 &  2.78 &  0.48 \\
  0.63 &  0.72 & 81.88 & 81.55 &  3.51 &  0.81 \\
 \hline
 \end{tabular}
 \\ \\
 \renewcommand{\arraystretch}{1.2}
 \begin{tabular}
 {|@{[}r@{,}r@{]}|c|r@{$\pm$}c@{$\pm$}c|}
 \hline
\multicolumn{6}{|c|}{$\roots \sim 200$} \\
 \hline
 \hline
 \multicolumn{2}{|c|}{} &
 \multicolumn{4}{|c|}{$\dsdcth$ (pb)} \\
 \cline{3-6}
 \multicolumn{2}{|c|}{$\cos\theta$} &
 \multicolumn{1}{|c|}{SM} &
 \multicolumn{3}{|c|}{Measurement} \\
 \hline
 \hline
 -0.72 & -0.54 &  1.57 &  1.73 &  0.38 &  0.08 \\
 -0.54 & -0.36 &  2.07 &  2.81 &  0.40 &  0.03 \\
 -0.36 & -0.18 &  2.66 &  2.86 &  0.45 &  0.03 \\
 -0.18 &  0.00 &  3.73 &  4.95 &  0.48 &  0.23 \\
  0.00 &  0.09 &  5.23 &  5.23 &  0.73 &  0.34 \\
  0.09 &  0.18 &  6.72 &  6.26 &  0.95 &  0.07 \\
  0.18 &  0.27 &  9.02 &  9.14 &  1.17 &  0.09 \\
  0.27 &  0.36 & 12.85 & 13.16 &  1.37 &  0.13 \\
  0.36 &  0.45 & 18.57 & 20.06 &  1.64 &  0.18 \\
  0.45 &  0.54 & 29.03 & 26.43 &  2.03 &  0.29 \\
  0.54 &  0.63 & 46.21 & 46.00 &  2.58 &  0.46 \\
  0.63 &  0.72 & 78.54 & 77.96 &  3.34 &  0.78 \\
 \hline
 \end{tabular}
 &
 \renewcommand{\arraystretch}{1.2}
 \begin{tabular}
 {|@{[}r@{,}r@{]}|c|r@{$\pm$}c@{$\pm$}c|}
 \hline
\multicolumn{6}{|c|}{$\roots \sim 202$} \\
 \hline
 \hline
 \multicolumn{2}{|c|}{} &
 \multicolumn{4}{|c|}{$\dsdcth$ (pb)} \\
 \cline{3-6}
 \multicolumn{2}{|c|}{$\cos\theta$} &
 \multicolumn{1}{|c|}{SM} &
 \multicolumn{3}{|c|}{Measurement} \\
 \hline
 \hline
 -0.72 & -0.54 &  1.61 &  1.50 &  0.55 &  0.09 \\
 -0.54 & -0.36 &  2.03 &  1.96 &  0.57 &  0.04 \\
 -0.36 & -0.18 &  2.58 &  2.98 &  0.62 &  0.03 \\
 -0.18 &  0.00 &  3.68 &  3.40 &  0.64 &  0.23 \\
  0.00 &  0.09 &  4.99 &  4.55 &  0.86 &  0.34 \\
  0.09 &  0.18 &  6.63 &  6.16 &  1.39 &  0.07 \\
  0.18 &  0.27 &  8.90 &  5.82 &  1.57 &  0.09 \\
  0.27 &  0.36 & 12.44 & 13.02 &  1.93 &  0.13 \\
  0.36 &  0.45 & 18.44 & 18.85 &  2.40 &  0.19 \\
  0.45 &  0.54 & 28.11 & 33.61 &  2.85 &  0.28 \\
  0.54 &  0.63 & 45.70 & 48.97 &  3.74 &  0.46 \\
  0.63 &  0.72 & 76.92 & 84.69 &  4.74 &  0.78 \\
 \hline
 \end{tabular}
 &
 \renewcommand{\arraystretch}{1.2}
 \begin{tabular}
 {|@{[}r@{,}r@{]}|c|r@{$\pm$}c@{$\pm$}c|}
 \hline
\multicolumn{6}{|c|}{$\roots \sim 205$} \\
 \hline
 \hline
 \multicolumn{2}{|c|}{} &
 \multicolumn{4}{|c|}{$\dsdcth$ (pb)} \\
 \cline{3-6}
 \multicolumn{2}{|c|}{$\cos\theta$} &
 \multicolumn{1}{|c|}{SM} &
 \multicolumn{3}{|c|}{Measurement} \\
 \hline
 \hline
 -0.72 & -0.54 &  1.58 &  1.94 &  0.37 &  0.07 \\
 -0.54 & -0.36 &  1.96 &  2.21 &  0.38 &  0.03 \\
 -0.36 & -0.18 &  2.56 &  3.01 &  0.43 &  0.03 \\
 -0.18 &  0.00 &  3.57 &  3.73 &  0.44 &  0.23 \\
  0.00 &  0.09 &  4.88 &  5.21 &  0.59 &  0.34 \\
  0.09 &  0.18 &  6.33 &  5.48 &  0.98 &  0.07 \\
  0.18 &  0.27 &  8.49 &  8.49 &  1.08 &  0.09 \\
  0.27 &  0.36 & 12.27 & 11.53 &  1.30 &  0.13 \\
  0.36 &  0.45 & 17.80 & 17.62 &  1.56 &  0.19 \\
  0.45 &  0.54 & 27.45 & 28.17 &  1.94 &  0.29 \\
  0.54 &  0.63 & 44.01 & 42.14 &  2.45 &  0.46 \\
  0.63 &  0.72 & 74.19 & 70.86 &  3.14 &  0.77 \\
 \hline
 \end{tabular}
 \\ \\
 &
 \renewcommand{\arraystretch}{1.2}
 \begin{tabular}
 {|@{[}r@{,}r@{]}|c|r@{$\pm$}c@{$\pm$}c|}
 \hline
\multicolumn{6}{|c|}{$\roots \sim 207$} \\
 \hline
 \hline
 \multicolumn{2}{|c|}{} &
 \multicolumn{4}{|c|}{$\dsdcth$ (pb)} \\
 \cline{3-6}
 \multicolumn{2}{|c|}{$\cos\theta$} &
 \multicolumn{1}{|c|}{SM} &
 \multicolumn{3}{|c|}{Measurement} \\
 \hline
 \hline
 -0.72 & -0.54 &  1.53 &  1.25 &  0.30 &  0.07 \\
 -0.54 & -0.36 &  1.95 &  2.25 &  0.32 &  0.03 \\
 -0.36 & -0.18 &  2.55 &  2.11 &  0.35 &  0.03 \\
 -0.18 &  0.00 &  3.41 &  3.85 &  0.36 &  0.23 \\
  0.00 &  0.09 &  4.84 &  5.84 &  0.49 &  0.34 \\
  0.09 &  0.18 &  6.34 &  6.30 &  0.81 &  0.07 \\
  0.18 &  0.27 &  8.30 &  7.59 &  0.89 &  0.09 \\
  0.27 &  0.36 & 11.82 & 10.86 &  1.06 &  0.12 \\
  0.36 &  0.45 & 17.70 & 16.84 &  1.29 &  0.19 \\
  0.45 &  0.54 & 26.95 & 25.22 &  1.60 &  0.28 \\
  0.54 &  0.63 & 43.65 & 42.40 &  2.03 &  0.46 \\
  0.63 &  0.72 & 73.37 & 72.69 &  2.60 &  0.77 \\
 \hline
 \end{tabular}
 &
 \end{tabular}
 \end{center}
 }
 %%%%%%%%%%%%%%%%%%%%%%%%%%%%%%%%%%%%
 %%%%%%%%%%%%%%%%%%%%%%%%%%%%%%%%%%%%
 \caption{\capsty{Differential cross-sections for non-radiative
          \eeee\ events at centre-of-mass energies from $\sim$189 to 207 GeV.
          The tables show the bins and the predicted Standard Model
          differential cross-sections and the measurements with 
          statistical and experimental systematic errors. }}
 \label{tab:ee:diff}
\end{table}
%. . . . . . . . . . . . . . . . . . . . . . . . . . . . . . . . . . .

As with the measurement of the forward-backward asymmetry, measurements
of the differential cross-sections require the determination of the
charge of the electron and positron, and a correction has to be applied
for mismeasurements of the charge. A bin by bin correction was applied to
%the differential cross-section -
the events detected in each polar angle bin $N_{\theta_i}$
which was largest in the very forward
and backward bins due to the large asymmetry in the cross-sections:
\begin{equation}
N^c_{\theta_i}=N_{\theta_i}\cdot(1+\xi_{\pm}) -
N_{\bar{\theta_i}}\cdot(\xi_{\pm}),
\end{equation}
\noindent
where $N_{\bar{\theta_i}}$ is the number of events detected in the opposite
hemisphere. The bin  $-0.72 < \cos\theta<-0.54 $ has the largest 
correction which  is about 25 to $30\%$. 
The differential cross-section is obtained as
\begin{equation}
\dsdcth_{i}=
    \frac{N^c_{\theta_i}}
   {{\cal{L}} \epsilon_{\theta_i} \cdot \Delta \cos\theta},
   \label{dsigmaee}
\end{equation}
\noindent
where $\epsilon_{\theta_i}$ is a bin correction factor dominated by the 
efficiency
evaluation and $\Delta \cos\theta$ is the bin width.
The $90^\circ$ region correction is taken into account by
adding $\sigma^{90}_+/\Delta \cos\theta$ and $\sigma^{90}_-/\Delta 
\cos\theta$  to the
corresponding bins.
The uncorrected $\cos{\theta}$ distribution is shown in 
Figure~\ref{fig:ee:ana}(right). The differential cross-sections for the 1998, 1999 and 2000 data are given
in Table \ref{tab:ee:diff}.

%% -------------------------------------------------------------------
\subsubsection*{Systematic errors}

\begin{table}[tp]
\begin{center}
 \begin{tabular}{|l|c||c|}
 \hline
 \multicolumn{3}{|c|}{\eeee} \\
 \hline
 \hline
                           &$\Delta\sigma/\sigma$ & $\Delta A_{FB}$ \\
 Source                    & (non-rad.)           & (non-rad.) \\
 \hline\hline
  Efficiency               &  76                  &   - \\
  Acceptance               &  48                  &  20 \\
  Charge misidentification &   -                  &  20 \\
  Backgrounds              &  20                  &  20 \\
  Luminosity               &  58                  &   - \\
  TPC sector instability   &  -                   &  20 \\
 \hline\hline
  Total uncertainty        & 109                  &  35 \\
 \hline
  Correlated               & 102                  &  35 \\
  Uncorrelated             &  38                  &  - \\
  \hline
 \end{tabular}
\caption{\capsty{Systematic uncertainties on the measured cross-sections 
                and forward-backward asymmetries for data taken
                at $\roots \sim$207 GeV. All numbers in units of 
                10$^{-4}$. The total uncertainty does not include
                the error due 
                to TPC sector instability which applies only to a part of the
                2000 data. The correlated error component includes errors 
                correlated between energies and channels and those correlated 
                with other LEP experiments.}}
\label{tab:ee:syst}
\end{center}
\end{table}
%%%%%%%%%%%%%%%%%%%%%%%%%%%%%%%%%%%%%%%%%%%%%%%%%%%%%%%

Apart from the luminosity, systematic uncertainties arise from the event
selection correlation, acceptance definition and
from the background subtraction.
The largest uncertainty comes from the
selection correlation ($\pm 0.7 \%$) due to the statistics of simulated events.
The simulated data for 2000 were produced exclusively with a
new tracking algorithm giving large correlations (which were measured
directly from the data)
and therefore the statistical error cannot be reduced.
%
%acceptance definition, through the determination of the efficiency correction.
The error on the cross-section due to
the acceptance definition
arises from the uncertainty on the absolute polar angle
determination. For events having fully reconstructed charged particle tracks this
uncertainty is
very small ($< 0.02^\circ$), while for those having the acceptance defined
by HPC clusters it is larger (about $ 0.04^\circ$). The latter sample has
an additional contribution due to the poorer precision ($\pm 0.2^\circ $)
on the cluster polar angle determination. The
total contribution to the acceptance definition depends on the fraction
of events with HPC-based acceptance and it ranges from $\pm 0.24 \%$ to
$\pm 0.30 \%$.
Beam energy spread effects on the acceptance were investigated and they were
found to be $\pm 0.28 \%$.

The corrections applied for the $\pm2^\circ$ polar angle fiducial cut
around $90^\circ$ in the analysis of the total and differential cross-sections
were computed at the different energies by using the program TOPAZ0 
and checked with  ALIBABA. 
No significant differences were found between the two generators. An error
of $\pm 0.25\%$ was assigned to the total cross-section correlated between all
energies.
The rate of events with two same-charged particle tracks was compatible between
data and simulation and consequently
the charge misassignment was determined from simulated events.
The charge misassignment error was determined by the statistics of the
simulated events and
%The charge misassignment error
it amounts to $\pm 0.002$ on the forward-backward
asymmetry, correlated among all energies.

The TPC failure in the last part of the 2000 run had a clear influence on the
Selection B efficiency: $63.0 \%$ instead of $75.5 \%$ obtained during the
first part of the 2000 run. However the overall effect on the 
cross-sections
and asymmetries was not very significant: the statistical error increase was
of the order of a few percent and only an additional systematic error of
$\pm 0.3 \%$ on the asymmetry was estimated because of the charge determination.

A breakdown of the systematic errors for data taken at $\roots \sim 207$ GeV
is given in Table~\ref{tab:ee:syst}.

%%%%%%%%%%%%%%%%%%%%%%%%%%%%%%%%%%%%%%%%%%%%%%%%%%%%%%%%%5

%%--------------------------------------------------------------------
%%-- ANALYSIS OF mm FINAL STATES -------------------------------------
%%--------------------------------------------------------------------
\subsection{$\boldmath{\mumu}$ final states}
\label{sec:mm}

An analysis of \mumu\ final states at \ee\ collision energies of 
$183$~GeV and above is presented. Data taken at energies of 
$183$~GeV and $189$~GeV \cite{ref:delphiff:183-189} have been
reanalysed to be consistent with data taken at higher energies.
Compared to the previous analysis, the new analysis benefited from an
increase in Monte Carlo statistics available in the whole LEP II data 
set. This made it possible to perform detailed comparisons of data and 
simulated events, which ultimately led to improvements in the estimation
of the systematic errors on the measurements.
For data taken at collision energies below 
$183$~GeV~\cite{ref:delphiff:130-172}, the statistical 
errors are so large that improvements in the systematic errors would be 
negligible. These data have not been reanalysed.

%% ------------------------------------------------------------------- 
\subsubsection{Analysis}
\label{sec:mm:analysis}

%%....................................................................
%%.. Summary .........................................................
%%....................................................................
%\fbox{
%\begin{minipage}{0.87\textwidth}
%\begin{itemize}
% \item Measurements $\sigma$, \Afb\ and \dsdcth
%
% \item Event selection
%
% \item \rootsp\ algorithm
%
% \item Efficiency
%
% \item Backgrounds
%
% \item Systematic errors - correlations
%
% \item 1 or 2 plots showing the quality of the data MC agreement
%       before the final selection. {\it{i.e.}} not just a final \rootsp\ plot
%
%
% \item Impact of failure of TPC S6
%
%
% \item Table giving
%  \begin{itemize}
%   \item Actual Energy
%
%   \item Luminosity analysed
%
%   \item Number of selected events
%
%   \item Efficiency
%
%   \item Background
%
%   \item Statistical relative error
%
%   \item Systematic relative error
%
%  \end{itemize}
%
% \item Need to cover different energies/years of data
%\end{itemize}
%\end{minipage}
%}
%%....................................................................
%%.. Contents ........................................................
%%....................................................................

%% ------------------------------------------------------------------- 
\subsubsection*{Run selection}

Runs in which important components of the detector, such as the muon chambers,
were not functioning satisfactorily were removed from the analysis.
%For the measurement of cross-sections and differential cross-sections, 
%runs in which there was no measurement of the luminosity were excluded.

%% ------------------------------------------------------------------- 
\subsubsection*{Event selection}

Events in the channel \eemm\ were selected by first imposing the following
kinematic requirements:
\begin{itemize}
 \item at least two reconstructed tracks from charged particles; 
 \item to allow for photon conversions or splitting of 
       tracks, events with up to 7 reconstructed tracks were accepted;
 \item the measured momenta of the two highest-momentum reconstructed 
       charged particles had to be greater than 15 GeV/$c$;
 \item the polar angles of the two highest-momentum reconstructed charged 
       particles had to lie between $14\mydeg$ and $166\mydeg$;
 \item to suppress backgrounds from \eett\ events, the momentum of the third
       highest-momentum charged particle had to be less than 5 GeV/$c$ in all 
       cases where the energy of the leading charged particle was less than 
       90\% of the beam energy.
\end{itemize}
Individual charged particles were identified as muons if:
\begin{itemize}
 \item there was at least one hit in the muon chambers associated to the 
       reconstructed track; or
 \item the associated energy deposits in the HCAL were consistent with 
       a minimum ionising particle; or 
 \item the associated energy in the electromagnetic calorimeters 
       was less than 1.5 \GeV.
\end{itemize}
The two highest-momentum charged particles had to be identified as muons. 
Events were rejected if:
\begin{itemize}
 \item either of the two highest-momentum charged particles had no associated 
       muon chambers hits and the energy per layer of the HCAL exceeded 
       5 GeV; or 
 \item the electromagnetic energy associated to either of the two 
       highest-momentum particles exceeded 10 GeV 
       and neither track had associated muon chamber hits - this removed 
       \eeee\ events.
\end{itemize}
To reduce cosmic ray backgrounds, the reconstructed tracks of the two
highest-momentum charged particles had to appear to come from the 
interaction region:
\begin{itemize}
 \item for tracks which had associated VD hits, the modulus of 
       the impact parameter relative to the beamspot in the plane transverse 
       to the beam axis had to be less than 0.1 cm, for at least one of the 
       tracks; 
 \item for tracks without hits in the VD, the modulus of the 
       impact parameter had to be less than 1.0 cm, for both tracks, 
       reflecting the worse resolution for these tracks;
 \item in all cases, the impact parameter with respect to the beamspot in the 
       direction parallel to the beam axis had to be less than 2.0 cm, 
       this cut being set by the length of the electron and positron bunches.
\end{itemize}

%% ------------------------------------------------------------------- 
\subsubsection*{$\boldmath{\rootsp}$ reconstruction}

Having selected suitable \eemm\ events, kinematic fits were used to determine
the invariant mass of the \mumu\ pair, $\rootsp$. The kinematic fits attempted 
to match the measured event to several possible event topologies, using energy
and momentum constraints to improve the determination of the 3-momenta of the
reconstructed muons.
Use of the kinematic fits improved the resolution of the measured
invariant masses of the \mumu\ pairs, compared to using the directly
measured momenta of all muons, or using the measured angles of the
muons, assuming a single photon emitted down the beampipe.
As well as using the tracks selected as muons, up to one reconstructed 
photon per event was included in the fit. Electromagnetic energy clusters,
not associated to any reconstructed track, were considered as photons if
they had an energy exceeding 5 GeV.

The following kinematic fits were tried in the order given: 
\begin{itemize}
 \item if a photon was found, the event was fitted assuming a seen photon
       and a possible additional unseen photon; 
 \item a fit in which there was assumed to be no photons radiated in the 
       event;
 \item a fit assuming a single photon emitted along the beampipe;
 \item a fit assuming a single photon emitted in any direction but unseen in
       the detector. 
\end{itemize}
In each case, if the fit gave a satisfactory $\chi^{2}$ the invariant mass 
of the \mumu\ pair was calculated from the fitted momenta of the particles.
Otherwise, the next fit was attempted. If none of the fits gave an acceptable 
$\chi^{2}$ the measured momenta of the muons were used to determine the 
invariant mass of the \mumu\ pair.
Events were divided into {\emph{inclusive}} and {\emph{non-radiative}}
classes according to the definitions given in the introduction to 
Section~\ref{sec:analysis}.

In the {\emph{non-radiative}} class, migrations into and out of the sample
were determined from the simulated events.
%% from the ratio of the 
%%purity\footnote{The purity is the ratio of the number of \mumu\ pairs in the 
%%sample which had generated values of $\rootsp$ inside the cuts defining the 
%%sample, compared to the total number of events reconstructed in the sample.} 
%%compared to the efficiency for reconstructing selected {\emph{non-radiative}}
%%events within the {\emph{non-radiative}} sample. 
The migrations amount to
a correction of between $1.6\%$ and $3.6\%$ on the measured cross-sections.

%% ------------------------------------------------------------------- 
\subsubsection*{Estimation of the selection efficiency and background}

%%%%%%%%%%%%%%%%%%%%%%%%%%%%%%%%%%%%%%%
\begin{figure}[tp]
\begin{center}
 \mbox{\epsfig{file=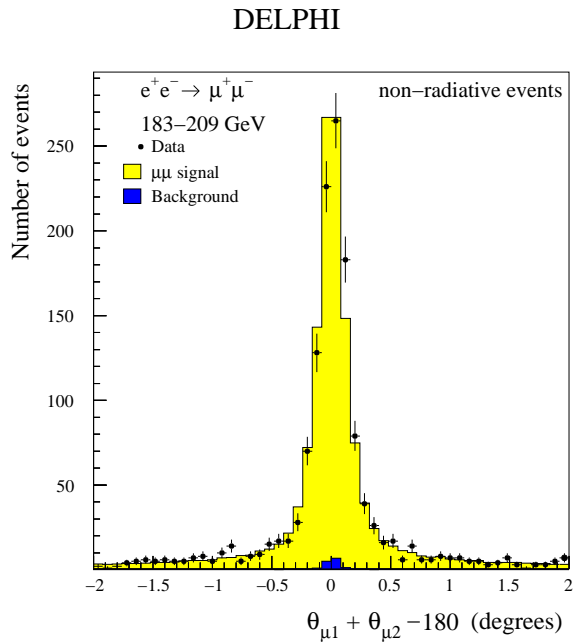,width=0.49\textwidth}}
\end{center}
\caption{\capsty{The distribution of $\theta_{\mu1}+\theta_{\mu2}-$180$\mydeg$
for the two highest momentum muons in data and simulation. 
For back-to-back events the sum should be  $$0$\mydeg$. 
This is broadened by radiation, resolution and beam
energy spread. Data from 1997-2000 are shown.}}
\label{fig:mm:qual}
\end{figure}
%%%%%%%%%%%%%%%%%%%%%%%%%%%%%%%%%%%%%%%

Selection efficiencies and backgrounds were determined from 
simulated events,
but critical components of the analysis, such as the efficiency for
reconstructing and identifying muons, were determined from the data.
To obtain good agreement between the data and the predictions of simulations
it was necessary to apply a number of corrections. 

Using \eemm\ events, collected at $\roots \sim \MZ$ at various times during 
1997 through to 2000, smearings and shifts in the mean values
for distributions of $1/p$ were computed for different charges of particle,
different ranges of $\cos\theta$, and different combinations of tracking 
detectors involved in the track reconstruction.
These corrections were applied to data taken at LEP II energies, 
improving the agreement between data and simulations in the distribution of 
the momenta of particles at these higher centre-of-mass energies.
Application of these corrections led to only small changes in the expected 
numbers of events selected in the {\emph{inclusive}} and 
{\emph{non-radiative}} classes. This reflects the fact that the 
relative weights given to measurements of angles in the kinematic fits 
were larger than the weights given to
the less precisely measured momenta of particles.

In the simulations of \eemm\ events, the momenta of the incoming electrons
and positrons were taken to be equal and opposite. However,
in the data the particles in the incoming bunches of electrons and positrons
had a momentum spread of $\sim 0.2$ GeV/$c$,
such that the energies of the electron and positron before collision were 
not necessarily equal.
%The predominant effect was that the centre-of-mass of the \mumu\ 
%system was not at rest in the detectors coordinate frame, even in the absence 
%of significant photon radiation.
This led to an observable broadening of the
acollinearity distribution at low acollinearities, in the data.
By applying boosts to the simulated \mumu\ events, after the full detector 
simulation, reflecting the momentum spread of the real beams, this effect 
was taken into account. This led to improved agreement between 
data and simulation in distributions which were used to study the 
performance of the kinematic fits. The agreement between data and simulation
after inclusion of the appropriate boosts is shown in Figure~\ref{fig:mm:qual}.
%In the simple case of a single photon emitted along the beampipe the 
%acollinearity of muons in \mumu\ events can be simply related to the 
%energy of the photon, which in turn can be related to the reconstructed
%$\sqsps$.
%Without modifying the kinematic fits the precision of the measurements of 
%the polar angles of muons was such that a large fraction
%of events without significant photon radiation would have failed the 4C 
%kinematic fit which assumed no photon radiation, instead these events would
%have been fitted under the assumption of single, low energy, photon emitted 
%along the beampipe. Modifying the 4C kinematic fit to take 
%into account the boost as an extra parameter, constrained by the expected
%dispersion of the boost, brought good agreement between the number of 
%events fitted in data and  simulation by this fit. The modified fitting 
%procedure was used in the measurements in the \eemm\ channel. 
Due to the relatively loose cuts in $\sqsps$, the application of the 
boosts in the simulation made a relatively small change to the expected 
number of events selected in the {\emph{non-radiative}} sample.

Corrections were also applied to bring agreement between the data and 
simulation for the numbers of muon chamber hits associated to reconstructed
tracks. These were determined from the high energy data themselves. Part of 
the corrections were to take into account unsimulated high voltage trips of 
the chambers which actually happened during the collection of the LEP II data 
set. Because the muon identification criteria select events from a union of
3 highly overlapping samples,
application of the correction had only a limited impact on the expected 
numbers of particles identified as muons.

After applying the corrections above, differences between data and
simulation were still found in the efficiency to reconstruct and
identify single muons. High purity samples of single muons with high
momentum were selected, and the efficiency to reconstruct and identify
a single muon was determined from the number of times a second muon
was found accompanying the first muon. The backgrounds from events
other than \eemm\ events, such as \eett\ and \eeWW\ which could lead
to only one muon and one other track in the detector, were negligible.
Averaging over all data from $\roots \sim 183$~GeV and above, it was
found that the efficiency for selecting \eemm\ events determined from
the simulation had to be reduced by $(1.08 \pm 0.49)\%$. The error on
this value, which is dominated by statistics of events selected in
data, is the most significant systematic error for the measurement of
the {\emph{non-radiative}} cross-sections, and is correlated between
measurements at all energies.

The efficiency for selecting \eemm\ events determined from the simulation 
was also corrected for the  efficiency for triggering DELPHI on these events.
This was estimated using the redundancy of the DELPHI 
trigger system in selected \eemm\ events.
%This was determined from the selected events. 
%DELPHI was able to trigger on single hard muons, such as those selected 
%in this analysis. By selecting events with a muon in both the forward and 
%backward hemispheres the efficiency could be determined by counting the 
%number of times only one of the muons gave a trigger compared to the number
%of time both muons gave a trigger. 
It was found to be stable over all years of 
data taking. Averaging over all years the efficiency was determined to be
$(99.82 \pm 0.07)\%$. No significant variation in efficiency was found over 
$\cos\theta$ within the acceptance of the analysis. 

Backgrounds from \eett, \eeWW, \eeZZ\ and $\ee \rightarrow \ee\ffbar$
processes were estimated from samples of simulated events. Backgrounds
from \eeee, \eeqq\ and two-photon collisions were estimated to be
negligible.  The background from cosmic rays inside the selected
sample was estimated by extrapolating the numbers of cosmic rays
failing the cosmic ray rejection cuts, from within a looser selection
on the impact parameters of the reconstructed tracks into the selected
\eemm\ region.  The estimated backgrounds are given in
Table~\ref{tab:mm:bg}.

\begin{figure}[tp]
\begin{center}
\begin{tabular}{cc}
 \mbox{\epsfig{file=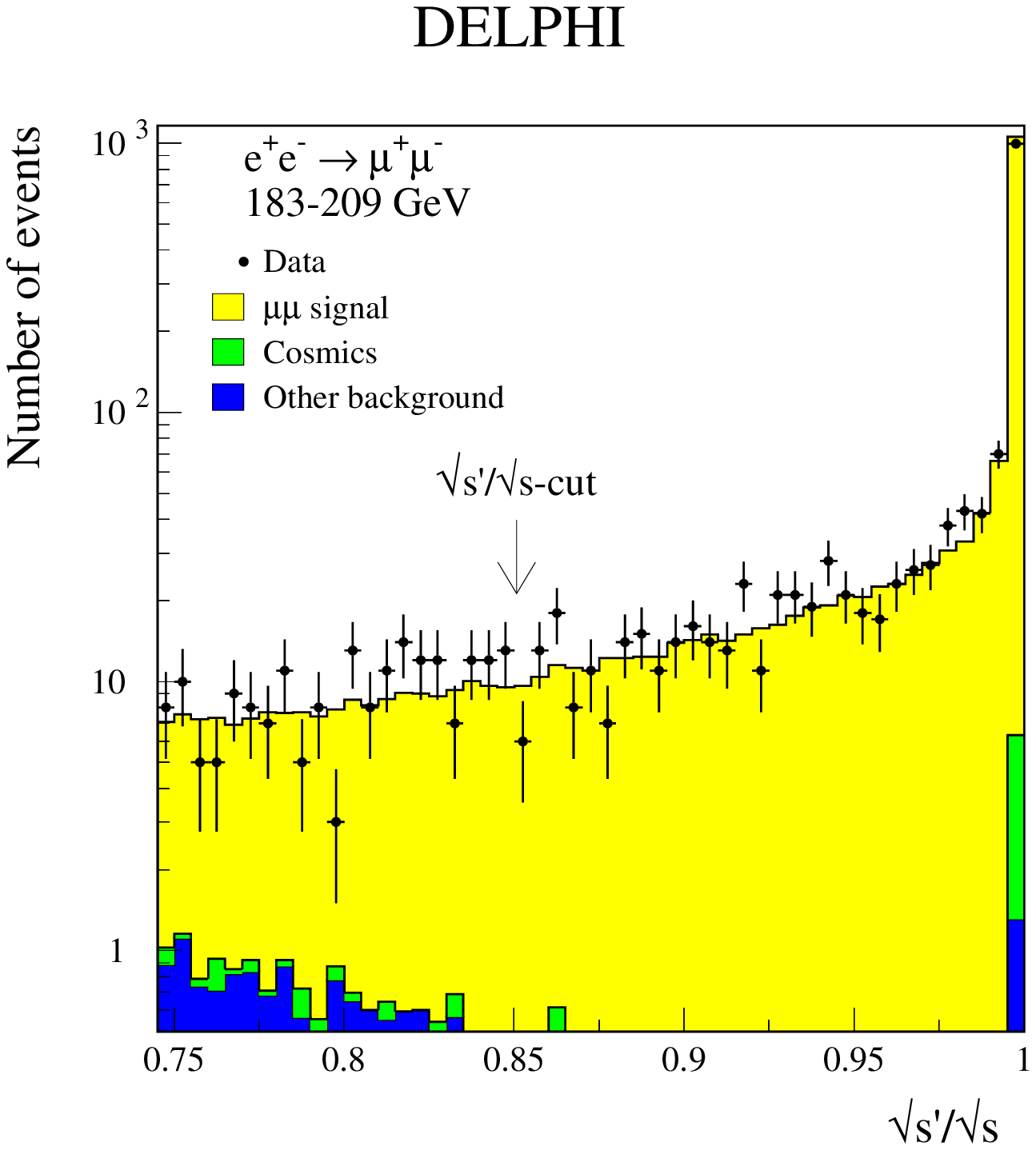,width=0.49\textwidth}} &
 \mbox{\epsfig{file=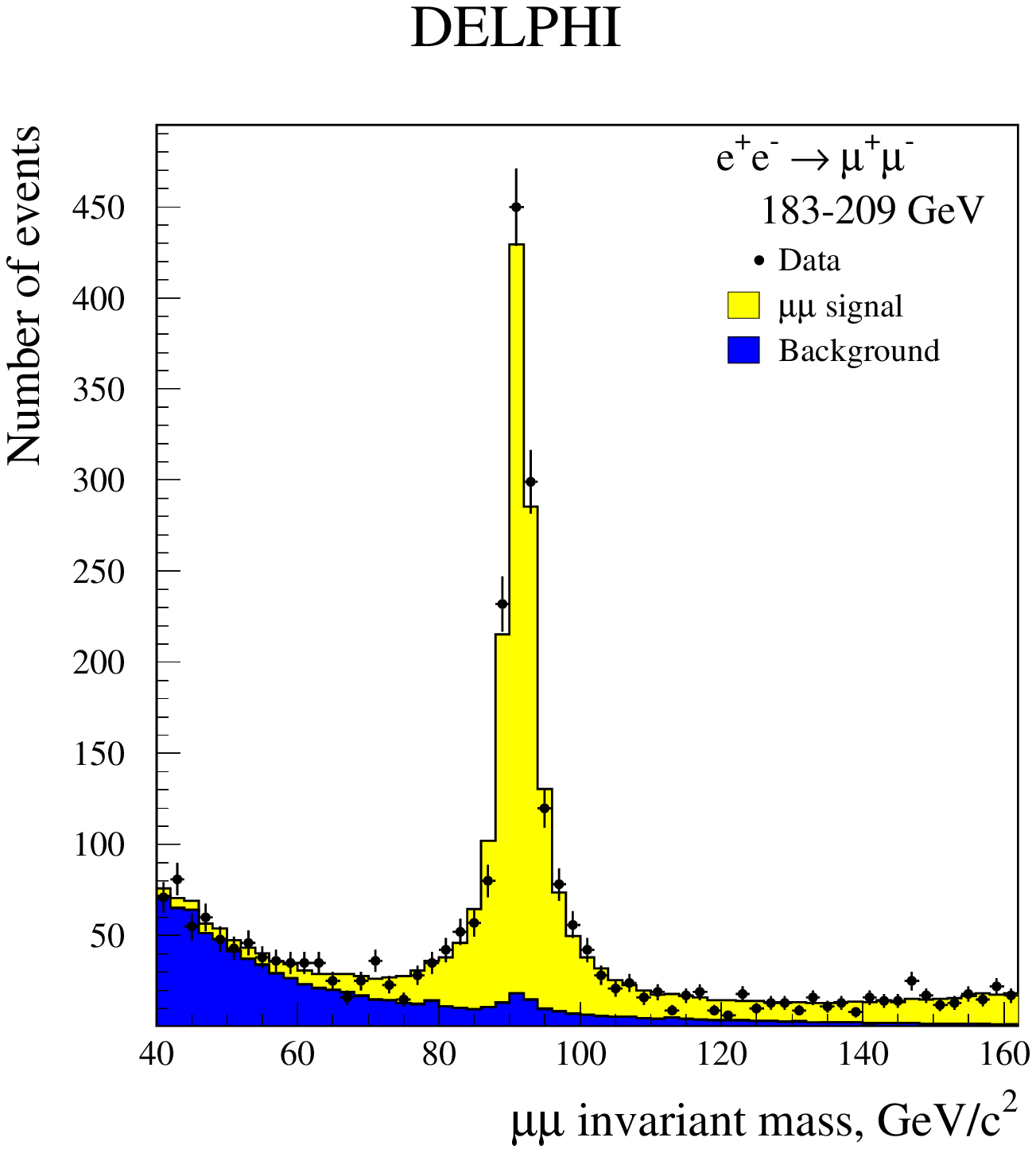,width=0.49\textwidth}} \\
\end{tabular}
\end{center}
\caption{\capsty{The reconstructed \sqsps\ distribution (left) and 
         the reconstructed \rootsp\ distribution (right) for \eemm\ events. 
         Both plots show data from 1997-2000. In the right hand plot all 
         selection criteria except for the cut on the \mumu\ invariant mass 
         were applied.}}
\label{fig:mm:sprm}
\end{figure}

The distributions of $\sqsps$ for events in the {\emph{non-radiative}} region
and $\rootsp$ in the {\emph{radiative return}} region are shown in 
Figure~\ref{fig:mm:sprm}. There is good agreement between data and simulation,
including in the low mass region dominated by backgrounds.

%%%%%%%%%%%%%%%%%%%%%%%%%%%%%%%%%%%%%%%%%%%%%%%%%%%%%%%%
\begin{table}[tp]
\begin{center}
\begin{tabular}{|l|c|c|}
\hline
\multicolumn{3}{|c|}{\eemm} \\
\hline
\hline
Background  & {\emph{Non-radiative}} & {\emph{Inclusive}} \\ 
source      & background             & background         \\
\hline\hline
$\eett$     & 0.08 $\pm$ 0.02        & 0.94 $\pm$ 0.03    \\
$\eeWW$     & 0.16 $\pm$ 0.07        & 3.56 $\pm$ 0.21    \\
$\eeZZ$     & 0.26 $\pm$ 0.06        & 1.01 $\pm$ 0.06    \\
$\ee \rightarrow \ee\ffbar$    & 0.10 $\pm$ 0.06        & 3.75 $\pm$ 0.59    \\
Cosmic rays & 0.44 $\pm$ 0.04        & 0.45 $\pm$ 0.04    \\
\hline\hline
Total       & 1.04 $\pm$ 0.12        & 9.72 $\pm$ 0.63    \\
\hline
\end{tabular}
\caption{\capsty{Residual background levels for data taken at 
         $\roots \sim$207 GeV
         (in \% relative to the number of selected \eemm\ candidates).}}
\label{tab:mm:bg}
\end{center}
\end{table}
%%%%%%%%%%%%%%%%%%%%%%%%%%%%%%%%%%%%%%%%%%%%%%%%%%%%%%%%

%%%%%%%%%%%%%%%%%%%%%%%%%%%%%%%%%%%%%%%
%%%%%%%%%%%%%%%%%%%%%%%%%%%%%%%%%%%%%%%

%% ------------------------------------------------------------------- 
\subsubsection{Results}
\label{sec:mm:res}

%%....................................................................
%%.. Contents ........................................................
%%....................................................................

In total $3684$ events were selected in the {\emph{inclusive}} sample
and $1595$ were selected in the {\emph{non-radiative}} class from data at the
183-207 GeV energy points. For the {\emph{non-radiative}} sample the average 
efficiency in the fiducial volume, taking into account
all corrections, was $93.25 \pm 0.48 \%$, and the background from
non \eemm\ collisions was estimated to increase from $(10 \pm 2)$fb at 
$\roots \sim 183$~GeV to  $(13 \pm 3)$fb at $\roots \sim 207$~GeV, which
is at most approximately $0.6\%$ of the signal cross-section.
In the {\emph{inclusive}} sample the total background amounted to, 
at most, $9\%$ of the signal cross-section. The dominant backgrounds came 
from four-fermion events which were misidentified as two-fermion events.
Cosmic ray events were estimated to account for between $0.31\%$ and $0.45\%$
of the selected events.  
The efficiency, background {\emph{etc.}} at each 
energy is given in Table~\ref{tab:mm:ana}.
The selected events were used to measure the total cross-sections,
forward-backward asymmetries and the differential cross-sections
for {\emph{non-radiative}} events.

%. . . . . . . . . . . . . . . . . . . . . . . . . . . . . . . . . . .
\begin{table}[p]
 {
%%\small
 \begin{center}
 \renewcommand{\arraystretch}{1.2}
 \begin{tabular}{|l|c|c|c|c|c|c|c|c|}
 \hline
 \multicolumn{9}{|c|}{\eemm} \\
 \hline
 \hline
                   & \multicolumn{8}{|c|}{Energy point (GeV)} \\
 \cline{2-9}
                   & 183    & 189    & 192    & 196    & 200    & 202    & 205    & 207    \\
 \hline
 \hline
 Energy (GeV)      & 182.65 & 188.56 & 191.60 & 195.53 & 199.53 & 201.65 & 204.85 & 206.55 \\
 Lumi ($\invpbarn$)& 52.54  & 156.38 & 25.79  & 73.98  & 83.14  & 40.51  & 75.55  & 137.07 \\
 No. Events        & 379    & 991    & 167    & 389    & 506    & 205    & 373    & 674    \\
 Efficiency (\%)   & 92.7   & 93.7   & 93.1   & 93.5   & 93.2   & 93.2   & 93.6   & 92.7   \\
 Background (\%)   & 0.7    & 0.9    & 0.8    & 0.8    & 0.8    & 0.9    & 1.1    & 1.0    \\
% Stat. error ($\%$)& ??.?   & ??.?   & ??.?   & ??.?   & ??.?   & ??.?   & ??.?   & ??.?   \\
% Syst. error ($\%$)& ??.?   & ??.?   & ??.?   & ??.?   & ??.?   & ??.?   & ??.?   & ??.?   \\
 \hline
 \end{tabular}
 \end{center}
 }
 \caption{\capsty{Details of LEP II analysis for the \eemm\ channel. The table 
  shows the actual centre-of-mass energy and luminosity analysed at each energy
  point, the number of events selected in the inclusive analysis and the 
  efficiencies for selecting events in the non-radiative samples and 
  the backgrounds selected in the non-radiative samples.}}
 \label{tab:mm:ana}
\end{table}
%. . . . . . . . . . . . . . . . . . . . . . . . . . . . . . . . . . . 

%. . . . . . . . . . . . . . . . . . . . . . . . . . . . . . . . . . . 
\begin{table}[p]
 %%%%%%%%%%%%%%%% mm %%%%%%%%%%%%%%%%
 %%%%%%%%%%%%%%%% mm %%%%%%%%%%%%%%%% This from John on 18/01/04
 {\small
 \begin{center}
 \renewcommand{\arraystretch}{1.2}
 \begin{tabular}{|c|c|c|c|c|}
 \hline
 \multicolumn{5}{|c|}{\eemm} \\
 \hline
 \hline
 \multicolumn{1}{|c|}{$\roots$} &
 \multicolumn{2}{|c|}{$\rootsp>75$ (GeV)} &
 \multicolumn{2}{|c|}{$\sqsps>0.85$} \\
 \cline{2-5}
 \multicolumn{1}{|c|}{(GeV)} &
 $\smu$ (pb) & $\Afbm$ & $\smu$ (pb) & $\Afbm$ \\
 \hline
 \hline
 130 & 
 $\begin{array}{c}
 24.30 \pm  3.20 \pm  0.83 \\ (20.04)
 \end{array}$ &
 $\begin{array}{c}
 0.450 \pm 0.120 \pm 0.002 \\ (0.339)
 \end{array}$ &
 $\begin{array}{c}
  9.70 \pm  1.90 \pm  0.36 \\ ( 8.11)
 \end{array}$ &
 $\begin{array}{c}
 0.670 \pm 0.150 \pm 0.003 \\ (0.719)
 \end{array}$ \\
 \hline
 136 & 
 $\begin{array}{c}
 17.00 \pm  2.60 \pm  0.58 \\ (17.07)
 \end{array}$ &
 $\begin{array}{c}
 0.560 \pm 0.130 \pm 0.002 \\ (0.340)
 \end{array}$ &
 $\begin{array}{c}
  6.60 \pm  1.60 \pm  0.25 \\ ( 7.00)
 \end{array}$ &
 $\begin{array}{c}
 0.740 \pm 0.160 \pm 0.003 \\ (0.699)
 \end{array}$ \\
 \hline
 161 & 
 $\begin{array}{c}
  9.30 \pm  1.10 \pm  0.32 \\ (10.32)
 \end{array}$ &
 $\begin{array}{c}
 0.390 \pm 0.110 \pm 0.002 \\ (0.335)
 \end{array}$ &
 $\begin{array}{c}
  3.60 \pm  0.70 \pm  0.13 \\ ( 4.43)
 \end{array}$ &
 $\begin{array}{c}
 0.430 \pm 0.160 \pm 0.003 \\ (0.629)
 \end{array}$ \\
 \hline
 172 & 
 $\begin{array}{c}
  8.90 \pm  1.10 \pm  0.31 \\ ( 8.74)
 \end{array}$ &
 $\begin{array}{c}
 0.550 \pm 0.100 \pm 0.002 \\ (0.332)
 \end{array}$ &
 $\begin{array}{c}
  3.60 \pm  0.70 \pm  0.12 \\ ( 3.79)
 \end{array}$ &
 $\begin{array}{c}
 0.940 \pm 0.140 \pm 0.003 \\ (0.610)
 \end{array}$ \\
 \hline
 183 & 
 $\begin{array}{c}
  8.28 \pm  0.46 \pm  0.07 \\ ( 7.56)
 \end{array}$ &
 $\begin{array}{c}
 0.299 \pm 0.051 \pm 0.003 \\ (0.329)
 \end{array}$ &
 $\begin{array}{c}
  3.61 \pm  0.28 \pm  0.03 \\ ( 3.30)
 \end{array}$ &
 $\begin{array}{c}
 0.588 \pm 0.064 \pm 0.001 \\ (0.596)
 \end{array}$ \\
 \hline
 189 & 
 $\begin{array}{c}
  7.17 \pm  0.24 \pm  0.06 \\ ( 7.01)
 \end{array}$ &
 $\begin{array}{c}
 0.336 \pm 0.032 \pm 0.003 \\ (0.327)
 \end{array}$ &
 $\begin{array}{c}
  3.07 \pm  0.15 \pm  0.02 \\ ( 3.07)
 \end{array}$ &
 $\begin{array}{c}
 0.600 \pm 0.039 \pm 0.001 \\ (0.589)
 \end{array}$ \\
 \hline
 192 & 
 $\begin{array}{c}
  7.37 \pm  0.61 \pm  0.07 \\ ( 6.76)
 \end{array}$ &
 $\begin{array}{c}
 0.275 \pm 0.078 \pm 0.003 \\ (0.327)
 \end{array}$ &
 $\begin{array}{c}
  2.82 \pm  0.36 \pm  0.02 \\ ( 2.97)
 \end{array}$ &
 $\begin{array}{c}
 0.636 \pm 0.098 \pm 0.001 \\ (0.586)
 \end{array}$ \\
 \hline
 196 & 
 $\begin{array}{c}
  5.89 \pm  0.32 \pm  0.06 \\ ( 6.45)
 \end{array}$ &
 $\begin{array}{c}
 0.297 \pm 0.052 \pm 0.004 \\ (0.326)
 \end{array}$ &
 $\begin{array}{c}
  2.76 \pm  0.21 \pm  0.02 \\ ( 2.84)
 \end{array}$ &
 $\begin{array}{c}
 0.586 \pm 0.061 \pm 0.001 \\ (0.582)
 \end{array}$ \\
 \hline
 200 & 
 $\begin{array}{c}
  6.95 \pm  0.33 \pm  0.07 \\ ( 6.16)
 \end{array}$ &
 $\begin{array}{c}
 0.334 \pm 0.044 \pm 0.003 \\ (0.324)
 \end{array}$ &
 $\begin{array}{c}
  3.08 \pm  0.21 \pm  0.02 \\ ( 2.71)
 \end{array}$ &
 $\begin{array}{c}
 0.548 \pm 0.056 \pm 0.001 \\ (0.578)
 \end{array}$ \\
 \hline
 202 & 
 $\begin{array}{c}
  5.70 \pm  0.44 \pm  0.06 \\ ( 6.02)
 \end{array}$ &
 $\begin{array}{c}
 0.365 \pm 0.070 \pm 0.004 \\ (0.324)
 \end{array}$ &
 $\begin{array}{c}
  2.46 \pm  0.27 \pm  0.02 \\ ( 2.65)
 \end{array}$ &
 $\begin{array}{c}
 0.544 \pm 0.090 \pm 0.001 \\ (0.577)
 \end{array}$ \\
 \hline
 205 & 
 $\begin{array}{c}
  5.46 \pm  0.31 \pm  0.06 \\ ( 5.81)
 \end{array}$ &
 $\begin{array}{c}
 0.304 \pm 0.053 \pm 0.004 \\ (0.323)
 \end{array}$ &
 $\begin{array}{c}
  2.35 \pm  0.19 \pm  0.02 \\ ( 2.56)
 \end{array}$ &
 $\begin{array}{c}
 0.642 \pm 0.061 \pm 0.001 \\ (0.574)
 \end{array}$ \\
 \hline
 207 & 
 $\begin{array}{c}
  5.49 \pm  0.23 \pm  0.06 \\ ( 5.70)
 \end{array}$ &
 $\begin{array}{c}
 0.373 \pm 0.039 \pm 0.006 \\ (0.323)
 \end{array}$ &
 $\begin{array}{c}
  2.47 \pm  0.15 \pm  0.02 \\ ( 2.52)
 \end{array}$ &
 $\begin{array}{c}
 0.558 \pm 0.048 \pm 0.001 \\ (0.573)
 \end{array}$ \\
 \hline
 \end{tabular}
 \end{center}
 }
 %%%%%%%%%%%%%%%%%%%%%%%%%%%%%%%%%%%%
 %%%%%%%%%%%%%%%%%%%%%%%%%%%%%%%%%%%%

 \caption{\capsty{Measured cross-sections and forward-backward asymmetries
          for {\it{inclusive}} and {\it{non-radiative}} \eemm\ events. The first uncertainty is statistical, the second systematic.
          Numbers in brackets are the theoretical predictions of ZFITTER, which are estimated to have a precision of $\pm 0.4\%$ on $\smu$ and $\pm 0.004$ on $\Afbm$. }}
 \label{tab:mm:res}
\end{table}
%. . . . . . . . . . . . . . . . . . . . . . . . . . . . . . . . . . . 

%% ------------------------------------------------------------------- 
\subsubsection*{Cross-sections}

The total cross-section for the {\emph{non-radiative}} and {\emph{inclusive}} 
samples of \eemm\ events was computed from
\begin{equation}
\smu = \eta_{4\pi} \frac{(N_{sel} - N_{bg})(1-f)}
                          {\epsilon {\mathcal{L}}}
%\smu = \eta_{4\pi} \frac{f\left(N_{\small{\mumu}} - N_{bg}\right)}
%                          {\epsilon {\mathcal{L}}}
\end{equation}
\noindent
where $N_{sel}$ is the numbers of events selected in each
sample, $N_{bg}$ is the predicted number of background events selected
in the sample for the luminosity $\mathcal{L}$. The efficiency for
selecting \eemm\ events in the sample is $\epsilon$, which includes
corrections for the efficiency to reconstruct and identify muons and
the trigger efficiency, obtained from the data. $f$ (applicable to the
{\emph{non-radiative}} cross-section only) is a factor for the
migration of \eemm\ events into the sample from lower centre-of-mass
energy. $\eta_{4\pi}$ is a correction factor required to extrapolate
the measurements from the polar angular acceptance to the full $4\pi$
acceptance, determined from simulated events.  The measured
cross-sections are given in Table~\ref{tab:mm:res} and shown in
Figure~\ref{fig:ana:sig-cmp}.

For the 207~GeV energy point separate measurements were made for the periods
before and after the failure of the sector of the TPC. These measured 
cross-sections were combined using the BLUE
technique~\cite{ref:ana:blue-sngl,*ref:ana:blue-multi} 
which performs a weighted average of the measurements taking into account
correlated errors.

%% ------------------------------------------------------------------- 
\subsubsection*{Forward-backward asymmetries}

The forward-backward asymmetry for each sample was computed from
\begin{equation}
 \Afb = \frac{N_{f} - c_{b} N_{b}}{N_{f} + c_{b} N_{b}}
\end{equation}
\noindent
where $N_{f}$ and $N_{b}$ are the numbers of \eemm\ events in 
which the $\mu^-$ is in 
the forward and
backward hemispheres defined as  $\cos\theta > 0$ and 
$\cos\theta < 0$ respectively.
Differences in the efficiencies for selecting events and the migration
of events into the {\emph{non-radiative}} samples in the
forward and backward hemispheres are accounted for by the factor $c_{b}$,
which was determined for each sample from simulated events.
These differences arise from the different fractions of {\emph{non-radiative}}
and {\emph{radiative return}} events in the forward and backward hemispheres
reflecting the very different forward-backward asymmetries of these subsamples.
$N_{f}$ and $N_{b}$ are computed from
\begin{equation}
 N_{f/b} = \eta_{2\pi} 
             \left(N^{f/b}_{\small{\mumu}} 
                              - N^{f/b}_{bg} + N^{f/b}_{\pm}\right),
\end{equation}
\noindent
where $N^{f/b}_{\small{\mumu}}$ and $N^{f/b}_{bg}$ are the numbers of events 
selected and the predicted backgrounds in each hemisphere. 
$\eta_{2\pi}$ is a factor to extrapolate the observed number of events
in the angular acceptance to $0 \leq |\cos\theta| \leq 1$.
$N^{f/b}_{\pm}$ account for the expected numbers of
events in each hemisphere with misidentified charges, such that an event in 
which the $\mu^{-}$ was produced in the 
forward hemisphere is reconstructed as an event with the $\mu^{-}$ in the
backward hemisphere and {\it{vice versa}}. There were approximately $0.5\%$ 
of events in which the charge of one of the particles
was misidentified, so that both particles were reconstructed with the same
charge. In these cases, the particle with the larger relative uncertainty on 
its momentum was considered to have its charge mismeasured and its charge was
inverted before the event was assigned to either the forward or backward 
hemisphere. From simulation this was found to identify the wrongly measured
charge in $93\pm5\%$ of cases.
The rate of events in which both particles had their charges mismeasured
was found to be negligible.
The measured forward-backward asymmetries are given in Table \ref{tab:mm:res} and shown in Figure~\ref{fig:ana:afb-cmp}.

For the 207~GeV energy point, separate measurements of the numbers of events 
in the forward and backward hemispheres were made for the periods before
and after the failure of the sector of the TPC. These measurements were 
added together and an average forward-backward asymmetry was computed.

%% ------------------------------------------------------------------- 
\subsubsection*{Differential cross-sections}

%. . . . . . . . . . . . . . . . . . . . . . . . . . . . . . . . . . . 
\begin{table}[p]
%%%%%%%%%%%%%%%% mm %%%%%%%%%%%%%%%%
%%%%%%%%%%%%%%%% mm %%%%%%%%%%%%%%%% This from John 28/01/04
 {\footnotesize
 \begin{center}
 \setlength{\tabcolsep}{1.0mm}
 \begin{tabular}{ccc}
 \multicolumn{3}{r}{\fbox{\hspace{7.08cm}
 $\eemm$
 \hspace{7.08cm}}} \\
 \\
 \renewcommand{\arraystretch}{1.2}
 \begin{tabular}
 {|@{[}r@{,}r@{]}|c|r@{$\pm$}
 c@{(}c@{)}@{$\pm$}c|}
 \hline
\multicolumn{7}{|c|}{$\roots \sim 183$} \\
 \hline
 \hline
 \multicolumn{2}{|c|}{} &
 \multicolumn{5}{|c|}{$\dsdcth$ (pb)} \\
 \cline{3-7}
 \multicolumn{2}{|c|}{$\cos\theta$} &
 \multicolumn{1}{|c|}{SM} &
 \multicolumn{4}{|c|}{Measurement} \\
 \hline
 \hline
 -0.97 & -0.80 &  0.47 & -0.01 &  0.00 &  0.24 &  0.00 \\
 -0.80 & -0.60 &  0.48 &  0.41 &  0.21 &  0.22 &  0.01 \\
 -0.60 & -0.40 &  0.57 &  1.00 &  0.32 &  0.24 &  0.01 \\
 -0.40 & -0.20 &  0.76 &  1.06 &  0.33 &  0.27 &  0.01 \\
 -0.20 &  0.00 &  1.05 &  1.14 &  0.35 &  0.33 &  0.01 \\
  0.00 &  0.20 &  1.44 &  1.69 &  0.43 &  0.39 &  0.02 \\
  0.20 &  0.40 &  1.92 &  1.79 &  0.43 &  0.44 &  0.02 \\
  0.40 &  0.60 &  2.52 &  3.34 &  0.59 &  0.51 &  0.03 \\
  0.60 &  0.80 &  3.22 &  3.30 &  0.60 &  0.58 &  0.03 \\
  0.80 &  0.97 &  4.00 &  4.07 &  0.72 &  0.71 &  0.04 \\
 \hline
 \end{tabular}
 &
 \renewcommand{\arraystretch}{1.2}
 \begin{tabular}
 {|@{[}r@{,}r@{]}|c|r@{$\pm$}
 c@{(}c@{)}@{$\pm$}c|}
 \hline
\multicolumn{7}{|c|}{$\roots \sim 189$} \\
 \hline
 \hline
 \multicolumn{2}{|c|}{} &
 \multicolumn{5}{|c|}{$\dsdcth$ (pb)} \\
 \cline{3-7}
 \multicolumn{2}{|c|}{$\cos\theta$} &
 \multicolumn{1}{|c|}{SM} &
 \multicolumn{4}{|c|}{Measurement} \\
 \hline
 \hline
 -0.97 & -0.80 &  0.46 &  0.41 &  0.13 &  0.14 &  0.01 \\
 -0.80 & -0.60 &  0.46 &  0.47 &  0.13 &  0.12 &  0.01 \\
 -0.60 & -0.40 &  0.54 &  0.54 &  0.14 &  0.14 &  0.01 \\
 -0.40 & -0.20 &  0.71 &  0.43 &  0.12 &  0.16 &  0.01 \\
 -0.20 &  0.00 &  0.98 &  1.21 &  0.21 &  0.19 &  0.02 \\
  0.00 &  0.20 &  1.33 &  1.25 &  0.21 &  0.22 &  0.02 \\
  0.20 &  0.40 &  1.78 &  2.03 &  0.26 &  0.24 &  0.02 \\
  0.40 &  0.60 &  2.33 &  2.19 &  0.28 &  0.28 &  0.02 \\
  0.60 &  0.80 &  2.98 &  2.86 &  0.32 &  0.32 &  0.03 \\
  0.80 &  0.97 &  3.70 &  3.90 &  0.40 &  0.39 &  0.04 \\
 \hline
 \end{tabular}
 &
 \renewcommand{\arraystretch}{1.2}
 \begin{tabular}
 {|@{[}r@{,}r@{]}|c|r@{$\pm$}
 c@{(}c@{)}@{$\pm$}c|}
 \hline
\multicolumn{7}{|c|}{$\roots \sim 192$} \\
 \hline
 \hline
 \multicolumn{2}{|c|}{} &
 \multicolumn{5}{|c|}{$\dsdcth$ (pb)} \\
 \cline{3-7}
 \multicolumn{2}{|c|}{$\cos\theta$} &
 \multicolumn{1}{|c|}{SM} &
 \multicolumn{4}{|c|}{Measurement} \\
 \hline
 \hline
 -0.97 & -0.80 &  0.45 & -0.01 &  0.00 &  0.33 &  0.00 \\
 -0.80 & -0.60 &  0.45 &  0.21 &  0.21 &  0.31 &  0.00 \\
 -0.60 & -0.40 &  0.53 &  0.83 &  0.42 &  0.33 &  0.01 \\
 -0.40 & -0.20 &  0.69 &  0.42 &  0.30 &  0.38 &  0.01 \\
 -0.20 &  0.00 &  0.94 &  1.09 &  0.49 &  0.45 &  0.01 \\
  0.00 &  0.20 &  1.28 &  0.67 &  0.40 &  0.54 &  0.01 \\
  0.20 &  0.40 &  1.71 &  2.98 &  0.80 &  0.60 &  0.02 \\
  0.40 &  0.60 &  2.24 &  2.27 &  0.69 &  0.68 &  0.02 \\
  0.60 &  0.80 &  2.87 &  2.30 &  0.70 &  0.77 &  0.02 \\
  0.80 &  0.97 &  3.57 &  3.22 &  0.90 &  0.94 &  0.03 \\
 \hline
 \end{tabular}
 \\ \\
 \renewcommand{\arraystretch}{1.2}
 \begin{tabular}
 {|@{[}r@{,}r@{]}|c|r@{$\pm$}
 c@{(}c@{)}@{$\pm$}c|}
 \hline
\multicolumn{7}{|c|}{$\roots \sim 196$} \\
 \hline
 \hline
 \multicolumn{2}{|c|}{} &
 \multicolumn{5}{|c|}{$\dsdcth$ (pb)} \\
 \cline{3-7}
 \multicolumn{2}{|c|}{$\cos\theta$} &
 \multicolumn{1}{|c|}{SM} &
 \multicolumn{4}{|c|}{Measurement} \\
 \hline
 \hline
 -0.97 & -0.80 &  0.44 &  0.68 &  0.24 &  0.19 &  0.01 \\
 -0.80 & -0.60 &  0.44 &  0.51 &  0.19 &  0.18 &  0.01 \\
 -0.60 & -0.40 &  0.51 &  0.65 &  0.22 &  0.19 &  0.01 \\
 -0.40 & -0.20 &  0.66 &  0.22 &  0.13 &  0.22 &  0.01 \\
 -0.20 &  0.00 &  0.90 &  0.84 &  0.26 &  0.26 &  0.01 \\
  0.00 &  0.20 &  1.22 &  1.71 &  0.37 &  0.31 &  0.02 \\
  0.20 &  0.40 &  1.63 &  1.33 &  0.32 &  0.35 &  0.01 \\
  0.40 &  0.60 &  2.14 &  2.08 &  0.39 &  0.39 &  0.02 \\
  0.60 &  0.80 &  2.74 &  2.92 &  0.46 &  0.45 &  0.02 \\
  0.80 &  0.97 &  3.41 &  2.75 &  0.49 &  0.54 &  0.03 \\
 \hline
 \end{tabular}
 &
 \renewcommand{\arraystretch}{1.2}
 \begin{tabular}
 {|@{[}r@{,}r@{]}|c|r@{$\pm$}
 c@{(}c@{)}@{$\pm$}c|}
 \hline
\multicolumn{7}{|c|}{$\roots \sim 200$} \\
 \hline
 \hline
 \multicolumn{2}{|c|}{} &
 \multicolumn{5}{|c|}{$\dsdcth$ (pb)} \\
 \cline{3-7}
 \multicolumn{2}{|c|}{$\cos\theta$} &
 \multicolumn{1}{|c|}{SM} &
 \multicolumn{4}{|c|}{Measurement} \\
 \hline
 \hline
 -0.97 & -0.80 &  0.43 &  0.53 &  0.20 &  0.18 &  0.01 \\
 -0.80 & -0.60 &  0.43 &  0.26 &  0.13 &  0.17 &  0.00 \\
 -0.60 & -0.40 &  0.49 &  0.83 &  0.23 &  0.18 &  0.01 \\
 -0.40 & -0.20 &  0.64 &  0.71 &  0.22 &  0.20 &  0.01 \\
 -0.20 &  0.00 &  0.86 &  1.10 &  0.28 &  0.24 &  0.01 \\
  0.00 &  0.20 &  1.17 &  1.44 &  0.32 &  0.28 &  0.01 \\
  0.20 &  0.40 &  1.56 &  1.66 &  0.33 &  0.32 &  0.02 \\
  0.40 &  0.60 &  2.04 &  2.72 &  0.42 &  0.36 &  0.02 \\
  0.60 &  0.80 &  2.61 &  2.80 &  0.43 &  0.41 &  0.02 \\
  0.80 &  0.97 &  3.25 &  3.18 &  0.50 &  0.50 &  0.03 \\
 \hline
 \end{tabular}
 &
 \renewcommand{\arraystretch}{1.2}
 \begin{tabular}
 {|@{[}r@{,}r@{]}|c|r@{$\pm$}
 c@{(}c@{)}@{$\pm$}c|}
 \hline
\multicolumn{7}{|c|}{$\roots \sim 202$} \\
 \hline
 \hline
 \multicolumn{2}{|c|}{} &
 \multicolumn{5}{|c|}{$\dsdcth$ (pb)} \\
 \cline{3-7}
 \multicolumn{2}{|c|}{$\cos\theta$} &
 \multicolumn{1}{|c|}{SM} &
 \multicolumn{4}{|c|}{Measurement} \\
 \hline
 \hline
 -0.97 & -0.80 &  0.43 &  1.43 &  0.48 &  0.26 &  0.02 \\
 -0.80 & -0.60 &  0.42 &  0.54 &  0.27 &  0.24 &  0.01 \\
 -0.60 & -0.40 &  0.49 &  0.25 &  0.18 &  0.25 &  0.00 \\
 -0.40 & -0.20 &  0.62 &  0.26 &  0.19 &  0.29 &  0.01 \\
 -0.20 &  0.00 &  0.84 &  0.56 &  0.28 &  0.34 &  0.01 \\
  0.00 &  0.20 &  1.14 &  0.71 &  0.32 &  0.40 &  0.01 \\
  0.20 &  0.40 &  1.52 &  2.45 &  0.58 &  0.45 &  0.02 \\
  0.40 &  0.60 &  1.99 &  1.45 &  0.44 &  0.51 &  0.02 \\
  0.60 &  0.80 &  2.55 &  1.87 &  0.50 &  0.58 &  0.02 \\
  0.80 &  0.97 &  3.18 &  3.14 &  0.71 &  0.71 &  0.03 \\
 \hline
 \end{tabular}
 \\ \\
 \renewcommand{\arraystretch}{1.2}
 \begin{tabular}
 {|@{[}r@{,}r@{]}|c|r@{$\pm$}
 c@{(}c@{)}@{$\pm$}c|}
 \hline
\multicolumn{7}{|c|}{$\roots \sim 205$} \\
 \hline
 \hline
 \multicolumn{2}{|c|}{} &
 \multicolumn{5}{|c|}{$\dsdcth$ (pb)} \\
 \cline{3-7}
 \multicolumn{2}{|c|}{$\cos\theta$} &
 \multicolumn{1}{|c|}{SM} &
 \multicolumn{4}{|c|}{Measurement} \\
 \hline
 \hline
 -0.97 & -0.80 &  0.42 &  0.67 &  0.24 &  0.19 &  0.01 \\
 -0.80 & -0.60 &  0.42 &  0.35 &  0.16 &  0.17 &  0.00 \\
 -0.60 & -0.40 &  0.47 &  0.20 &  0.12 &  0.18 &  0.00 \\
 -0.40 & -0.20 &  0.60 &  0.34 &  0.16 &  0.20 &  0.01 \\
 -0.20 &  0.00 &  0.81 &  0.57 &  0.20 &  0.24 &  0.01 \\
  0.00 &  0.20 &  1.10 &  0.95 &  0.27 &  0.28 &  0.01 \\
  0.20 &  0.40 &  1.47 &  1.33 &  0.31 &  0.32 &  0.01 \\
  0.40 &  0.60 &  1.92 &  1.40 &  0.32 &  0.37 &  0.02 \\
  0.60 &  0.80 &  2.46 &  2.16 &  0.40 &  0.42 &  0.02 \\
  0.80 &  0.97 &  3.06 &  4.23 &  0.61 &  0.51 &  0.03 \\
 \hline
 \end{tabular}
 & &
 \renewcommand{\arraystretch}{1.2}
 \begin{tabular}
 {|@{[}r@{,}r@{]}|c|r@{$\pm$}
 c@{(}c@{)}@{$\pm$}c|}
 \hline
\multicolumn{7}{|c|}{$\roots \sim 207$} \\
 \hline
 \hline
 \multicolumn{2}{|c|}{} &
 \multicolumn{5}{|c|}{$\dsdcth$ (pb)} \\
 \cline{3-7}
 \multicolumn{2}{|c|}{$\cos\theta$} &
 \multicolumn{1}{|c|}{SM} &
 \multicolumn{4}{|c|}{Measurement} \\
 \hline
 \hline
 -0.97 & -0.80 &  0.42 &  0.47 &  0.15 &  0.14 &  0.01 \\
 -0.80 & -0.60 &  0.41 &  0.47 &  0.13 &  0.13 &  0.00 \\
 -0.60 & -0.40 &  0.47 &  0.42 &  0.13 &  0.13 &  0.00 \\
 -0.40 & -0.20 &  0.59 &  0.54 &  0.14 &  0.15 &  0.01 \\
 -0.20 &  0.00 &  0.80 &  0.71 &  0.16 &  0.18 &  0.01 \\
  0.00 &  0.20 &  1.08 &  1.00 &  0.20 &  0.21 &  0.01 \\
  0.20 &  0.40 &  1.44 &  1.39 &  0.23 &  0.24 &  0.01 \\
  0.40 &  0.60 &  1.88 &  2.61 &  0.32 &  0.27 &  0.02 \\
  0.60 &  0.80 &  2.41 &  1.93 &  0.28 &  0.31 &  0.02 \\
  0.80 &  0.97 &  3.01 &  2.65 &  0.36 &  0.38 &  0.02 \\
 \hline
 \end{tabular}
 \\ \\
 \end{tabular}
 \end{center}
 }
%%%%%%%%%%%%%%%%%%%%%%%%%%%%%%%%%%%%
 %%%%%%%%%%%%%%%%%%%%%%%%%%%%%%%%%%%%

 \caption{\capsty{Differential cross-sections for non-radiative
          \eemm\ events at centre-of-mass energies from 183 to 207 GeV. The 
          tables show the bins, the predictions of the Standard Model 
          and the measurements. The errors quoted are the statistical and 
          experimental systematic errors. The statistical errors are shown as 
          the measured errors and, in brackets, the expected errors, computed 
          from the square root of the observed and expected numbers 
          of events respectively.}}
 \label{tab:mm:diff}
\end{table}
%. . . . . . . . . . . . . . . . . . . . . . . . . . . . . . . . . . . 

The differential cross-sections, \dsdcth\, for the
{\emph{non-radiative}} samples of \eemm\ events were computed in 10
bins, $i$, of $\cos\theta$ in the range $-0.970 < \cos\theta < 0.970$
using
\begin{equation}
%% \frac{d\sigma_{i}}{d \cos\theta} 
\dsdcth_{i}
           = \frac{\eta_{i} f^{i} 
               \left(N^{i}_{\small{\mumu}} - N^{i}_{bg} + N_{\pm}^{i}\right)}
                          {\epsilon_{i} {\mathcal{L}} \Delta_{i}},
\end{equation}
\noindent
where $N^{i}_{\small{\mumu}}$ and $N^{i}_{bg}$ are the number of
observed events and the expected number of background events in each
bin of $\cos\theta$, where $\theta$ is the polar angle of the
negatively charged muon with respect to the beam axis. $\epsilon_{i}$
is the efficiency in each bin of $\cos\theta$, which was corrected, in
all bins, for the global track reconstruction and muon identification
correction factor determined for the total cross-sections and the
trigger efficiency.  The bin width in $\cos\theta$ is $\Delta_{i}$.
The migration factor between {\emph{non-radiative}} and
{\emph{radiative return}} events, $f$, was computed individually
for each bin from simulated events

Migrations between neighbouring bins of $\cos\theta$ were found to be negligible.
The expected number of events with mismeasured charges in each bin was accounted for
by $N_{\pm}^{i}$.

Measurements in the outer bins of $\cos\theta$ were not corrected to
the full angular acceptance.  Corrections were made for the
experimental cuts on the polar angle of both muons by the factor $\eta$
which was computed from simulated events and found to be significant
only in the bins with largest $|\cos\theta|$.
%However, the predictions of \ZFITTER\ correspond to 
%a cut on the polar angle of the $\mu^{-}$ only, whereas the efficiencies were 
%computed for the experimental cuts on the polar angles of the $\mu^{+}$ and
%$\mu^{-}$ as in the measurements of the total cross-section. The results
%were extrapolated to a cut on the $\mu^{-}$ only by the correction $\eta$ 
%which was computed from simulated events and found to be significant only in
%the bins with largest $|\cos\theta|$. 
The measured differential cross-sections are given in Table~\ref{tab:mm:diff}.

For the 207~GeV energy point separate measurements of the differential 
cross-sections in each bin were made for the periods
before and after the failure of the sector of the TPC. The measurements
were combined using the BLUE technique, taking into account
correlated systematic errors.

%% ------------------------------------------------------------------- 
\subsubsection*{Systematic errors}

%%
%% NB
%%  - Uncertainties from S6 were found to be neglible except for
%%    charge misid - but this is already taken into account in the 
%%    charge misid error for A0U
%%

%%%%%%%%%%%%%%%%%%%%%%%%%%%%%%%%%%%%%%%%%%%%%%%%%%%%%%%
%%%%%%%%%%%%%%%%%%%%%%%%%%%%%%%%%%%%%%%%%%%%%%%%%%%%%%%
\begin{table}[tp]
\begin{center}
 \begin{tabular}{|l|c|c||c|c|}
 \hline
 \multicolumn{5}{|c|}{\eemm} \\
 \hline
 \hline
                           &$\Delta\sigma/\sigma$ &$\Delta\sigma/\sigma$&
                              $\Delta A_{FB}$ &$\Delta A_{FB}$\\
 Source                    & (non-rad.) & (inclus.)&
                              (non-rad.)&(inclus.) \\
 \hline\hline
   Efficiency              &  51 &  51 &   9 &  20 \\
   \rootsp\ reconstruction &  14 &  20 &   2 &   0 \\
  Charge misidentification &   - &   - &   6 &   4 \\
  QED                      &   2 &   9 &   1 &   0 \\
  Backgrounds              &  12 &  70 &   6 &  51 \\
  Luminosity               &  55 &  55 &   - &   - \\
 \hline\hline
  Total uncertainty        &  77 & 105 &  12 &  55 \\
 \hline
  Correlated               &  75 & 103 &  11 &  53 \\
  Uncorrelated             &  16 &  17 &   5 &  17 \\
 \hline
 \end{tabular}
\caption{\capsty{Systematic uncertainties on the measured cross-sections 
         and forward-backward asymmetries for data taken
         at $\roots \sim$ 207 GeV. All numbers in units of 10$^{-4}$.
Separate measurements were made before and after the TPC sector failure, and 
then combined. There is thus no explicit systematic uncertainty related to 
the TPC sector instability. 
         The correlated error component includes errors 
         correlated between energies and channels and those correlated 
         with other LEP experiments.}}
\label{tab:mm:syst}
\end{center}
\end{table}
%%%%%%%%%%%%%%%%%%%%%%%%%%%%%%%%%%%%%%%%%%%%%%%%%%%%%%%
%%%%%%%%%%%%%%%%%%%%%%%%%%%%%%%%%%%%%%%%%%%%%%%%%%%%%%%

For measurements of the {\emph{non-radiative}} cross-sections the
largest single systematic error comes from the determination of the
selection efficiency which is dominated by the correction factor for
muon identification and track reconstruction efficiencies. The
relative error on the correction factor is $0.49\%$ and is correlated
between all energies and between all bins in the estimation
of the differential cross-sections. Additional, smaller sources of
error on the efficiency come from the statistics of simulated event
samples and the estimated trigger efficiency.

%Comparison of the efficiencies and ISR migration factors in the forward and 
%backward hemispheres in the simulated \eemm\ samples led to an error of
%$0.0004$ on the {\emph{non-radiative} forward-backward asymmetry

The uncertainty on the ISR migration factors arising from 
misreconstruction of \rootsp\ has two components: an uncorrelated component
from the statistics of simulated \eemm\ events; and an uncertainty estimated 
from switching off the simulation of the beam energy spread in the simulated 
sample, which was found by comparing the change in the migration factor from 
the standard analysis. This was determined for a single energy and applied
as a correlated error over all energies. These uncertainties were estimated 
bin-by-bin for the differential cross-sections.

The uncertainty on the angular acceptance corrections comes from the
statistics of the simulated \eemm\ samples. The uncertainty on the 
forward-backward asymmetry and differential cross-sections due to charge 
misidentification comes from
the estimation of the efficiency for correctly identifying which particle
had the mismeasured charge in such events. The uncertainty amounts to, at most,
$0.0006$ on the {\emph{non-radiative}} forward-backward asymmetries. This 
introduces an anticorrelation between bins in the forward and backward 
hemispheres in the differential cross-sections.

The uncertainties arising from missing higher orders in the QED radiative
corrections in the simulation were estimated by taking half the difference
in the measurements obtained using \KK\ simulated events in which the highest 
order corrections available in the program were either included or excluded.
%The \KK\ generator used to simulate \eemm\ events uses the coherent exclusive
%exponentiation scheme~\cite{ref:mc:kk} to generate ISR and FSR and take into 
%account the interference between these sources of radiation. This scheme
%is based on calculations of second order QED corrections. 
%The events generated are unweighted, however, events have auxiliary weights
%computing the probability to generate the same event at lower orders in QED.
%To estimate the uncertainty arising from missing higher orders the 
%analysis was remade using events weighted with the probability determined
%at one lower order in the exponentiation scheme. The error arising for
%the different measurements was taken to be half the difference in the final 
%measurements.

Uncertainties on the backgrounds come from the statistics of the simulated
samples and from the number of cosmic rays failing vertex-selection cuts used 
to estimate the residual cosmic ray background, which are uncorrelated 
between energy points. The final source comes from the theoretical precision 
on the total cross-section of the different simulated background samples, 
which is correlated  between energy points and is significant for the 
measurement of the {\emph{inclusive}} cross-sections and forward-backward 
asymmetries.

Uncertainties from the evaluation of the luminosity were included in the 
measurements of the cross-sections.
The systematic errors are summarised in Table~\ref{tab:mm:syst}.

%%--------------------------------------------------------------------
%%-- ANALYSIS OF tt FINAL STATES -------------------------------------
%%--------------------------------------------------------------------
\subsection{$\boldmath{\tautau}$ final states}
\label{sec:tt}

The analysis of tau pair production in 1997-98 LEP runs was presented
in \cite{ref:delphiff:183-189}.  Here we present a complete
reanalysis of those data together with the new analysis of data taken
in 1999-2000. Several aspects of the analysis have changed with
respect to the previous publication: improved track reconstruction
algorithm, optimised event selection procedure and a better description
of signal and background processes with new event generators. All the
data taken in the 1997-2000 runs were analysed with a homogeneous
procedure which is described below.  The data taken above the $\Zzero$
resonance in the 1995-96 runs~\cite{ref:delphiff:130-172} were not
reanalysed since the improvements of the new analysis are
negligible compared to the large statistical uncertainty of those
measurements.

%% -------------------------------------------------------------------
%\subsubsection{Signal definition}
%\label{sec:tt:sigdef}
%%%....................................................................
%%%.. Summary .........................................................
%%%....................................................................
%%\fbox{
%%\begin{minipage}{0.87\textwidth}
%%\begin{itemize}
%% \item Specific signal definitions
%%\end{itemize}
%%\end{minipage}
%%}
%
%%%....................................................................
%%%.. Contents ........................................................
%%%....................................................................
%
%
%The definition of the $\tau$ pair signal was similar to that for 
%the muon channel. Only the genuine $\eettg$ process was accepted.
%Production of a $\tau$ pair accompanied by another fermion pair
%was considered as a four-fermion background and subtracted
%using the WPHACT generator.
%
%The effective centre-of-mass energy $\rootsp$ was defined as
%an invariant mass of the two tau leptons after 
%the final state radiation. The definitions of non-radiative and
%inclusive event samples were $\sqsps>0.85$ and
%$\rootsp>$75 GeV respectively.

%% ------------------------------------------------------------------- 
\subsubsection{Analysis}
\label{sec:tt:analysis}

%%....................................................................
%%.. Summary .........................................................
%%....................................................................
%\fbox{
%\begin{minipage}{0.87\textwidth}
%\begin{itemize}
% \item Measurments $\sigma$, \Afb\ and \dsdcth
%
% \item Event selection
%
% \item \rootsp\ algorithm
%
% \item Efficiency
%
% \item Backgrounds
%
% \item Systematic errors - correlations
%
% \item 1 or 2 plots showing the quality of the data MC agreement
%       before the final selection. {\it{i.e.}} not just a final \rootsp\ plot
%
% \item Table giving
%  \begin{itemize}
%   \item Actual Energy
%
%   \item Luminosity analysed
%
%   \item Number of selected events
%
%   \item Efficiency
%
%   \item Background
%
%   \item Statistical relative error
%
%   \item Systemtic relative error
%
%  \end{itemize}
%
% \item Need to cover different energies/years of data
%\end{itemize}
%\end{minipage}
%}
%%....................................................................
%%.. Contents ........................................................
%%....................................................................

%% . . . . . . . . . . . . . . . . . . . . . . . . . . . . . . . . . . 
\subsubsection*{Run selection}
\label{sec:tt:run}

Runs in which 
detector components essential for the analysis were not
fully operational were rejected to ensure a high quality 
of the data used in the analysis. This included the requirement that the
TPC, HPC and Forward Electromagnetic Calorimeters (FEMC) 
had a performance close to optimal.
In addition, from the two pairs of detectors
VD and Inner Detector (ID), and the forward tracking chambers, FCA and FCB, 
at least one of the two detectors in both pairs had to be operational.
About 2.5\% of total integrated luminosity was rejected
by this procedure.

%% . . . . . . . . . . . . . . . . . . . . . . . . . . . . . . . . . . 
\subsubsection*{Track and photon selection}
\label{sec:tt:track}

The $\tau$-pair selection was based on the reconstructed 
kinematic properties of the events. Therefore it was
important that only well reconstructed tracks and photons
were used in the analysis. To ensure this the following 
selection was applied to the tracks found by the reconstruction
program:

\begin{itemize}

\item The particle momentum had to be at least 100 MeV/$c$;

\item The error of the momentum reconstruction had to be
less than the absolute value of the momentum;

\item The track extrapolation had to originate from 
the region around the nominal interaction point 
(less than 5 cm in the plane perpendicular to the beam, less than 10 cm 
along the z axis);

\item The track had to be seen either in VD or in ID;

\item The track had to be seen in at least one tracking device beyond
VD and ID \mbox{(namely TPC, OD, FCA, FCB).}

\end{itemize}

The last condition was not applied to the tracks 
close to the azimuthal boundaries of the TPC sectors,
nor was it applied to the tracks pointing to the broken
TPC sector for the last period of 2000 data taking.
In these cases, if the last condition was not fulfilled the track
was still accepted if it had a measured RICH tracking point
or if there was any deposition in an electromagnetic or hadron
calorimeter within 10 degrees from the track direction.

An unassociated electromagnetic cluster was accepted as a photon
if it had more than 0.5 GeV of energy deposition and
was found more than one degree from the nearest track.
Identified electron-positron pairs compatible with a photon conversion
were considered as photons rather than charged particles.

The event selection was based on the tracks and photons 
accepted by these procedures.

%% . . . . . . . . . . . . . . . . . . . . . . . . . . . . . . . . . . 
\subsubsection*{Event selection}
\label{sec:tt:cuts}

The selection of $\tau$-pairs was similar to the one 
presented in \cite{ref:delphiff:183-189}. Several cuts were changed
to optimise the product of efficiency and purity. 

Events with fewer than 2 tracks were rejected.
The charged particle tracks were grouped into two jets using 
the LUCLUS algorithm \cite{ref:mc:pythia_orig}.
Tracks in each jet were 
considered as decay products of a $\tau$-candidate. 
The most energetic charged particle in each jet was called 
{\it the leading charged particle} and its track was called
{\it the leading track}.
The momentum direction of the $\tau$-candidate was approximated
by the vector sum of momenta of its charged decay products.

%The reduced centre-of-mass energy, 
%$\rootsp$, was estimated from momentum directions
%of the two $\tau$ candidates assuming a single ISR photon
%emitted along the beam: 

%\begin{equation}
%\label{eq:tt:sprm}
%\frac{s^{\prime}}{s} = 
%           \frac{\sin{\theta_1}+\sin{\theta_2}-|\sin{(\theta_1+\theta_2)}|}
%                {\sin{\theta_1}+\sin{\theta_2}+|\sin{(\theta_1+\theta_2)}|},
%\end{equation}

%\noindent where $\theta_i$ are the polar angles of the
%tau-candidate momenta. Under the assumption of a single collinear
%ISR photon the $\tau$-candidate momenta can be estimated from: 

%\begin{equation}
%\label{eq:tt:ptau}
%P^{\tau}_i = 2 \cdot \frac{\sin{\theta_j}}
%{\sin{\theta_1}+\sin{\theta_2}+|\sin{(\theta_1+\theta_2)}|}P_{beam}~~,
%\end{equation}

%\noindent where $j$ denotes the $\tau$ candidate opposite to the 
%candidate $i$.

Under the assumption that a single ISR photon is emitted along the
beam direction, the $\tau$-candidate momenta can be estimated
(neglecting the $\tau$ mass) from:
\begin{equation}
\label{eq:tt:ptau}
P^{\tau}_i = 2 \cdot \frac{\sin{\theta_j}}
{\sin{\theta_1}+\sin{\theta_2}+|\sin{(\theta_1+\theta_2)}|}P_{beam}~~,
\end{equation}
where $\theta_i$ are the polar angles of the
tau-candidate momenta, and $j$ denotes the $\tau$ candidate opposite to the 
candidate $i$. The reduced centre-of-mass energy, $\rootsp$, is then
given by:
\begin{equation}
\label{eq:tt:sprm}
\frac{s^{\prime}}{s} = 
           \frac{\sin{\theta_1}+\sin{\theta_2}-|\sin{(\theta_1+\theta_2)}|}
                {\sin{\theta_1}+\sin{\theta_2}+|\sin{(\theta_1+\theta_2)}|}.
\end{equation}

The $\tau$ pairs were selected using the set of cuts listed below:

\begin{itemize}

\item To reject multi-hadronic events the number of selected tracks $N_{ch}$
had to be less than seven: $N_{ch} < 7$. 
For events in which both jets contained more than one track
the invariant mass of each jet (computed using the charged particles,
which were assumed to be pions)
had to be less than 2.5 GeV/$c^2$;

\item The background from WW pairs, ZZ pairs and the $\ee \rightarrow \ee \ffbar$ process
was suppressed by requiring that the acoplanarity 
%(i.e. acolinearity in the r-$\varphi$ plane) %UNCOMMENT IF NECESSARY!
of the two leading tracks 
had to be less than 12 degrees: $\theta_{ACOP} < 12^\circ$;

\item To suppress background from two-photon collisions 
the visible energy (computed as a scalar sum of
charged particle momenta and photon energies) had to be larger than
20\% of the nominal centre-of-mass energy: 
$\Evis > 0.2\sqrt{s}$. The cut value was increased
to $0.225\sqrt{s}$ for $\ee$ and $\mumu$ candidates
(events with exactly two tracks 
both satisfying loose electron identification 
or muon identification criteria);

\item Additional suppression of 
two-photon  background was achieved using
a cut on the transverse momentum P$_T$ (computed as the 
projection onto the r-$\phi$ plane of the vector sum
of all charged particle momenta): P$_T < 2.5~\GeV/c$.
The cut value was increased to $8~\GeV/c$ for $\ee$ and $\mumu$ candidates;

\item The background from $\eemm$ and partially from $\eeee$
events was rejected using the ``radial momentum'' variable,
$\prad   = \sqrt{(p_1/P^\tau_1)^2 + (p_2/P^\tau_2)^2}$, 
where $p_i$ are the momenta of leading charged
particles and $P^\tau_i$ are the $\tau$ momenta 
estimated using eqn.~\ref{eq:tt:ptau}.
The selection cut was $\prad   < 0.95$ for $\mumu$ candidates
and $\prad   < 1.2$ for all other events;

\item The bulk of the $\eeee$ background was
rejected using the depositions in the electromagnetic
calorimeters HPC and FEMC. The ``reduced electromagnetic energies''
$e_i$ were defined for each jet as $e_i = E_i/E^\tau_i$,
where $E_i$ are the total electromagnetic energy depositions
in cones of half angle 30$^\circ$ around the two leading
charged particles and $E^\tau_i$ are the estimated $\tau$ energies, 
computed from the 
estimated momenta $P^{\tau}_i$ of eqn.~\ref{eq:tt:ptau}.
For the events contained within the HPC acceptance
both reduced electromagnetic energies had to satisfy
the condition $e_i < 0.85$. In addition, the distance
between the point with coordinates ($e_1,e_2$) 
and the point (1,1) expected for $\eeee$ events had to be 
more than 0.5. In the forward part of the detector the $\eeee$
background was significantly higher than in the barrel part.
Therefore the selection cuts were slightly tighter for the events
contained within the FEMC acceptance and an additional cut was 
applied: the energy deposited 
beyond the first layer of the HCAL had to be non-zero;

\item The background from cosmic-ray events was rejected using the
impact parameters of the two leading tracks in the r-$\phi$
plane ($R^{IMP}_{1,2}$) and along the z axis ($Z^{IMP}_{1,2}$).
The values of $Z^{IMP}_{1,2}$ had to be not simultaneously larger
than +3 cm or simultaneously smaller than -3 cm and 
at least one of $R^{IMP}_{1}$ or $R^{IMP}_{2}$ had to be less than 3 mm in 
absolute value. In addition, for the two-track events,
the difference $|Z^{IMP}_{1} - Z^{IMP}_{2}|$
had to be less than 3 cm;

\item The remaining background from $\eeee$, $\eemm$ and cosmic-ray events
was reduced by rejecting back-to-back events with very low acollinearity
between the two leading tracks: $\theta_{ACOL} < 0.3^\circ$.

\end{itemize}

After the selection procedure the residual background
level was relatively low (10-20\%) for {\emph{non-radiative}} events and
for {\emph{radiative return}} events. 
However the region of intermediate values of
reconstructed $\rootsp$ was dominated by background from $\ggee$ events.
Therefore all $\ee$ candidate events were rejected in the 
region $\sqrt{1/3} < \sqsps < 0.85$.

\begin{figure}[tbp]
\begin{center}
\begin{tabular}{cc}
 \mbox{\epsfig{file=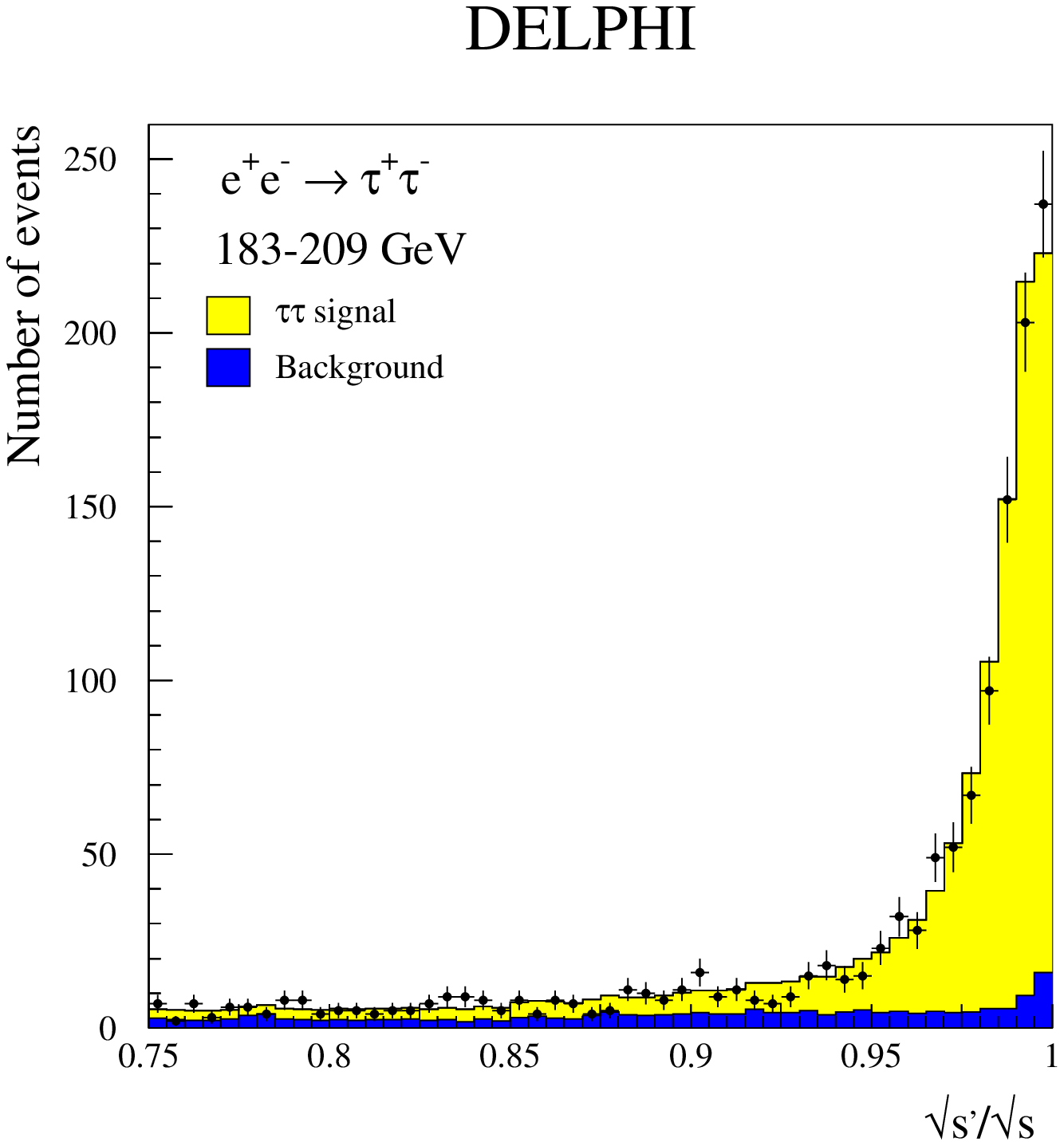,width=0.49\textwidth}} &
  \mbox{\epsfig{file=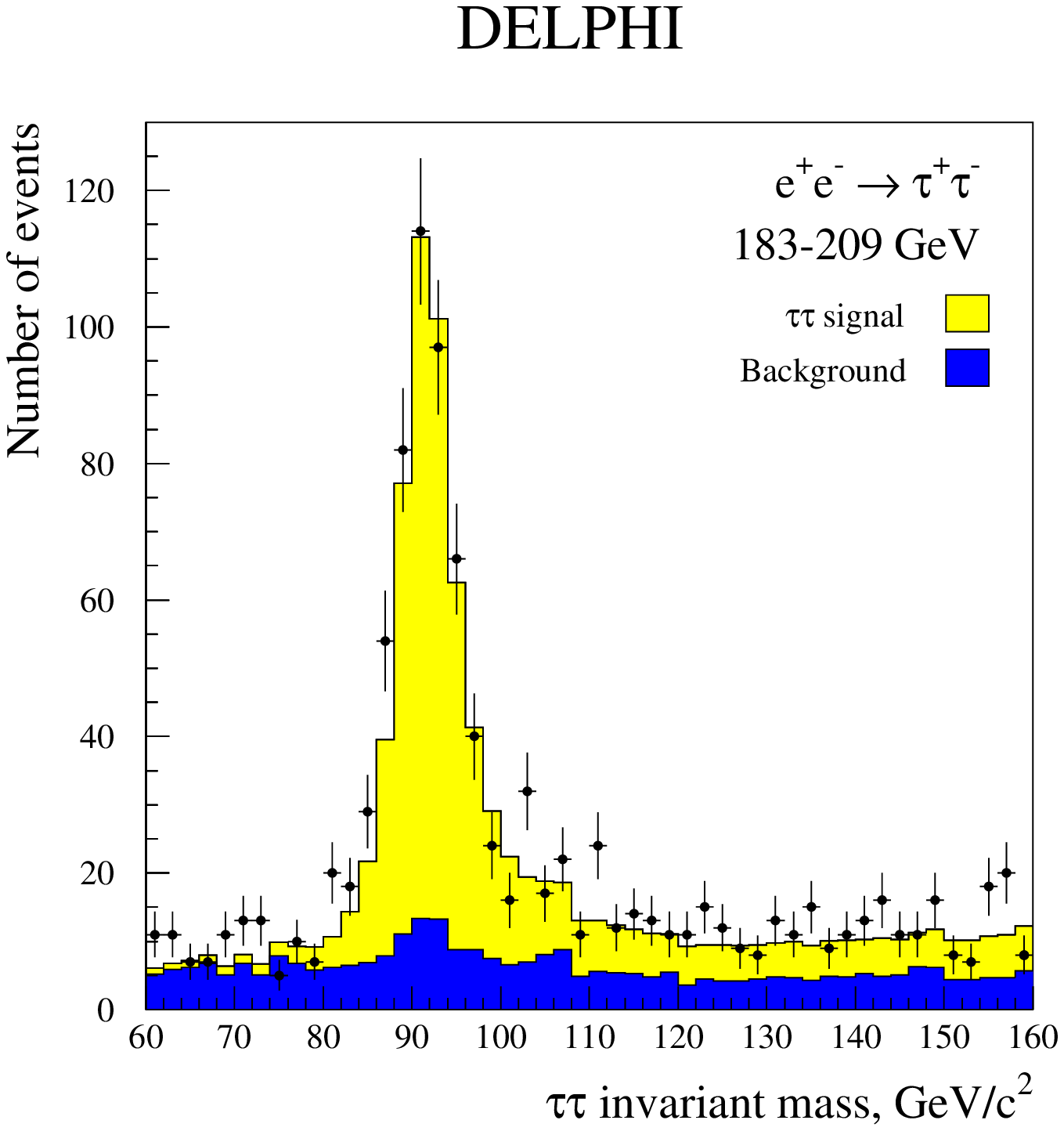,width=0.49\textwidth}} \\
\end{tabular}
\end{center}
\caption{\capsty{The reconstructed \sqsps\ distribution (left) and 
         the reconstructed \rootsp\ distribution (right) for \eett\ events. 
         Both plots show data from 1997-2000.}}
\label{fig:tt:sprm}
\end{figure}

Events were associated to the {\it{non-radiative}} or {\it{inclusive}} 
samples according to the value of $\rootsp$ estimated using 
eqn.~\ref{eq:tt:sprm}. The distribution of the reduced centre-of-mass 
energy, for selected $\tau$ pair events from centre-of-mass energies of 
183 to 207~$\GeV$, is shown in Figure \ref{fig:tt:sprm}. The ratio $\sqsps$ is shown 
for the {\emph{non-radiative}} sample, while for the events with a
hard ISR photon the value of $\rootsp$,
which approximates the invariant mass of the final state leptons, is shown.

%%%%%%%%%%%%%%%%%%%%%%%%%%%%%%%%%%%%%%%
%%%%%%%%%%%%%%%%%%%%%%%%%%%%%%%%%%%%%%%

%% . . . . . . . . . . . . . . . . . . . . . . . . . . . . . . . . . . 
\subsubsection*{Estimation of the selection efficiency and residual background}
\label{sec:tt:calib}

Unlike in the cases of $\eeee$ and $\eemm$, the selection efficiency
of $\eett$ events cannot be estimated or verified from real data.
The estimation of the selection efficiency is entirely based on simulations.
To ensure good agreement between real and simulated data an extensive study
was performed using test samples of real and simulated events.
Where necessary, the simulation was corrected by introducing
calibration constants, smearing distributions, etc.

The ($\theta$-dependent) energy scale and
energy resolution of the electromagnetic
calorimeters were calibrated using well reconstructed $\eeee$ events
selected from high energy runs. Non-linearities were checked
using $\eeee$ events selected from $\Zzero$ runs and $\ggee$ events
at high energies. The corresponding re-scaling and smearing
were applied to electromagnetic energies in simulated events.
Also, a small forward-backward asymmetry of the electromagnetic calorimeters
was found and corrected for.

The absolute scale and resolution of charged-particle momentum
measurements were calibrated using $\eemm$ events from $\Zzero$ runs.
The real data needed a small $\theta$-dependent correction
to the measured 1/$p_T$, while simulated data needed 
a smearing of the 1/$p_T$ distribution. The effect of this calibration
was checked with {\emph{non-radiative}} $\eemm$ events from high energy data.
A rather small (about 2$\sigma$) discrepancy was found and
taken as a systematic uncertainty.

The momentum dependence of the muon chamber efficiency was studied using
test samples of muons selected with very tight HCAL criteria.
The $\eemm$ events from $\Zzero$ and high energy runs were used
to select muons with $\sim$45 GeV/$c$ and $\sim$100 GeV/$c$ momenta, while
$\ggmm$ and $\tau \rightarrow \mu\nu\nu$ events from high energy
runs were used for momentum regions of 0--45 and 45--100 GeV/$c$.
Both the mean efficiency and momentum dependence were corrected
in the simulation. 

Other calibrations included the electron punch-through in the
first layer of the HCAL, the specific energy loss (dE/dx) measurements from the TPC 
and the LEP energy spread. After applying 
all necessary corrections and calibrations the selection 
efficiency was calculated from simulated data as the fraction
of events generated above the nominal $\rootsp$ cut which pass
the selection criteria.

\begin{figure}[tbp]
\begin{center}
\begin{tabular}{cc}
 \mbox{\epsfig{file=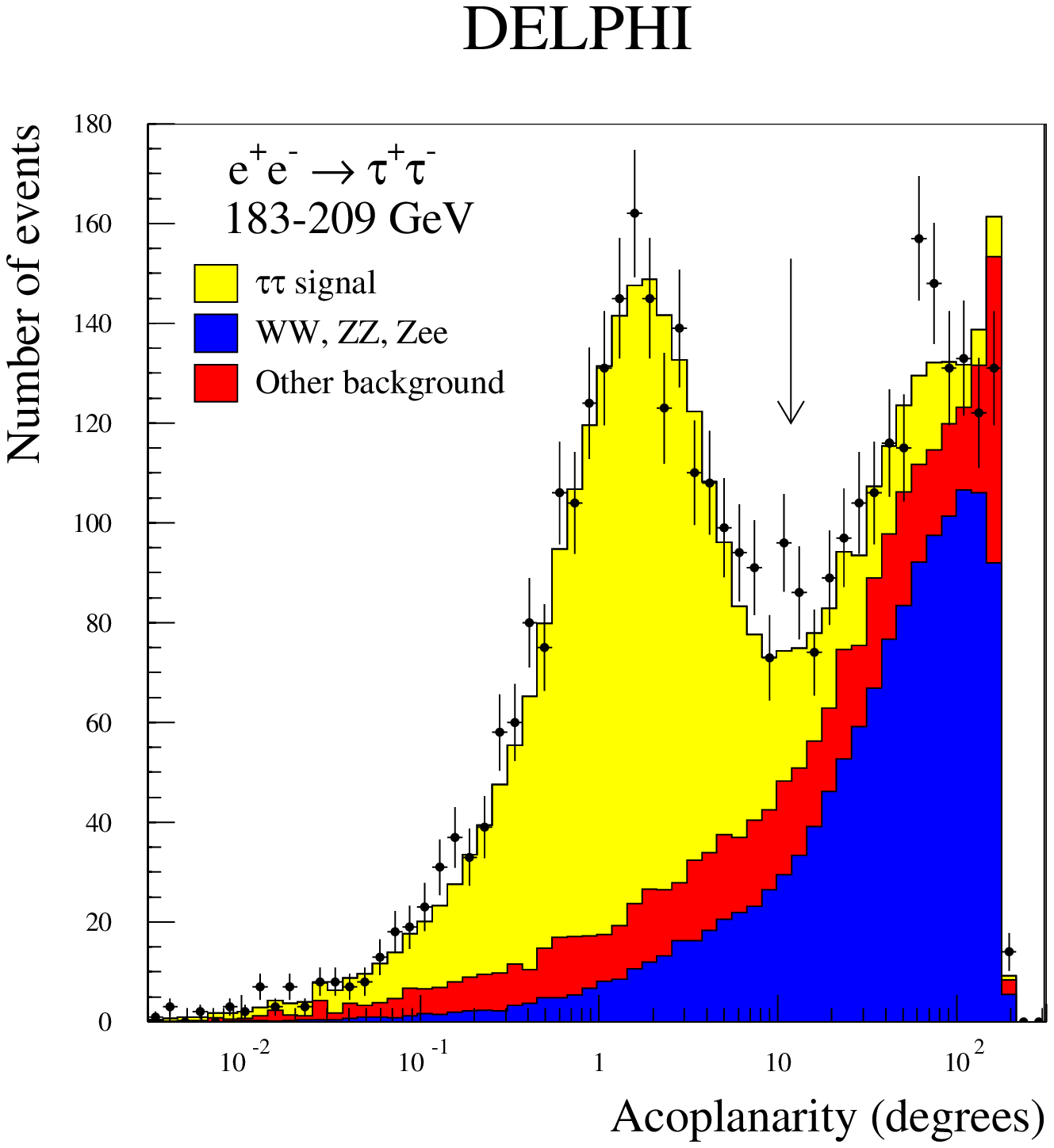,width=0.49\textwidth}} &
  \mbox{\epsfig{file=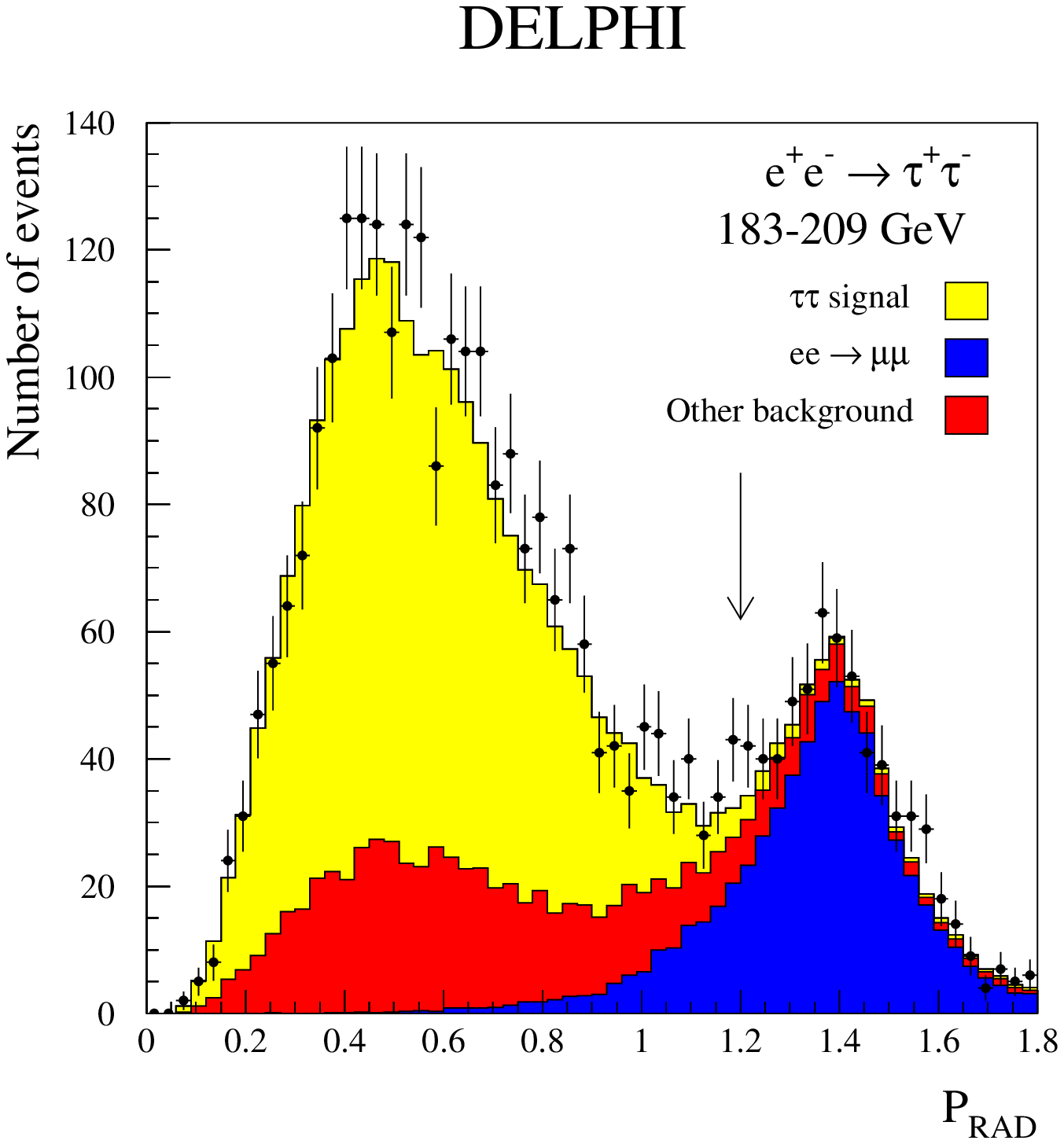,width=0.49\textwidth}} \\
\end{tabular}
\end{center}
\caption{\capsty{The acoplanarity distribution (left) and the distribution 
         of the variable $\prad$ (right). The acoplanarity distribution is 
         sensitive to the four-fermion backgrounds while the $\prad$ 
         distribution is sensitive to the \eemm background.
         Both plots show data from 1997-2000. The dominant backgrounds have 
         been normalised to obtain good agreement between simulation and data 
in the region dominated by background.}}
\label{fig:tt:bkg}
\end{figure}

The residual background level was normalised using the real data. 
For each type of background a selection cut most sensitive 
to this background source was chosen and all other cuts were
applied to the real and simulated data.
The events failing the ``sensitive cut'' were then used 
to normalise the background level predicted by simulation
to the real data. As an example the distributions of
acoplanarity (sensitive to four-fermion background)
and radial momentum (sensitive to $\eemm$) are presented
in Figure~\ref{fig:tt:bkg}.
The full statistics of the 1997-2000 runs are shown.
Arrows show the selection cuts; events to the left of the arrows
were selected as $\tau$ candidates and events to the right
were used for background level normalisation.

Due to this normalisation procedure the simulation
is used only to predict the shapes of distributions,
while the absolute background level is estimated
from the real data. The size of the corrections
applied to the simulation was of the order of 5-15\%.
The estimated residual backgrounds from different channels
are presented in Table~\ref{tab:tt:bg} for {\emph{non-radiative}} and 
{\emph{inclusive}} samples selected from 2000 data. 
For other years the background levels were similar.
The estimation of background uncertainties is
discussed in section \ref{sec:tt:syst}.

The feed-through of \eett\ events from lower values of \sqsps\ into the
{\emph{non-radiative}} sample was estimated from simulated events
to be approximately $4.4\%$, almost independent of \roots.

%%%%%%%%%%%%%%%%%%%%%%%%%%%%%%%%%%%%%%%
%%%%%%%%%%%%%%%%%%%%%%%%%%%%%%%%%%%%%%%

%%%%%%%%%%%%%%%%%%%%%%%%%%%%%%%%%%%%%%%%%%%%%%%%%%%%%%%%
\begin{table}[tp]
\begin{center}
\begin{tabular}{|l|c|c|}
\hline
\multicolumn{3}{|c|}{\eett} \\
\hline
\hline
Background  & {\emph{Non-radiative}} & {\emph{Inclusive}} \\ 
source      & background             & background         \\
\hline\hline
$\eeee$     &   4.50 $\pm$ 0.54      &    3.37 $\pm$ 0.37 \\
$\eemm$     &   1.05 $\pm$ 0.15      &    1.18 $\pm$ 0.16 \\
$\eeqq$     &   0.19 $\pm$ 0.03      &    1.40 $\pm$ 0.14 \\
$\ggtt$     &   0.02 $\pm$ 0.01      &    0.23 $\pm$ 0.05 \\
$\ggee$     &   0.11 $\pm$ 0.05      &    1.89 $\pm$ 0.31 \\
$\ggmm$     &   0.07 $\pm$ 0.04      &    0.23 $\pm$ 0.06 \\
$\ggqq$     &   0.00 $\pm$ 0.04      &    0.10 $\pm$ 0.06 \\
$\eeWW$     &   2.93 $\pm$ 0.50      &    3.15 $\pm$ 0.54 \\
$\eeZZ$     &   1.09 $\pm$ 0.19      &    2.04 $\pm$ 0.35 \\
$\ee \rightarrow \ee\ffbar$    &   2.19 $\pm$ 0.38      &    5.50 $\pm$ 0.94 \\
Cosmic rays &   0.28 $\pm$ 0.22      &    0.24 $\pm$ 0.15 \\
\hline\hline
Total       &  12.43 $\pm$ 0.89      &   19.34 $\pm$ 1.26 \\
\hline
\end{tabular}
\caption{\capsty{Residual background level for 2000 data
(in \% relative to the number of selected $\tau$-pair candidates).}}
\label{tab:tt:bg}
\end{center}
\end{table}
%%%%%%%%%%%%%%%%%%%%%%%%%%%%%%%%%%%%%%%%%%%%%%%%%%%%%%%%

%%%%%%%%%%%%%%%%%%%%%%%%%%%%%%%%%%%%%%%%%%%%%%%%%%%%%%%
\begin{table}[tp]
 {
%%\small
 \begin{center}
 \renewcommand{\arraystretch}{1.2}
 \begin{tabular}{|l|c|c|c|c|c|c|c|c|}
 \hline
 \multicolumn{9}{|c|}{\eett} \\
 \hline
 \hline
                   & \multicolumn{8}{|c|}{Energy point (GeV)} \\
 \cline{2-9}
                   & 183    & 189    & 192    & 196    & 200    & 202    & 205    & 207    \\
 \hline
 \hline
 Energy (GeV)      & 182.66 & 188.63 & 191.58 & 195.51 & 199.51 & 201.64 & 204.89 & 206.56 \\
 Lumi ($\invpbarn$)& 53.11  & 152.67 & 25.14  & 75.99  & 82.65  & 40.40  & 81.49  & 136.39 \\
 No. Events        & 228    & 551    & 79     & 232    & 269    & 119    & 273    & 401    \\
 Efficiency (\%)   & 52.1   & 52.2   & 52.0   & 52.0   & 52.0   & 52.0   & 51.7   & 51.7   \\
% Trigger Efficiency& ??.?   & ??.?   & ??.?   & ??.?   & ??.?   & ??.?   & ??.?   & ??.?   \\
 Background (\%)   & 12.2   & 11.8   & 14.6   & 12.6   & 13.6   & 13.8   & 12.3   & 12.5   \\
% Stat. error ($\%$)& ??.?   & ??.?   & ??.?   & ??.?   & ??.?   & ??.?   & ??.?   & ??.?   \\
% Syst. error ($\%$)& ??.?   & ??.?   & ??.?   & ??.?   & ??.?   & ??.?   & ??.?   & ??.?   \\
 \hline

 \end{tabular}
 \end{center}
 }
 \caption{\capsty{Details of LEP II analysis for \eett\ channel. The
  table shows the actual centre-of-mass energy and luminosity analysed
  at each energy point, the number of events selected in the
  inclusive analysis and the efficiencies for selecting
  events in the non-radiative samples and the backgrounds
  selected in the non-radiative samples.}}
 \label{tab:tt:ana}
\end{table}
%%%%%%%%%%%%%%%%%%%%%%%%%%%%%%%%%%%%%%%%%%%%%%%%%%%%%%%

%% ------------------------------------------------------------------- 
%\pagebreak
\subsubsection{Results}
\label{sec:tt:res}

%%....................................................................
%%.. Summary .........................................................
%%....................................................................
%\fbox{
%\begin{minipage}{0.87\textwidth}
%\begin{itemize}
% \item Table of results with statistical and total systematic errors
%       as in Table \ref{tab:tt:res}.
%\end{itemize}
%\end{minipage}
%}

%%....................................................................
%%.. Contents ........................................................
%%....................................................................

%%%%%%%%%%%%%%%%%%%%%%%%%%%%%%%%%%%%%%%%%%%%%%%%%%%%%%%
\begin{table}[p]
 %%%%%%%%%%%%%%%% tt %%%%%%%%%%%%%%%%
 {\small
 \begin{center}
 \renewcommand{\arraystretch}{1.2}
 \begin{tabular}{|c|c|c|c|c|}
 \hline
 \multicolumn{5}{|c|}{\eett} \\
 \hline
 \hline
 \multicolumn{1}{|c|}{$\roots$} &
 \multicolumn{2}{|c|}{$\rootsp>75$ (GeV)} &
 \multicolumn{2}{|c|}{$\sqsps>0.85$} \\
 \cline{2-5}
 \multicolumn{1}{|c|}{(GeV)} &
 $\stau$ (pb) & $\Afbt$ & $\stau$ (pb) & $\Afbt$ \\
 \hline
 \hline
 130 & 
 $\begin{array}{c}
 22.20 \pm  4.60 \pm  1.56 \\ (20.30)
 \end{array}$ &
 $\begin{array}{c}
 0.310 \pm 0.170 \pm 0.020 \\ (0.337)
 \end{array}$ &
 $\begin{array}{c}
 10.20 \pm  3.10 \pm  0.72 \\ ( 8.31)
 \end{array}$ &
 $\begin{array}{c}
 0.730 \pm 0.170 \pm 0.020 \\ (0.719)
 \end{array}$ \\
 \hline
 136 & 
 $\begin{array}{c}
 17.70 \pm  3.90 \pm  1.24 \\ (17.29)
 \end{array}$ &
 $\begin{array}{c}
 0.260 \pm 0.190 \pm 0.020 \\ (0.338)
 \end{array}$ &
 $\begin{array}{c}
  8.80 \pm  3.00 \pm  0.62 \\ ( 7.17)
 \end{array}$ &
 $\begin{array}{c}
 0.490 \pm 0.230 \pm 0.020 \\ (0.699)
 \end{array}$ \\
 \hline
 161 & 
 $\begin{array}{c}
 11.70 \pm  1.80 \pm  0.82 \\ (10.44)
 \end{array}$ &
 $\begin{array}{c}
 0.390 \pm 0.120 \pm 0.020 \\ (0.332)
 \end{array}$ &
 $\begin{array}{c}
  5.10 \pm  1.20 \pm  0.36 \\ ( 4.54)
 \end{array}$ &
 $\begin{array}{c}
 0.920 \pm 0.080 \pm 0.020 \\ (0.628)
 \end{array}$ \\
 \hline
 172 & 
 $\begin{array}{c}
 11.20 \pm  1.80 \pm  0.79 \\ ( 8.83)
 \end{array}$ &
 $\begin{array}{c}
 0.190 \pm 0.140 \pm 0.020 \\ (0.329)
 \end{array}$ &
 $\begin{array}{c}
  4.50 \pm  1.10 \pm  0.32 \\ ( 3.89)
 \end{array}$ &
 $\begin{array}{c}
 0.130 \pm 0.200 \pm 0.020 \\ (0.610)
 \end{array}$ \\
 \hline
 183 & 
 $\begin{array}{c}
  8.73 \pm  0.70 \pm  0.20 \\ ( 7.64)
 \end{array}$ &
 $\begin{array}{c}
 0.400 \pm 0.074 \pm 0.012 \\ (0.326)
 \end{array}$ &
 $\begin{array}{c}
  3.29 \pm  0.38 \pm  0.07 \\ ( 3.39)
 \end{array}$ &
 $\begin{array}{c}
 0.671 \pm 0.080 \pm 0.012 \\ (0.596)
 \end{array}$ \\
 \hline
 189 & 
 $\begin{array}{c}
  7.23 \pm  0.38 \pm  0.17 \\ ( 7.08)
 \end{array}$ &
 $\begin{array}{c}
 0.448 \pm 0.047 \pm 0.012 \\ (0.324)
 \end{array}$ &
 $\begin{array}{c}
  3.11 \pm  0.22 \pm  0.07 \\ ( 3.15)
 \end{array}$ &
 $\begin{array}{c}
 0.697 \pm 0.048 \pm 0.011 \\ (0.589)
 \end{array}$ \\
 \hline
 192 & 
 $\begin{array}{c}
  6.16 \pm  0.89 \pm  0.15 \\ ( 6.83)
 \end{array}$ &
 $\begin{array}{c}
 0.415 \pm 0.134 \pm 0.012 \\ (0.324)
 \end{array}$ &
 $\begin{array}{c}
  2.50 \pm  0.48 \pm  0.06 \\ ( 3.04)
 \end{array}$ &
 $\begin{array}{c}
 0.578 \pm 0.150 \pm 0.011 \\ (0.586)
 \end{array}$ \\
 \hline
 196 & 
 $\begin{array}{c}
  5.97 \pm  0.51 \pm  0.14 \\ ( 6.52)
 \end{array}$ &
 $\begin{array}{c}
 0.156 \pm 0.080 \pm 0.012 \\ (0.323)
 \end{array}$ &
 $\begin{array}{c}
  2.89 \pm  0.30 \pm  0.06 \\ ( 2.91)
 \end{array}$ &
 $\begin{array}{c}
 0.465 \pm 0.083 \pm 0.011 \\ (0.582)
 \end{array}$ \\
 \hline
 200 & 
 $\begin{array}{c}
  6.50 \pm  0.50 \pm  0.15 \\ ( 6.22)
 \end{array}$ &
 $\begin{array}{c}
 0.226 \pm 0.073 \pm 0.012 \\ (0.322)
 \end{array}$ &
 $\begin{array}{c}
  2.61 \pm  0.27 \pm  0.06 \\ ( 2.78)
 \end{array}$ &
 $\begin{array}{c}
 0.540 \pm 0.080 \pm 0.011 \\ (0.578)
 \end{array}$ \\
 \hline
 202 & 
 $\begin{array}{c}
  5.74 \pm  0.68 \pm  0.14 \\ ( 6.08)
 \end{array}$ &
 $\begin{array}{c}
 0.314 \pm 0.110 \pm 0.012 \\ (0.321)
 \end{array}$ &
 $\begin{array}{c}
  2.55 \pm  0.38 \pm  0.06 \\ ( 2.72)
 \end{array}$ &
 $\begin{array}{c}
 0.464 \pm 0.122 \pm 0.011 \\ (0.576)
 \end{array}$ \\
 \hline
 205 & 
 $\begin{array}{c}
  6.94 \pm  0.52 \pm  0.16 \\ ( 5.86)
 \end{array}$ &
 $\begin{array}{c}
 0.434 \pm 0.070 \pm 0.012 \\ (0.320)
 \end{array}$ &
 $\begin{array}{c}
  2.80 \pm  0.28 \pm  0.06 \\ ( 2.62)
 \end{array}$ &
 $\begin{array}{c}
 0.709 \pm 0.068 \pm 0.011 \\ (0.574)
 \end{array}$ \\
 \hline
 207 & 
 $\begin{array}{c}
  5.95 \pm  0.38 \pm  0.14 \\ ( 5.76)
 \end{array}$ &
 $\begin{array}{c}
 0.347 \pm 0.060 \pm 0.012 \\ (0.320)
 \end{array}$ &
 $\begin{array}{c}
  2.53 \pm  0.21 \pm  0.06 \\ ( 2.58)
 \end{array}$ &
 $\begin{array}{c}
 0.666 \pm 0.059 \pm 0.011 \\ (0.572)
 \end{array}$ \\
 \hline
 \end{tabular}
 \end{center}
 }
 %%%%%%%%%%%%%%%%%%%%%%%%%%%%%%%%%%%%

 \caption{\capsty{Measured cross-sections and forward-backward asymmetries
          for {\it{inclusive}} and {\it{non-radiative}} \eett\ events. The 
first uncertainty is statistical, the second systematic.
          Numbers in brackets are the theoretical predictions of ZFITTER, which are estimated to have a precision of $\pm 0.4\%$ on $\stau$ and $\pm0.004$ on $\Afbt$. }}
 \label{tab:tt:res}
\end{table}
%%%%%%%%%%%%%%%%%%%%%%%%%%%%%%%%%%%%%%%%%%%%%%%%%%%%%%%

%% . . . . . . . . . . . . . . . . . . . . . . . . . . . . . . . . . . 
\subsubsection*{Cross-sections}
\label{sec:tt:cross}

The total cross-sections of $\tau$ pair production for the 
{\it{non-radiative}} and {\it{inclusive}} samples were calculated as follows:

\begin{equation}
\sigma_{\tau\tau} = \frac{(N_{sel} - N_{bg})(1 - f)}
                         {\epsilon {\mathcal L}}.
\end{equation}

\noindent Here ${\mathcal L}$ is the total integrated luminosity for
the corresponding energy point, $N_{sel}$ is the number of events
selected in the corresponding sample, $f$ (applicable to the
{\it{non-radiative}} cross-section only) is the fraction of
feed-through events estimated from the $\tau$-pair simulation and
$\epsilon$ is the selection efficiency, computed within the full
$4\pi$ acceptance, which takes into account the effect of selection
cuts (including the cut on $\rootsp$) and the trigger efficiency.  The
trigger efficiency was estimated from the real data using the
redundancy of the DELPHI trigger system. Its typical value was about
99.9\% with typical uncertainty of about 0.05\%. For the 2000 data
taken with one TPC sector broken the trigger efficiency value was
98.9$\pm$0.9\% (for the whole detector).  The values of the selection
efficiency are summarised in Table~\ref{tab:tt:ana}.  The residual
number of background events $N_{bg}$ was estimated from simulation and
normalised to the real data as discussed in the previous section. The
measured cross-sections are presented in Table~\ref{tab:tt:res} and
shown in Figure~\ref{fig:ana:sig-cmp}.

%% . . . . . . . . . . . . . . . . . . . . . . . . . . . . . . . . . . 
\subsubsection*{Forward-backward asymmetries}
\label{sec:tt:afb}

%%%%%%%%%%%%%%%%%%%%%%%%%%%%%%%%%%%%%%%%%%%%%%%%%%%%%%%
\begin{table}[p]
%%%%%%%%%%%%%%%% tt %%%%%%%%%%%%%%%%
%%%%%%%%%%%%%%%% tt %%%%%%%%%%%%%%%% This from John 28/01/04
 {\footnotesize
 \begin{center}
 \setlength{\tabcolsep}{1.0mm}
 \begin{tabular}{ccc}
 \multicolumn{3}{r}{\fbox{\hspace{7.12cm}
 $\eett$
 \hspace{7.12cm}}} \\
 \\
 \renewcommand{\arraystretch}{1.2}
 \begin{tabular}
 {|@{[}r@{,}r@{]}|c|r@{$\pm$}
 c@{(}c@{)}@{$\pm$}c|}
 \hline
\multicolumn{7}{|c|}{$\roots \sim 183$} \\
 \hline
 \hline
 \multicolumn{2}{|c|}{} &
 \multicolumn{5}{|c|}{$\dsdcth$ (pb)} \\
 \cline{3-7}
 \multicolumn{2}{|c|}{$\cos\theta$} &
 \multicolumn{1}{|c|}{SM} &
 \multicolumn{4}{|c|}{Measurement} \\
 \hline
 \hline
 -0.96 & -0.80 &  0.48 & -0.17 &  0.00 &  0.50 &  0.03 \\
 -0.80 & -0.60 &  0.49 &  0.54 &  0.37 &  0.36 &  0.03 \\
 -0.60 & -0.40 &  0.59 &  0.42 &  0.28 &  0.32 &  0.02 \\
 -0.40 & -0.20 &  0.78 &  0.44 &  0.28 &  0.36 &  0.02 \\
 -0.20 &  0.00 &  1.08 &  1.22 &  0.48 &  0.46 &  0.04 \\
  0.00 &  0.20 &  1.47 &  1.51 &  0.53 &  0.52 &  0.05 \\
  0.20 &  0.40 &  1.97 &  1.86 &  0.55 &  0.55 &  0.06 \\
  0.40 &  0.60 &  2.58 &  2.30 &  0.61 &  0.64 &  0.07 \\
  0.60 &  0.80 &  3.30 &  3.76 &  0.94 &  0.89 &  0.12 \\
  0.80 &  0.96 &  4.08 &  4.77 &  1.40 &  1.31 &  0.15 \\
 \hline
 \end{tabular}
 &
 \renewcommand{\arraystretch}{1.2}
 \begin{tabular}
 {|@{[}r@{,}r@{]}|c|r@{$\pm$}
 c@{(}c@{)}@{$\pm$}c|}
 \hline
\multicolumn{7}{|c|}{$\roots \sim 189$} \\
 \hline
 \hline
 \multicolumn{2}{|c|}{} &
 \multicolumn{5}{|c|}{$\dsdcth$ (pb)} \\
 \cline{3-7}
 \multicolumn{2}{|c|}{$\cos\theta$} &
 \multicolumn{1}{|c|}{SM} &
 \multicolumn{4}{|c|}{Measurement} \\
 \hline
 \hline
 -0.96 & -0.80 &  0.47 &  0.19 &  0.20 &  0.27 &  0.03 \\
 -0.80 & -0.60 &  0.47 &  0.13 &  0.14 &  0.21 &  0.03 \\
 -0.60 & -0.40 &  0.56 &  0.35 &  0.16 &  0.18 &  0.03 \\
 -0.40 & -0.20 &  0.73 &  0.74 &  0.21 &  0.21 &  0.03 \\
 -0.20 &  0.00 &  1.00 &  0.78 &  0.24 &  0.26 &  0.04 \\
  0.00 &  0.20 &  1.36 &  1.78 &  0.34 &  0.30 &  0.04 \\
  0.20 &  0.40 &  1.83 &  2.31 &  0.35 &  0.31 &  0.05 \\
  0.40 &  0.60 &  2.39 &  2.59 &  0.37 &  0.36 &  0.06 \\
  0.60 &  0.80 &  3.06 &  2.61 &  0.48 &  0.51 &  0.11 \\
  0.80 &  0.96 &  3.78 &  3.14 &  0.67 &  0.73 &  0.13 \\
 \hline
 \end{tabular}
 &
 \renewcommand{\arraystretch}{1.2}
 \begin{tabular}
 {|@{[}r@{,}r@{]}|c|r@{$\pm$}
 c@{(}c@{)}@{$\pm$}c|}
 \hline
\multicolumn{7}{|c|}{$\roots \sim 192$} \\
 \hline
 \hline
 \multicolumn{2}{|c|}{} &
 \multicolumn{5}{|c|}{$\dsdcth$ (pb)} \\
 \cline{3-7}
 \multicolumn{2}{|c|}{$\cos\theta$} &
 \multicolumn{1}{|c|}{SM} &
 \multicolumn{4}{|c|}{Measurement} \\
 \hline
 \hline
 -0.96 & -0.80 &  0.46 &  0.60 &  0.74 &  0.69 &  0.02 \\
 -0.80 & -0.60 &  0.46 &  0.22 &  0.41 &  0.52 &  0.03 \\
 -0.60 & -0.40 &  0.54 &  0.45 &  0.41 &  0.45 &  0.01 \\
 -0.40 & -0.20 &  0.71 &  0.44 &  0.41 &  0.50 &  0.02 \\
 -0.20 &  0.00 &  0.97 &  0.85 &  0.60 &  0.63 &  0.03 \\
  0.00 &  0.20 &  1.32 &  1.48 &  0.76 &  0.72 &  0.03 \\
  0.20 &  0.40 &  1.76 &  1.77 &  0.77 &  0.77 &  0.04 \\
  0.40 &  0.60 &  2.30 &  1.67 &  0.77 &  0.88 &  0.05 \\
  0.60 &  0.80 &  2.95 &  2.57 &  1.18 &  1.25 &  0.07 \\
  0.80 &  0.96 &  3.65 &  0.25 &  0.74 &  1.75 &  0.09 \\
 \hline
 \end{tabular}
 \\ \\
 \renewcommand{\arraystretch}{1.2}
 \begin{tabular}
 {|@{[}r@{,}r@{]}|c|r@{$\pm$}
 c@{(}c@{)}@{$\pm$}c|}
 \hline
\multicolumn{7}{|c|}{$\roots \sim 196$} \\
 \hline
 \hline
 \multicolumn{2}{|c|}{} &
 \multicolumn{5}{|c|}{$\dsdcth$ (pb)} \\
 \cline{3-7}
 \multicolumn{2}{|c|}{$\cos\theta$} &
 \multicolumn{1}{|c|}{SM} &
 \multicolumn{4}{|c|}{Measurement} \\
 \hline
 \hline
 -0.96 & -0.80 &  0.45 &  0.35 &  0.34 &  0.39 &  0.01 \\
 -0.80 & -0.60 &  0.45 &  0.36 &  0.27 &  0.29 &  0.03 \\
 -0.60 & -0.40 &  0.52 &  1.02 &  0.33 &  0.25 &  0.01 \\
 -0.40 & -0.20 &  0.68 &  0.73 &  0.29 &  0.28 &  0.02 \\
 -0.20 &  0.00 &  0.92 &  1.31 &  0.41 &  0.35 &  0.03 \\
  0.00 &  0.20 &  1.25 &  1.03 &  0.37 &  0.40 &  0.03 \\
  0.20 &  0.40 &  1.68 &  1.38 &  0.40 &  0.43 &  0.04 \\
  0.40 &  0.60 &  2.19 &  2.24 &  0.50 &  0.50 &  0.05 \\
  0.60 &  0.80 &  2.81 &  2.85 &  0.71 &  0.70 &  0.07 \\
  0.80 &  0.96 &  3.48 &  2.92 &  0.91 &  0.98 &  0.08 \\
 \hline
 \end{tabular}
 &
 \renewcommand{\arraystretch}{1.2}
 \begin{tabular}
 {|@{[}r@{,}r@{]}|c|r@{$\pm$}
 c@{(}c@{)}@{$\pm$}c|}
 \hline
\multicolumn{7}{|c|}{$\roots \sim 200$} \\
 \hline
 \hline
 \multicolumn{2}{|c|}{} &
 \multicolumn{5}{|c|}{$\dsdcth$ (pb)} \\
 \cline{3-7}
 \multicolumn{2}{|c|}{$\cos\theta$} &
 \multicolumn{1}{|c|}{SM} &
 \multicolumn{4}{|c|}{Measurement} \\
 \hline
 \hline
 -0.96 & -0.80 &  0.44 &  0.32 &  0.32 &  0.36 &  0.01 \\
 -0.80 & -0.60 &  0.44 &  0.81 &  0.35 &  0.28 &  0.03 \\
 -0.60 & -0.40 &  0.51 &  0.50 &  0.23 &  0.24 &  0.01 \\
 -0.40 & -0.20 &  0.65 &  0.49 &  0.24 &  0.27 &  0.02 \\
 -0.20 &  0.00 &  0.88 &  0.98 &  0.35 &  0.33 &  0.03 \\
  0.00 &  0.20 &  1.20 &  0.93 &  0.34 &  0.38 &  0.03 \\
  0.20 &  0.40 &  1.60 &  1.96 &  0.44 &  0.41 &  0.04 \\
  0.40 &  0.60 &  2.09 &  1.86 &  0.44 &  0.47 &  0.05 \\
  0.60 &  0.80 &  2.68 &  2.20 &  0.61 &  0.66 &  0.07 \\
  0.80 &  0.96 &  3.32 &  2.89 &  0.87 &  0.92 &  0.08 \\
 \hline
 \end{tabular}
 &
 \renewcommand{\arraystretch}{1.2}
 \begin{tabular}
 {|@{[}r@{,}r@{]}|c|r@{$\pm$}
 c@{(}c@{)}@{$\pm$}c|}
 \hline
\multicolumn{7}{|c|}{$\roots \sim 202$} \\
 \hline
 \hline
 \multicolumn{2}{|c|}{} &
 \multicolumn{5}{|c|}{$\dsdcth$ (pb)} \\
 \cline{3-7}
 \multicolumn{2}{|c|}{$\cos\theta$} &
 \multicolumn{1}{|c|}{SM} &
 \multicolumn{4}{|c|}{Measurement} \\
 \hline
 \hline
 -0.96 & -0.80 &  0.44 & -0.13 &  0.00 &  0.52 &  0.01 \\
 -0.80 & -0.60 &  0.43 &  0.83 &  0.50 &  0.39 &  0.03 \\
 -0.60 & -0.40 &  0.50 &  0.42 &  0.31 &  0.34 &  0.01 \\
 -0.40 & -0.20 &  0.64 &  0.60 &  0.36 &  0.38 &  0.02 \\
 -0.20 &  0.00 &  0.86 &  1.54 &  0.61 &  0.47 &  0.03 \\
  0.00 &  0.20 &  1.17 &  0.65 &  0.42 &  0.54 &  0.03 \\
  0.20 &  0.40 &  1.56 &  2.46 &  0.70 &  0.58 &  0.03 \\
  0.40 &  0.60 &  2.04 &  2.01 &  0.65 &  0.66 &  0.05 \\
  0.60 &  0.80 &  2.62 &  1.11 &  0.69 &  0.93 &  0.07 \\
  0.80 &  0.96 &  3.24 &  2.28 &  1.12 &  1.30 &  0.08 \\
 \hline
 \end{tabular}
 \\ \\
 \renewcommand{\arraystretch}{1.2}
 \begin{tabular}
 {|@{[}r@{,}r@{]}|c|r@{$\pm$}
 c@{(}c@{)}@{$\pm$}c|}
 \hline
\multicolumn{7}{|c|}{$\roots \sim 205$} \\
 \hline
 \hline
 \multicolumn{2}{|c|}{} &
 \multicolumn{5}{|c|}{$\dsdcth$ (pb)} \\
 \cline{3-7}
 \multicolumn{2}{|c|}{$\cos\theta$} &
 \multicolumn{1}{|c|}{SM} &
 \multicolumn{4}{|c|}{Measurement} \\
 \hline
 \hline
 -0.96 & -0.80 &  0.43 & -0.12 &  0.00 &  0.36 &  0.01 \\
 -0.80 & -0.60 &  0.42 &  0.48 &  0.23 &  0.27 &  0.02 \\
 -0.60 & -0.40 &  0.48 &  0.44 &  0.23 &  0.23 &  0.01 \\
 -0.40 & -0.20 &  0.62 &  0.53 &  0.26 &  0.26 &  0.01 \\
 -0.20 &  0.00 &  0.83 &  0.48 &  0.28 &  0.33 &  0.02 \\
  0.00 &  0.20 &  1.13 &  1.29 &  0.39 &  0.37 &  0.02 \\
  0.20 &  0.40 &  1.50 &  1.30 &  0.37 &  0.40 &  0.03 \\
  0.40 &  0.60 &  1.97 &  2.25 &  0.49 &  0.46 &  0.04 \\
  0.60 &  0.80 &  2.52 &  2.87 &  0.66 &  0.64 &  0.06 \\
  0.80 &  0.96 &  3.13 &  4.99 &  1.08 &  0.90 &  0.07 \\
 \hline
 \end{tabular}
 & &
 \renewcommand{\arraystretch}{1.2}
 \begin{tabular}
 {|@{[}r@{,}r@{]}|c|r@{$\pm$}
 c@{(}c@{)}@{$\pm$}c|}
 \hline
\multicolumn{7}{|c|}{$\roots \sim 207$} \\
 \hline
 \hline
 \multicolumn{2}{|c|}{} &
 \multicolumn{5}{|c|}{$\dsdcth$ (pb)} \\
 \cline{3-7}
 \multicolumn{2}{|c|}{$\cos\theta$} &
 \multicolumn{1}{|c|}{SM} &
 \multicolumn{4}{|c|}{Measurement} \\
 \hline
 \hline
 -0.96 & -0.80 &  0.43 &  0.30 &  0.23 &  0.27 &  0.01 \\
 -0.80 & -0.60 &  0.42 &  0.00 &  0.15 &  0.21 &  0.02 \\
 -0.60 & -0.40 &  0.48 &  0.50 &  0.18 &  0.18 &  0.01 \\
 -0.40 & -0.20 &  0.61 &  0.60 &  0.20 &  0.20 &  0.01 \\
 -0.20 &  0.00 &  0.82 &  0.62 &  0.22 &  0.25 &  0.02 \\
  0.00 &  0.20 &  1.11 &  1.66 &  0.32 &  0.28 &  0.02 \\
  0.20 &  0.40 &  1.48 &  1.61 &  0.32 &  0.31 &  0.03 \\
  0.40 &  0.60 &  1.93 &  1.54 &  0.32 &  0.35 &  0.04 \\
  0.60 &  0.80 &  2.48 &  1.72 &  0.42 &  0.49 &  0.06 \\
  0.80 &  0.96 &  3.07 &  4.32 &  0.81 &  0.69 &  0.07 \\
 \hline
 \end{tabular}
 \\ \\
 \end{tabular}
 \end{center}
 }
%%%%%%%%%%%%%%%%%%%%%%%%%%%%%%%%%%%%
%%%%%%%%%%%%%%%%%%%%%%%%%%%%%%%%%%%%

 \caption{\capsty{Differential cross-sections for non-radiative
         \eett events at centre-of-mass energies from 183 to 207 GeV. The 
          tables show the bins, the predictions of the Standard Model 
          and the measurements. The errors quoted are the statistical and 
          experimental systematic errors. The statistical errors are shown as 
          the measured errors and, in brackets, the expected errors, computed 
          from the square root of the observed and expected numbers 
          of events respectively.}}

 \label{tab:tt:diff}
\end{table}
%%%%%%%%%%%%%%%%%%%%%%%%%%%%%%%%%%%%%%%%%%%%%%%%%%%%%%%

The forward-backward asymmetry of $\tau$ pair production was 
calculated as

\begin{equation}
A^{\tau\tau}_{FB} = 
\frac{\sigma_F - \sigma_B}{\sigma_F + \sigma_B}
\end{equation}

The forward cross-section $\sigma_F$ ($0^\circ<\theta_{\tau^-}<90^\circ$)
and the backward cross-section $\sigma_B$ 
($90^\circ<\theta_{\tau^-}<180^\circ$)
were determined in a similar way to the determination
of the total cross-section $\sigma_{\tau\tau}$.
The $\tau$ charge and the direction of the $\tau$ momentum 
were estimated from the charges and momenta of its
charged decay products. 
To reduce charge misidentification,
only events with at least one jet consisting of exactly one 
charged particle ({\emph{1-N topology}} events)
were used in this analysis. 
The charge of that single particle was assigned
to the corresponding $\tau$ and an opposite charge
was assigned to the other $\tau$ regardless of the charge of the multi-track
jet.
%Unlike \eemm\ events where the efficiency for correctly identifying the
%tracks with the wrongly assigned charge is high, the efficiency in the
%\eett\ channel is low, due to additional sources of charge misidentification
%in this channel compared to \eemm, such as tracks being unreconstructed. 
The {\emph{1-1 topology}} events with two reconstructed tracks of the same 
charge ({\emph{like-sign events}}) were excluded from the analysis. 
For the period of broken TPC sector in 2000 the tracks
detected only in the VD and ID were not used for the charge identification.

Misidentification of the $\tau$ charge results in
a reduction of the absolute value of $A_{FB}$.
The probability of charge misidentification
was estimated from the real data using the rate of the like-sign events.
The corresponding correction to $A_{FB}$ was found to be
$+0.0068 \pm 0.0034$ (accidentally the same for {\emph{non-radiative}}
and  {\emph{inclusive}} samples). The corrected values of $A_{FB}$
are summarised in Table~\ref{tab:tt:res} and shown in 
Figure~\ref{fig:ana:afb-cmp}.

\subsubsection*{Differential cross-sections}
\label{sec:tt:xdif}

In addition to the measurements of the forward-backward asymmetries,
differential cross-sections, $\dsdcth$ for $\tau$-pair production were 
determined.
The region of angular acceptance of this analysis 
($-0.96<\cos{\theta}<0.96$) was divided into ten bins
with respect to the polar angle of the momentum direction of
the negatively charged tau lepton. The cross-section
for every bin was determined similarly to the
determination of the total cross-section and the correction
for the charge misidentification effect was applied.
The measured differential cross-sections are presented in
Table~\ref{tab:tt:diff}.

%% . . . . . . . . . . . . . . . . . . . . . . . . . . . . . . . . . . 
\subsubsection*{Systematic errors}
\label{sec:tt:syst}

%%%%%%%%%%%%%%%%%%%%%%%%%%%%%%%%%%%%%%%%%%%%%%%%%%%%%%%
\begin{table}[tp]
\begin{center}
\begin{tabular}{|l|c|c||c|c|}
\hline
\multicolumn{5}{|c|}{\eett} \\
\hline
\hline
       &$\Delta\sigma/\sigma$ &$\Delta\sigma/\sigma$&
         $\Delta A_{FB}$ &$\Delta A_{FB}$\\
Source & (non-rad.) & (inclus.)&(non-rad.)&(inclus.) \\
\hline\hline
Track Selection         & 72 &  35&  34&  48   \\
Event Selection         &127 & 104&  51&  59   \\
Detector Calibration    & 91 &  86&  58&  50   \\
Background Level        &102 & 157&  39&  43   \\
Light pair subtraction  &  9 &  12&   4&   3   \\
Trigger Efficiency      &  9 &   9&   0&   0   \\
$\rootsp$ Reconstruction& 25 &   3&  14&   1   \\
Feed-through            & 32 &   2&  17&   1   \\
Charge Misidentification&  0 &   0&  34&  34   \\
Simulation Statistics   & 21 &  19&  22&  22   \\
QED                     &  7 &  14&   1&  17   \\
Tau Polarisation        & 33 &  14&  16&  23   \\
Tau branchings          &  6 &  11&   0&   0   \\
Luminosity              & 55 &  55&   -&   -   \\
\hline\hline                                   
Total uncertainty       &216 & 220& 105& 113   \\
\hline                                         
Correlated              &208 & 213&  97& 106   \\
Uncorrelated            & 59 &  53&  41&  39   \\
\hline\hline
TPC sector instability  &127 &  90&  41&  16   \\
\hline
\end{tabular}
\caption{\capsty{Systematic uncertainties for the 2000 data. 
         All numbers in units of 10$^{-4}$. The total uncertainties do not 
         include the error due to TPC sector instability which applies only 
         to a part of the 2000 data. The correlated error component includes 
         errors correlated between energies and channels and those correlated 
         with other LEP experiments.}}
\label{tab:tt:syst}
\end{center}
\end{table}
%%%%%%%%%%%%%%%%%%%%%%%%%%%%%%%%%%%%%%%%%%%%%%%%%%%%%%%

A breakdown of the systematic uncertainties for cross-sections
and asymmetries of {\emph{non-radiative}} and {\emph{inclusive}} samples 
is presented in Table~\ref{tab:tt:syst}. The numbers
are given for the results of 2000; for other years
the uncertainties were of a similar size.

The most important sources of systematic uncertainty 
were the choice of track- and event-selection cuts, 
background level and detector calibration. 
To estimate the effect of track selection every selection cut
was varied in a wide range and the full analysis chain 
was repeated. If the observed change of result exceeded
the expected statistical fluctuation (1$\sigma$) then the quadratic
difference between the change and the expected fluctuation
was taken as the systematic error,
otherwise no systematic error was assigned.
%In other case the full
%observed change of the result was taken as a conservative estimate of 
%the systematic error.
The largest uncertainty came 
from the impact parameters, $Z^{IMP}_{1,2}$, and from the combination of 
detectors which was used in the track fit. 

The influence of event-selection cuts was checked in a similar way.
Every event-selection cut was varied typically by the
experimental resolution on the corresponding variable.
The largest contribution came from the cuts on $\Evis$,
$\prad$ and electromagnetic energy. To reduce the effect of statistical
fluctuations the full statistics of 1997-2000 were used to estimate the 
uncertainty from track and event selections and therefore
these systematic errors were common and correlated for the data of
different years.

The uncertainty of the detector calibration was mainly
due to the limited statistics of test samples selected
from real data of $\Zzero$ and high energy runs
to estimate various corrections to the detector
simulation. This part of the uncertainty was uncorrelated
between different years.
The second part of the uncertainty was due to residual disagreements 
in high energy data between real and simulated distributions used for
calibration (correlated between different years). The largest contribution
to the detector calibration uncertainty was the calibration
of the HPC response.

The uncertainty of the residual background level consisted of three parts:
the precision of the event generators (correlated);
the statistics of the simulated background events (uncorrelated); 
and the uncertainty
of the background level normalisation procedure.
The normalisation uncertainty was determined from the statistics 
of real data events used for the normalisation (uncorrelated)
and from residual data-simulation disagreements in the 
distributions of variables 
sensitive to particular background types (correlated).
The uncertainties of the background from different sources are presented 
in Table~\ref{tab:tt:bg}.

Other (smaller) sources of systematic uncertainties were: statistics
of simulated signal events; QED uncertainties computed by changing the
order of perturbation theory to which QED corrections are included in
the \KK\ generator; subtraction of $\tau$ pairs accompanied by light
pair production; reconstruction of $\rootsp$; estimation of the
feed-through event fraction; beam energy spread; trigger efficiency;
world averages of $\tau$-decay branching ratios; and $\tau$-polarisation 
at LEP II energies.

An uncertainty specific to the measurements of the forward-backward 
asymmetries and differential cross-sections originated from the 
estimation of the charge misidentification probability.
The systematic errors on the total {\emph{non-radiative}} cross-section
were assigned also to the differential cross-section.
A correlation between the $\cos{\theta}$ bins was assumed 
for all sources except the statistics of simulated signal
and simulated background. The charge misidentification 
uncertainty was assumed to be anti-correlated between the 
bins with $\cos{\theta}<0$ and $\cos{\theta}>0$.

A relatively small additional systematic uncertainty for the period
of the broken TPC sector came from $\rootsp$ reconstruction,
trigger efficiency and charge identification. It was taken
into account in the procedure of averaging of the two periods
of 2000.

%% . . . . . . . . . . . . . . . . . . . . . . . . . . . . . . . . . . 
\subsubsection*{Cross-check with Z$^0$ data}

For an additional cross-check of the data quality we have used 
the data taken in 1997-2000 during the LEP
runs near the peak of the Z resonance. After the usual run selection
the total integrated luminosity was about 11 pb$^{-1}$. 
Tau pairs were selected from this sample using a set of cuts
very similar to the ones used for the high-energy analysis,
and the total cross-section and forward-backward asymmetry
were calculated using the same technique as for the high energies.

For the combined 1997-2000 sample the statistical uncertainties
were $\Delta \sigma/\sigma = \pm 1.2\%$ for the cross-section
and $\Delta A_{FB} = \pm 0.012$ for the asymmetry estimation.
Both values are similar to, or below, the systematic errors 
of the measurements at high energy.

The measured value of the forward-backward asymmetry was 
in good agreement with the Standard Model expectation,
while the cross-section was found to be 1.1 standard deviations
below the expected value (only the statistical error was taken into account).
The year-by-year studies also did not show any significant
departure from the expectations. These cross-checks give additional
confidence that the measurements performed at the high energies
are valid within the quoted uncertainties.

%%--------------------------------------------------------------------
%%-- ANALYSIS OF qq FINAL STATES -------------------------------------
%%--------------------------------------------------------------------
\subsection{$\boldmath{\qqbar}$ final states}
\label{sec:qq}

   The analysis of the $\eeqqg$ process followed closely the 
procedure applied previously to the data collected near 183 
and 189 GeV and published in~\cite{ref:delphiff:183-189}. 
It benefited however from several improvements. In particular, 
more accurate generators were used for the simulation of the 
signal and background final states, and the enlarged Monte Carlo data 
sample allowed an improved understanding and control of 
residual differences between real and simulated distributions.
 
   These improvements were implemented in the analysis of the 
latest data sets (collected above 190 GeV) and stimulated a
reanalysis of the published data collected near 183 and 189 GeV. 
The magnitude of the changes as a result of these improvements 
did not justify reanalysing the data collected at 130-172 GeV 
~\cite{ref:delphiff:130-172}, which have larger statistical 
uncertainties.

   Two cross-sections were computed at each collision energy:
an {\em inclusive} cross-section corresponding to reduced 
centre-of-mass energies larger than 10\% of the collision 
energy, and a {\em non-radiative} cross-section corresponding 
to reduced centre-of-mass energies exceeding 85\% of the 
collision energy.

%\section{$\eeqqg$}

%% ------------------------------------------------------------------- 
%% ------------------------------------------------------------------- 
%% ------------------------------------------------------------------- 
\subsubsection{Analysis}
\label{sec:qq:analysis}

%% ------------------------------------------------------------------- 
\subsubsection*{Run selection}
\label{sec:qq:run}

  The cross-section computations are based on data samples 
collected during running periods where the subdetectors of 
prime importance for the event selection were close enough 
to their nominal operating conditions. A small fraction of the
data collected was discarded for this reason, mainly due to 
inefficiencies of the TPC, of the forward chambers (FCA-FCB) 
or of the FEMC. They correspond to 
about 1.4\% of the total integrated luminosity.

\subsubsection*{Track selection}
\label{sec:qq:track}

%%%%%%%%%%%%%%%%%%%%%%%%%%%%%%%%%%%%%%%
\begin{figure}[tbp]
\begin{center}
\begin{tabular}{cc}
 \mbox{\epsfig{file=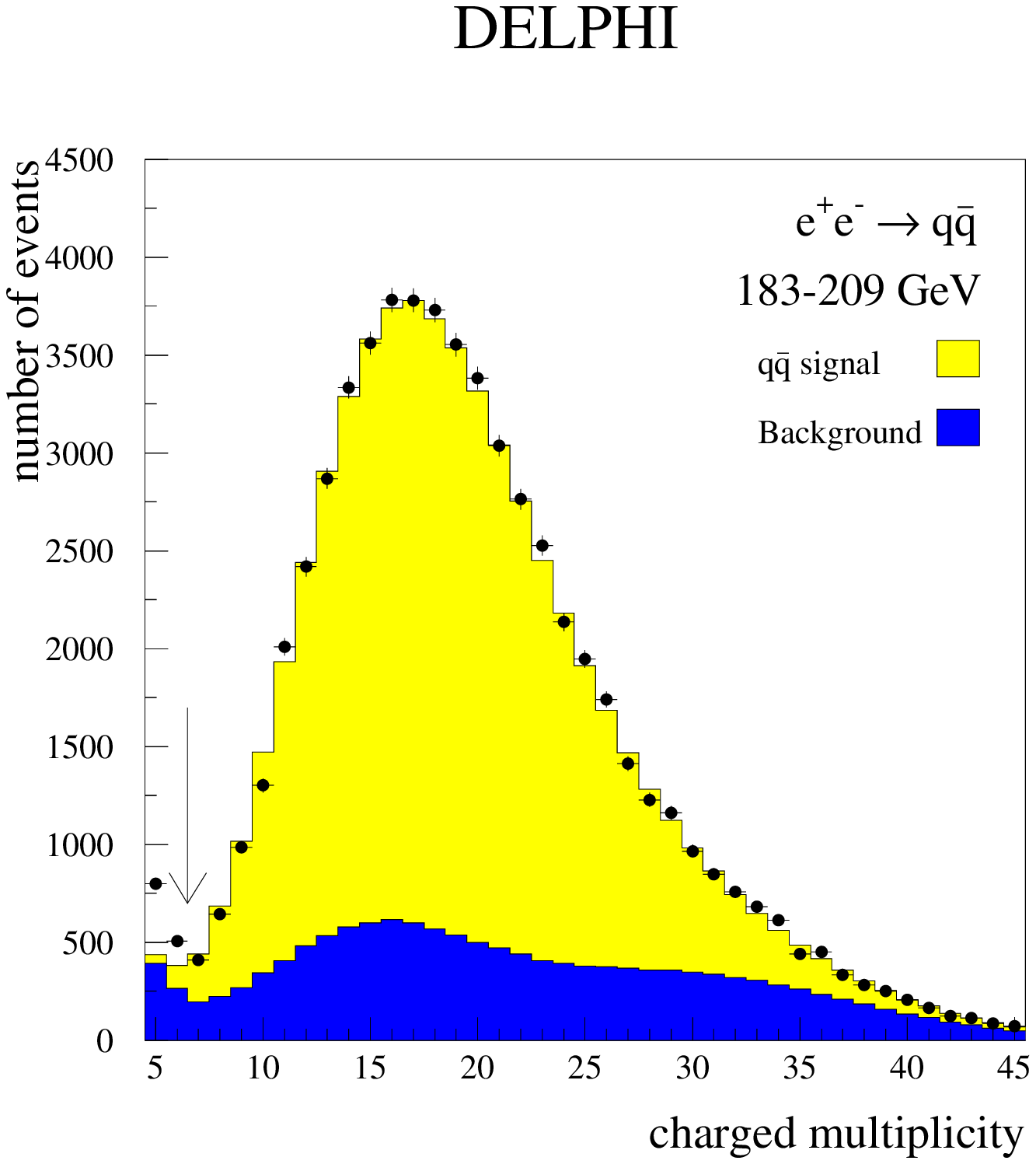,width=0.49\textwidth}} &
  \mbox{\epsfig{file=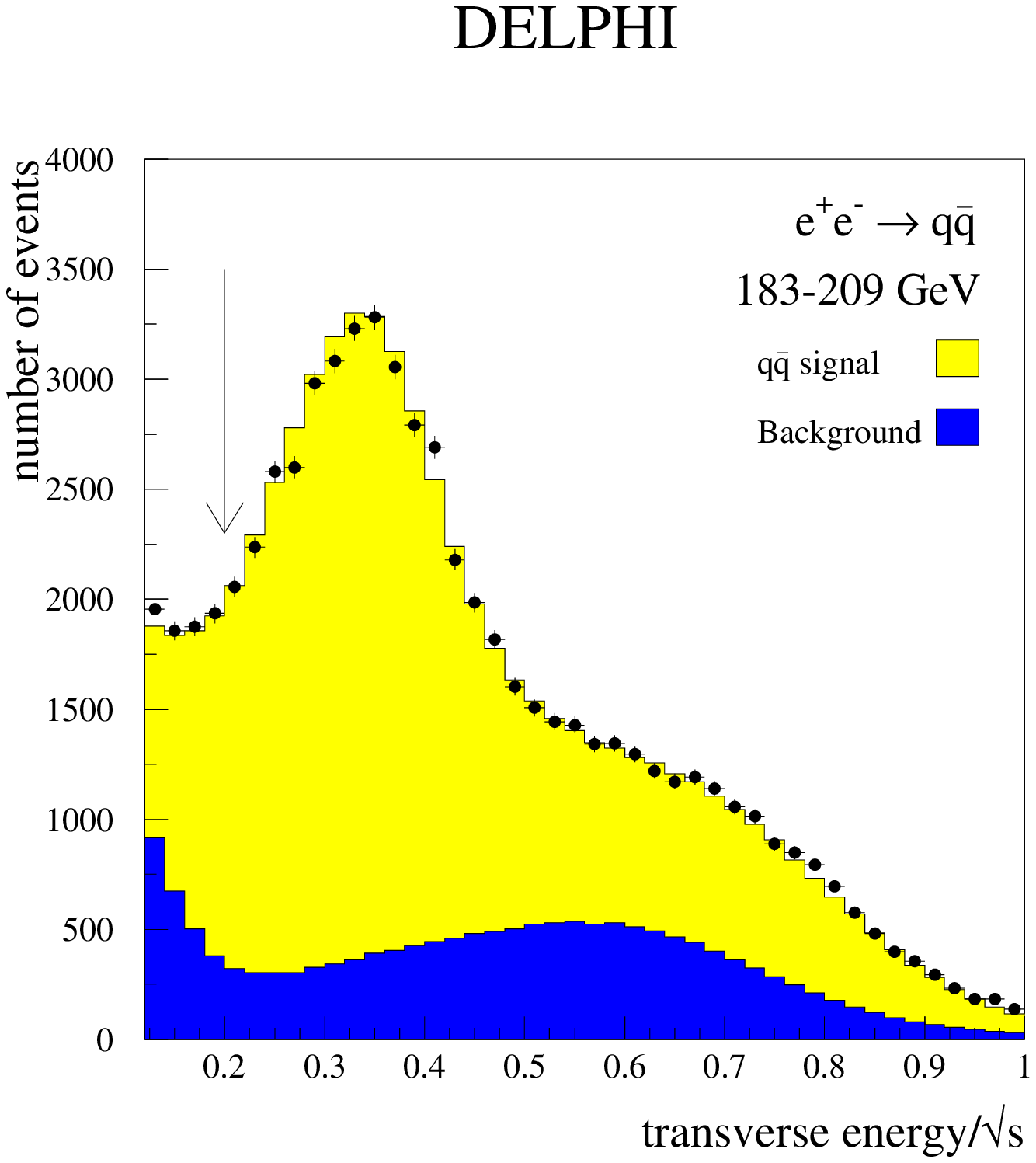,width=0.49\textwidth}} \\
\end{tabular}
\end{center}
\caption{\capsty{Charged multiplicity (left) and 
         transverse energy (right) distributions for \eeqq\ events. 
         The arrows indicate the cut values applied. The distributions 
         combine data collected from 1997 to 2000.}}
\label{fig:qq:nch}
\end{figure}
%%%%%%%%%%%%%%%%%%%%%%%%%%%%%%%%%%%%%%%

   The selection of $\qqbar$ events relied mainly on charged particles
for which the tracks were required to originate from the vertex
(i.e. their transverse and longitudinal impact parameters with respect
to the nominal interaction point had to be below 4 cm and 10 cm
respectively), to have a momentum greater than 400 MeV/$c$ measured
with a resolution better than 100\%, and to have a reconstructed track
length of at least 30 cm. For a small fraction of particle
trajectories which were only reconstructed in the VD and ID, the
length was required to exceed only 10 cm.

%% ------------------------------------------------------------------- 
\subsubsection*{Event selection}
\label{sec:qq:event}

   The event selection was mainly oriented towards an efficient 
rejection of the backgrounds due to low multiplicity events and 
two-photon collisions. For this purpose, final states were accepted 
if they contained at least 7 selected tracks and if the total energy
carried by the selected tracks exceeded 10\% of the collision energy.  
In order to suppress the contamination by two-photon 
collisions further, the event total transverse energy was computed, based 
on the transverse momentum of each selected charged particle and on the 
electromagnetic showers reconstructed in the HPC and the FEMC
with a shower energy above 500 MeV. Events were rejected if their 
total transverse energy was below 20\% of the collision energy. 
%%GM adds this...otherwise Fig7 is not referenced
Figure~\ref{fig:qq:nch} shows the distributions of charged multiplicity and 
transverse energy for candidate \eeqq\ events combined over the years 
1997 to 2000.
%%%
Finally, a large fraction of the residual Bhabha events was discarded 
by requiring the quadratic sum of the total energy 
(i.e. $E_{rad}=\sqrt{E_{\rm F}^2 +E_{\rm B}^2}$)
seen in each end-cap of the FEMC (computed with showers of at 
least 1 GeV) to be less than 90\% of the beam energy. 
The sample of events satisfying the whole set of selection criteria 
described previously was thus mainly contaminated by four-fermion 
final states.  

   The selection of {\emph{non-radiative}} final states included the 
additional requirement that the reduced centre-of-mass energy was 
larger than 85\% of the collision energy (i.e. $\rootsp\,>\,0.85\,\roots$).

%% ------------------------------------------------------------------- 
\subsubsection*{$\boldmath{\rootsp}$ determination}
\label{sec:qq:sprime}

%%%%%%%%%%%%%%%%%%%%%%%%%%%%%%%%%%%%%%%
\begin{figure}[tbp]
\begin{center}
\begin{tabular}{cc}
 \mbox{\epsfig{file=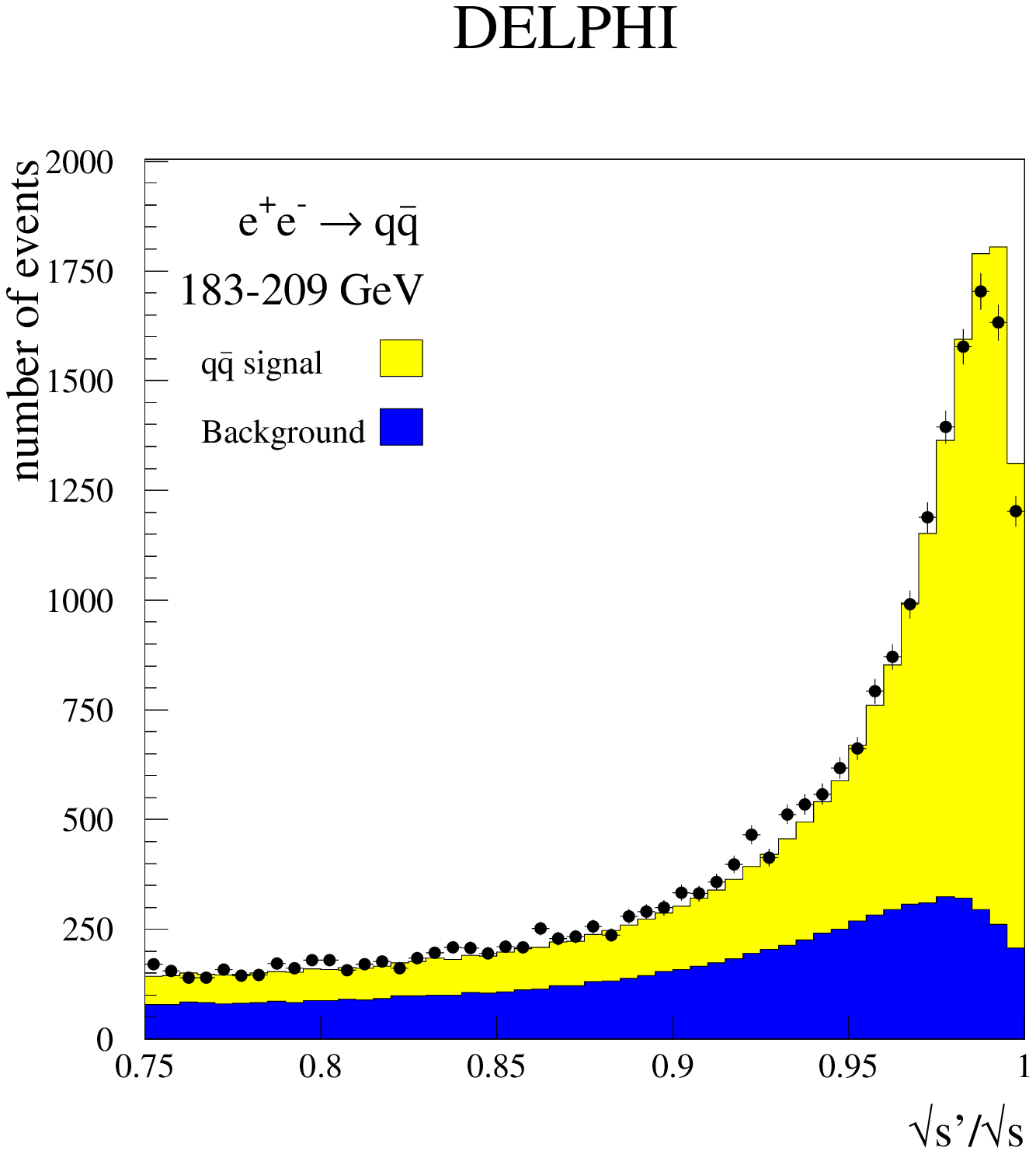,width=0.49\textwidth}} &
  \mbox{\epsfig{file=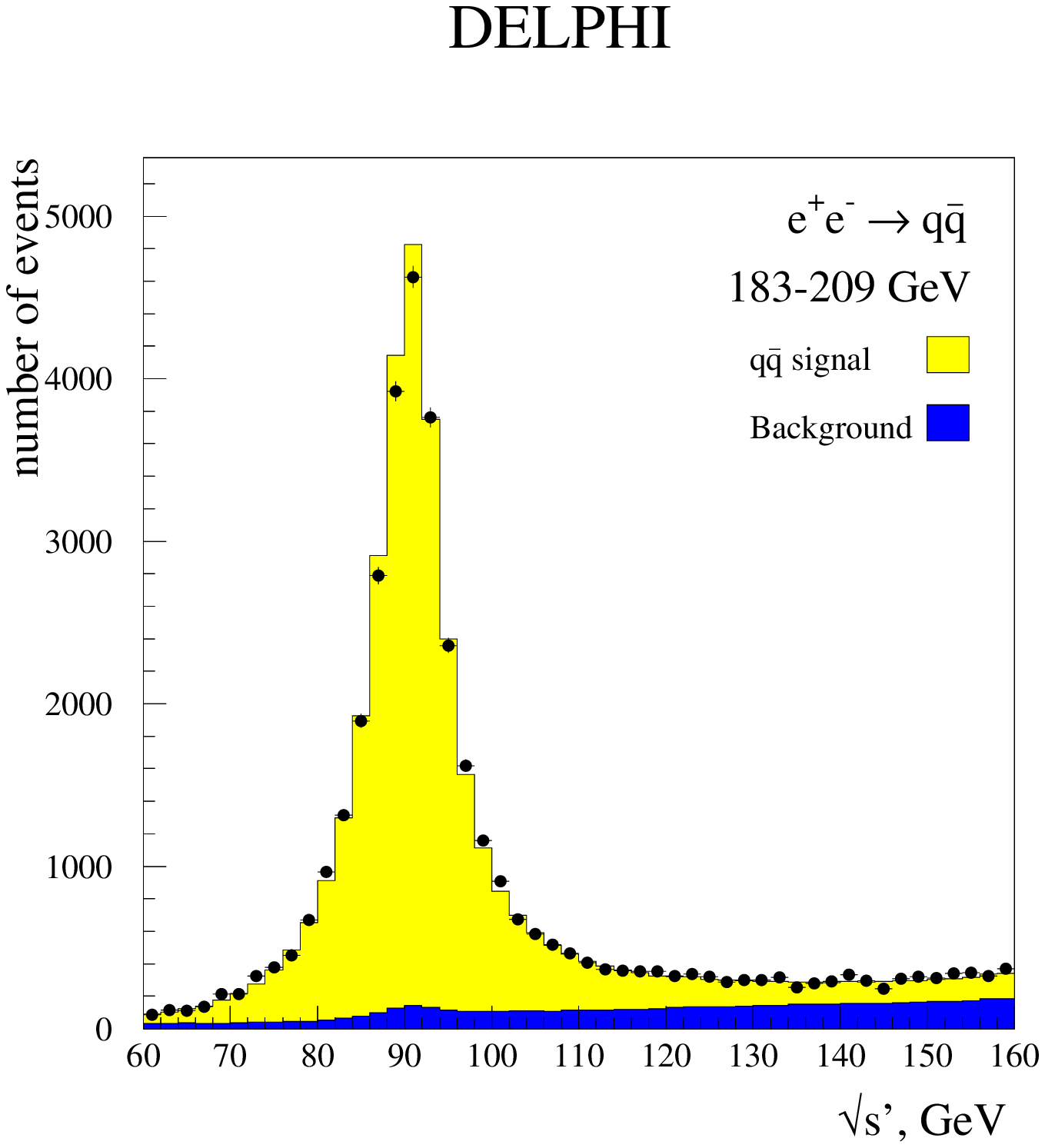,width=0.49\textwidth}} \\
\end{tabular}
\end{center}
\caption{\capsty{Reconstructed \sqsps\ (left) and \rootsp\ (right)
         distributions for \eeqq\ events. Data are from 1997-2000.}}
\label{fig:qq:sprm}
\end{figure}
%%%%%%%%%%%%%%%%%%%%%%%%%%%%%%%%%%%%%%%

   The determination of the reduced energy of each event ($\rootsp$)
was mainly based on hadronic jets. A first step consisted in 
reconstructing jets according to the DURHAM clustering 
algorithm~\cite{ref:ana:durham}, using the selected charged 
particles as well as the neutral particles with energy larger 
than 1.5~GeV. A constrained fit 
was then performed, imposing energy and momentum conservation 
and assuming that an ISR photon was emitted along the beam lines. 
The free parameters of the fit were the jet directions and the ISR 
photon energy ($E_{\gamma}$). The value of $\rootsp$ was then 
derived from the fitted value of $E_{\gamma}$ according to the 
following expression: $s' = s - 2 E_{\gamma} \roots$. 

  The quality of the agreement between the real and simulated 
distributions of $\rootsp$, which is essential for an accurate 
determination of the {\emph{non-radiative}} event selection 
efficiency, was reasonable, as illustrated in Figure~\ref{fig:qq:sprm}. 
The differences observed for $\rootsp$ values close to $\roots$ 
and to the Z mass reflect residual discrepancies between the 
real and simulated $\rootsp$ resolutions. They have minor
consequences on the analysis, which is mainly sensitive to 
differences occurring near the cut value of $0.85\,\roots$.  

%% ------------------------------------------------------------------- 
\subsubsection*{Estimation of selection efficiency and backgrounds}
\label{sec:qq:eff}

%%
%% NB.
%%  - trigger efficiency taken to be >99.99%
%%

%%%%%%%%%%%%%%%%%%%%%%%%%%%%%%%%%%%%%%%
\begin{table}[tp]
 {
%%\small
 \begin{center}
 \renewcommand{\arraystretch}{1.2}
 \begin{tabular}{|l|c|c|c|c|c|c|c|c|}
 \hline
 \multicolumn{9}{|c|}{\eeqq} \\
 \hline
 \hline
                   & \multicolumn{8}{|c|}{Energy point (GeV)} \\
 \cline{2-9}
                   & 183   & 189   & 192   & 196   & 200   & 202   & 205   & 207    \\
 \hline
 \hline
 Energy (GeV)        & 182.7 & 188.6 & 191.6 & 195.5 & 199.5 & 201.6 & 204.9 & 206.5 \\
 Lumi ($\invpbarn$)&  53.5 & 155.0 &  25.1 &  76.1 &  83.0 &  40.3 &  81.9 & 136.9 \\
 No. Events &  5859 &  15582 &   2433 &  7241 &  7452 &  3573 &  6792 &  10982  \\
 Efficiency ($\%$)   & 92.3  & 92.3  & 92.4  & 92.3  & 92.5  & 92.4  & 92.1  &  92.2 \\
 Background ($\%$)   &  26.7  & 30.3  & 31.2  & 32.3  & 34.4  & 35.2  & 36.6  &  37.6 \\
% Stat. uncertainty ($\%$)&  3.1  &  2.0  &  5.1  &  3.0  &  3.0  &  4.5  &  3.3  &   2.6  \\
% Syst. uncertainty ($\%$)&  1.7  &  1.8  &  1.9  &  1.9  &  1.9  &  2.0  &  2.0  &   2.0  \\
 \hline

% - inclusive  No. Events        &  5859  & 15582  &  2433  &  7241  &  7452  &  3573  &  6792  &  10982 \\

 \end{tabular}
 \end{center}
 }
 \caption{\capsty{The table 
  shows the actual centre-of-mass energy and luminosity analysed at each energy
  point, the number of events selected in the inclusive samples
  and the efficiencies and backgrounds for the non-radiative 
  selection.}}
 \label{tab:qq:ana}
\end{table}
%%%%%%%%%%%%%%%%%%%%%%%%%%%%%%%%%%%%%%%

   The selection efficiency was derived at each collision energy 
from the simulated sample of signal events satisfying the criteria 
above. The sample was generated with the KK program~\cite{ref:mc:kk}, the event hadronisation being performed with the 
PYTHIA algorithm~\cite{ref:mc:pythia,ref:mc:pythia_orig}. 
The selection efficiency values obtained 
for the {\emph{inclusive}} sample varied from about 84\% to 
78\% for increasing values of the collision energy. It remained 
near 92\% at all collision energies for the {\emph{non-radiative}} 
sample. Values of the selection efficiency at various centre-of-mass energies 
may be 
found in Table~\ref{tab:qq:ana}. 
The fraction of events generated below 85\% of $\roots$ and 
reconstructed above the cut value represented less than 10\% 
of the {\emph{non-radiative}} sample.

   The residual background events contaminating the selected
{\emph{non-radiative}} ({\emph{inclusive}}) sample originated mainly
from WW pairs, which amounted to more than 90\% (70\%) of the total
background. The {\emph{inclusive}} sample contained also significant
contributions from \eeZZ, $\ee \rightarrow \ee\ffbar$ and two-photon
collision events. Other backgrounds, like those containing lepton
pairs of opposite charge, amounted only to a few percent of the total
residual background.

\begin{figure}[t]
\begin{center}
\vspace*{-1.8cm}
 \mbox{\epsfig{file=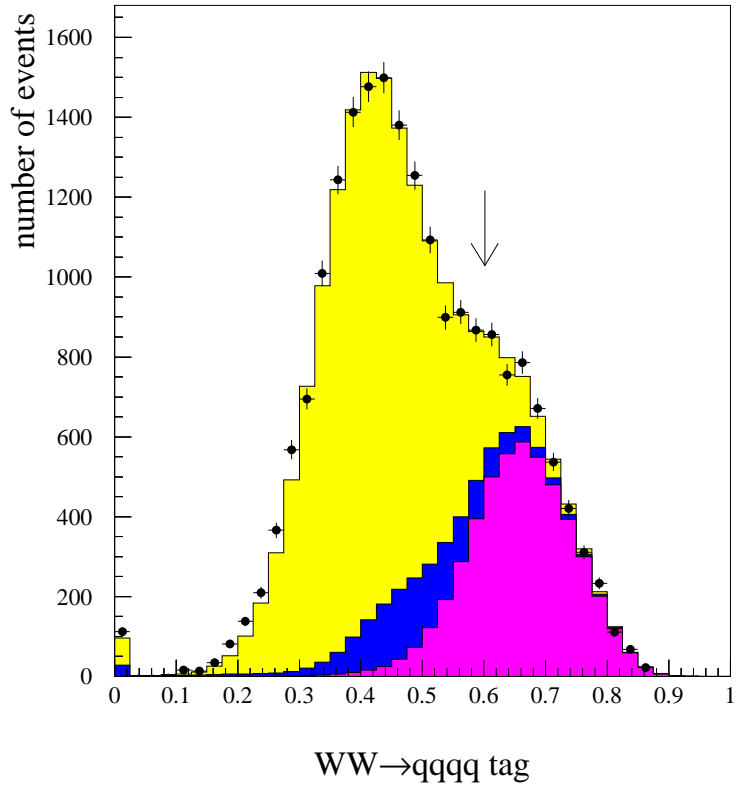,width=0.7\textwidth}} 
\end{center}
\caption{\capsty{Distribution of the multi-variable selection 
parameter used to tag WW events in the {\emph{non-radiative}} 
sample. The light area represents the signal simulation, 
the darker area stands for the $WW\rightarrow qqqq$ 
background, and the darkest area displays the 
remaining backgrounds. 
The distributions combine data collected from 1997 to 2000.}}
\label{fig:qq:ww}
\end{figure}

   In order to assess the magnitude of the residual contamination 
of the {\emph{non-radiative}} sample by WW events, which accounted
for up to 1/3 of the sample size, a dedicated study of the accuracy 
of the WW generator program (WPHACT) was performed. 
Variables reflecting the event shapes were 
%%%combined in a 
multiplied to create a
multi-variable selection parameter (see Figure~\ref{fig:qq:ww}), 
which was 
applied to the data in order to reject a substantial fraction of 
the background. 
The variables included: the energies of the most and least energetic jets; 
the minimal interjet angle; the broadening of the narrowest jet; the
value of the clusterization parameter of the LUCLUS~\cite{ref:mc:pythia_orig}
algorithm at which the event changed from 3 to 4 jets.   
About 60-70\% of the residual WW events 
were discarded in this way, while
the signal selection efficiency dropped by a few percent.
The cross-sections computed before and after applying these 
additional rejection criteria differed by only small amounts (0.2-0.3 pb), 
fully compatible with the statistical uncertainty 
affecting the comparisons. 
 
%%%%%%%%%%%%%%%%%%%%%%%%%%%%%%%%%%%%%%%
%%%%%%%%%%%%%%%%%%%%%%%%%%%%%%%%%%%%%%%

\subsubsection*{Forward region simulation}
\label{sec:qq:forw}

    The simulated rate of tracks belonging to multi-hadronic final 
states and reconstructed in the forward regions of the detector 
(i.e. with a polar angle between 9$^{\circ}$ and 30$^{\circ}$ or 
between 150$^{\circ}$ and 171$^{\circ}$)  
underestimated the observed rate by up to 10\%. This deficit was 
regarded as a consequence of the approximations made in the  
description of the forward material of the detector at the simulation 
level.

   In order to account for the deficit, an artificial inefficiency was 
applied to the real tracks reconstructed in the forward regions. It 
consisted in discarding randomly some of these tracks, with 
weights extracted 
from the ratio between the real and simulated track polar angle
distributions in four different momentum bins. These corrections 
were extracted from the calibration data collected each year near 
the Z resonance peak and applied to the high-energy data of 
the same year. 
The corresponding change of the {\emph{inclusive}} and 
{\emph{non-radiative}} cross-sections was $-1.2$\% on average.
The associated systematic uncertainty was estimated as
one third of the change (i.e. 0.4\%). 

%In the previously published analysis of 183~GeV and 189~GeV 
%data a similar excess was also present but of a significantly 
%smaller size.

   An alternative treatment was used as a cross-check. It 
consisted in assigning weights to the simulated forward tracks 
in order to reproduce the excess of the real ones. This approach 
resulted in values of the {\emph{inclusive}} cross-section well 
consistent with those obtained with the previous method, which 
was however preferred for the cross-section determination 
as it was significantly simpler to implement in the $\rootsp$ 
reconstruction algorithm. 

\subsubsection*{Impact of the TPC instability in the year 2000}
\label{sec:qq:s6}
   
   The failure of one of the TPC sectors in the year 2000 translated 
into a small loss in the reconstruction efficiency for tracks traversing 
the sector affected, which was estimated to occur in about 40\% of 
multihadronic events. According to the event simulation, the corresponding 
drop of the event selection efficiency for the {\emph{inclusive}} 
sample was moderate ($-$0.6\%). Several cross-checks were 
performed in order to test how accurately the simulation reproduced 
the modified track and jet reconstructions.
While the rate of real tracks was reproduced quite well, residual 
differences were found between real and simulated jet parameters,
which were accounted for by a specific systematic uncertainty. The 
latter was estimated as $\pm0.1\%$ for the {\emph{inclusive}} sample selection 
efficiency and $\pm0.5\%$ for the {\emph{non-radiative}} sample selection 
efficiency.

\subsubsection*{Subtraction of interference between ISR and FSR}
\label{sec:qq:ifi}

%\fbox{
%\begin{minipage}{0.87\textwidth}
%\begin{itemize}
% \item Explain how IFI was subtracted
% \item NB. for the purposes of making calculations for these corrections
%       $\rootsp$ was taken as \Mff.
%\end{itemize}
%\end{minipage}
%}

Computations performed with the ZFITTER package showed 
that the inclusion of the ISR-FSR interference in the computation of the 
{\emph{non-radiative}} cross-section diminishes its magnitude by almost 
$0.5\%$. 

Since the \KK\ generator used for the simulation of multi-hadronic final 
states did not account for interference between initial and final 
state photon radiation, and since the expected change in the cross-section 
is not uniform, peaking at $|\cos\theta|=1$, where the efficiency is smallest,
the quark polar angle distribution was reweighted
by the relative difference in the differential cross-sections computed 
from ZFITTER, with and without ISR-FSR interference. The change in the 
selection efficiency was computed, and found to be negligible. Thus 
the measured {\emph{non-radiative}}
cross-sections were corrected by the full $0.5\%$ correction.

%%%%%%%%%%%%%%%%%%%%%%%%%%%%%%%%%%%%%%%

%% ------------------------------------------------------------------- 
\subsubsection{Results}
\label{sec:qq:res}

\subsubsection*{Determination of the cross-sections}
\label{sec:qq:xsec}

   The $\eeqq$ cross-sections were derived from the expression:
\begin{equation}
 \sqq = \frac{(N_{sel}-N_{bg})~(1-f)}{\epsilon~\mathcal{L}}
\end{equation}
\noindent
where $N_{sel}$ stands for the number of events selected, 
$N_{bg}$ is the number of background events expected, 
$\epsilon$ is the selection efficiency and ${\mathcal{L}}$ stands for the 
integrated luminosity. The term $(1-f)$ applies only to 
the {\emph{non-radiative}} cross-section, the parameter $f$ 
expressing the feed-up by events produced at a 
reduced centre-of-mass energy below 85\% of $\roots$. 

   The cross-section values found at each energy point are presented
in Table~\ref{tab:qq:res} and shown in Figure~\ref{fig:ana:sig-cmp},
together with the Standard Model predictions computed with the ZFITTER
program. The {\emph{non-radiative}} cross-sections at energies from
130~GeV to 172~GeV supersede those of~\cite{ref:delphiff:130-172}
because of the correction for ISR-FSR interference mentioned in
Section~\ref{sec:qq:analysis}.

%%%%%%%%%%%%%%%%%%%%%%%%%%%%%%%%%%%%%%%
\begin{table}[tp]
 %%%%%%%%%%%%%%%% qq %%%%%%%%%%%%%%%%
 {\small
 \begin{center}
 \renewcommand{\arraystretch}{1.2}
 \begin{tabular}{|c|c|c||c|c|c|}
 \hline
 \multicolumn{6}{|c|}{\eeqq} \\
 \hline
 \hline
 \multicolumn{1}{|c|}{$\roots$} &
 \multicolumn{1}{|c|}{$\sqsps>0.10$ } &
 \multicolumn{1}{|c||}{$\sqsps>0.85$} &
 \multicolumn{1}{|c|}{$\roots$} &
 \multicolumn{1}{|c|}{$\sqsps>0.10$ } &
 \multicolumn{1}{|c|}{$\sqsps>0.85$} \\
 \cline{2-3}
 \cline{5-6}
 \multicolumn{1}{|c|}{(GeV)} &
 $\sqq$ (pb) & $\sqq$ (pb) &
 \multicolumn{1}{|c|}{(GeV)} &
 $\sqq$ (pb) & $\sqq$ (pb) \\
 \hline
 \hline
 130 & 
 $\begin{array}{c}
 328.4 \pm  11.3 \pm   3.7 \\ (328.1)
 \end{array}$ &
 $\begin{array}{c}
  82.4 \pm   5.2 \pm   2.6 \\ ( 82.5)
 \end{array}$ &
 192 & 
 $\begin{array}{c}
  92.9 \pm   2.4 \pm   1.0 \\ ( 93.4)
 \end{array}$ &
 $\begin{array}{c}
  22.1 \pm   1.1 \pm   0.3 \\ ( 21.2)
 \end{array}$ \\
 \hline
 136 & 
 $\begin{array}{c}
 259.6 \pm  10.0 \pm   3.1 \\ (272.0)
 \end{array}$ &
 $\begin{array}{c}
  65.3 \pm   4.7 \pm   2.1 \\ ( 66.4)
 \end{array}$ &
 196 & 
 $\begin{array}{c}
  91.1 \pm   1.4 \pm   0.9 \\ ( 88.7)
 \end{array}$ &
 $\begin{array}{c}
  21.2 \pm   0.6 \pm   0.3 \\ ( 20.1)
 \end{array}$ \\
 \hline
 161 & 
 $\begin{array}{c}
 158.3 \pm   4.4 \pm   2.0 \\ (151.4)
 \end{array}$ &
 $\begin{array}{c}
  41.0 \pm   2.1 \pm   1.3 \\ ( 35.1)
 \end{array}$ &
 200 & 
 $\begin{array}{c}
  85.2 \pm   1.3 \pm   0.9 \\ ( 84.2)
 \end{array}$ &
 $\begin{array}{c}
  19.5 \pm   0.6 \pm   0.3 \\ ( 19.0)
 \end{array}$ \\
 \hline
 172 & 
 $\begin{array}{c}
 125.5 \pm   4.2 \pm   1.9 \\ (125.1)
 \end{array}$ &
 $\begin{array}{c}
  30.4 \pm   1.9 \pm   1.0 \\ ( 28.7)
 \end{array}$ &
 202 & 
 $\begin{array}{c}
  84.2 \pm   1.9 \pm   0.9 \\ ( 82.0)
 \end{array}$ &
 $\begin{array}{c}
  18.9 \pm   0.8 \pm   0.3 \\ ( 18.5)
 \end{array}$ \\
 \hline
 183 & 
 $\begin{array}{c}
 107.6 \pm   1.7 \pm   1.0 \\ (106.0)
 \end{array}$ &
 $\begin{array}{c}
  25.5 \pm   0.8 \pm   0.3 \\ ( 24.2)
 \end{array}$ &
 205 & 
 $\begin{array}{c}
  77.8 \pm   1.3 \pm   0.9 \\ ( 78.9)
 \end{array}$ &
 $\begin{array}{c}
  17.7 \pm   0.6 \pm   0.3 \\ ( 17.8)
 \end{array}$ \\
 \hline
 189 & 
 $\begin{array}{c}
  96.9 \pm   1.0 \pm   0.9 \\ ( 97.3)
 \end{array}$ &
 $\begin{array}{c}
  22.6 \pm   0.5 \pm   0.3 \\ ( 22.1)
 \end{array}$ &
 207 & 
 $\begin{array}{c}
  74.7 \pm   1.0 \pm   0.8 \\ ( 77.3)
 \end{array}$ &
 $\begin{array}{c}
  17.0 \pm   0.4 \pm   0.3 \\ ( 17.4)
 \end{array}$ \\
 \hline
 \end{tabular}
 \end{center}
 }
 %%%%%%%%%%%%%%%%%%%%%%%%%%%%%%%%%%%%

 \caption{\capsty{Measured values of the {\it{inclusive}} and 
          {\it{non-radiative}}
          cross-sections for the process $\eeqq$ at collision energies ranging 
          from 130 to 207 GeV. The uncertainties include the statistical and 
          all systematic contributions. Values in parentheses 
          are the Standard Model predictions computed with the ZFITTER 
          program, and are estimated to have a precision of $\pm0.26\%$.}} 
 \label{tab:qq:res}
\end{table}

\subsubsection*{Systematic uncertainty}
\label{sec:qq:syst}

   Most of the systematic uncertainties on the selection efficiency 
resulted from small residual discrepancies between real and 
simulated distributions. Because of these differences, the selection 
cuts were expected to have slightly different effects on the 
simulated distributions than on the real ones. The associated   
uncertainties were estimated from the observed changes of 
the cross-section when varying the cut values. They added 
up to a total contribution of 1.0\%. 

   The systematic uncertainty related to the ISR modelling was 
estimated by comparing the selection efficiencies extracted from  
two different simulation samples, produced either with the KK 
or with the PYTHIA generators. 
The observed difference in the selection efficiency was very 
small, i.e. 0.1\%, and was converted into a systematic uncertainty of 
identical size. 

   The uncertainty related to the accuracy of the fragmentation 
modelling in the Monte Carlo was derived from the comparison 
between a sample where the hadronisation was performed with 
PYTHIA and a sample where it was done with ARIADNE (version 
4.08)~\cite{ref:th:ariadne}
%%% text added after EPJ%%%
, with parameters tuned to DELPHI data~\cite{ref:mc:tuning}.
%%% 
The difference between the two 
values of the selection efficiency (0.1-0.25\%) was taken as
the corresponding systematic uncertainty.

   A 0.7\% systematic uncertainty was assigned to the 
{\emph{non-radiative}} cross-section as a consequence of the cut 
on $\rootsp$. Its value follows from the resolution attributed 
to the reconstructed value of $\rootsp$.

    The systematic uncertainty on the residual backgrounds accounts
for the theoretical precision of the generators used and for the
accuracy on the selection efficiency associated to each final
state. The uncertainty affecting the {\emph{non-radiative}}
cross-section was completely dominated by the precision on the WW
background estimation, while the {\emph{inclusive}} cross-section was
also affected by the uncertainties related to the subtraction of
residual contaminations due to \eeZZ, $\ee \rightarrow \ee\ffbar$ and
two-photon collision final states.

   The quadratic combination of the contributions to the 
systematic uncertainties described above translate into 
a total uncertainty on the selection efficiency amounting 
to 1.1\% for the {\emph{inclusive}} cross-section and 1.3\% 
for the {\emph{non-radiative}} cross-sections.
As for the residual backgrounds, a 0.5 pb (0.2 pb) total 
uncertainty was assigned to the {\emph{inclusive}} 
({\emph{non-radiative}}) cross-section. 
The correlation between systematic uncertainties assigned 
to cross-sections measured at different energy points was
estimated as 80\% when coming from the event selection, 
and as 100\% when originating from the residual background 
subtraction.

The breakdown of the systematic uncertainties on the $\eeqq$ 
cross-sections is provided in Table~\ref{tab:qq:syst}.

%%%%%%%%%Put in a copy of Table 7
\begin{table}[tp]
\begin{center}
 \begin{tabular}{|l|c||c|}
 \hline
 \multicolumn{3}{|c|}{\eeqq} \\
 \hline
 \hline
                           &$\Delta\sigma/\sigma$ & $\Delta\sigma/\sigma$ \\
 Source                    & (non-rad.)           & (inclus.) \\
 \hline\hline
  Forward corrections      &  40                  &  40 \\
  Selection cuts           & 100                  & 100 \\ 
  ISR modelling            &  10                  &  10 \\
  Fragmentation            &  25                  &  10 \\
  $\rootsp$ cut            &  70                  &   - \\
  Backgrounds              & 120                  &  70 \\
  Luminosity               &  55                  &  55 \\
  TPC sector instability   &  50                  &  10 \\
 \hline\hline
  Total uncertainty        & 186                  & 140 \\
 \hline
  Correlated               & 169                  & 124 \\
  Uncorrelated             &  79                  &  65 \\
  \hline
 \end{tabular}
\caption{\capsty{Breakdown of the systematic uncertainties on 
         the $\eeqq$ cross-sections for data taken at $\roots \sim 207~\GeV$. All numbers in units of $10^{-4}$. The total uncertainties do not include the uncertainty due to the TPC 
         instability, which applies only to a part of the data collected in 2000.
The correlated error component includes errors 
         correlated between energies and channels and those correlated 
         with other LEP experiments.   }}
\label{tab:qq:syst}
\end{center}
\end{table}

\subsection{Summary of results}
\label{sec:results}

%%....................................................................
%%.. Summary .........................................................
%%....................................................................
%\fbox{
%\begin{minipage}{0.87\textwidth}
%\begin{itemize}
% \item R ratios
%
% \item Comments about agreement between measurements and theoretical 
%       predictions for all channels
%
% \item Summary plots of cross-sections and asymmetries for
%       all channels
%
%\end{itemize}
%\end{minipage}
%}

%%....................................................................
%%.. Contents ........................................................
%%....................................................................

%%%%%%%%%%%%%%%%%%%%%%%%%%%%%%%%%%%%%%%
\begin{figure}[p]
\begin{center}
 \begin{tabular}{cc}
  \mbox{\epsfig{file=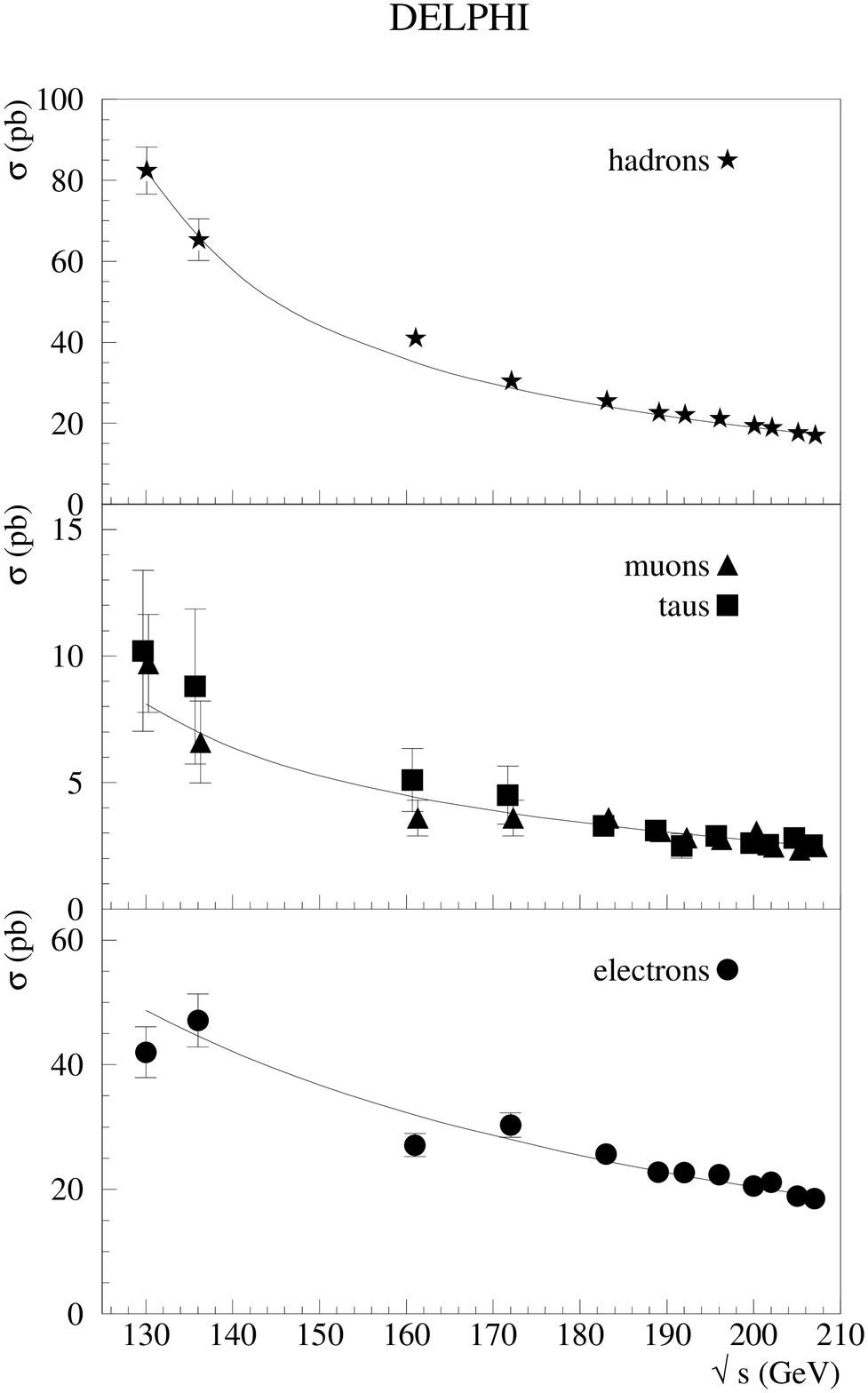,width=0.49\textwidth}}
  \mbox{\epsfig{file=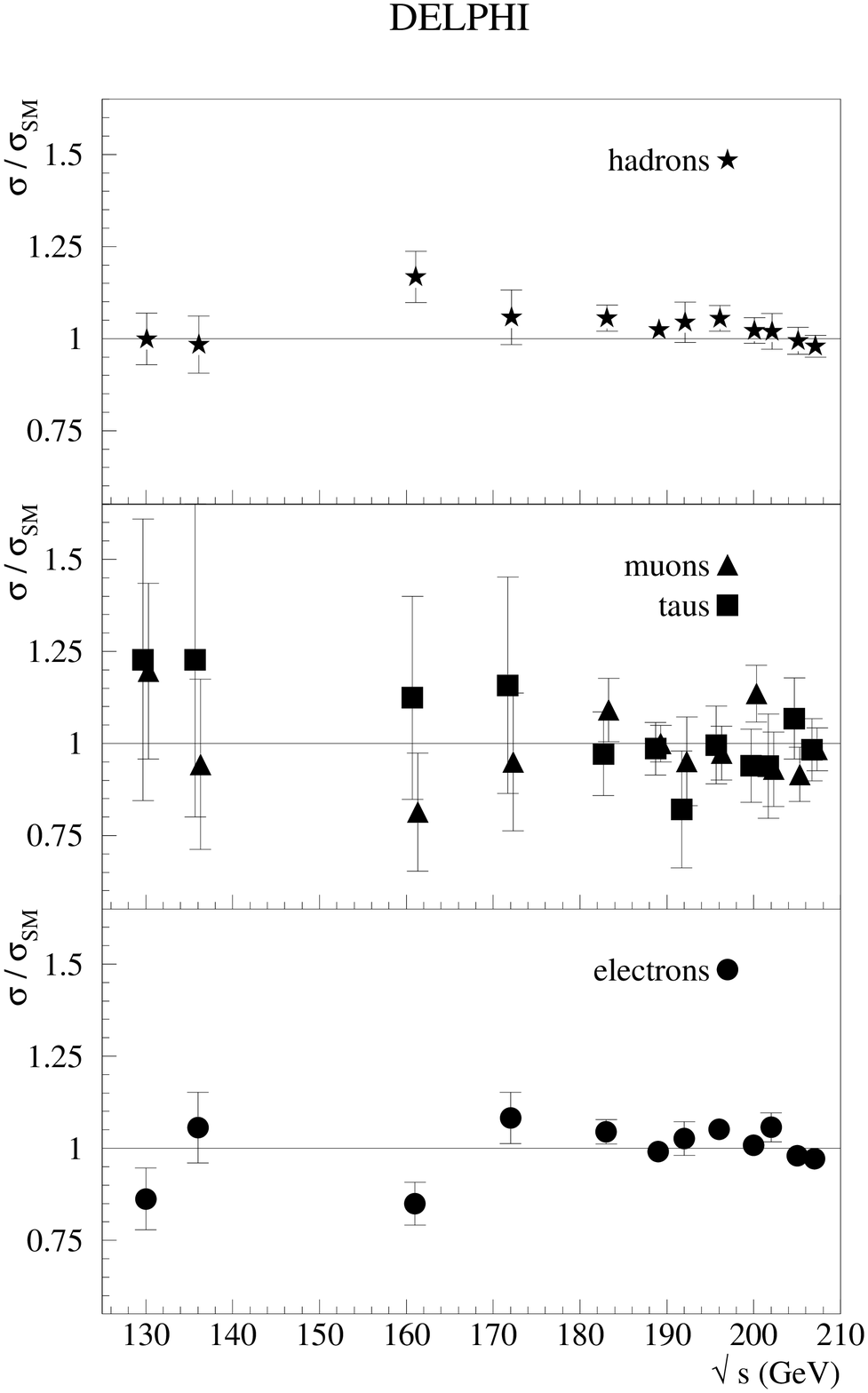,width=0.49\textwidth}}
 \end{tabular}
\end{center}
\caption{\capsty{Comparisons of the measurements of cross-sections to 
         predictions of the Standard Model from \BHWIDE\ and \ZFITTER\ for
         {\it{non-radiative}} samples of \ee, \mumu, \tautau\ and \qqbar\ 
         final states. The measurements are compared to the predictions shown 
         as curves (left) and as values of $\mathcal{R}$ the ratio of the 
         measurement to the predictions (right). For clarity the results
         for \mumu\ and \tautau\ final states are slightly displaced 
         from the actual centre-of-mass energy of the data. The errors 
         include the statistical and experimental systematic errors.}}
\label{fig:ana:sig-cmp}
\end{figure}
%%%%%%%%%%%%%%%%%%%%%%%%%%%%%%%%%%%%%%%

%%%%%%%%%%%%%%%%%%%%%%%%%%%%%%%%%%%%%%%
\begin{figure}[tp]
\begin{center}
 \begin{tabular}{cc}
  \mbox{\epsfig{file=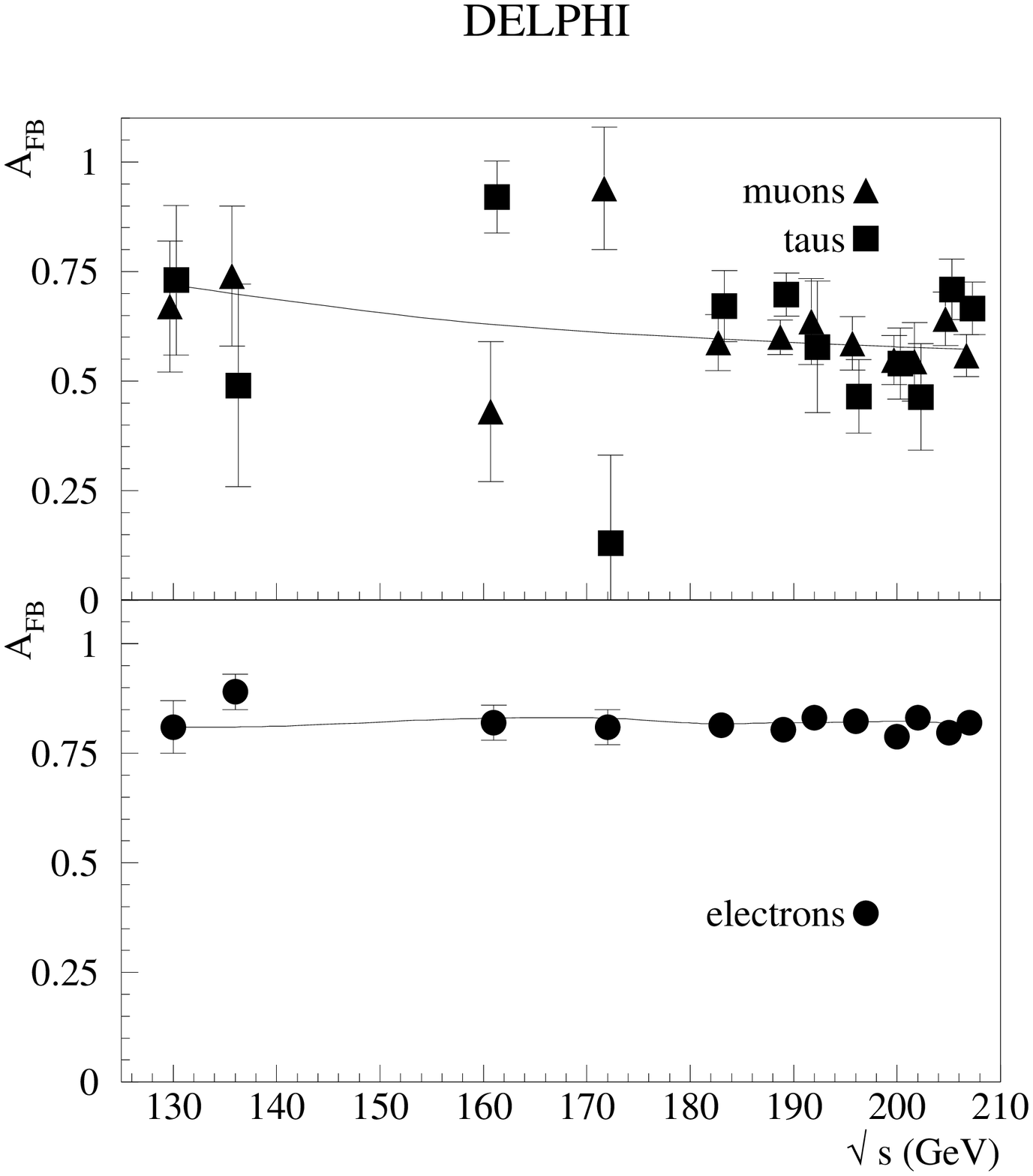,width=0.49\textwidth}}
  \mbox{\epsfig{file=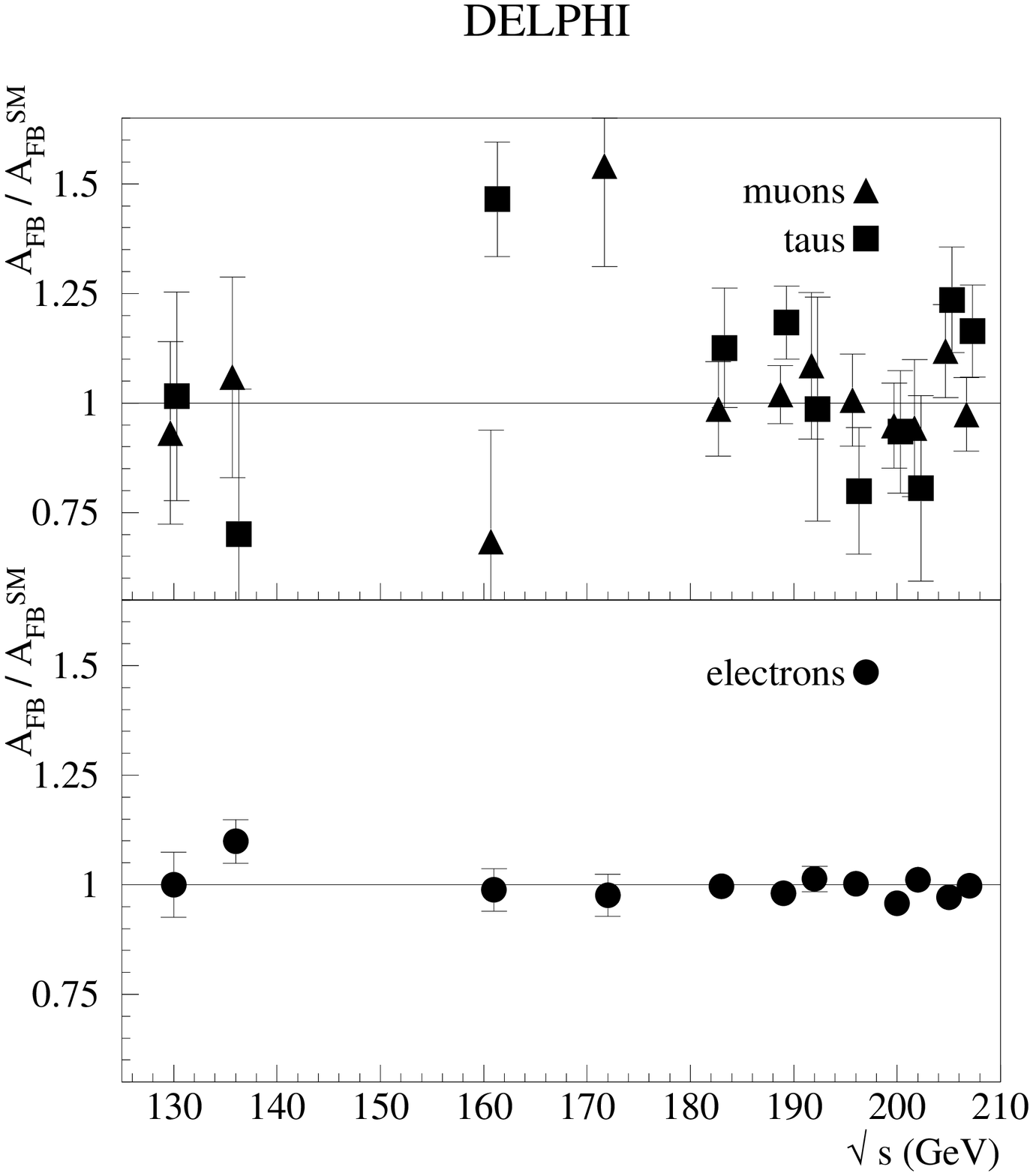,width=0.49\textwidth}}
 \end{tabular}
\end{center}
\caption{\capsty{Comparisons of the measurements of forward-backward 
         asymmetries to predictions of the Standard Model from 
\BHWIDE\ and \ZFITTER\ for {\it{non-radiative}} samples of \ee, \mumu\ and 
         \tautau\ final states. The measurements are compared to the 
          predictions
         shown as curves (left) and as values of $\mathcal{R}$ the ratio of 
         the measurement to the predictions (right). For clarity the results
         for \mumu\ and \tautau\ final states are slightly displaced 
         from the actual centre-of-mass energy of the data. The errors 
         include the statistical and experimental systematic errors.}}
\label{fig:ana:afb-cmp}
\end{figure}
%%%%%%%%%%%%%%%%%%%%%%%%%%%%%%%%%%%%%%%

The measurements of cross-sections and forward-backward asymmetries for 
{\it{non-radiative}} samples of events are compared to theoretical predictions 
in Figures~\ref{fig:ana:sig-cmp} and~\ref{fig:ana:afb-cmp}. In both cases the
measurements are compared directly to the predictions and the ratios
of the measurements to the prediction are shown, 
\begin{displaymath}
 {\mathcal{R}}=\frac{\mathcal{O}}{{\mathcal{O}}_{SM}},
\end{displaymath}
where $\mathcal{O}$ is the measured observable and ${\mathcal{O}}_{SM}$ 
is the prediction for the observable.
A useful guide to the compatibility of the different sets of measurements
with the theoretical predictions is to average the ratios of the measured 
results to the predictions over all centre-of-mass energies, 
$\left\langle{\mathcal{R}}\right\rangle$. Such averages have been calculated 
for the measurements made on {\it{non-radiative}} samples of events:
\begin{itemize}
 \item cross-sections for \ee, \mumu, \tautau, \qqbar\ final states and \lplm, 
       an average of the combination of the \mumu and \tautau\ final 
       states; and
 \item forward-backward asymmetries for \ee, \mumu, \tautau\ and \lplm\ final
       states; and
 \item bin-by-bin averages of the differential cross-sections for
       \mumu\ and \tautau\ final states.
\end{itemize}
The averages were made using the BLUE 
technique~\cite{ref:ana:blue-sngl,*ref:ana:blue-multi}
so that correlations between systematic errors in the different averages
were taken into account. The combination method makes it easy to
identify the statistical and systematic uncertainties on 
$\left\langle {\mathcal{R}} \right\rangle$, for each set of measurements. 
For the cross-section and forward-backward asymmetry measurements the averages
are made over all centre-of-mass energies from 130 to 207 GeV. Data taken at
lower energies, from 130 to 172 GeV, are taken 
from~\cite{ref:delphiff:130-172}, except for the \qqbar\ final state
where the updated values of Table~\ref{tab:qq:res} are used. The 
experimental systematic errors for the 
lower energy data, with the exception of luminosity errors, are taken to be 
correlated amongst themselves for each measurement, and uncorrelated with the 
errors at higher energies.
Theoretical uncertainties on the cross-sections and asymmetries are taken 
from~\cite{ref:th:lepffwrkshp}.
For the differential cross-section the averages are made over energies from
183-207 GeV, and for the \mumu\ and \tautau\ final states the statistical error
is taken from the expected error. The relative theoretical uncertainties on the
differential cross-sections are taken to be the same as the relative 
uncertainties on the cross-sections as given in~\cite{ref:th:lepffwrkshp}, and
assumed to be fully correlated between bins.

%In addition it is possible to compute the $\chi^2$ of the sets of measurements
%to the predictions and the $\chi^{2}$ of the individual values of 
%${\mathcal{R}}$ at each energy to the average.

%%%%%%%%%%%%%%%%%%%%%%%%%%%%%%%%%%%%%%%%%%%%%%%%%%%%%%%%%%%%%%
%%%%%%%%%%%%%%%%%%%%%%%%%%%%%%%%%%%%%%%%%%%%%%%%%%%%%%%%%%%%%% This from John 18/1/04
\begin{table}[tbp]
\begin{center}
\begin{tabular}{|c|c|c|}
\hline
          & Final   &                                     \\
 Meas.    & States  & $\langle{\mathcal{R}}\rangle\pm$(stat)$\pm$(syst)$\pm$(theory) \\
\hline\hline
 $\sigma$ & \ee     & $1.0006\pm0.0086\pm0.0077\pm0.0200$   \\
\cline{2-3}
          & \mumu   & $0.9961\pm0.0244\pm0.0062\pm0.0040$   \\
\cline{2-3}
          & \tautau & $0.9852\pm0.0341\pm0.0203\pm0.0040$   \\
\cline{2-3}
          & \qqbar  & $1.0256\pm0.0103\pm0.0130\pm0.0026$   \\
\cline{2-3}
          & \lplm   & $0.9930\pm0.0200\pm0.0074\pm0.0040$   \\
\hline
 $\Afb$   & \ee     & $0.9896\pm0.0061\pm0.0043\pm0.0244$   \\
\cline{2-3}
          & \mumu   & $1.0120\pm0.0335\pm0.0010\pm0.0068$   \\
\cline{2-3}
          & \tautau & $1.1121\pm0.0405\pm0.0156\pm0.0068$   \\
\cline{2-3}
          & \lplm   & $1.0494\pm0.0259\pm0.0059\pm0.0068$   \\
\hline
\end{tabular}
\end{center}
\caption{\capsty{Average values of $\mathcal{R}$ for cross-sections and
         forward-backward asymmetries for \ee, \mumu, \tautau\ and \qqbar\
         final states. The errors on the averages,
         $\langle{\mathcal{R}}\rangle$, are the statistical, experimental
         systematic and theoretical uncertainties.}}
\label{tab:ana:rvals}
\end{table}
%%%%%%%%%%%%%%%%%%%%%%%%%%%%%%%%%%%%%%%%%%%%%%%%%%%%%%%%%%%%%%
%%%%%%%%%%%%%%%%%%%%%%%%%%%%%%%%%%%%%%%%%%%%%%%%%%%%%%%%%%%%%%

Results for the averages of $\left\langle{\mathcal{R}}\right\rangle$
for cross-sections and forward-backward asymmetries are given in
Table~\ref{tab:ana:rvals}. They indicate satisfactory agreement
between the data and the predictions, with the largest deviation from
the expectation being approximately 2.6 standard deviations. In most
cases the uncertainties on the averages are dominated by statistical
errors, except for the cross-section and forward-backward asymmetry
for \eeee\, where the theoretical errors dominate, and for the
cross-section for \eeqq\, where the experimental systematic error
dominates.

%%%%%%%%%%%%%%%%%%%%%%%%%%%%%%%%%%%%%%%%%%%%%%%%%%%%%%%%%%%%%%
%%%%%%%%%%%%%%%%%%%%%%%%%%%%%%%%%Updated with John's new Table 22 28/01/04
\begin{table}[tbp]
\begin{center}
\begin{tabular}{c}
 \begin{tabular}{|@{[}r@{,}r@{]}|r|}
 \hline
 \multicolumn{3}{|c|}{\eeee} \\
 \hline\hline
 \multicolumn{2}{|c|}{$\cos\theta$} &
  $\langle{\mathcal{R}}\rangle\pm$
   (stat)$\pm$(syst)$\pm$(theory) \\
 \hline\hline
 -0.72 & -0.54 & 1.106 $\pm$ 0.091 $\pm$ 0.049 $\pm$ 0.020 \\
 -0.54 & -0.36 & 1.077 $\pm$ 0.074 $\pm$ 0.014 $\pm$ 0.020 \\
 -0.36 & -0.18 & 0.951 $\pm$ 0.060 $\pm$ 0.008 $\pm$ 0.020 \\
 -0.18 &  0.00 & 1.113 $\pm$ 0.049 $\pm$ 0.026 $\pm$ 0.020 \\
  0.00 &  0.09 & 1.005 $\pm$ 0.047 $\pm$ 0.027 $\pm$ 0.020 \\
  0.09 &  0.18 & 0.999 $\pm$ 0.055 $\pm$ 0.007 $\pm$ 0.020 \\
  0.18 &  0.27 & 0.958 $\pm$ 0.046 $\pm$ 0.007 $\pm$ 0.020 \\
  0.27 &  0.36 & 0.943 $\pm$ 0.038 $\pm$ 0.007 $\pm$ 0.020 \\
  0.36 &  0.45 & 1.030 $\pm$ 0.033 $\pm$ 0.008 $\pm$ 0.020 \\
  0.45 &  0.54 & 0.974 $\pm$ 0.026 $\pm$ 0.008 $\pm$ 0.020 \\
  0.54 &  0.63 & 1.007 $\pm$ 0.021 $\pm$ 0.008 $\pm$ 0.020 \\
  0.63 &  0.72 & 0.999 $\pm$ 0.016 $\pm$ 0.008 $\pm$ 0.020 \\
 \hline
 \end{tabular}
\\

\\
 \begin{tabular}{|@{[}r@{,}r@{]}|r|}
 \hline
 \multicolumn{3}{|c|}{\eemm} \\
 \hline\hline
 \multicolumn{2}{|c|}{$\cos\theta$} &
  $\langle{\mathcal{R}}\rangle\pm$
   (stat)$\pm$(syst)$\pm$(theory) \\
 \hline\hline
 -0.97 & -0.80 & 1.166 $\pm$ 0.150 $\pm$ 0.009 $\pm$ 0.004 \\
 -0.80 & -0.60 & 0.964 $\pm$ 0.138 $\pm$ 0.008 $\pm$ 0.004 \\
 -0.60 & -0.40 & 1.092 $\pm$ 0.127 $\pm$ 0.008 $\pm$ 0.004 \\
 -0.40 & -0.20 & 0.761 $\pm$ 0.112 $\pm$ 0.007 $\pm$ 0.004 \\
 -0.20 &  0.00 & 1.031 $\pm$ 0.098 $\pm$ 0.007 $\pm$ 0.004 \\
  0.00 &  0.20 & 1.004 $\pm$ 0.085 $\pm$ 0.007 $\pm$ 0.004 \\
  0.20 &  0.40 & 1.067 $\pm$ 0.072 $\pm$ 0.007 $\pm$ 0.004 \\
  0.40 &  0.60 & 1.083 $\pm$ 0.063 $\pm$ 0.006 $\pm$ 0.004 \\
  0.60 &  0.80 & 0.933 $\pm$ 0.056 $\pm$ 0.006 $\pm$ 0.004 \\
  0.80 &  0.97 & 1.003 $\pm$ 0.054 $\pm$ 0.006 $\pm$ 0.004 \\
 \hline
 \end{tabular}
\\

\\
 \begin{tabular}{|@{[}r@{,}r@{]}|r|}
 \hline
 \multicolumn{3}{|c|}{\eett} \\
 \hline\hline
 \multicolumn{2}{|c|}{$\cos\theta$} &
  $\langle{\mathcal{R}}\rangle\pm$
   (stat)$\pm$(syst)$\pm$(theory) \\
 \hline\hline
 -0.96 & -0.80 & 0.397 $\pm$ 0.291 $\pm$ 0.025 $\pm$ 0.004 \\
 -0.80 & -0.60 & 0.757 $\pm$ 0.221 $\pm$ 0.031 $\pm$ 0.004 \\
 -0.60 & -0.40 & 0.970 $\pm$ 0.166 $\pm$ 0.023 $\pm$ 0.004 \\
 -0.40 & -0.20 & 0.898 $\pm$ 0.144 $\pm$ 0.022 $\pm$ 0.004 \\
 -0.20 &  0.00 & 0.972 $\pm$ 0.133 $\pm$ 0.023 $\pm$ 0.004 \\
  0.00 &  0.20 & 1.122 $\pm$ 0.111 $\pm$ 0.021 $\pm$ 0.004 \\
  0.20 &  0.40 & 1.106 $\pm$ 0.089 $\pm$ 0.020 $\pm$ 0.004 \\
  0.40 &  0.60 & 0.967 $\pm$ 0.078 $\pm$ 0.021 $\pm$ 0.004 \\
  0.60 &  0.80 & 0.875 $\pm$ 0.086 $\pm$ 0.022 $\pm$ 0.004 \\
  0.80 &  0.96 & 1.027 $\pm$ 0.098 $\pm$ 0.021 $\pm$ 0.004 \\
 \hline
 \end{tabular}

\end{tabular}
\end{center}
\caption{\capsty{Bin-by-bin average values of $\mathcal{R}$ for differential
         cross-sections for \ee, \mumu\ and \tautau\ final states. 
         The errors on the averages, $\langle{\mathcal{R}}\rangle$, are 
         the statistical, experimental 
         systematic and theoretical uncertainties.}}
\label{tab:ana:rdsdc}
\end{table}
%%%%%%%%%%%%%%%%%%%%%%%%%%%%%%%%%%%%%%%%%%%%%%%%%%%%%%%%%%%%%%

%%%%%%%%%%%%%%%%%%%%%%%%%%%%%%%%%%%%%%%
\begin{figure}[tp]
\begin{center}
  \mbox{\epsfig{file=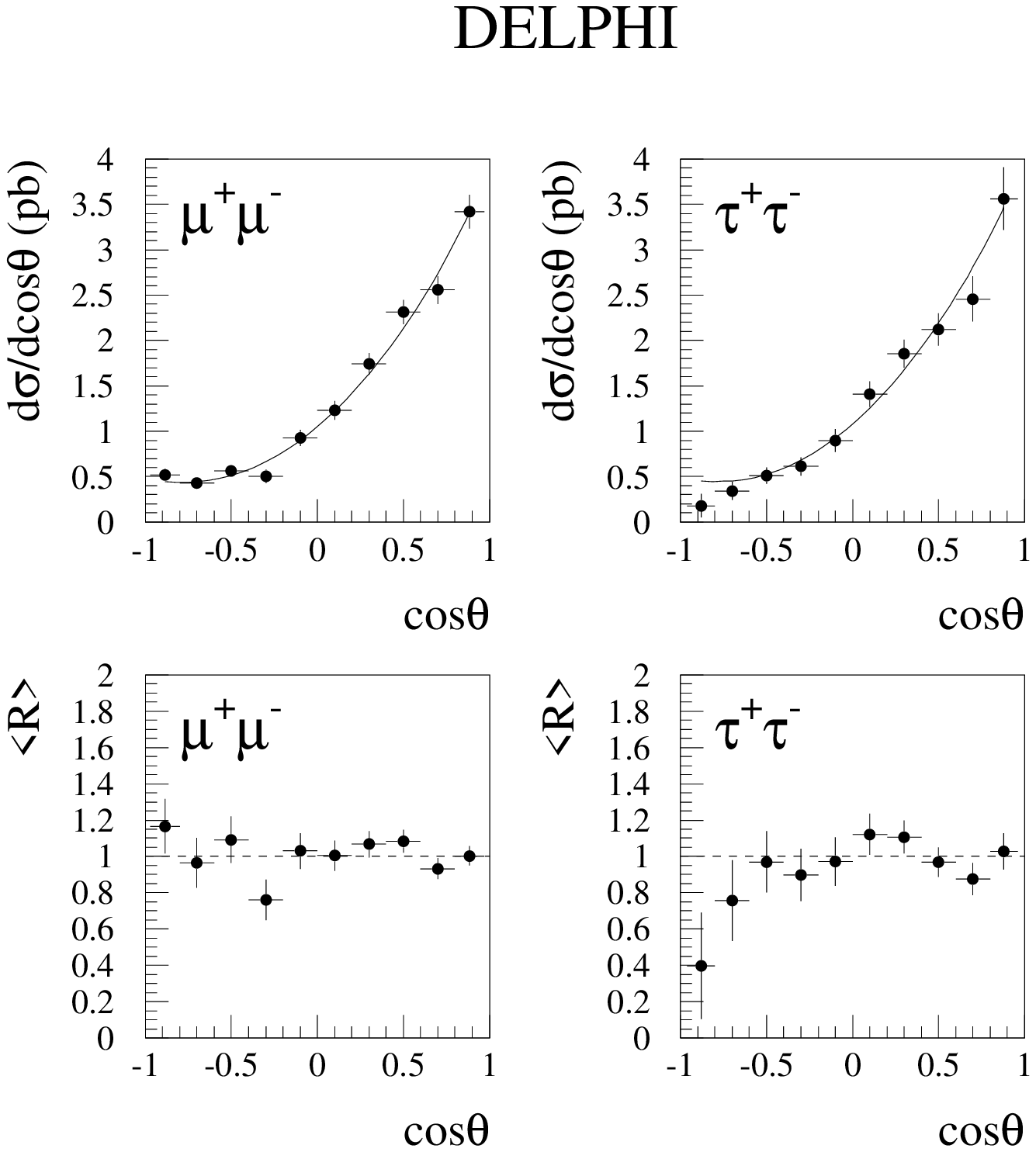,width=0.94\textwidth}}
\end{center}
\caption{\capsty{Bin-by-bin averages, $\langle{\mathcal{R}}\rangle$, for
         \mumu\ and \tautau\ final states. 
         The lower plots show the values of $\langle{\mathcal{R}}\rangle$
         while the upper plots show the 
SM \dsdcth\ multiplied by $\langle{\mathcal{R}}\rangle$
%%the $\langle{\mathcal{R}}\rangle$ expressed
%%         as measurements of \dsdcth\ 
at the luminosity weighted centre-of-mass
         energy for each measurement channel.}}
\label{fig:ana:rdsdc}
\end{figure}
%%%%%%%%%%%%%%%%%%%%%%%%%%%%%%%%%%%%%%%

Results for the bin-by-bin averages of ratios of the measured differential 
cross-sections to the predictions are given in Table~\ref{tab:ana:rdsdc}.
These are shown in Figure~\ref{fig:ana:rdsdc}, in which the averaged ratios
have been expressed as a differential cross-section at the luminosity weighted 
centre-of-mass energy for the \mumu\ and \tautau\ final states by multiplying
the averaged ratios, $\langle{\mathcal{R}}\rangle$, by the predicted 
differential cross-sections in each bin at the appropriate centre-of-mass 
energy. The results show good agreement with the expectation.

Overall the data show good agreement between the measurements and the 
predictions of the Standard Model from \ZFITTER\ and \BHWIDE.
In the following section the measurements are interpreted in a variety of 
models which allow for physics beyond the Standard Model. Since each of these 
models predicts specific behaviours for either the differential cross-section, 
or alternatively cross-sections and forward-backward asymmetries, as a function
of centre-of-mass energy, the predictions of the models are compared directly
to the individual measurements rather than to the averaged 
$\langle{\mathcal{R}}\rangle$ values given in this section.

%%--------------------------------------------------------------------
%%-- INTERPRETATION --------------------------------------------------
%%--------------------------------------------------------------------
\section{Interpretation}
\label{sec:interp}

%%....................................................................
%%.. Summary .........................................................
%%....................................................................
%\fbox{
%\begin{minipage}{0.87\textwidth}
%\begin{itemize}
% \item General review of interpretations
%
% \item Discussion of issues
% \begin{itemize}
%  \item Radiative corrections
%  \item Fitting
%  \item Limit setting
% \end{itemize}
%
%\end{itemize}
%\end{minipage}
%}

%%....................................................................
%%.. Contents ........................................................
%%....................................................................

The results of the measurements of cross-sections, forward-backward
asymmetries and angular distributions presented in
Section~\ref{sec:analysis} have been interpreted in a variety of
scenarios. The data were used to determine the parameters of the
S-matrix formalism for \eeff, as well as to investigate a variety of
models which include explicit forms of physics beyond the Standard
Model\footnote{In the comparison of the data with the models, it has
been assumed that the one-loop electroweak corrections to the
observables are those given by the Standard Model. It has been pointed
out~\cite{ref:1loop:thry2002} that this may not always be the case.}:
models with \Zprime\ bosons, contact interactions,
%the exchange of \sneut\ in \rpviol\ SUSY 
models which include the exchange of gravitons in large extra
dimensions and models which consider  
possible $s$ or $t$ channel sneutrino $\snul$
exchange in R-parity violating supersymmetry.
For the S-matrix ansatz and the search for \Zprime\ bosons the LEP II
data reported here were used together with measurements from LEP I. For other
studies, the LEP II data alone were used.

In the following subsections, each of these interpretations is discussed.
In all cases the theoretical basis of the model is summarised and the
relevant parameters are identified. The methods of fitting parameters of
the models to the data are discussed. 
%%%after EPJ referee%%%
The theoretical uncertainties on the Standard Model
predictions for cross-sections and asymmetries, as given in 
Section~\ref{sec:analysis}, are explicitly included in the fits.
%%%
In most cases, the relevant papers
discuss the predictions of the models only at Born level, however, to confront
the models with data it is necessary to take into account QED radiative 
corrections. The methods used to implement radiative corrections are 
described for each model. In the 
absence of evidence for physics beyond the Standard Model, limits on the
physical parameters of the models can be derived. There is no universally
agreed way to compute limits, particularly when measurements lie close
to a physical boundary. Where possible results of the interpretations
are first quoted as measurements, and the methods used to derive limits 
from these results are discussed.

%%%After EPJ referee%%%%
In principle the introduction of processes beyond the Standard Model
will affect the cross-section for the channel \eeee\  at small
angles, on which the measurement of the luminosity is based. However
for values of the parameters of the models discussed in this section,
the changes to the cross-section in this region are negligible
compared with the experimental and theoretical uncertainties discussed
in Section~\ref{sec:lumi}.
%%%

%%--------------------------------------------------------------------
%%-- SMATRIX FITS ----------------------------------------------------
%%--------------------------------------------------------------------

\subsection{S-matrix ansatz}
\label{sec:smat}

%% ------------------------------------------------------------------- 
\subsubsection{Theory}
\label{sec:smat:thry}

%%....................................................................
%%.. Summary .........................................................
%%....................................................................
%\fbox{
%\begin{minipage}{0.87\textwidth}
%\begin{itemize}
% \item Summary of theory papers
% \item What are the important parameters
% \item Choices of other parameters for the particular models
%\end{itemize}
%\end{minipage}
%}

The S-matrix 
formalism~\cite{ref:smat:thry1991,*ref:smat:thry1992} 
is a rigorous semi model-independent approach to 
describe the cross-sections and the forward-backward asymmetries in \ee\ 
annihilations.
In this model, the cross-sections can be parametrised as follows:
\begin{equation}
\sigma^0_{\mathrm{a}}(s)=\frac{4}{3}\pi\alpha^2(\hbar c)^2
\left[
      \frac{\gfa}{s}
     +\frac{{\jfa} (s-\MZbar^2 c^4) + {\rfa} \, s}
           {(s-\MZbar^2 c^4)^2 + \MZbar^2 c^4 \GZbar^2}
\right]
\,\,\,{\mathrm{with}}\,\,
\begin{array}{l}
\mathrm{a}={\mathrm{tot,fb}}\\
\mathrm{f=had,e,\mu,\tau}\,,
\end{array}
\label{eqn:smat:eq1}
\end{equation}
while the forward-backward asymmetries are given by:
\begin{equation}
A^0_{\mathrm{FB}}(s)=\frac{3}{4}
                   \frac{\sigma^0_{\mathrm{fb}}(s)}
                        {\sigma^0_{\mathrm{tot}}(s)}\,.
\end{equation}
%%where $\roots$ is the centre-of-mass energy.

The mass $\MZbar$ and width $\GZbar$ of the \Zzero\ 
in the S-matrix formalism are related to the values obtained
from Standard Model fits, \MZ\ and \GZ, in which an
$s$ dependent width term is included by
\begin{equation}
\begin{array}{lllll@{}c@{}r}
  \MZ & \equiv  & \MZbar\sqrt{1+\GZbar^2/\MZbar^2} & \approx & \MZbar & + & 34.20~\MeV /c^2\phantom{\,,} \\[3mm]
  \GZ & \equiv  & \GZbar\sqrt{1+\GZbar^2/\MZbar^2} & \approx & \GZbar & + &  0.94~\MeV\,.\phantom{/c^2}
\end{array}
\end{equation}

The parameters $\rf$ and $\jf$ respectively scale the \Zzero\ exchange and the 
$\gamma$\Zzero\ 
interference contributions to the total cross-section and forward-backward
asymmetries; 
%%GM adds this and removes detailed formulae
they are functions of the effective vector and axial-vector couplings of 
the fermions.
The contribution $\gf$ of the pure $\gamma$ exchange was fixed to 
the value predicted by QED in all fits. 
%%The \Zzero\ exchange term, the 
%%$\gamma$\Zzero\ interference term and the photon exchange term are given by:
%%\begin{equation}
%%\begin{array}{l@{}l@{}l@{}l}
%%\displaystyle
%%\rtotf & = & \kappa^2
%%             \left[\ahate^2+\vhate^2\rule{0mm}{4mm}\right]
%%             \left[\ahatf^2+\vhatf^2\right]
%%            -2\kappa\,\vhate\,\vhatf C_{Im}         \\[5mm]
%%\jtotf & = & 2\kappa\,\vhate\,\vhatf \left(C_{Re}+C_{Im}\right) \\[5mm]
%%\gtotf & = & Q^2_{\mathrm{e}}Q^2_{\mathrm{f}}\left|F_A(s)\right|^2 \\[5mm]
%%\rfbf  & = & 4\kappa^2\ahate\,\vhate\,\ahatf\,\vhatf
%%            -2\kappa\,\ahate\,\ahatf C_{Im}         \\[5mm]
%%\jfbf  & = & 2\kappa\,\ahate\,\ahatf \left(C_{Re}+C_{Im}\right) \\[5mm]
%%\gfbf  & = & 0 \,,
%%\end{array}
%%\end{equation}
%%with the following definitions:
%%\begin{equation}
%%\begin{array}{l}
%%\displaystyle
%%\kappa    = \dfrac{G_F\MZ^2}{2\sqrt{2\,}\pi\alpha} \approx 1.50\\[5mm]
%%C_{Im}    = \dfrac{\GZ}{\MZ}  \left.Q_{\mathrm{e}}Q_{\mathrm{f}}\right.
%%                                {\mathrm{Im}} \left\{F_A(s)\right\} \\[5mm]
%%C_{Re}    =                   \left.Q_{\mathrm{e}}Q_{\mathrm{f}}\right.
%%                                {\mathrm{Re}} \left\{F_A(s)\right\} \\[5mm]
%%F_A(s)  = \dfrac{\alpha(s)}{\alpha} \,,
%%\end{array}
%%\end{equation}
%%where $\alpha(s)$ is the complex fine-structure constant, and 
%%$\alpha\equiv\alpha(0)$.
The photonic virtual and bremsstrahlung corrections are included through 
the convolution of eqn.~\ref{eqn:smat:eq1} with the photonic flux function.

%%....................................................................
%%.. Contents ........................................................
%%....................................................................

%% ------------------------------------------------------------------- 
\subsubsection{Results}
\label{sec:smat:res}

%%....................................................................
%%.. Summary .........................................................
%%....................................................................
%\fbox{
%\begin{minipage}{0.87\textwidth}
%\begin{itemize}
% \item Which data is fitted
% \item How radiative corrections to new physics are handled
% \item Treatment of correlated errors
% \item Parameters for fit
% \item How limits are extracted from fit
% \item Quote confidence level for limits
% \item Plot showing sensitivity of data to new physics \\
%       and/or \\
%       Likelihood/$\chi^{2}$ curves/surfaces \\
%       and/or \\
%       Plot showing limits/excluded region of parameters
% \item Table of results including fit results
%\end{itemize}
%\end{minipage}
%}
%%....................................................................
%%.. Contents ........................................................
%%....................................................................

%=========================================================================%

Published measurements from LEP I \cite{ref:delphils:91,*ref:delphils:95,ref:delphils:00} and the runs above the Z in 1995-97 \cite{ref:delphiff:130-172}
and the results described in Section~\ref{sec:analysis} were
analysed in the framework of the S-matrix approach, achieving a substantial 
improvement in the precision of the $\gamma$\Zzero\ interference compared to 
the accuracy obtained from the \Zzero\ data alone \cite{ref:delphils:91,*ref:delphils:95,ref:delphils:00}, and updating the results presented in 
reference~\cite{ref:delphiff:130-172}.

Fits to the hadronic and leptonic cross-sections and leptonic forward-backward 
asymmetries were carried out in this framework using the corresponding 
branch of the 
\mbox{ZFITTER}/\mbox{SMATASY}\ 6.36~\cite{ref:th:zfitter,ref:smat:thry1991,*ref:smat:thry1992,ref:smat:thry1995} 
program.
%%The fits included hadronic and leptonic DELPHI measurements performed near 
%%the \Zzero\
%%resonance~\cite{ref:delphils:91,*ref:delphils:95,ref:delphils:00}, and
%%non-radiative hadronic, \mumu\ and \tautau\ measurements at higher
%%energies (Section~\ref{sec:analysis}).
Data on \eeee\ at LEP II are not used in the
fit due to the large $t$-channel contribution to the measurements, which is 
not described by the S-matrix formalism.
In the fits, LEP I and LEP II measurements were assumed to be uncorrelated.
Results for the mass and the width are quoted in terms of \MZ\ and \GZ.

The results of the fits are presented in Table~\ref{tab:smat:smatrix_fit}. 
Using the LEP I data only, the $\chisq$ amounted to 162.0 ($ndof=161$) for 
the 16-parameter fit (i.e. without assuming lepton universality), and to 176.1 
($ndof=169$) for the 8-parameter fit (where lepton universality was assumed), 
the number of fitted points being 177.
Using the combined LEP I and LEP II data, the $\chisq$ amounted to 
245.1 ($ndof=221)$ and 256.1 ($ndof=229$) for the 16-parameter fit and the 
8-parameter fit, respectively, the number of fitted points being 237.
The correlation coefficients between the free parameters of the 16- and 
8-parameter fits for the LEP I and LEP I + LEP II data are shown in 
Tables~\ref{tab:smat:smatrix_corr_16_lep1}, \ref{tab:smat:smatrix_corr_16_lep1+lep2}, 
\ref{tab:smat:smatrix_corr_8_lep1} and \ref{tab:smat:smatrix_corr_8_lep1+lep2}.
The data support the hypothesis of lepton universality. Overall, the 
measurements are in good agreement with the Standard Model predictions.

\begin{figure}[tp]
\begin{center}
 \mbox{\epsfig{figure=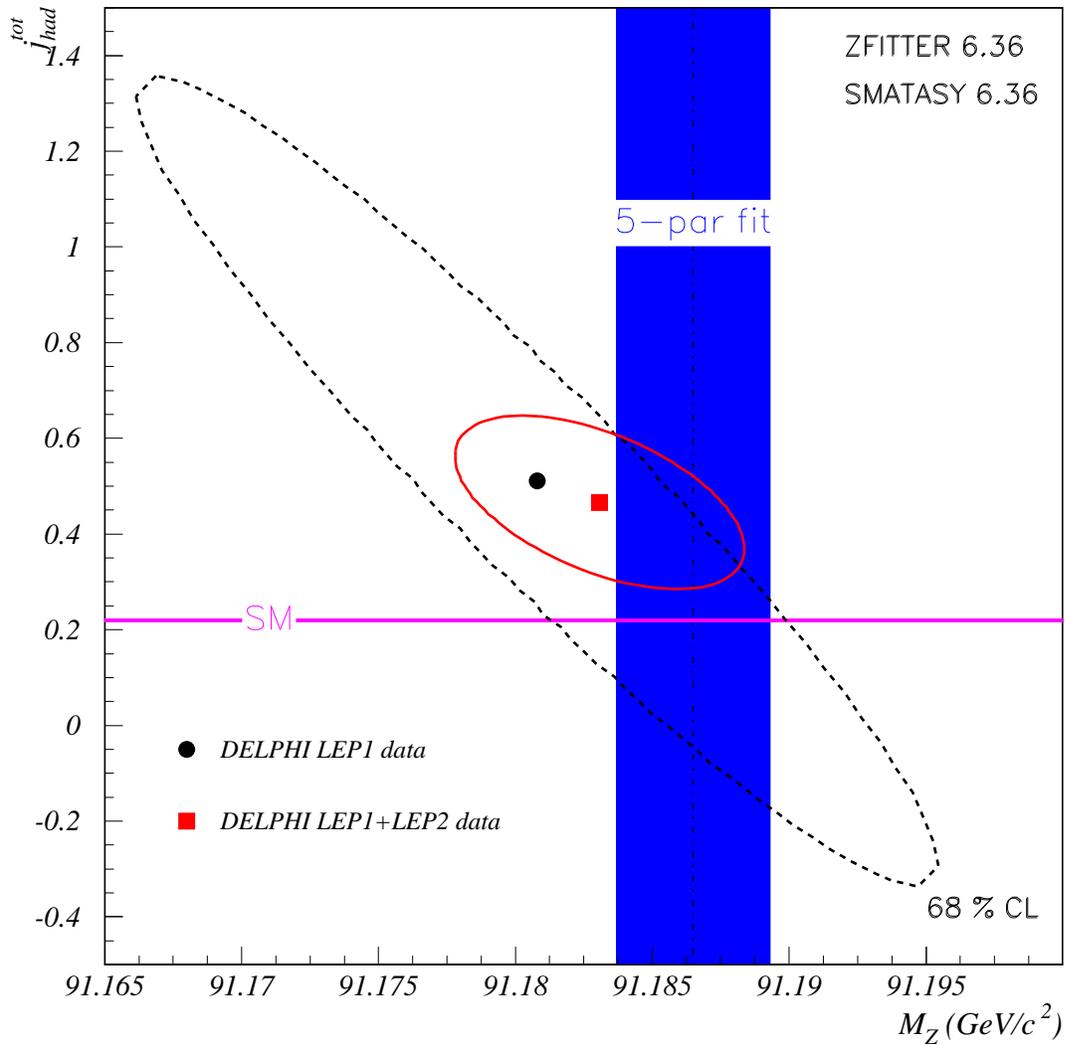,width=\linewidth}}
\end{center}
\caption{\capsty{Contour plot in the (\MZ,$\jtoth$) plane. The dashed
curve shows the region accepted at the 68\% confidence level from an
8-parameter fit to data taken at the energies around the \Zzero; the
solid curve shows the region accepted at the same confidence level
when the high energy data are also included in the fit. The band
labelled {\emph{5-par fit}} is the fit to \MZ\ obtained from a
5-parameter, Standard Model, fit to LEP I data. The narrow band
labelled {\emph{SM}} is the Standard Model expectation for $\jtoth$,
as calculated using ZFITTER.}}
\label{fig:smat:contplot}
\end{figure}

The correlations between the parameters \MZ\ and $\jtoth$ for
8-parameter fits to LEP I and LEP I + LEP II data are shown in
Figure~\ref{fig:smat:contplot}. It can be seen that a significant
improvement on the precision on the hadronic interference parameter,
$\jtoth$, is obtained when the high energy data are included in the
fit. The fitted value of \MZ\ is consistent with the value obtained
from a 5 parameter, Standard Model, fit to LEP I
data~\cite{ref:delphils:00}, $\MZ=91.1863\pm0.0028~\GeV/c^2$, and the
value of $\jtoth$ is consistent with, though somewhat above, the
Standard Model expectation $\jtoth=0.2201^{+0.0024}_{-0.0039}$, where
the uncertainties correspond to uncertainties in the input parameters.

\begin{table}[p]
\begin{center}
\renewcommand{\arraystretch}{1.5}
\small
\begin{tabular}{|c||c|c|c|c||c|}
\hline
 & \multicolumn{2}{|c|}{LEP I} & \multicolumn{2}{|c||}{LEP I + LEP II} & SM \\
\hline\hline
 $\MZ$    & $91.1936 \pm 0.0112$  & $91.1808 \pm 0.0094$  & $91.1844 \pm 0.0036$  & $91.1831 \pm 0.0034$  & --      \\[1 mm]
 $\GZ$    &  $2.4861 \pm 0.0048$  &  $2.4886 \pm 0.0046$  &  $2.4894 \pm 0.0041$  &  $2.4893 \pm 0.0041$  & 2.497   \\[1 mm]
 $\rtoth$ &  $2.9490 \pm 0.0110$  &  $2.9543 \pm 0.0107$  &  $2.9567 \pm 0.0096$  &  $2.9564 \pm 0.0095$  & 2.966   \\[1 mm]
\hline\hline
 $\rtote$ & $0.14092 \pm 0.00095$ &                       & $0.14129 \pm 0.00091$ &                       &         \\[1 mm]
 $\rtotm$ & $0.14274 \pm 0.00072$ &                       & $0.14301 \pm 0.00067$ &                       &         \\[1 mm]
 $\rtott$ & $0.14161 \pm 0.00100$ &                       & $0.14204 \pm 0.00096$ &                       &         \\[1 mm]
\hline
 $\rtotl$ &                       & $0.14230 \pm 0.00062$ &                       & $0.14240 \pm 0.00058$ & 0.1427  \\[1 mm]
\hline\hline
 $\jtoth$ &$-0.21    \pm 0.64$    & $0.51    \pm 0.55$    & $0.44    \pm 0.12$    & $0.47    \pm 0.12$    & 0.22    \\[1 mm]
\hline\hline
 $\jtote$ &$-0.094   \pm 0.075$   &                       &$-0.042   \pm 0.048$   &                       &         \\[1 mm]
 $\jtotm$ & $0.056   \pm 0.042$   &                       & $0.029   \pm 0.019$   &                       &         \\[1 mm]
 $\jtott$ & $0.004   \pm 0.046$   &                       &$-0.013   \pm 0.026$   &                       &         \\[1 mm]
\hline
 $\jtotl$ &                       & $0.047   \pm 0.037$   &                       & $0.010   \pm 0.015$   & 0.004   \\[1 mm]
\hline\hline
 $\rfbe$  & $0.00306 \pm 0.00091$ &                       & $0.00298 \pm 0.00090$ &                       &         \\[1 mm]
 $\rfbm$  & $0.00275 \pm 0.00051$ &                       & $0.00286 \pm 0.00049$ &                       &         \\[1 mm]
 $\rfbt$  & $0.00406 \pm 0.00072$ &                       & $0.00428 \pm 0.00070$ &                       &         \\[1 mm]
\hline
 $\rfbl$  &                       & $0.00304 \pm 0.00038$ &                       & $0.00327 \pm 0.00037$ & 0.00273 \\[1 mm]
\hline\hline
 $\jfbe$  & $0.802   \pm 0.075$   &                       & $0.805   \pm 0.075$   &                       &         \\[1 mm]
 $\jfbm$  & $0.711   \pm 0.037$   &                       & $0.802   \pm 0.024$   &                       &         \\[1 mm]
 $\jfbt$  & $0.707   \pm 0.047$   &                       & $0.832   \pm 0.031$   &                       &         \\[1 mm]
\hline
 $\jfbl$  &                       & $0.725   \pm 0.027$   &                       & $0.811   \pm 0.018$   & 0.799   \\[1 mm]
\hline
\end{tabular}
\end{center}
\caption{\capsty{Results of the 16- and 8-parameter fits to the LEP I only and
combined LEP I + LEP II data. Also shown are the Standard Model 
predictions for the fit parameters. \MZ\ (~$\GZ$) are measured in units 
of GeV/$c^2$ (GeV), 
all other quantities are dimensionless.}}
\label{tab:smat:smatrix_fit}
\end{table}

\begin{table}[p]
\begin{center}
\rotatebox{90}{
\renewcommand{\arraystretch}{1.5}
\begin{tabular}{|l||r|r|r|r|r|r|r|r|r|r|r|r|r|r|r|r|}
\hline
\rule{0 mm}{3.5 mm}
  & \mco{$\GZ$}
  & \mco{$\rtoth$} & \mco{$\rtote$} & \mco{$\rtotm$} &\mco{$\rtott$}
  & \mco{$\jtoth$} & \mco{$\jtote$} & \mco{$\jtotm$} &\mco{$\jtott$}
  & \mco{$\rfbe$}  & \mco{$\rfbm$}  & \mco{$\rfbt$}
  & \mco{$\jfbe$}  & \mco{$\jfbm$}  & \mco{$\jfbt$}                 \\[0.5 mm]
\hline
\hline
$\MZ$    & -0.50 & -0.46 & -0.29 & -0.32 & -0.25 & -0.96 & -0.80 & -0.70 & -0.64 &  0.13 &  0.24 &  0.16 & -0.03 &  0.00 &  0.00 \\[0.5 mm]
$\GZ$    &       &  0.90 &  0.52 &  0.67 &  0.49 &  0.53 &  0.40 &  0.38 &  0.35 & -0.06 & -0.11 & -0.07 &  0.04 &  0.04 &  0.03 \\[0.5 mm]
$\rtoth$ &       &       &  0.53 &  0.68 &  0.49 &  0.49 &  0.37 &  0.35 &  0.32 & -0.05 & -0.10 & -0.06 &  0.04 &  0.05 &  0.04 \\[0.5 mm]
$\rtote$ &       &       &       &  0.39 &  0.28 &  0.30 &  0.26 &  0.22 &  0.20 &  0.07 & -0.06 & -0.04 &  0.08 &  0.03 &  0.02 \\[0.5 mm]
$\rtotm$ &       &       &       &       &  0.36 &  0.34 &  0.26 &  0.33 &  0.23 & -0.03 & -0.05 & -0.04 &  0.03 &  0.08 &  0.03 \\[0.5 mm]
$\rtott$ &       &       &       &       &       &  0.26 &  0.20 &  0.19 &  0.25 & -0.03 & -0.05 &  0.00 &  0.02 &  0.02 &  0.09 \\[0.5 mm]
$\jtoth$ &       &       &       &       &       &       &  0.78 &  0.70 &  0.63 & -0.12 & -0.24 & -0.16 &  0.03 &  0.00 &  0.01 \\[0.5 mm]
$\jtote$ &       &       &       &       &       &       &       &  0.57 &  0.52 & -0.08 & -0.20 & -0.13 &  0.10 &  0.00 &  0.00 \\[0.5 mm]
$\jtotm$ &       &       &       &       &       &       &       &       &  0.46 & -0.09 & -0.15 & -0.12 &  0.02 & -0.04 &  0.00 \\[0.5 mm]
$\jtott$ &       &       &       &       &       &       &       &       &       & -0.08 & -0.16 & -0.08 &  0.02 &  0.00 & -0.04 \\[0.5 mm]
$\rfbe $ &       &       &       &       &       &       &       &       &       &       &  0.04 &  0.03 &  0.09 &  0.00 &  0.00 \\[0.5 mm]
$\rfbm $ &       &       &       &       &       &       &       &       &       &       &       &  0.05 & -0.01 &  0.20 &  0.00 \\[0.5 mm]
$\rfbt $ &       &       &       &       &       &       &       &       &       &       &       &       &  0.00 &  0.00 &  0.18 \\[0.5 mm]
$\jfbe $ &       &       &       &       &       &       &       &       &       &       &       &       &       &  0.00 &  0.00 \\[0.5 mm]
$\jfbm $ &       &       &       &       &       &       &       &       &       &       &       &       &       &       &  0.00 \\[0.5 mm]
\hline
\end{tabular}
}
\end{center}
\caption{\capsty{Correlation matrix for the 16-parameter fit to LEP I data.}}
\label{tab:smat:smatrix_corr_16_lep1}
\end{table}

\begin{table}[p]
\begin{center}
\rotatebox{90}{
\renewcommand{\arraystretch}{1.5}
\begin{tabular}{|l||r|r|r|r|r|r|r|r|r|r|r|r|r|r|r|r|}
\hline
\rule{0 mm}{3.5 mm}
  & \mco{$\GZ$}
  & \mco{$\rtoth$} & \mco{$\rtote$} & \mco{$\rtotm$} &\mco{$\rtott$}
  & \mco{$\jtoth$} & \mco{$\jtote$} & \mco{$\jtotm$} &\mco{$\jtott$}
  & \mco{$\rfbe$}  & \mco{$\rfbm$}  & \mco{$\rfbt$}
  & \mco{$\jfbe$}  & \mco{$\jfbm$}  & \mco{$\jfbt$}                 \\[0.5 mm]
\hline
\hline
$\MZ$    &  0.03 &  0.06 &  0.01 &  0.04 &  0.02 & -0.55 & -0.35 & -0.15 & -0.16 &  0.04 &  0.08 &  0.05 &  0.00 & -0.05 & -0.06 \\[0.5 mm]
$\GZ$    &       &  0.87 &  0.44 &  0.61 &  0.42 &  0.02 & -0.02 &  0.02 &  0.02 &  0.01 &  0.02 &  0.02 &  0.03 &  0.05 &  0.05 \\[0.5 mm]
$\rtoth$ &       &       &  0.45 &  0.62 &  0.43 &  0.00 & -0.03 &  0.01 &  0.02 &  0.01 &  0.02 &  0.02 &  0.03 &  0.05 &  0.04 \\[0.5 mm]
$\rtote$ &       &       &       &  0.31 &  0.22 &  0.01 &  0.04 &  0.01 &  0.01 &  0.12 &  0.01 &  0.01 &  0.07 &  0.03 &  0.02 \\[0.5 mm]
$\rtotm$ &       &       &       &       &  0.30 & -0.01 & -0.02 &  0.08 &  0.01 &  0.01 &  0.03 &  0.02 &  0.02 &  0.10 &  0.03 \\[0.5 mm]
$\rtott$ &       &       &       &       &       &  0.00 & -0.01 &  0.01 &  0.09 &  0.01 &  0.01 &  0.04 &  0.01 &  0.03 &  0.13 \\[0.5 mm]
$\jtoth$ &       &       &       &       &       &       &  0.23 &  0.12 &  0.14 & -0.02 & -0.05 & -0.03 &  0.00 &  0.05 &  0.06 \\[0.5 mm]
$\jtote$ &       &       &       &       &       &       &       &  0.06 &  0.06 &  0.03 & -0.03 & -0.02 &  0.12 &  0.02 &  0.03 \\[0.5 mm]
$\jtotm$ &       &       &       &       &       &       &       &       &  0.04 & -0.01 &  0.07 & -0.01 &  0.00 &  0.34 &  0.02 \\[0.5 mm]
$\jtott$ &       &       &       &       &       &       &       &       &       & -0.01 & -0.01 &  0.08 &  0.00 &  0.02 &  0.39 \\[0.5 mm]
$\rfbe $ &       &       &       &       &       &       &       &       &       &       &  0.01 &  0.01 &  0.09 &  0.00 &  0.00 \\[0.5 mm]
$\rfbm $ &       &       &       &       &       &       &       &       &       &       &       &  0.02 &  0.00 &  0.12 &  0.00 \\[0.5 mm]
$\rfbt $ &       &       &       &       &       &       &       &       &       &       &       &       &  0.00 &  0.00 &  0.11 \\[0.5 mm]
$\jfbe $ &       &       &       &       &       &       &       &       &       &       &       &       &       &  0.00 &  0.00 \\[0.5 mm]
$\jfbm $ &       &       &       &       &       &       &       &       &       &       &       &       &       &       &  0.01 \\[0.5 mm]
\hline
\end{tabular}
}
\end{center}
\caption{\capsty{Correlation matrix for the 16-parameter fit to 
          LEP I + LEP II data.}}
\label{tab:smat:smatrix_corr_16_lep1+lep2}
\end{table}

\begin{table}[p]
\begin{center}
\renewcommand{\arraystretch}{1.5}
\begin{tabular}{|c||c|c|c|c|c|c|c|}
\hline
\rule{0 mm}{5 mm}
          & $\GZ$ & $\rtoth$ & $\rtotl$ & $\jtoth$ & $\jtotl$ & $\rfbl$ & $\jfbl$ \\[1 mm]
\hline\hline
 $\MZ$    & -0.42 &    -0.39 &    -0.32 &    -0.95 &    -0.83 &    0.27 &     0.03 \\[1 mm]
 $\GZ$    &       &     0.90 &     0.74 &     0.46 &     0.38 &   -0.09 &     0.05 \\[1 mm]
 $\rtoth$ &       &          &     0.75 &     0.43 &     0.35 &   -0.08 &     0.05 \\[1 mm]
 $\rtotl$ &       &          &          &     0.35 &     0.33 &   -0.04 &     0.09 \\[1 mm]
 $\jtoth$ &       &          &          &          &     0.81 &   -0.26 &    -0.03 \\[1 mm]
 $\jtotl$ &       &          &          &          &          &   -0.20 &    -0.04 \\[1 mm]
 $\rfbl$  &       &          &          &          &          &         &     0.17 \\[1 mm]
\hline
\end{tabular}
\end{center}
\caption{\capsty{Correlation matrix for the 8-parameter fit to LEP I data.}}
\label{tab:smat:smatrix_corr_8_lep1}
\end{table}
\begin{table}[p]
\begin{center}
\renewcommand{\arraystretch}{1.5}
\begin{tabular}{|c||c|c|c|c|c|c|c|}
\hline
\rule{0 mm}{5 mm}
          & $\GZ$ & $\rtoth$ & $\rtotl$ & $\jtoth$ & $\jtotl$ & $\rfbl$ & $\jfbl$ \\[1 mm]
\hline\hline
 $\MZ$    &  0.02 &     0.05 &     0.03 &    -0.53 &    -0.29 &    0.09 &    -0.10 \\[1 mm]
 $\GZ$    &       &     0.87 &     0.70 &     0.03 &     0.02 &    0.03 &     0.07 \\[1 mm]
 $\rtoth$ &       &          &     0.71 &     0.01 &     0.01 &    0.03 &     0.07 \\[1 mm]
 $\rtotl$ &       &          &          &     0.00 &     0.07 &    0.06 &     0.12 \\[1 mm]
 $\jtoth$ &       &          &          &          &     0.22 &   -0.06 &     0.08 \\[1 mm]
 $\jtotl$ &       &          &          &          &          &    0.05 &     0.35 \\[1 mm]
 $\rfbl$  &       &          &          &          &          &         &     0.10 \\[1 mm]
\hline
\end{tabular}
\end{center}
\caption{\capsty{Correlation matrix for the 8-parameter fit to LEP I + LEP II 
data.}}
\label{tab:smat:smatrix_corr_8_lep1+lep2}
\end{table}

%%%%%%%%%%%%%%%%%%%%%%%%%%%%%%%%%%%%%%%%%%%%%%%%%%%%%%%%%%%%%%%%%%%%%%%%%%%%

%%--------------------------------------------------------------------
%%-- ZPRIME FITS -----------------------------------------------------
%%--------------------------------------------------------------------
\subsection{$\boldmath{\Zprime}$ Bosons}
\label{sec:zp}

%% ------------------------------------------------------------------- 
\subsubsection{Theory}
\label{sec:zp:thry}

%%....................................................................
%%.. Summary .........................................................
%%....................................................................
%\fbox{
%\begin{minipage}{0.87\textwidth}
%\begin{itemize}
% \item Summary of theory papers
% \item What are the important parameters
% \item Choices of other parameters for the particular models
%\end{itemize}
%\end{minipage}
%}

%%....................................................................
%%.. Contents ........................................................
%%....................................................................
Many theories which are more general than the Standard Model 
predict the existence of additional heavy gauge bosons. The
consequences of several of these models were investigated, complemented by a
model-independent fit to the leptonic data.

%\paragraph{Specific ${\Zprime}$ Models }
The existence of an additional heavy gauge boson ${\Zprime}$ can be
parametrised by the mass of the boson ${\MZp}$ and by its couplings 
to fermions. In addition, a possible mixing between ${\Zprime}$ and 
the standard Z, represented by a mixing angle $\thtzzp$, has to be 
taken into account~\cite{ref:zp:thry1991,ref:zp:thry1992}. 
In order to deal with a restricted number of free parameters, it is useful to
consider specific ${\Zprime}$ models with well-defined couplings. Popular
models are:

\begin{itemize}
 \item The $E_6$ model~\cite{ref:zp:thry1984,ref:zp:thry1986}
is based on a symmetry breaking of the $E_6$ GUT where two gauge groups 
$U(1)_\chi$ and $U(1)_\psi$ are introduced,
%The gauge eigenstate $\mathrm{Z}^{0'}$ will be a mixed state of these two groups:
%\begin{displaymath}
\begin{eqnarray}
J^{\mu}_{\ssize{\Zprime}} = J^{\mu}_{\chi}\cos{\Theta_6} +J^{\mu}_{\psi}\sin{\Theta_6}\,.
%\mathrm{Z}^{0'} = Z_{\chi}\cos{\Theta_6} + Z_{\psi}\sin{\Theta_6}\,.
 \label{eq:zp:e6}
\end{eqnarray}
%\end{displaymath}
The free parameter, $\Theta_6$, of this model is the mixing of the $\chi$ and $\psi$ fields
to form the $\Zprime$. Usual choices of $\Theta_6$ are 
$\Theta_6 = 0,\, \pi/2,\, -\arctan \sqrt{5/3}$ 
( $\chi$, $\psi$ and $\eta$ model);

 \item The L-R model~\cite{ref:zp:thry1974,ref:zp:thry1975}
includes a right-handed $SU(2)_R$ extension
to the Standard Model gauge group $SU(2)_L \otimes U(1)$. 
The free parameter
$\alpha_{LR}$ describes the coupling of the heavy bosons to fermions.
$\alpha_{LR}$ can take values between 
$\sqrt{2/3} \le \alpha_{LR} \le \sqrt{\cot^2
\theta_W - 1}$ ($\sim 1.53$), where $\theta_W$ is the weak mixing angle.

\end{itemize}
%........................................................................%
%\paragraph{Model independent approach}
In a more general approach, the ${\Zprime}$ boson is directly
described in terms of its couplings $a_f'$ and $v_f'$
~\cite{ref:zp:thry1997}.
Off the ${\Zprime}$ resonance, pair production is only sensitive to the
normalised couplings $a_f^N$ and $v_f^N$. As a consequence, the couplings and
the mass of the ${\Zprime}$ boson cannot be measured independently.
The normalised couplings are:
\begin{equation}
a_f^N = a_f' \sqrt{ \frac{s}{\MZpsq-s} } \ , \quad
v_f^N = v_f' \sqrt{ \frac{s}{\mathrm{M}_{\ssize{\Zprime}}^{2}-s} } \ .
%a_f^N = a_f' \sqrt{ \frac{s}{\mathrm{M}_{\ssize{\Zprime}}^{2}-s} } \ , \quad
%v_f^N = v_f' \sqrt{ \frac{s}{\mathrm{M}_{\ssize{\Zprime}}^{2}-s} } \ .
\end{equation}
%

%% ------------------------------------------------------------------- 
\subsubsection{Results}
\label{sec:zp:res}

%%....................................................................
%%.. Summary .........................................................
%%....................................................................
%\fbox{
%\begin{minipage}{0.87\textwidth}
%\begin{itemize}
% \item Which data is fitted
% \item How radiative corrections to new physics are handled
% \item Treatment of correlated errors
% \item Parameters for fit
% \item How limits are extracted from fit
% \item Quote confidence level for limits
% \item Plot showing sensitivity of data to new physics \\
%       and/or \\
%       Likelihood/$\chi^{2}$ curves/surfaces \\
%       and/or \\
%       Plot showing limits/excluded region of parameters
% \item Table of results including fit results
%\end{itemize}
%\end{minipage}
%}

%%....................................................................
%%.. Contents ........................................................
%%....................................................................

The {\em non-radiative} hadronic total cross-sections and $\mumu$ and
$\tautau$ total cross-sections and asymmetries presented here together
with existing data from LEP
I~\cite{ref:delphils:91,*ref:delphils:95,ref:delphils:00} and LEP II at
$\roots \sim$ 130-172 GeV~\cite{ref:delphiff:130-172} were used to fit
the data to models including additional \Zprime\ bosons.

Fits were made to the mass of $\Zprime$, ${\MZp}$, the mass of the Z, $\MZ$,
and to the mixing angle between the two bosonic fields, $\thtzzp$, for 4
different models referred to as $\chi$, $\psi$, $\eta$ and 
L-R as described in \ref{sec:zp:thry}. 
The theoretical predictions came from the ZEFIT package (version 6.10)
~\cite{ref:zp:thry1991} 
together with the ZFITTER program (version 6.10).
The program ZEFIT provides predictions for the
cross-sections and forward-backward asymmetries for each model as a function 
of $\MZ$, ${\MZp}$, $\thtzzp$, the masses of the Higgs boson, $\MH$, and the
top quark, $\MT$, 
the strong coupling constant, $\alpha_s$, 
and the $\Zprime$-model parameters $\Theta_6$ or $\alpha_{LR}$.
For the L-R model $\alpha_{LR}$ was set to $1.1$.
In order to reduce the number of free parameters, the following input
parameters were used: $\MT = 175~\GeV/c^{2}$, $\MH = 150~\GeV/c^{2}$, 
$\alpha_s$ = 0.118.

%the Higgs was set to $M_{H}$ = 150 $\GeVcsq$ and the coupling $\alpha_s$
%to the value 0.118.
% Varying the values of these parameters has a negligible influence on 
% the fit results.

The correlations between the experimental errors were taken into account when 
the $\chi^2$ was calculated between the predictions and the measurements.
The most important correlation is between the errors on the
luminosity for the cross-section measurements.
Correlations between LEP I and LEP II measurements are very small and
therefore neglected. 

\begin{figure}[p]
 \begin{center}
   \mbox{\epsfig{file=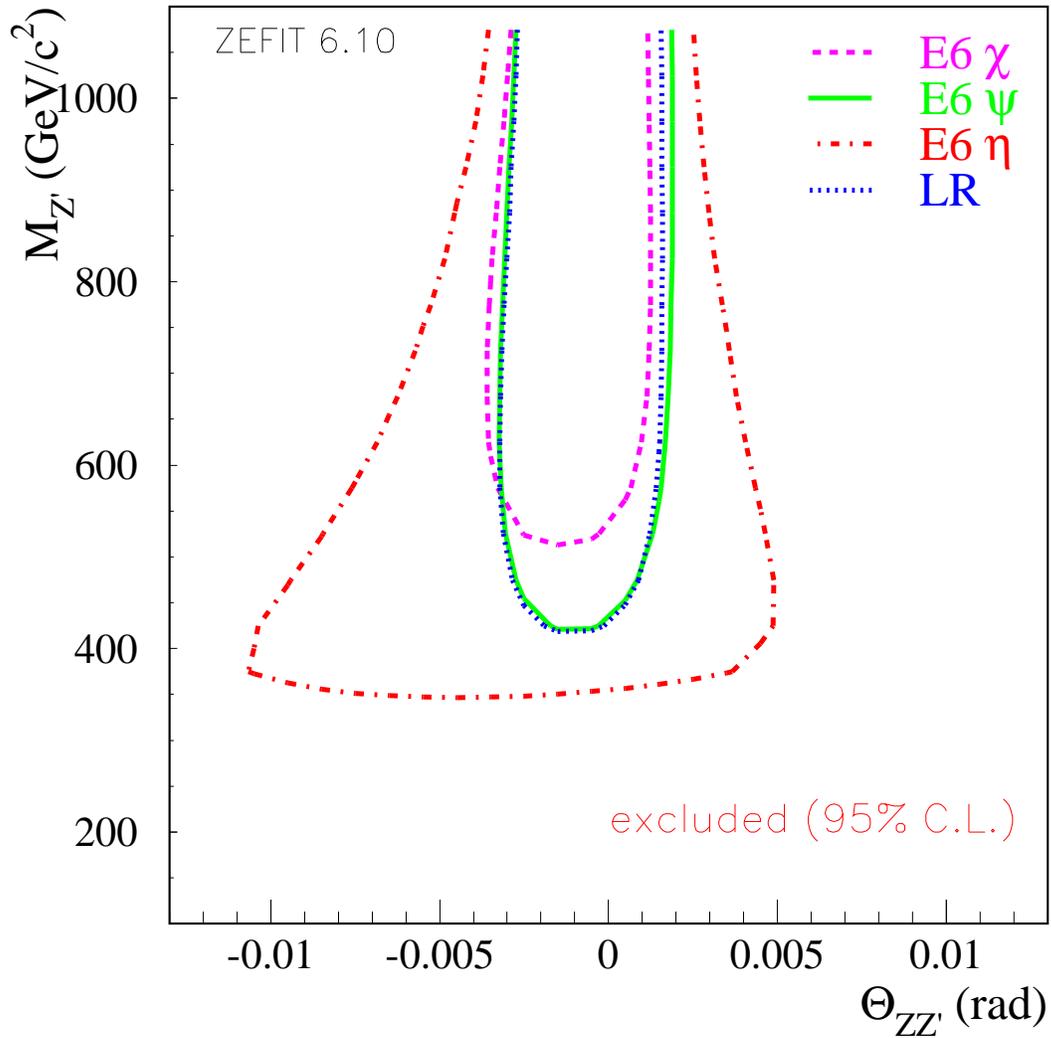,width=\linewidth}}
  \caption[]{\capsty
           {The allowed domains in the \thtzzp$ - $\MZp  plane
           for the $\chi$, $\psi$, $\eta$ and L-R
           models.
           Values outside the contours are excluded at 95\% confidence level.}}
  \label{moddep_Mmix_limits}
 \end{center}
\end{figure}

No evidence was found for the existence of a $\Zprime$--boson in any of the
models. The fitted value of $\MZ$ was found to be in agreement with the
Standard Model value. 
Two-dimensional exclusion contours at 95\% confidence level, obtained 
with $\chi^2 > \chi^2_{min} + 5.99$~\cite{ref:zp:minuit} 
in the ${\MZp}$--$\thtzzp$ plane,
were made. The allowed regions for $\MZp$ and $\thtzzp$ are shown in
Figure~\ref{moddep_Mmix_limits}. The one-dimensional limits at 95\%
confidence level for both ${\MZp}$ and $\thtzzp$, 
obtained with $\chi^2 > \chi^2_{min}$ + 3.84, are shown in Table~\ref{tab_masslimits_moddep}. 
%%%%%%%%%%%%%%%%%%%%%%%Repeat caveat mentioned in footnote
We repeat that these limits are based on the assumption that the 
electroweak one-loop corrections are those of the SM. In particular
for the L-R symmetric model, the authors of
reference~\cite{ref:1loop:thry2002} have shown that inclusion of the
extra gauge bosons in the corrections can affect the limits on the
mixing angle.
%%%%%%%%%%%%%%%%%%%%%%%%%%%%%%
The limits for the $\Zprime$ mass range from 360 to 545 $\GeV/c^{2}$, an increase of
between 50 and 125 $\GeV/c^{2}$ on the limits
presented in~\cite{ref:delphiff:183-189}, depending on the model.

In addition the Sequential Standard Model \cite{ref:zpr:sqsm} has been considered.
This model proposes the existence of a $\Zprime$
with exactly the same coupling to fermions as the Standard Model Z.
A limit of \mbox{$\MZp > 1305 \: \GeV/c^{2}$} is found at $95\%$
confidence level, an increase of 595 $\GeV/c^{2}$ on the limit
presented in~\cite{ref:delphiff:183-189}.

%%%%%%%%%%%%%%
%%%%%%%%%%%%%%

%%%%%%%%%%%%%%
\begin{table}[p]
\begin{center}
\renewcommand{\arraystretch}{1.5}
\begin{tabular}{|cc|c|c|c|c|}
\hline
 \multicolumn{2}{|c|}{Model}          & $\chi$  & $\psi$ & $\eta$ & L-R    \\
\hline \hline
 $\MZplim$           & ($\GeV/c^{2}$) &~~545~~&~~475~~&~~360~~&~~455~~ \\ \hline
$| \ \thtzzplim \ |$ & (mrad)         & 3.1     & 2.7    & 9.2    & 2.8    \\ 
\hline
\end{tabular}
\end{center}
\caption{\capsty 
           {95\% confidence level lower limits on the $\Zprime$ mass and 
           upper limits on the Z$\Zprime$ mixing angle within the $\chi$, 
           $\psi$, $\eta$ and L-R models. 
                }} 
  \label{tab_masslimits_moddep}
\end{table}
%%%%%%%%%%%%%%

Model-independent fits were performed to the leptonic cross-sections and
forward--backward asymmetries, for the leptonic couplings of a $\Zprime$,
$a^{N}_{l'}$ and $v^{N}_{l'}$, normalised for the overall coupling scale
and the mass of the $\Zprime$~\cite{ref:zp:thry1997}.

Several values of the mass of the ${\Zprime}$
were considered (i.e. 300, 500 and 1000 $\GeV/c^{2}$), and the
Z${\Zprime}$--mixing was neglected.
%Figure \ref{fig_mod_ind} shows the values of the couplings $a_f'$ and $v_f'$
%which are compatible with the DELPHI data with a confidence level of 95\%.
The 95\% confidence level limits on the normalised couplings are
$|a^{N}_{l'}| < 0.19$ and $|v^{N}_{l'}| < 0.19$, an increase of $0.04$
and a decrease of $0.03$, respectively, on limits given in
\cite{ref:delphiff:183-189}.

%%--------------------------------------------------------------------
%%-- CONTACT FITS ----------------------------------------------------
%%--------------------------------------------------------------------
\subsection{Contact interactions}
\label{sec:cntc}

%% ------------------------------------------------------------------- 
\subsubsection{Theory}
\label{sec:cntc:thry}

%%....................................................................
%%.. Summary .........................................................
%%....................................................................
%\fbox{
%\begin{minipage}{0.87\textwidth}
%\begin{itemize}
% \item Summary of theory papers~\cite{ref:cntc:thryELP,ref:cntc:thryKroha}
% \item What are the important parameters
% \item Choices of other parameters for the particular models
%\end{itemize}
%\end{minipage}
%}

%%....................................................................
%%.. Contents ........................................................
%%....................................................................

%%%%%%%%%%%%%%%%%%%%%%%%%%%
\begin{table}[tp]
 \begin{center}
  \begin{tabular}{|c|c|c|c|c|}
   \hline
   Model      & $\eta_{LL}$ & $\eta_{RR}$ & $\eta_{LR}$ & $\eta_{RL}$ \\
   \hline\hline
   LL$^{\pm}$ &   $\pm 1$   &      0      &      0      &      0      \\
   \hline
   RR$^{\pm}$ &      0      &   $\pm 1$   &      0      &      0      \\
   \hline
   VV$^{\pm}$ &   $\pm 1$   &   $\pm 1$   &   $\pm 1$   &   $\pm 1$   \\
   \hline
   AA$^{\pm}$ &   $\pm 1$   &   $\pm 1$   &   $\mp 1$   &   $\mp 1$   \\
   \hline
   LR$^{\pm}$ &      0      &      0      &   $\pm 1$   &      0      \\
   \hline
   RL$^{\pm}$ &      0      &      0      &      0      &   $\pm 1$   \\
   \hline
   V0$^{\pm}$ &   $\pm 1$   &   $\pm 1$   &      0      &      0      \\
   \hline
   A0$^{\pm}$ &      0      &      0      &  $\pm 1$    &   $\pm 1$   \\
   \hline
  \end{tabular}
 \end{center}
 \caption{\capsty{Choices of $\eta_{ij}$ for different contact interaction 
                  models, LL, RR {\emph{etc}}.}}
 \label{tab:cntc:def}
\end{table}
%%%%%%%%%%%%%%%%%%%%%%%%%%%

Contact interactions between initial and final state fermionic currents provide
a rather general description of the low energy behaviour of new physics 
with a characteristic high energy scale. Following 
reference~\cite{ref:cntc:thryELP}, these interactions are parameterised by
an effective Lagrangian, $\cal{L}$$_{eff}$, of the form:
%%%
\begin{equation}
 \mbox{$\cal{L}$}_{eff} = \frac{g^{2}}{(1+\delta)\Lambda^{2}} 
                          \sum_{i,j=L,R} \eta_{ij} 
                           \overline{e}_{i} \gamma_{\mu} e_{i}
                            \overline{f}_{j} \gamma^{\mu} f_{j},
 \label{eqn:cntclag}
\end{equation}
%%%
where $g^{2}/{4\pi}$ is taken to be 1 by convention, $\delta=1 (0)$ for 
$f=e (f \neq e)$, $\eta_{ij}=\pm 1$ or $0$,
$\Lambda$ is the scale of the contact interactions%
\footnote{The choice of $g^{2}$ is somewhat arbitrary; if the coupling
          constant was taken to be $\alpha_{s}$ much lower limits on
          $\Lambda$ would be obtained.}, $e_{i}$ and $f_{j}$ are left
          or right-handed spinors. This effective Lagrangian is added
          to the Standard Model Lagrangian.  By assuming different
          helicity couplings between the initial-state and final-state
          currents and either constructive or destructive interference
          with the Standard Model (according to the choice of each
          $\eta_{ij}$) a set of different models can be defined from
          this Lagrangian~\cite{ref:cntc:thryKroha}. The values of
          $\eta_{ij}$ for the models investigated in this paper are
          given in Table~\ref{tab:cntc:def}.  The differential
          cross-section for scattering the outgoing fermion at an
          angle $\theta$ with respect to the incident $e^{-}$
          direction is given
          by~\cite{ref:cntc:thryALEPH,ref:cntc:thryOPAL}
\begin{equation}
\dsdcth
%%  \frac{d\sigma}{d\cos \theta} 
= 
          \frac{\pi \alpha^{2}}{2s} N_{c}^{f} 
           \left\{
            \begin{array}[c]{l}
             \left[ \left|\overline{A}_{LR}^{ee}\right|^{2} +
                     \left|\overline{A}_{RL}^{ee}\right|^{2} \right]
                 (\frac{s}{t})^{2} \delta \: + 
             \\
              \hspace{3mm}
              \left[ \left|A_{LR}^{ef}\right|^{2} +
                    \left|A_{RL}^{ef}\right|^{2} \right]
                 (\frac{t}{s})^{2} \: + 
             \\
               \hspace{6mm}
               \left[ \left|A_{LL}^{ef}\right|^{2} +
                    \left|A_{RR}^{ef}\right|^{2} \right]
                 (\frac{u}{s})^{2}
            \end{array}
           \right\} ,
\end{equation}
where $s$, $t$ and $u$ are the Mandelstam variables and $N_{c}^{f}$ is
the number of colours for fermion~$f$. The $A_{ij}$ and 
$\overline{A}_{ij}$ are helicity amplitudes for the scattering 
process~\cite{ref:cntc:thryELP}. When the helicity amplitudes are squared, 3 
sets of terms arise: the first set contains purely Standard Model terms; 
the second set of terms derive from the interference between contact 
interactions and the Standard Model, these terms are proportional to 
$1/\Lambda^{2}$; the final set of terms are due to contact interactions alone
and are proportional to $1/\Lambda^{4}$. For the purpose of fitting contact
interaction models to the data, a new parameter 
$\epsilon=1/\Lambda^{2}$ is defined, with $\epsilon=0$ in the limit that 
there are no contact interactions. This parameter is allowed to 
take both positive and negative values in the fits. It is worth noting that 
there is a symmetry between models with $\eta_{ij}=+1$ and those with 
$\eta_{ij}=-1$. The predicted differential cross-section in the 
constructive (+) models is the same as the destructive ($-$) models for 
$\epsilon^{-}=-\epsilon^{+}$. Therefore, starting from a model with 
constructive interference with the Standard Model, the region $\epsilon \ge 0$
represents physical values of $1/\Lambda^2$, while the region $\epsilon \le 0$
represents physical values for the equivalent model with destructive 
interference.

%% ------------------------------------------------------------------- 
\subsubsection{Results}
\label{sec:cntc:res}

%%....................................................................
%%.. Summary .........................................................
%%....................................................................
%\fbox{
%\begin{minipage}{0.87\textwidth}
%\begin{itemize}
% \item Which data is fitted
% \item How radiative corrections to new physics are handled
% \item Treatment of correlated errors
% \item Parameters for fit
% \item How limits are extracted from fit
% \item Quote confidence level for limits
% \item Plot showing sensitivity of data to new physics \\
%       and/or \\
%       Likelihood/$\chi^{2}$ curves/surfaces \\
%       and/or \\
%       Plot showing limits/excluded region of parameters
% \item Table of results including fit results
%\end{itemize}
%\end{minipage}
%}
%%....................................................................
%%.. Contents ........................................................
%%....................................................................

Cross-section and forward-backward asymmetry measurements from all
LEP II centre-of-mass energies were compared to each of the contact 
interaction models mentioned above%
\footnote{For leptonic final 
          states, models with only $\eta_{LR}=\pm 1$ are
          equivalent to models with only $\eta_{RL}=\pm 1$.}
using $\chi^{2}$ fits,
considering separately the $e^{+}e^{-}$, $\mu^{+}\mu^{-}$ and 
$\tau^{+}\tau^{-}$ final states, and a simultaneous fit to all three final 
states, assuming lepton universality in the contact interactions to obtain
$\chi^2$ curves as a function of $\epsilon$. 
The predicted tree-level results were corrected for QED radiation.
Correlations between the
systematic uncertainties were taken into account in the fits. Making use of 
the symmetry, mentioned above, between models with constructive and destructive
interference with the Standard Model, it is possible to fit pairs of models 
by allowing $\epsilon$ to take both positive and negative values.

The resulting $\chi^{2}$ as a function of $\epsilon$ were not always
Gaussian-like parabolae. The values of $\epsilon$ extracted from the points
with minimum $\chi^{2}$ from each of the fits were all compatible with the 
Standard Model expectation $\epsilon=0~\TeV^{-2}$,
{\it{i.e.}} the differences in $\chi^{2}$ between the best fit points
and the Standard Model point were always less than nine, except in the
fit for the A0 model in \eett, in which case the difference in $\chi^{2}$ 
was found to be 11.2.

Errors on $\epsilon$ were derived by finding the 
points above and below the best fit points for which the $\chi^{2}$ 
increased by nine above the minimum value, which would correspond to a three 
sigma uncertainty for Gaussian-like parabolae. These three sigma uncertainties 
were divided by three, to give one sigma errors. 
$95\%$ confidence level lower limits on $\Lambda$, by integrating under 
the likelihood curves, ${\mathcal{L}}(\epsilon)$, were obtained from the 
$\chi^{2}$ fits using
\begin{displaymath}
 {\mathcal{L}}(\epsilon) = \exp -\frac{1}{2}\chi^{2}(\epsilon)
\end{displaymath}
over the physically allowed region of $\epsilon$ for each choice of model%
\footnote{Integrating under the likelihood curve for $\epsilon$ is equivalent 
to obtaining a Bayesian limit assuming a prior uniform in $\epsilon$ over the 
physically allowed regions and zero for the unphysical regions.}.
The fitted values of $\epsilon$ with upper and lower errors and the limits
on $\Lambda$ for models with constructive, $\Lambda^{+}$, and destructive,
$\Lambda^{-}$, interference with the Standard Model are given in 
Table~\ref{tab:cntc:res}.

%%%%%%%%%%%%%%%%%%%%
\begin{table}[p]
 \begin{center}
 \renewcommand{\arraystretch}{1.1}
  \begin{tabular}{cc}
  \renewcommand{\arraystretch}{1.5}
  \begin{tabular}{|c|r|c|c|}
   \hline
   \multicolumn{4}{|c|}{\boldmath $e^{+}e^{-} \rightarrow e^{+}e^{-}$ \unboldmath} \\
   \hline
   Model  & $\epsilon^{+\sigma_{+}}_{-\sigma_{-}} (\TeV^{-2})$ &
  $\Lambda^{-} (\TeV)$ & $\Lambda^{+} (\TeV)$ \\
   \hline
   \hline
   LL &  0.0071$^{+ 0.0166}_{- 0.0135}$ &    6.8 &    5.3 \\
   \hline
   RR &  0.0106$^{+ 0.0157}_{- 0.0148}$ &    6.8 &    5.2 \\
   \hline
   VV &  0.0024$^{+ 0.0033}_{- 0.0038}$ &   13.9 &   11.7 \\
   \hline
   AA &  0.0176$^{+ 0.0148}_{- 0.0292}$ &    4.6 &    4.8 \\
   \hline
   RL &  0.0035$^{+ 0.0156}_{- 0.0133}$ &    6.7 &    5.7 \\
   \hline
   LR &  0.0035$^{+ 0.0156}_{- 0.0133}$ &    6.7 &    5.7 \\
   \hline
   V0 &  0.0038$^{+ 0.0077}_{- 0.0069}$ &   10.1 &    8.0 \\
   \hline
   A0 &  0.0058$^{+ 0.0060}_{- 0.0080}$ &    9.8 &    8.5 \\
   \hline
  \end{tabular}
  &
  \renewcommand{\arraystretch}{1.5}
  \begin{tabular}{|c|r|c|c|}
   \hline
   \multicolumn{4}{|c|}{\boldmath $e^{+}e^{-} \rightarrow \mu^{+}\mu^{-}$ \unboldmath} \\        
   \hline
   Model  & $\epsilon^{+\sigma_{+}}_{-\sigma_{-}} (\TeV^{-2})$ &
  $\Lambda^{-} (\TeV)$ & $\Lambda^{+} (\TeV)$ \\
   \hline
   \hline
   LL &  0.0019$^{+ 0.0093}_{- 0.0100}$ &    7.6 &    7.3 \\
   \hline
   RR &  0.0016$^{+ 0.0103}_{- 0.0109}$ &    7.2 &    7.0 \\
   \hline
   VV & -0.0006$^{+ 0.0040}_{- 0.0034}$ &   12.9 &   12.2 \\
   \hline
   AA &  0.0028$^{+ 0.0045}_{- 0.0057}$ &   10.9 &   10.1 \\
   \hline
   RL & -0.2377$^{+ 0.0919}_{- 0.0139}$ &    2.0 &    6.3 \\
   \hline
   LR & -0.2377$^{+ 0.0919}_{- 0.0139}$ &    2.0 &    6.3 \\
   \hline
   V0 & -0.0011$^{+ 0.0057}_{- 0.0044}$ &   11.5 &   10.9 \\
   \hline
   A0 & -0.2396$^{+ 0.0866}_{- 0.0067}$ &    2.0 &    9.0 \\
   \hline
  \end{tabular}
  \\
  \\
  \renewcommand{\arraystretch}{1.5}
  \begin{tabular}{|c|r|c|c|}
   \hline
   \multicolumn{4}{|c|}{\boldmath $e^{+}e^{-} \rightarrow \tau^{+} \tau^{-}$ \unboldmath} \\
   \hline
   Model  & $\epsilon^{+\sigma_{+}}_{-\sigma_{-}} (\TeV^{-2})$ &
  $\Lambda^{-} (\TeV)$ & $\Lambda^{+} (\TeV)$ \\
   \hline
   \hline
   LL & -0.0194$^{+ 0.0137}_{- 0.0166}$ &    4.6 &    7.9 \\
   \hline
   RR & -0.0213$^{+ 0.0150}_{- 0.0189}$ &    4.4 &    7.6 \\
   \hline
   VV & -0.0127$^{+ 0.0057}_{- 0.0053}$ &    7.1 &   15.8 \\
   \hline
   AA &  0.0029$^{+ 0.0060}_{- 0.0073}$ &    9.4 &    8.8 \\
   \hline
   RL & -0.1974$^{+ 0.0678}_{- 0.0220}$ &    2.1 &    7.9 \\
   \hline
   LR & -0.1974$^{+ 0.0678}_{- 0.0220}$ &    2.1 &    7.9 \\
   \hline
   V0 & -0.0134$^{+ 0.0083}_{- 0.0069}$ &    6.7 &   11.8 \\
   \hline
   A0 & -0.2223$^{+ 0.0720}_{- 0.0098}$ &    2.1 &   11.6 \\
   \hline
  \end{tabular}
 &
  \renewcommand{\arraystretch}{1.5}
  \begin{tabular}{|c|r|c|c|}
   \hline
   \multicolumn{4}{|c|}{\boldmath $e^{+}e^{-} \rightarrow l^{+} l^{-}$ \unboldmath} \\
   \hline
   Model  & $\epsilon^{+\sigma_{+}}_{-\sigma_{-}} (\TeV^{-2})$ &
  $\Lambda^{-} (\TeV)$ & $\Lambda^{+} (\TeV)$ \\
   \hline
   \hline
   LL & -0.0017$^{+ 0.0068}_{- 0.0071}$ &    8.2 &    9.1 \\
   \hline
   RR & -0.0015$^{+ 0.0073}_{- 0.0077}$ &    7.9 &    8.7 \\
   \hline
   VV & -0.0012$^{+ 0.0023}_{- 0.0024}$ &   13.7 &   16.5 \\
   \hline
   AA &  0.0011$^{+ 0.0042}_{- 0.0039}$ &   12.1 &   10.6 \\
   \hline
   RL & -0.0071$^{+ 0.0090}_{- 0.0097}$ &    6.5 &    8.7 \\
   \hline
   LR & -0.0071$^{+ 0.0090}_{- 0.0097}$ &    6.5 &    8.7 \\
   \hline
   V0 & -0.0007$^{+ 0.0035}_{- 0.0037}$ &   11.5 &   12.6 \\
   \hline
   A0 & -0.0035$^{+ 0.0045}_{- 0.0048}$ &    9.2 &   12.2 \\
   \hline
  \end{tabular}
 \end{tabular}
 \end{center}
%%%%%%%%%%%%%%%%%%%%%%%%%%%%%%%%%%%%%%%%%%%%%%%%%%%%%%%%%%%%

 \caption[]{\capsty{ 
               {Fitted values of $\epsilon$
               and 95\% confidence lower limits on the scale,
               $\Lambda$, of contact interactions in the models 
                 discussed in the text, for $\eeee$, $\eemm$, $\eett$  
               and $\eell$, a simultaneous fit to the above, assuming 
               lepton universality in the contact interactions. 
               The errors on $\epsilon$ come from the statistical and 
               systematic errors on the determination of the cross-sections
               and asymmetries.}}}
\label{tab:cntc:res}
\end{table}
%%%%%%%%%%%%%%%%%%%%

%%--------------------------------------------------------------------
%%-- GRAVITY FITS ----------------------------------------------------
%%--------------------------------------------------------------------
\subsection{Gravity in large extra dimensions}
\label{sec:grav}

%% ------------------------------------------------------------------- 
\subsubsection{Theory}
\label{sec:grav:thry}

%%....................................................................
%%.. Summary .........................................................
%%....................................................................
%\fbox{
%\begin{minipage}{0.87\textwidth}
%\begin{itemize}
% \item Summary of theory papers
% \item What are the important parameters
% \item Choices of other parameters for the particular models
%\end{itemize}
%\end{minipage}
%}

%%....................................................................
%%.. Contents ........................................................
%%....................................................................
The large difference between the electroweak scale 
($\mathrm{M_{EW}} \sim 10^{2} - 10^{3}~\GeV$)  and the scale at which 
quantum gravitational effects become strong, the Planck scale
($\mathrm{M_{Pl}} \sim 10^{19}~\GeV$), leads to the well known 
``hierarchy problem''. A 
solution, not relying on supersymmetry or technicolour, that involves an 
effective Planck scale, $\mathrm{M_{D}}$, of 
$\mathcal{O}$(TeV) is achieved by introducing $n$ 
compactified dimensions, into which spin 2 gravitons propagate, in addition to
the 4 dimensions of standard space-time~\cite{ref:grav:thryadd}. 
The Planck mass seen in the 4 uncompactified dimensions, $\mathrm{M_{Pl}}$, 
can be expressed in terms of $\mathrm{M_{D}}$, the effective Planck scale in 
the $n+4$ dimensional theory,
\begin{displaymath}
{\mathrm{M_{Pl}}}^{2} ~ \sim ~ {\mathrm{R}}^n {\mathrm{M_{D}}}^{n+2}, 
\end{displaymath}
where $\mathrm{R}$ is the size of the extra dimensions.  With
$\mathrm{M_{D}}=1$~TeV, the case where $n=1$ is excluded as Newtonian
gravitation would be modified at solar system distances, whereas $n=2$
corresponds to a radius for extra dimensions of $\mathcal{O}$(1 mm),
which is excluded by recent gravitational experiments which test the
inverse square law of gravitational attraction down to
$\mathcal{O}$(100$\mu$m)~\cite{ref:grav:hoyle2}. There
are also severe limits from astrophysics for a small number ($n=2,3$)
of large extra dimensions, however, higher numbers of dimensions are
not ruled out. In addition it is possible to construct models which
evade gravity and astrophysics bounds with a slight modification of
the extra dimension scenario~\cite{ref:grav:evasion}.

In high energy collisions at LEP and other colliders, new channels
not present in the Standard Model would be available in which gravitons 
could be produced or exchanged.
Virtual graviton exchange would affect the differential cross-section for
$\eeff$, with the largest contributions seen at low angles with respect to the 
incoming electron or positron. 
Embedding the model into a string model, and identifying the effective
Planck scale, $\mathrm{M_{D}}$, with the string scale, $\mathrm{M_{s}}$,
the differential cross-section for $\eeff$ with the inclusion of the spin 2 
graviton can be expressed as~\cite{ref:grav:thryhew}
\begin{displaymath}
\dsdcth~=~\left.\dsdcth\right|_{SM}
     +~C^{f}_{1}(s,\cos\theta)\left[\frac{\lambda}{\mathrm{M^4_s}}\right]~
      +~C^{f}_{2}(s,\cos\theta)\left[\frac{\lambda}{\mathrm{M^4_s}}\right]^2,
%\label{eqn:grav:eeff}
\end{displaymath}
while the differential cross-section for the process \eeee\ can be 
expressed as \cite{ref:grav:thryriz}
\begin{displaymath}
\dsdcth~=~\left.\dsdcth\right|_{SM} 
     -~C^{e}_{1}(s,t)\left[\frac{\lambda}{\mathrm{M^4_s}}\right]~
      +~C^{e}_{2}(s,t)\left[\frac{\lambda}{\mathrm{M^4_s}}\right]^2,
%\label{eqn:grav:eeee}
\end{displaymath}
\noindent
with $\theta$, as usual, being the polar angle of the outgoing fermion 
with respect to 
the direction of the incoming electron. The functions $C^{f}_{1}$, $C^{f}_{2}$,
$C^{e}_{1}$ and $C^{e}_{2}$ are known~\cite{ref:grav:thryhew,ref:grav:thryriz}.
For \eeff\ the expansion of $\cos\theta$ extends up to the fourth power.
The dimensionless parameter $\lambda$, of $\mathcal{O}$(1), 
is not explicitly calculable
without full knowledge of the underlying quantum gravitational theory. It can 
be either positive or negative \cite{ref:grav:thryhew,ref:grav:thrygiu}. For 
the purposes of the fits, two cases, $\lambda = \pm 1$, are considered.
This parametrisation has no explicit dependence on the number of 
extra dimensions, $n$.

\eject

%% ------------------------------------------------------------------- 
\subsubsection{Results}
\label{sec:grav:res}

%%....................................................................
%%.. Summary .........................................................
%%....................................................................
%\fbox{
%\begin{minipage}{0.87\textwidth}
%\begin{itemize}
% \item Which data is fitted
% \item How radiative corrections to new physics are handled
% \item Treatment of correlated errors
% \item Parameters for fit
% \item How limits are extracted from fit
% \item Quote confidence level for limits
% \item Plot showing sensitivity of data to new physics \\
%       and/or \\
%       Likelihood/$\chi^{2}$ curves/surfaces \\
%       and/or \\
%       Plot showing limits/excluded region of parameters
% \item Table of results including fit results
%\end{itemize}
%\end{minipage}
%}
%%....................................................................
%%.. Contents ........................................................
%%....................................................................

%%%%%%%%%%%%%%%%%%%%%%%%%%%%
\begin{table}[tp]
\begin{center}
\begin{tabular}{|c|c|c|c|} 
\hline
            & $\epsilon_{fit}$ &           & $\mathrm{M_{s}}$(TeV) \\ 
Final State & ( TeV$^{-4}$)    & $\lambda$ & $[95\% {\mathrm{C.L.}}]$ \\
\hline
\hline
\ee
            & $+0.54^{+0.67}_{-0.69}$  &
                        $\begin{array}{c} -1 \\ +1 \end{array}$ & 
                         $\begin{array}{c} 0.986  \\ 0.874 \end{array}$  \\ 
\hline
\mumu
            & $+0.84^{+1.92}_{-1.92}$  &
                        $\begin{array}{c} -1 \\ +1 \end{array}$ & 
                         $\begin{array}{c} 0.739  \\ 0.695 \end{array}$  \\ 
\hline
\tautau
            & $-3.04^{+3.28}_{-3.28}$ &       
                        $\begin{array}{c} -1 \\ +1 \end{array}$ & 
                         $\begin{array}{c} 0.578  \\  0.690 \end{array}$ \\
\hline
\lplm
            & $ +0.44^{+0.63}_{-0.63}$ &
                        $\begin{array}{c} -1 \\ +1 \end{array}$ & 
                         $\begin{array}{c} 0.998 \\ 0.898 \end{array}$   \\ 
\hline
\end{tabular}

\end{center}
\caption{\capsty{$95\%$ confidence level lower limits on $\mathrm{M_{s}}$ 
         in models of gravity in extra dimensions for $\ee$, $\mumu$ and 
         $\tautau$ final states, and for $\lplm$, a simultaneous fit to the 
         final states above.}}
\label{tab:grav:res}
\end{table}
%%%%%%%%%%%%%%%%%%%%%%%%%%%%

%%%%%%%%%%%%%%%%%%%%%%%%%%%%
\begin{figure}[tp]
\begin{center}
 \mbox{\epsfig{file=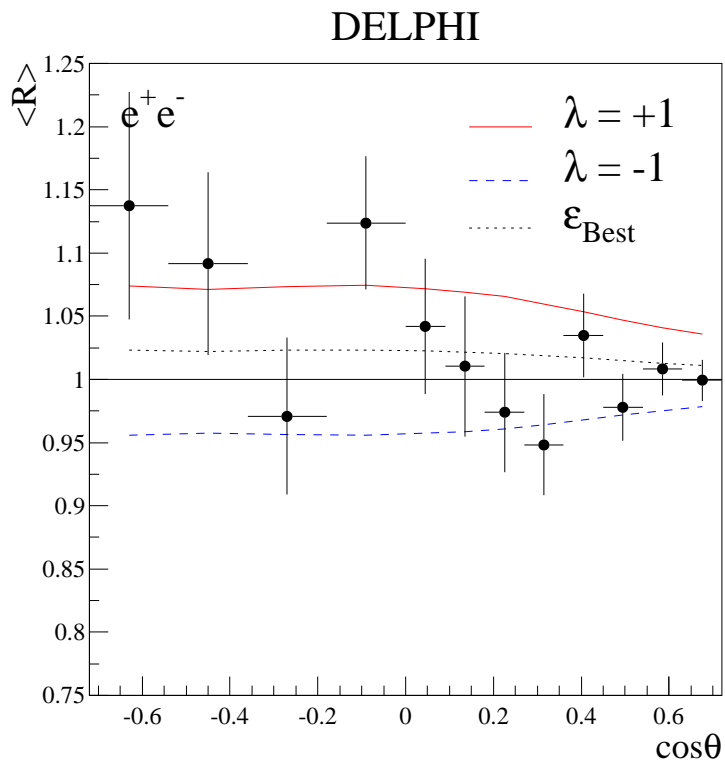,width=0.66\textwidth}}
\caption{\capsty{The deviations of the measured values of 
         $\dsdcth$ for $\eeee$ from the predictions of the Standard Model, 
         averaged over all energies. Superimposed
         are the predicted deviations at the luminosity weighted 
centre-of-mass energy for gravity in large extra dimensions for the 
         $95\%$ confidence limits,  
         $\mathrm{M_{s}}=$0.986 TeV for $\lambda=-1$ and
         $\mathrm{M_{s}}=$0.874 TeV for $\lambda=+1$.}}
\label{fig:grav:fit}
\end{center}
\end{figure}
%%%%%%%%%%%%%%%%%%%%%%%%%%%%

$\chi^{2}$ fits to the measured differential cross--sections for \ee, \mumu\ 
and \tautau\ final states reported in this paper for the parameter 
$\epsilon = \lambda / \mathrm{M^4_s}$ were performed. 
In order to fit the prediction to the data, the angular distributions given 
in \cite{ref:grav:thryhew,ref:grav:thryriz}
were corrected to account for radiative effects, dominated by initial state 
radiation. The corrections take the form of a numerical version of the 
radiator function contained in \ZFITTER, which are computed assuming only 
Standard Model processes.
Separate radiator functions were derived for each angular bin in several 
bins of $\sqsps$. The tree-level new physics cross-section in a 
given bin was subsequently convoluted with the appropriate numerical 
radiator function to produce the corrected cross-section in that bin. 

For the \mumu\ and \tautau\ final states the statistical errors expected from 
the Standard Model were used. The correlations between the systematic errors 
between bins of $\cos\theta$ and between channels and energies were 
taken into account in the fit. Errors on the parameter $\epsilon$ were 
determined in the same way as in Section~\ref{sec:cntc:res}.

The values of $\epsilon$ obtained are compatible with the Standard Model, 
i.e. $\epsilon=0$ TeV$^{-4}$. 
Table~\ref{tab:grav:res} shows the fitted values of $\epsilon$ and
$95\%$ confidence level lower limits on $\mathrm{M_{s}}$. These limits
were obtained using a method equivalent to that used to
extract the limits on the scale, $\Lambda$, of contact interactions, 
as described in Section~\ref{sec:cntc:res}.

The deviations of the measured angular distributions from the Standard Model 
predictions averaged over all energies in the channel \eeee\ are shown in 
Figure~\ref{fig:grav:fit} together with the expected deviations as a 
function of $\cos\theta$ for the fitted best value of $\epsilon$ and for
the $95\%$ lower limits on $\mathrm{M_{s}}$ for $\lambda=\pm 1$.
%%%%%%%%%%%%%%%%%%%%%%%% Peter's sneutrino section 28/04/04

\subsection{Sneutrino exchange models}
\label{sec:sneutrino}
These models consider possible $s$- or $t$-channel sneutrino $\snul$
exchange in R-parity violating supersymmetry \cite{ref:susy:SUSY}, which 
can affect
the channel $\eell$.
The purely leptonic part of the R-parity violating superpotential
has the form
\begin{displaymath}
 \lambda_{ijk} L_{L}^{i} L_{L}^{j} \overline{E}_{R}^{k}
\end{displaymath}
where $ijk$ are family indices, $L_{L}$ represents a left-handed leptonic 
superfield doublet and $\overline{E}_{R}$ corresponds to the right-handed
singlet superfield of charged leptons. The coupling $\lambda_{ijk}$ is only
non-zero for combinations involving at least two generations and 
for $i \lt j$. 

 For the channel $\eeee$ there are possible contributions from the
$s$-channel production and $t$-channel exchange of 
either $\snumu$ ($\lambda_{121} \neq 0$) 
or $\snutau$ ($\lambda_{131} \neq 0$).
For the channels $\eemm$ and $\eett$ there is no $s$-channel contribution
if only one of the $\lambda_{ijk}$'s is non-zero. 
For $\eemm$ there are 
$t$-channel contributions from either $\snue$ ($\lambda_{121} \neq 0$),
$\snumu$ ($\lambda_{122} \neq 0$) or 
from $\snutau$ ($\lambda_{132}$ or $\lambda_{231} \neq 0$). 
If both $\lambda_{131} \neq 0$ and $\lambda_{232} \neq 0$ then the
$s$-channel production of $\snutau$ is possible. 
For  $\eett$ there are
$t$-channel contributions from either $\snue$ ($\lambda_{131} \neq 0$),
$\snumu$ ($\lambda_{123}$ or $\lambda_{231} \neq 0$) or 
from $\snutau$ ( $\lambda_{133} \neq 0$).
If both $\lambda_{121} \neq 0$ and $\lambda_{233} \neq 0$ then the
$s$-channel production of $\snumu$ is possible. 

 All these possibilities are considered here. For a given scenario
the $s$- or $t$-channel sneutrino exchange amplitude contribution is 
added to the Standard Model contribution as appropriate. If there is no
sneutrino exchange for a specific channel then the prediction for that 
channel is just the SM value.

In the case of $s$-channel sneutrino graphs, if the sneutrino mass,
\msneut, is equal to the centre-of-mass energy of the $\ee$ beams,
resonant sneutrino production occurs, which can lead to a large change
in the cross-section. A lesser change in the cross-section will occur
for $\msneut < \roots$ due to the process of {\it radiative
return}. There is some sensitivity to \msneut\ just above \roots\
due to the finite width of the particle.  It is assumed here that the
sneutrino width is 1 GeV.

%........................................................................%
\subsubsection{Fits to models of sneutrino exchange}

The total cross-section and forward-backward asymmetry values for
the channels $\eeee$, $\eemm$ and $\eett$,
at each centre-of-mass energy, were used in the fits. 
The theoretical prediction consisted
of Improved Born Approximation Standard Model terms, 
plus sneutrino exchange, plus
interference terms. 
%The IBA predictions were convoluted with
%QED radiative corrections, which were computed using the {\tt DYMU3}
%package \cite{DYMU3}.

\begin{figure}[p]
 \begin{center}
  \epsfig{file=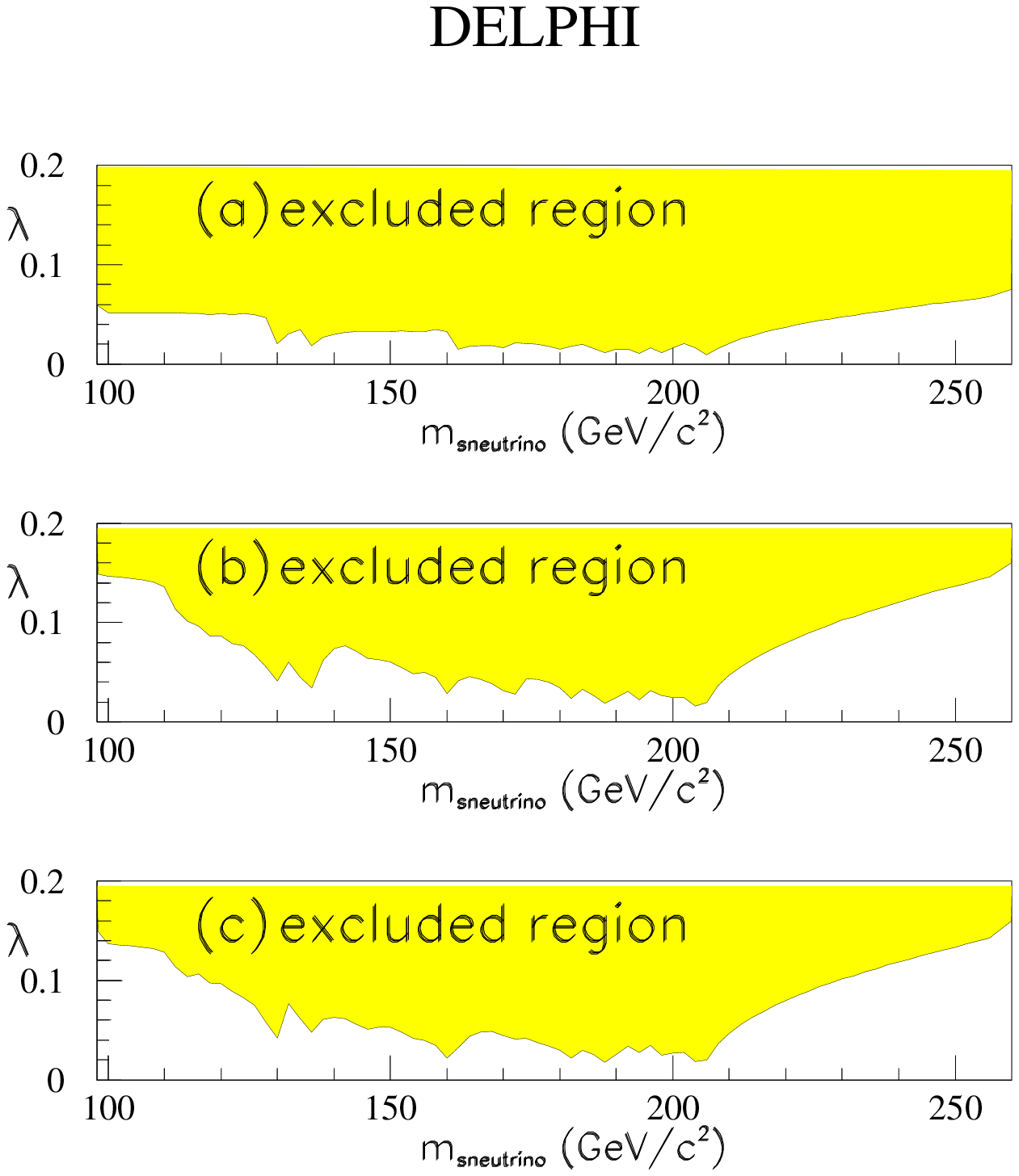,height=16.0cm,width=\textwidth}
%file=exclall.eps,width=\textwidth}
 \end{center}
 \caption{\capsty{ The $95\%$ exclusion limits 
               for a) $\lambda_{121} (or \lambda_{131}$), as
               a function of $\msneut$ obtained from the $\eeee$ channel; 
               b) $\lambda_{131} = \lambda_{232} = \lambda$, as
               a function of $\msneut$ obtained from the $\eemm$ channel; 
               c) $\lambda_{121} = \lambda_{233} = \lambda$, as
               a function of $\msneut$ obtained from the $\eett$ channel.  
               The sneutrino width is taken to be 1 GeV.}}
 \label{fig:exclall}
\end{figure}

 All the fits considered result in values of $\lambda$ which are compatible
with zero; so results are expressed as $95\%$ confidence limits.
The first fits considered are to those terms which modify the $\eeee$
channel. These involve the $s$- and $t$-channel exchange of 
a $\snumu$ ( $\lambda_{121} \neq 0$) or $\snutau$ ( $\lambda_{131} \neq 0$).
The resulting $95 \%$ limits on $\lambda$, as a function of $\msneut$,
are given in Figure \ref{fig:exclall}a.
The best limits on $\lambda$ are obtained for the case
where \msneut\ is close to the actual centre-of-mass energy of the
LEP collisions, but the radiative return process
gives some sensitivity between these points. It can be seen that
$\lambda$ greater than approximately 0.05 can be excluded for \msneut\ in the 
LEP II range of energies at the $95\%$ confidence level. 

 For the case that only one $\lambda$ value is non-zero there are only 
$t$-channel sneutrino effects for $\eemm$ and $\eett$.
The values of $\lambda$ obtained for the $\eemm$ 
channel 
%(i.e. for $\lambda_{121}$, $\lambda_{122}$, $\lambda_{132}$ or $\lambda_{231}$)
and for the $\eett$ channel
%(i.e. for $\lambda_{131}$, $\lambda_{123}$, $\lambda_{232}$ or $\lambda_{133}$)
are all consistent with zero, so results are expressed as $95\%$
confidence limits in Table \ref{tab:susy}.

%%%%%%%%%%%%%%%%
%\begin{table}[tbp]
\begin{table}[bp]
 \begin{center}
  \begin{tabular}{|c|c|c|}
   \hline
                    & $\msneut=100 \: \GeV/c^2$ & $\msneut=200 \: \GeV/c^2$ \\
     coupling       &        ($95\%$ c.l.)       &        ($95\%$ c.l.)      \\
   \hline
%    $\lambda_{121}$, $\lambda_{122}$, $\lambda_{132}$ or $\lambda_{231}$ &
     $\lambda$ (t-chann. $\snul$ in $\eemm$) &
           0.19            &           0.26            \\
%    $\lambda_{131}$, $\lambda_{123}$, $\lambda_{232}$ or  $\lambda_{133}$ & 
     $\lambda$ (t-chann. $\snul$ in $\eett$) &
           0.56            &           0.63            \\
   \hline
  \end{tabular}
 \end{center}
 \caption{\capsty{$\em$ Limits on the various couplings $\lambda$  in
               $t$-channel sneutrino exchange in $\eemm$  and
               $\eett$.}}
 \label{tab:susy}
\end{table}
%%%%%%%%%%%%%%%%

For the fits assuming that $\lambda_{131} = \lambda_{232} = \lambda$,
the resulting $95 \%$ confidence limits on $\lambda$, as a function of
$\msneut$, are given in Figure \ref{fig:exclall}b. A similar exclusion
pattern to that obtained from the $\eeee$ channel is obtained. Values of $\lambda$ greater than approximately 0.07 can be excluded
for \msneut\ in most of the LEP II range of energies at the $95\%$
confidence level.  The exclusion contour for $\lambda_{121} =
\lambda_{233} = \lambda$ is shown in Figure \ref{fig:exclall}c, from
which it can be seen that again a similar exclusion pattern is
obtained.

%%%%%%%%%%%%%%%%%%%%
%%%%%%%%%%%%%%%%%%%%

%%%%%%%%%%%%%%%%%%%%%%%%%%%%%%%%%%%%%%%%%%%%%%%%%%%%%%%%%%%%%%%%%%%%%%%%%%%%
%%%%%%%%%%%%%%%%%%%%%%%%%%%%%%%%%%%%%%%%%%%%%%%%%%%%%%%%%%%%%%%%%%%%%%%%%%%%

%%--------------------------------------------------------------------
%%-- CONCLUSIONS -----------------------------------------------------
%%--------------------------------------------------------------------
\section{Conclusions}   
\label{sec:conclusions}

Analyses of cross-sections and forward-backward asymmetries and 
differential cross-sections in the \eeee, \eemm\ and \eett channels, 
at centre-of-mass energies from 183 to 207 \GeV, have been 
presented, along with measurements of cross-sections for \eeqq, in 
Sections~\ref{sec:ee} to~\ref{sec:qq}. 
The results of the measurements are presented in 
Tables~\ref{tab:ee:res}, \ref{tab:mm:res}, \ref{tab:tt:res} 
and~\ref{tab:qq:res} for cross-sections and asymmetries and in 
Tables~\ref{tab:ee:diff}, \ref{tab:mm:diff} 
and~\ref{tab:tt:diff} for differential cross-sections.
Results of the measurements of cross-sections and asymmetries
from previous analyses~\cite{ref:delphiff:130-172} at centre-of-mass
energies of 130 to 172 GeV are included in the tables.

To compare data to the predictions of the Standard Model,
averages of the ratios of measurements for {\emph{non-radiative} samples 
of events to the predictions
of the Standard Model have been made over all centre-of-mass energies.
The results of these averages are presented in Tables~\ref{tab:ana:rvals}
and~\ref{tab:ana:rdsdc}. The precision obtained on the averaged ratios for
the cross-sections are $\pm2.3\%$ for \eeee, dominated by theoretical 
uncertainties, $\pm2.5\%$ and $\pm4.0\%$ for \eemm\ and \eett, respectively, 
dominated by statistical uncertainties and $\pm1.7\%$ for \eeqq, dominated by
experimental systematic uncertainties. The results are consistent with the
expectations of the Standard Model. Full details can be found in 
Section~\ref{sec:results}.

The measurements for {\emph{non-radiative}} samples have been used to
fit the parameters of the S-matrix model of \eeff\ and to search for
physics beyond the Standard Model in a number of models. In all cases
correlations between the experimental uncertainties have been taken
into account along with theoretical uncertainties.  Using LEP I
data~\cite{ref:delphils:00} alone, the best fit to the parameter
$\jtoth$ of the S-matrix Model is $0.51 \pm 0.55$. Including the LEP
II measurements the best fit is $0.47 \pm 0.12$, consistent with,
though somewhat above, the expectation of the Standard Model
$\jtoth=0.22$.  A complete set of results and description of the
analysis can be found in Section~\ref{sec:smat}.  LEP I and LEP II
data have also been used to search for \Zprime\ bosons, yielding
limits on the mass of such bosons which range from $360$ to $1305$
$\GeV/c^2$, depending on the model assumed. Full sets of results are
presented in Section~\ref{sec:zp}.

Using measurements for {\emph{non-radiative}} samples of leptons from
the full range of centre-of-mass energies, limits have been derived on
the scales of contact interactions and the scale associated with the
propagation of gravitons in models with large extra dimensions; these
are described in Sections~\ref{sec:cntc} and~\ref{sec:grav}. For
contact interactions the limits range from $6.5$ to $16.5$ \TeV,
assuming lepton universality.  For models of gravity in large extra
dimensions the limits on the scales are found to be $0.898~\TeV$ or
$0.998~\TeV$, respectively, in the case when the new physics
interferes either constructively or destructively with the Standard
Model processes. Full details can be found in
Section~\ref{sec:grav}. In supersymmetric theories with R-parity
violation, exchange of sneutrinos can affect the total and
differential cross-sections for $\eell$. In
Section~\ref{sec:sneutrino} it is shown that limits at the level of
$0.05$ can be set on the R-parity violating couplings for sneutrino
masses in the LEP II range of energies.

%         Modified on 04-06-1999 by dimartino
%-------------------------------------------------------------------
\subsection*{Acknowledgements}
\vskip 3 mm
 We are greatly indebted to our technical 
collaborators, to the members of the CERN-SL Division for the excellent 
performance of the LEP collider, and to the funding agencies for their
support in building and operating the DELPHI detector.\\
We acknowledge in particular the support of \\
Austrian Federal Ministry of Education, Science and Culture,
GZ 616.364/2-III/2a/98, \\
FNRS--FWO, Flanders Institute to encourage scientific and technological 
research in the industry (IWT), Belgium,  \\
FINEP, CNPq, CAPES, FUJB and FAPERJ, Brazil, \\
Czech Ministry of Industry and Trade, GA CR 202/99/1362,\\
Commission of the European Communities (DG XII), \\
Direction des Sciences de la Mati$\grave{\mbox{\rm e}}$re, CEA, France, \\
Bundesministerium f$\ddot{\mbox{\rm u}}$r Bildung, Wissenschaft, Forschung 
und Technologie, Germany,\\
General Secretariat for Research and Technology, Greece, \\
National Science Foundation (NWO) and Foundation for Research on Matter (FOM),
The Netherlands, \\
Norwegian Research Council,  \\
State Committee for Scientific Research, Poland, SPUB-M/CERN/PO3/DZ296/2000,
SPUB-M/CERN/PO3/DZ297/2000, 2P03B 104 19 and 2P03B 69 23(2002-2004)\\
FCT - Funda\c{c}\~ao para a Ci\^encia e Tecnologia, Portugal, \\
Vedecka grantova agentura MS SR, Slovakia, Nr. 95/5195/134, \\
Ministry of Science and Technology of the Republic of Slovenia, \\
CICYT, Spain, AEN99-0950 and AEN99-0761,  \\
The Swedish Research Council,      \\
Particle Physics and Astronomy Research Council, UK, \\
Department of Energy, USA, DE-FG02-01ER41155. \\
EEC RTN contract HPRN-CT-00292-2002. \\

%=========================================================================%

%%----------------------------------------------------------------------------
%%-- Bibliography ------------------------------------------------------------
%%----------------------------------------------------------------------------

\end{document}